\documentclass{acta_2022}

\usepackage{amssymb}
\usepackage{amsmath}

\usepackage[longnamesfirst]{natbib} \usepackage{har2nat}
\setcitestyle{comma,aysep={}}
\shortcites{samplepreprint} %

\usepackage{amssymb,amsmath}
\usepackage{newtxtext}
\usepackage{newtxmath} 
\DeclareSymbolFont{CMlargesymbols}{OMX}{cmex}{m}{n}
\DeclareMathDelimiter{(}{\mathopen} {operators}{"28}{CMlargesymbols}{"00}
\DeclareMathDelimiter{)}{\mathclose}{operators}{"29}{CMlargesymbols}{"01}
\DeclareMathAlphabet\mathcal{OMS}{cmsy}{m}{n}
\SetMathAlphabet\mathcal{bold}{OMS}{cmsy}{b}{n}
\pagerange{\pageref{firstpage}--\pageref{lastpage}}
\allowdisplaybreaks
\doi{XXXXXXXX}  %
\usepackage[UKenglish]{babel} %
\newcommand{\ignore}[1]{}
\usepackage[twoside,
        paperheight=247mm,
        paperwidth=174mm,
        marginratio={3:5,2:3},
        left=18mm,
        top=20mm]{geometry}
\numberwithin{figure}{section}
\numberwithin{table}{section}
\usepackage{graphics,graphicx}
\usepackage[cmyk]{xcolor}
\usepackage{enumitem} %

\newtheorem{theorem}{Theorem}[section]

\title[Optimal experimental design]{Optimal experimental design:\\Formulations and computations}

\author[X.~Huan, J.~Jagalur and Y.~Marzouk]{Xun Huan\\
  University of Michigan, 1231 Beal Ave, Ann Arbor, MI 48109,
  USA \\ Email: {xhuan@umich.edu} \\
  \and
  Jayanth~Jagalur\\
  Center for Applied Scientific Computing, Lawrence Livermore National
  Laboratory, \\ 7000 East Ave, Livermore, CA 94550, USA
  \\ Email: {jagalur1@llnl.gov} \\
  \and
  Youssef~Marzouk\\
  Massachusetts Institute of Technology, \\ 77 Massachusetts Ave,
  Cambridge, MA 02139, USA \\ Email: {ymarz@mit.edu}}

\usepackage[textsize=tiny]{todonotes}
\usepackage{bm}
\usepackage{mathtools}

\usepackage{subfig} %

\definecolor{darkred}{rgb}{.7,0,0}

\definecolor{darkgreen}{rgb}{.15,.55,0}

\definecolor{darkblue}{rgb}{0,0,0.7}

\newcommand{\design}{\xi}
\newcommand{\pdf}{p}
\newcommand{\policy}{\pi}
\newcommand{\param}{\theta}
\newcommand{\Param}{\Theta}
\newcommand{\paramset}{\bm{\Theta}}
\newcommand{\info}{{\sf i}}
\newcommand{\Info}{I}

\newcommand{\Var}{\mathbb{V}\text{ar}}
\newcommand{\Cov}{\mathbb{C}\text{ov}}
\newcommand{\DKL}{D_{\text{KL}}}
\newcommand{\tr}{\text{tr}}
\newcommand{\argmax}{\operatornamewithlimits{argmax}}
\newcommand{\argmin}{\operatornamewithlimits{argmin}}
\newcommand{\normdist}{\mathcal{N}}

\newcommand{\CA}{\mathcal{A}}
\newcommand{\CF}{\mathcal{F}}

\newcommand{\CS}{\mathcal{S}}

\newcommand{\EE}{\mathbb{E}}
\newcommand{\PP}{\mathbb{P}}
\newcommand{\RR}{\mathbb{R}}

\newcommand{\hg}{\hat{g}}

\DeclareMathOperator\supp{supp}

\newcommand{\rev}[1]{#1}

\newcommand{\revnew}[1]{#1}

\newcommand{\revnewnew}[1]{#1}

\newcommand{\revarx}[1]{#1}

\newcommand{\revarxtwo}[1]{#1}

\DeclarePairedDelimiter\norm{\lVert}{\rVert}
\newcommand{\lr}[1]{ \left( #1 \right)}

\newcommand{\Ubar}{\bar{U}}

\begin{document}

\label{firstpage}
\maketitle

\vspace{8.75em}

\begin{abstract}

Questions of `how best to acquire data' are essential to modeling and prediction in the natural and social sciences, engineering applications, and beyond. Optimal experimental design (OED) formalizes these questions and creates computational methods to answer them.
This article presents a systematic survey of modern OED, from its foundations in classical design theory to current research involving OED for complex models.
We begin by reviewing criteria used to formulate an OED problem and thus to encode the goal of performing an \revnew{experiment.}
We emphasize the flexibility of the Bayesian and decision-theoretic approach, which encompasses information-based criteria that are well-suited to nonlinear and non-Gaussian statistical models. 
We then discuss methods for estimating or bounding the values of these design criteria; this endeavor can be quite challenging due to strong nonlinearities, high parameter dimension, large per-sample costs, or settings where the model is implicit.
A complementary set of computational issues involves optimization methods used to find a design; we discuss such methods in the discrete (combinatorial) setting of observation \revnew{selection and in} settings where an exact design can be continuously parameterized.
Finally we present emerging methods for sequential OED that build non-myopic design policies, rather than explicit designs; these methods naturally adapt to the outcomes of past experiments in proposing new experiments, while seeking coordination among all experiments to be performed.
Throughout, we highlight important open questions and challenges.

\vspace{0.6em}

\hspace{17em}Secondary 62L05, 94A17, 65M32

\end{abstract}

\tableofcontents 

\vspace{1em}
\section{Introduction}
\label{s:introduction}
Acquiring data to {inform models} and guide decisions is
an essential part of scientific inquiry, engineering design, and even
policy making. Seldom can we construct a useful model in isolation
from data. Rather, data must be used to infer parameters of models, to
assess whether models can provide useful predictions, and to spur a
wide variety of model improvements. In this setting, it is natural to
consider how to acquire data \emph{efficiently}. Experimental data and
field measurements can be costly or time-consuming to acquire; other
information sources, similarly, may be expensive to query. Yet we
face a multitude of choices in designing such queries. Where to place
a sensor? What experimental conditions to impose? What quantity to
observe? Do measurements need to be very precise, or would a noisier
measurement suffice? When should measurements be made?  And how much data
should be collected? More broadly, what \emph{combination} of
observations would be most informative or useful---and how should we
precisely define notions of `informative' or `useful' in the first
place?  Crucially, these notions must allow experimental choices to be
made \emph{before} data are acquired.

\emph{Optimal experimental design} (OED) aims to answer these
questions, through mathematical formulations that formalize and tailor
them to the ultimate goals of acquiring data. A model developer may
have many possible goals, and hence there are many possible criteria
for what comprises a good experimental design. Another essential
aspect of OED involves numerical algorithms, e.g., for evaluating
suitable design criteria, for optimizing over possible experimental
configurations, and possibly doing so in a `closed-loop' sequential
fashion. Collectively, these endeavors lie at the intersection of many
fields: statistical inference and decision theory,
information theory, Monte Carlo methods, continuous and combinatorial optimization, dynamic programming and stochastic control, and even reinforcement learning.

Every OED problem has two essential ingredients: an \emph{experiment}, which is the source of data, and a mathematical \emph{model}. The role of the model is to simulate what might happen in candidate experiments, and to assess how the results of such experiments might improve the model and its predictions. Formally, the model is a statistical model; in some applications, evaluating this statistical model also involves significant numerical simulation.
An underlying presumption of OED is that it is \emph{worthwhile} to perform many calculations involving the model, in order to plan experiments that are more efficient and effective. These calculations might be far less expensive (by some metric) than performing experiments, or there might be other impediments to experimentation that make finding an optimal design in advance, or building an online optimal design policy for online settings, worth the~effort.

The notion of an `experiment' should be construed quite broadly, and
certainly not limited to laboratory experiments in the traditional
sense. A high-fidelity simulation of a complex model, used to produce
data to calibrate the parameters of a simpler model, constitutes an
experiment. Arranging a network of sensors, or planning the path of an
airborne vehicle carrying instruments, also constitutes designing an
experiment. Application domains in which OED is used are similarly
vast. OED has long been an essential part of statistical modeling,
from the design of clinical trials \citep{Berry_2010} to the design of
computer experiments \citep{Sacks_1989}; the latter is closely related
to the classical problem of design for regression
\citep{elfving1952optimum,kieferWolfowitz1959optimum,kiefer1961optimum}.
But OED can also be applied to problems involving parameter inference
in ordinary or partial differential equations
\citep{huan2013simulation}, to a wide range of inverse \revnew{problems
\citep{Haber_2008,Alexanderian_2016,Ruthotto2018,Alexanderian_2021,Helin_2022}, and}
to myriad other `complex' models of data-generating processes---in
astronomy \citep{Loredo_2010}, systems biology \citep{Liepe_2013},
aeroelasticity \citep{Riley_2019}, reliability testing
\citep{Weaver_2021}, \revarx{nuclear physics \citep{melendez2021designing},} and beyond.

The evolution of experimental design has a rich and fascinating history.
Early twentieth-century approaches to experimental design were
motivated by agricultural experiments and similar
applications. Statistical methods in this setting often involved
hypothesis testing, and a good design was one that maximized the
sensitivity of the test.
Notable works by Fisher and his collaborators during this period
introduced concepts such as balance, orthogonality, blocking, and
aliasing \citep{fisher1936design,craig1936design,yates1933principles,yates1937design,yates1940lattice,bose1939construction,bose1939partially}.
Wald recognized that these ideas were relevant to many other fields,
and in a seminal paper \citep{wald1943efficient} introduced formal
notions of the efficiency of a design. With this framework, he was
able to explain the success of designs based on Latin
squares, commonly used in agricultural experimentation.
Following Wald's work, many core ideas of modern OED emerged towards
the middle of the twentieth century
\citep{elfving1952optimum, Lindley_1956,kiefer1958nonrandomized,stone1959application,kieferWolfowitz1959optimum,kiefer1959optimum,kiefer1961optimum},
some of them influenced by the emerging discipline of decision theory.
Kiefer distinguished various design criteria by giving them meaningful
names, and thus originated current `alphabetic optimality'
\mbox{terminology} \citep{kiefer1958nonrandomized}.  Much later, Kiefer also
showed inter-relationships among various design criteria by
introducing more general notions of `universal' optimality
\citep{kiefer1974general}.
Lindley's work in the same period \citep{Lindley_1956} aligns
more closely with Bayesian statistical thinking, and deserves special
mention in light of the current popularity of information theoretic design
criteria. These approaches remained largely intractable half a
century ago, but with advances in computing power and
algorithms, these more general and arguably more rigorous
design criteria have become feasible to implement.
For a more detailed history of OED, we refer readers to
\citet{wynn1984jack} and to the texts by
\citet{fedorov1972theory,Shah_1989}, and \citet{Pukelsheim_2006}.

In recent years, interest in OED has expanded from the statistics literature into the uncertainty quantification and applied mathematics communities, where, as noted earlier, an animating goal has been to advance OED methodologies for `complex' models. Here many forms of complexity are relevant: high-dimensional parameters, computationally intensive likelihood functions that involve the evaluation of ordinary or partial differential equations, strong nonlinearity in the dependence of observables on parameters, and the associated non-Gaussianity of posterior distributions in the Bayesian setting. Additional complexities can arise due to data-generating processes that evolve in time, which present the opportunity to design and implement experiments adaptively, i.e., where the results of previous experiments influence the next; this is the setting of \textit{sequential} optimal experimental design.
Another thread relevant to modern OED has come from the computer science literature, where much attention has been paid to the optimization of \emph{set functions}; methods here underpin a combinatorial approach to OED, where the problem is cast as choosing a subset of a given `ground set' of feasible candidate experiments.
And in recent years, tools from machine learning and in particular
deep learning have become quite useful for OED, for instance by offering new ways of evaluating complex design criteria in high dimensional, non-Gaussian settings. Of particular interest are expressive machine learning models and learning algorithms for the underlying density (or density ratio) estimation tasks. We will discuss all of these threads, and many more, in the ensuing sections.

\medskip

We mention here several other excellent surveys that may be of
interest to the reader. \citet{Steinberg_1984} and
\citet{Pukelsheim_2006} (updated from the original 1993 version)
provide comprehensive reviews of non-Bayesian OED for linear models.
\citet{Ford_1989} discuss design for nonlinear models, while
\citet{DasGupta_1995} discusses Bayesian designs, mostly for linear
models. \citet{Atkinson_2007} provide extensive coverage of linear optimal
design theory, with some extension to nonlinear and Bayesian methods.
The paper of \citet{Chaloner_1995} is an influential review
from a statistical perspective, highlighting features of the Bayesian
and decision-theoretic approach to OED. We view the Bayesian approach
as very natural in the setting of design, and will largely adopt such
a perspective here. \citet{Clyde_2001} presents an overview of
Bayesian OED formulations with various choices of utility.

More recent reviews have placed a greater emphasis on computation.
\citet{Ryan_2016} discuss Bayesian formulations of OED and then survey
computational algorithms for realizing Bayesian designs, emphasizing
Monte Carlo methods for estimating design criteria and for searching
through design space. \citet{Alexanderian_2021} reviews OED for
Bayesian inverse problems, emphasizing formulations and algorithms for
the infinite-dimensional (function space)
setting. \citet{Rainforth_2023} provide a survey highlighting recent machine
learning and reinforcement learning tools in OED, including sequential
design. \citet{Strutz_2023} present a review of variational OED
methodologies and their application to geophysical and in particular
seismological problems. There may be other recent reviews of which we
are unaware, and we apologize for such omissions.

\subsection{Scope and organization of this article}

This article aims to provide a broad, comprehensive survey of methodologies for optimal experimental design. Our perspective covers both formulations---i.e.,~the many ways of \textit{posing} an OED problem---and computations---i.e., numerical algorithms for \textit{finding} optimal designs, or tractable approximations of optimal designs, for a range of problem settings.
The second topic in particular is multi-faceted---we will draw upon Monte Carlo methods, algorithms for inference and density estimation, and a variety of optimization approaches---but naturally enjoys a close interplay with the first.
Our goal is to guide readers who are new to OED from the basic ideas to the research frontier, and to illuminate open issues and challenges at that frontier. Indeed, we believe that the present moment is ripe for a survey of the field: the problems that remain are quite challenging, and a synthesis of ideas and approaches from many different intellectual communities (some of which have been rather disconnected) is needed to make progress.

The article is organized as follows.
We begin in Section~\ref{s:criteria} by describing possible design
criteria for OED, each encoding a different notion of what
constitutes a `good' experiment. The applicability of these criteria
ranges from the rather specific (e.g., linear-Gaussian problems) to
the very general. Many are rooted in information-theoretic \revnew{and/or}
decision-theoretic formulations. We also mention alternative design
heuristics that have been used in practice, and clarify distinctions
between the OED problem and several related but different problems,
such as Bayesian optimization.

In Section~\ref{s:estimation} we turn to the first step of
computation: numerical algorithms for estimating or bounding the
values of these design criteria. We survey Monte Carlo schemes, as
well as variational approximations and density estimation methods, for
this purpose. Some of these algorithms are applicable to so-called
`implicit' Bayesian models, where evaluations of the likelihood
function or prior density may~be unavailable. We also discuss
challenges associated with high-dimensional parameters and data, and
dimension reduction schemes that mitigate these challenges.
In Section~\ref{s:optimization} we discuss strategies and algorithms
for efficiently optimizing design criteria in various settings. On one
hand, we address problems where an exact design of interest is
represented by continuous (real-valued) variables. On the~other hand,
we discuss settings where the design optimization problem is discrete
and combinatorial, e.g., a form of subset selection, and continuous
relaxations thereof.

Sections~\ref{s:criteria}--\ref{s:optimization} focus on the
all-at-once (`batch') design of experiments---that is, `fixed' or
static designs that cannot be adjusted as the outcomes of experiments
are realized. In contrast, Section~\ref{s:sequential} turns to the
problem of sequential OED, where design decisions are naturally
adapted according to the outcomes of previous experiments, while
taking into consideration the information to be gained from future
experiments. We present sequential OED in a fully Bayesian setting,
leveraging the formalism of Markov decision processes. We then
highlight recent computational methods for solving this challenging
problem, which make use of dynamic programming, various reinforcement
learning techniques, and information bounds.

We close in  Section~\ref{s:outlook} with a discussion of open questions and
opportunities for future work.

\vspace{1em}
\section{Optimal design criteria}
\label{s:criteria}
We begin by addressing the central question of any OED formulation: In what sense should one deem a candidate design to be `good?' More specifically, by what considerations should one candidate design be deemed better than another? Answering these questions is essential to the notion of \emph{optimal} design. The answers are formalized by choosing a quantitative design criterion---a function that can then be maximized in order to identify the corresponding optimal design. In this section, we will discuss a wide variety of design criteria and the goals they encode.

To formulate these criteria, we must first specify a \emph{statistical model} for the observations $Y$ obtained under a candidate design.
We need such a model in order to predict the outcomes of candidate experiments, and to relate these outcomes to the ensuing estimation or prediction tasks that motivated the acquisition of data in the first place.
We will initially consider parametric statistical models. Any such model is a family of
probability distributions for $Y$, indexed by (unknown) model
parameters $\param \in \paramset \subseteq \mathbb{R}^p$ and by the
choice of design $\design \in \Xi$:
\begin{equation}
\label{eq:statmodel}
  \mathcal{M} = \{ \revnew{\textit{\textsf{F}}_{Y ; \param, \design}} :  \param \in \paramset, \ \design \in \Xi \}.
\end{equation}

This model encodes, for any given value of $\param$ and $\design$, a complete probabilistic description of the resulting observations, via the cumulative distribution function $\revnew{\textit{\textsf{F}}_{Y; \param, \design}}$.
Further, if every element of $\mathcal{M}$ is absolutely continuous with
respect to Lebesgue measure, we can also write the statistical model
as a family of conditional probability density functions, i.e.,
$\mathcal{M} = \{ \pdf_{Y \vert \param, \design}( y \vert  \param,
\design) :  \param \in \paramset, \ \design \in \Xi \}$. We will make
this simplifying assumption from here on, with the understanding that
the \rev{conditional} density of $Y$ could be replaced by a conditional probability measure $\mu(dy \vert \param, \design)$ as needed.

The parameters $\param$ in the model are unknown and hence the object of estimation or inference. On the other hand, $\design$ can be controlled by the experimenter.
For~instance, if candidate $Y$ \revnew{corresponded} to observations of a spatiotemporal process, $\design$ might represent the spatial and temporal coordinates of a chosen {set} of observations. In a regression model, $\design$ \revnew{encodes} the values of the covariates (i.e., independent variables) at which observations are obtained.
Different ways of representing $\design$ will give rise to different optimization problems, which we will discuss in Section~\ref{s:optimization}.

The \revnew{statistical model} immediately yields a likelihood function $\param \mapsto \pdf_{Y | \param, \design}(y \vert \param, \design)$, and the corresponding \rev{symmetric, positive semi-definite} $p \times p$ Fisher information matrix
\begin{equation}
  \label{eq:fisherinfo}
  F(\param, \design) \coloneqq \mathbb{E}_{Y \vert \param, \design}[\nabla_\param \log \pdf_{Y | \param, \design} (Y \vert \param, \design) \otimes \nabla_\param \log \pdf_{Y | \param, \design}(Y \vert \param, \design)],
\end{equation} 
which is a central object in estimation theory \citep{Lehmann_1998}. Below we will discuss various design criteria based on the Fisher information matrix.

Alternatively, one can take a Bayesian approach and endow the unknown parameters, now denoted as $\Param$, with a prior distribution. Let this distribution have (Lebesgue) density $\pdf_{\Param}$ on $\mathbb{R}^p$. We will always assume that the prior density is functionally independent of the design $\design$. The posterior density of $\Param$ is then given by Bayes' rule: 
\begin{equation} \label{eq:Bayes}
    \pdf_{\Param | Y, \design}( \param | y, \design) = \frac{\pdf_{Y | \Param, \design}(y | \param, \design) \, \pdf_{\Param}( \param )}{\pdf_{Y|\design}(y | \design)}.
\end{equation}
In the Bayesian paradigm, the prior and posterior distributions represent, respectively, our states of knowledge before and after having observed $Y=y$. The marginal density of the observations,
\begin{align*}
  \pdf_{Y|\design}(y|\design) = \int \pdf_{Y | \Param, \design}(y|\param, \design)\,  \pdf_{\Param}( \param )\, \mathrm{d} \param,
\end{align*}
appearing in the denominator of \eqref{eq:Bayes}, is called the evidence or marginal likelihood. We will also call this distribution the \textit{prior predictive}, as it reflects our probabilistic prediction of future values of $Y$ given only the prior on $\Param$, the statistical model, and a chosen design.
Having observed a particular value of the data, say $y^\ast$, at some chosen design $\design$, the \textit{posterior predictive} density of the data $Y$ for a new design $\design^+$~is
\begin{align*}
  \pdf_{Y|\design^+, y^\ast, \design}(y \vert \design^+, y^\ast, \design ) = \int \pdf_{Y | \Param, \design}(y \vert \param, \design^+)\, \pdf_{\Param | Y, \design}(\param \vert y^\ast, \design)\, \mathrm{d}\param.
\end{align*}

Many design criteria discussed below will explicitly take advantage of this Bayesian formulation of the inference problem. Indeed, we will see that it is useful to have the ability to incorporate prior information on $\Param$---in general, but especially in nonlinear design settings---and that the Bayesian approach to OED is natural for decision-theoretic reasons as well.

\subsection{Design criteria for the linear-Gaussian setting}
\label{s:lineardesign}

A rich variety of design criteria, both Bayesian and non-Bayesian, have been developed for linear-Gaussian models. This class of statistical models has numerous practical applications: certainly linear regression, but also \emph{linear inverse problems}, where observations depend \emph{indirectly} on the parameters through the action of some linear operator. Canonical linear inverse problems include deconvolution, computerized tomography, and source inversion, among many others \citep{kaipio2006statistical}. Design criteria in this setting are often quite explicit and tractable, and also serve as a building block for certain nonlinear \rev{design approaches.}

We specify a general linear-Gaussian model as
\begin{equation}
  \label{eq:linearGaussianmodel}
Y = G \param + \rev{\mathcal{E}},
\end{equation}
where $ G \in \mathbb{R}^{n \times p}$ represents the linear `forward' operator, mapping parameters to data in $\mathbb{R}^n$, and $\rev{\mathcal{E}}$ is a Gaussian random variable with full-rank covariance matrix $\Gamma_{Y \vert \Param} \in \mathbb{R}^{n \times n}$ that does not depend on $\param$. We let $\rev{\mathcal{E}}$ have mean zero; choosing otherwise would not affect the developments below. In general, both $G$ and $\Gamma_{Y \vert \Param}$ can depend on the design $\design$; that is, we have $G(\design)$ and $\Gamma_{Y \vert \Param}(\design)$.
The  linear-Gaussian model can be summarized as $Y | \param, \design \,  \sim  \, \normdist \left (G(\design) \param, \Gamma_{Y \vert \Param}(\design) \right )$.

\subsubsection{Classical alphabetic optimality}
\label{sss:classical}

The Fisher information matrix associated with \eqref{eq:linearGaussianmodel} is
\begin{equation}
  \label{eq:fisherinfolinear}
  F(\param, \design)  = F(\design) = G(\design)^\top \Gamma_{Y \vert \Param}(\design)^{-1} G(\design).
\end{equation}
It is thus independent of the value of the parameters $\param$; this
property of linear models greatly simplifies the construction of
design criteria. Note also that the inverse of the Fisher information
matrix, $F^{-1}$, \rev{when it exists,} is the covariance of the least-squares estimate 
and hence the maximum likelihood estimate, \rev{$\hat{\param}(y)$,} of $\param$.
Many classical design criteria are therefore chosen to be scalar-valued functionals of the matrix $F$. Indeed, we must somehow `scalarize' $F$ to produce a useful optimization objective, and various scalarizations encode different goals. We recall several of these so-called `alphabetic optimality' criteria as follows.

\revarxtwo{Suppose that $F$ is invertible, which is invariably the situation of interest in classical design and what we shall assume in the remainder of this subsection. Then \textit{A-optimal} design seeks
$$
\design^\ast \in  \arg \min_{\design \in \Xi} \tr \left ( F ^{-1} (\design) \right ),
$$
which can be interpreted as minimizing the average variance of the $p$ components of $\hat{\param}$. Note that this optimization problem is not equivalent to $\arg \max_{\design \in \Xi} \tr \bigl ( F(\design) \bigr )$.}

\textit{D-optimal} design, similarly, seeks
$$
\design^\ast \in \arg \min_{\design \in \Xi} \log \det \left ( F ^{-1} (\design) \right),
$$
which minimizes the volume (in $\mathbb{R}^p$) of the smallest $100(1-\alpha)\%$ confidence ellipsoid for $\param$, for any confidence level $1-\alpha$. A useful distinguishing feature of D-optimal designs is that they are invariant under linear reparameterization (and hence rescaling) of $\param$; that is, if ${\param}' = M\param$ for some invertible matrix $M$, then a design that is D-optimal for $\param$ is also D-optimal for $\param'$. This is not true, in general, for other optimality criteria.

While the A- and D-optimality criteria explicitly involve all the
eigenvalues of $F$, E-optimal designs minimize the maximum eigenvalue
of $F^{-1}(\design)$, $\lambda_{\text{max}} (F^{-1}(\design))$
(equivalently, maximize $\lambda_{\text{min}} (F(\design)) = 1 /
\lambda_{\text{max}} (F^{-1}(\design))$). Such designs thus minimize
the variance of $ q^\top \hat{\param}$ among all $q \in \mathbb{R}^p$
satisfying a norm constraint, that is, they minimize $\max_{\|q\| = w } \Var(q^\top \hat{\param}) = \max_{\|q\| = w } q^\top F(\design)^{-1} q$ for any $w > 0$.

A-, D-, and E-optimality are perhaps the essential building-block
design criteria for linear models, but there are numerous others. Some
focus on the estimation of a linear combination of the parameters
$\param$, or a subset of the elements of $\param$. Suppose, for
example, that we are primarily interested in $A^\top \param$, where $A \in \mathbb{R}^{p \times s}$,  $s < p$, and $A$ has rank $s$. Then $\text{D}_{\text{A}}$-optimality generalizes D-optimality by seeking 
\begin{equation}
\label{eq:DAoptimality}
\design^\ast \in \arg \min_{\design \in \Xi} \log \det \left ( A^\top F^{-1}(\design) A \right ).
\end{equation}
This objective is justified by noting that $A^\top F^{-1}(\design) A$ is the covariance matrix of $A^\top \hat{\param}$. If we put $A = [I_s \ \  0]^\top$, then the design criterion \eqref{eq:DAoptimality} focuses on the first $s$ elements of $\param$; this is called $\text{D}_{\text{S}}$-optimality.

An analogous generalization of A-optimality, \revnew{for some matrix $L \in \mathbb{R}^{p \times p}$,} is called L-optimality
\revnew{\citep{Atkinson_2007}:}
\begin{equation*}
\revnew{  \design^\ast \in \arg \min_{\design \in \Xi}\tr  (  F^{-1}(\design) L) \, .}
\end{equation*}
If $L$ is symmetric and has rank $s \leq p$, then it can be expressed as $L = A A^\top$ for $A \in \mathbb{R}^{p \times s}$. Then, using the cyclic property of the trace, the L-optimality objective can be rewritten as $\tr (A^\top F^{-1}(\design) A )$.  If $A^\top$ is a row vector in this setting
\rev{(i.e., $s=1$),}
then the criterion seeks to minimize the variance of a single linear combination of the parameters, and it is called c-optimality. 

Other criteria instead seek to control the variance of predictions of the linear model. Consider, specifically, a regression model on some compact domain $\mathcal{X}$, where each row of $G$ is given by the evaluation of a feature vector $f: \mathcal{X} \to \mathbb{R}^p$; that is, \rev{the $i$th row of $G$ is $G(i, :) = f^\top(x_i)$} for some $x_i \in \mathcal{X}$ that is in the support of the design $\design$. The G-optimality criterion considers the variance of the predicted response at any point $x \in \mathcal{X}$, $f^\top(x) F^{-1}(\design) f(x)$, and seeks a design $\design$ that will minimize the maximum value of this variance, that is,
\begin{equation}
\design^\ast \in \argmin_{\design \in \Xi} \max_{x \in \mathcal{X}}
f^\top(x) F^{-1}(\design) f(x) .
\label{e:Gopt}
\end{equation}
Variants of this criterion that target the \textit{average} variance
of the predicted response over a region, rather than its maximum, are
called I-optimality or V-optimality. For a much more comprehensive
discussion of classical alphabetic optimality criteria and their
properties, we refer to \citet{Hedayat_1981} and \citet{Shah_1989} as well as \citet[Chapter 10]{Atkinson_2007}.

We should note here that the assumption of normality on $Y$ is
generally not needed for these criteria to apply, and for the
statistical interpretations given above to hold. Rather, we need only
that $\mathbb{E}[Y] = G \param $ and $\Cov(Y)=
\Gamma_{Y \vert \Param}$. The best~linear unbiased estimator still
follows from the least squares solution in this setting,
and its performance is bounded by the Fisher information matrix (cf.\ Cram\'{e}r--Rao bounds). In many situations (including the usual setting for G-optimality described above), it is further assumed that $\Gamma_{Y \vert \Param} = \sigma^2 I$, that is, the observational errors are uncorrelated and have constant variance.

\subsubsection{Bayesian alphabetic optimality}
\label{s:bayesalphopt}
In the Bayesian setting, the parameters $\param$ of the linear model \eqref{eq:linearGaussianmodel} are endowed with~a~prior distribution. Since these parameters are now modeled as random variables, we write them as uppercase $\Param$ and require that $\rev{\mathcal{E}}$ and $\Param$ be independent. Choosing a conjugate Gaussian prior, $\Param \sim \normdist(\mu_{\Param}, \Gamma_{\Param})$, we obtain a posterior distribution that is again Gaussian; it can be written in closed form as  $\Param | y, \design \,  \sim \normdist(\mu_{\Param | Y}(y, \design), \Gamma_{\Param \vert Y}(\design))$, where
\begin{eqnarray}
  \Gamma_{\Param | Y}(\design) & \coloneqq & \left ( G(\design)^\top \Gamma_{Y \vert \Param}(\design)^{-1} G(\design) + \Gamma_{\Param}^{-1} \right )^{-1} \label{e:LG_post_cov} \\
  \mu_{\Param | Y}(y, \design)  & \coloneqq & \Gamma_{\Param | Y}(\design) \left ( G(\design)^\top \Gamma_{Y \vert \Param}(\design)^{-1} y + \Gamma_{\Param}^{-1} \mu_{\Param} \right ) .\label{e:LG_post_mean}
\end{eqnarray}
The posterior mean therefore depends on the realized value of the data $y$, but the posterior covariance matrix does not.

The Bayesian analogue to classical alphabetic optimality uses design criteria that are functions of the \emph{posterior covariance matrix}. Prior knowledge---and more generally the `balance' of information between the prior and the likelihood, where the latter may be affected by the number of observations---will therefore affect the optimal design, since the design criteria depend on the dispersion (shape and scale) of the posterior.

For instance, Bayesian A-optimality seeks designs that minimize the trace of the posterior covariance matrix,
$$
\design^\ast \in \arg \min_{\design \in \Xi}  \tr \left (\Gamma_{\Param | Y}(\design) \right) = \tr \left (  ( F(\design) + \Gamma_{\Param}^{-1} )^{-1} \right ).
$$
Note that, in contrast with classical A-optimality, this objective no longer requires $F$ to be invertible, as long as the prior covariance $\Gamma_{\Param}$ is chosen \revnew{to have full rank.} The same is true of all other Bayesian alphabetic optimality criteria, making these criteria well-suited to designs with fewer than $p$ support points, and to ill-posed inverse problems generally \citep{Haber_2008}.
Bayesian D-optimality seeks to minimize the log-determinant of the posterior covariance matrix,
\begin{equation}
\label{e:bayesdopt}
  \design^\ast \in \arg \min_{\design \in \Xi} \log \det \left ( \Gamma_{\Param | Y}(\design) \right ). 
\end{equation}
Similarly, Bayesian $\text{D}_{\text{A}}$-optimality seeks to minimize
the log-determinant of the covariance of the posterior predictive
distribution of \rev{$A^\top \param$, for some matrix $A \in
  \mathbb{R}^{p \times s}$ with rank $s < p$; hence the goal is to
  minimize $\log \det \left ( A^\top  \Gamma_{\Param | Y}(\design) A
  \right )$.} See \citet{Attia2018} for an application of this
criterion, and of a Bayesian analogue of classical
L-optimality. Bayesian E-optimal design minimizes the maximum
eigenvalue of $\Gamma_{\Param | Y}(\design)$, and so on. For an
extensive discussion of~Bayesian alphabetic optimality criteria and
their interpretations, we refer the reader to \citet{Chaloner_1995}
and \citet{DasGupta_1995}. We will also revisit several of these criteria from the more general viewpoint of decision theory in Section~\ref{ss:decision_theoretic}; the \rev{decision-theoretic} formulation lets us derive many Bayesian alphabetic optimality criteria from specific utility functions and, crucially, enables generalization to nonlinear models.

In the Bayesian setting, it is also natural to consider linear models with unknown variance parameters, e.g., $\Gamma_{Y | \Param} = \sigma^2 I$ with unknown $\sigma^2$, and to endow these variance parameters with suitable priors. In general, this extension modifies the optimality criteria discussed above---with some exceptions, such as Bayesian A-optimality using a conjugate inverse Gamma prior for $\sigma^2$; for more information, see \citet{Chaloner_1995}.

\subsubsection{Designs as probability measures}
\label{sss:design_measure}
So far, we have deliberately remained somewhat non-specific in describing how the design $\design$ enters the statistical models \eqref{eq:statmodel} or \eqref{eq:linearGaussianmodel}, other than to think of $\design$ as representing all the \rev{chosen} `coordinates' or locations of the observations on some continuous domain $\mathcal{X}$, or the indices \rev{of observations selected} from some countable set of candidates. One rather elegant way of formalizing these examples is to cast the design as a \emph{probability measure} on some domain $\mathcal{X}$. This viewpoint, \rev{originating in \citet{kieferWolfowitz1959optimum},} is widely adopted in the classical literature on optimal design.

To explain this perspective, let us first consider the discrete case,
where $\mathcal{X}$ is a countable and perhaps finite set of
observation indices, $\mathcal{X} = \{1, 2, 3, \ldots \}$. Suppose
that we wish to choose $n$ observations in total. If $n_i$
observations are taken at each point $i=1, 2, 3, \ldots$ and $\sum_i
n_i = n$, then we can write $\design_i = n_i / n$ and consider
$(\design_i)_{i}$ to be a probability measure
\citep{Chaloner_1995}. This class of `quantized' measures, with
integer $n_i$ and hence weights that are multiples of $1/n$, is called
an \textit{exact design}. If each point can only be observed
once---i.e., if the selection is without replacement---then we further
require $n_i \in \{0, 1\}$. It is often useful to relax the integer
constraint, such that $\design$ is any probability measure over the
set of candidate indices; in this case, $\design$ is called a
\textit{continuous} or approximate design. Then \rev{$\Xi$ is} the set of all probability mass functions over $\mathcal{X}$. 

Now consider the setting of continuous observation indices, on a compact set $\mathcal{X} \subseteq \mathbb{R}^d$, for  $d \geq 1$. This setting allows candidate observations to be indexed by some continuous coordinates, e.g., angles for a tomography problem, real-valued spatial coordinates for a sensor placement problem, \rev{or any other covariates in a generic regression problem.} The notion of a continuous \rev{\emph{design}} extends naturally: $\design$ is simply a probability measure on $\mathcal{X}$, and $\Xi$ is the set of all such probability measures. An exact design here would be a finite mixture of Dirac measures with quantized weights---that is, a measure supported on a finite collection of points in $\mathcal{X}$ with mixture weights that are multiples of $1/n$, i.e., $\design = \frac1n \sum_{i=1}^n \delta_{x_i}$, for $x_i \in \mathcal{X}$.

A nice consequence of this general viewpoint is that many quantities relevant to the preceding design criteria can be written as integrals with respect to $\design$. Consider a linear regression model with features $f: \mathcal{X} \to \mathbb{R}^p$ and uncorrelated observational errors. If the design $\design$ is supported on $n$ equally weighted points $x_i \in \mathcal{X}$, $i=1, \ldots, n$, then the Fisher information matrix of the model is
\begin{align}
  F(\design) = \frac{1}{\sigma^2}\sum_{i=1}^n f(x_i) f(x_i)^\top .
  \label{e:FIregression}
\end{align}
This expression follows from \eqref{eq:fisherinfolinear} by setting
\begin{align*}
  G
= \left [ f(x_1)^\top;  f(x_2)^\top; \ldots; f(x_n)^\top \right ] \in
\mathbb{R}^{n \times p}
\end{align*}
and $\Gamma_{Y | \Param} = \sigma^2 I_n$. 
If the design has continuous support, then we simply have instead
$$
F(\design) = \revnew{\frac{n}{\sigma^2}} \int_{\mathcal{X}} f(x) f(x)^\top \design(\mathrm{d}x),
$$
\revnew{where the $n$ above ensures that the scaling of the integral is consistent with \eqref{e:FIregression}.} 

It is natural to wonder how to reconcile this \revarx{continuous-support} viewpoint with the existence of classical optimal designs supported on a \textit{finite} set of points in $\mathcal{X}$ \citep{Atwood_1969}. In other words, when optimizing \rev{some} design criterion over the set of all probability measures on \rev{the infinite set} $\mathcal{X}$, why should a minimizer be supported only on a finite number of locations? \rev{In fact, as} explained in \citet{Atwood_1969} and \citet{Kiefer_1960}, under conditions satisfied by any of the design criteria in Section~\ref{sss:classical}, there exists an optimal design supported on at most $p(p+1)/2$ points. Intuition for the result follows from Carath\'{e}odory's theorem, compactness of $\mathcal{X}$, and the fact that $F$ is a $p \times p$ symmetric matrix: the optimal information matrix $F$ can always be expressed as a convex combination of \rev{at most} $p(p+1)/2$ rank-one matrices, \rev{each produced by a single-point design.} The need for $F$ to be full rank, on the other hand, imposes a lower bound of $p$ on the number of points in an optimal design. \rev{In the case of Bayesian alphabetic optimality criteria for linear models, similar upper bounds for the number of support points in an optimal design have also been derived \citep{chaloner1984optimal}. In the nonlinear setting, however, the Fisher information matrix depends on the parameter $\param$ (see Section~\ref{sss:nonlinear}). A common approach, discussed below, is then to average a local design criterion over a distribution on $\param$. Because now (infinitely) many information matrices $F(\param, \design)$ are involved, upper bounds on the number of support points do not in general hold \citep{Atkinson_2007,chaloner1989optimal}.}

\rev{An important theme in classical design theory has been the construction of so-called `equivalence theorems' for continuous designs. The first such result was due to \citet{kiefer1960equivalence}, who showed that any continuous D-optimal design is also G-optimal, and vice-versa. This result was substantially generalized by \citet{kiefer1974general} and \citet{whittle1973some}. The resulting `general equivalence theorem' relies on the fact that the design criteria to be minimized are convex functionals $\phi$ of the probability measure $\design$. Under this condition, with some further assumptions on the regularity of $\phi$, the equivalence theorem provides multiple equivalent conditions for the optimality of a design $\design^\ast$, some of which correspond to verifying that the directional derivatives of $\phi$ (with respect to feasible designs $\design$) are zero at $\design^\ast$. The latter are useful to check optimality of a given design measure (in the continuous case), and have been employed in algorithms \citep{wynn1972results}. Variants of the general equivalence theorem have been established for Bayesian alphabetic optimality in linear models \citep{chaloner1984optimal,pilz1991bayesian} and for certain local optimality criteria averaged over the prior in nonlinear models \citep{chaloner1989optimal}. There is a vast array of results along this theme, which we will not attempt to survey here. Instead we refer the reader to the comprehensive framework in \citet{Pukelsheim_2006}, which tackles design optimality for linear models using tools of convex analysis, and the historical perspective in \citet{wynn1984jack}.} 

\rev{Since our interest is largely in nonlinear problems (as well as infinite-dimensional linear problems), we will generally resort to numerical methods for optimizing $\design$, which} must in any case \rev{employ} \emph{tractable} finite-dimensional parameterizations of candidate designs. Moreover, \emph{only an exact design can be realized in an experiment}, and hence some notion of `rounding' is \revarx{usually} needed if a problem is initially solved from a continuous design perspective. We will discuss these considerations further in Section~\ref{s:optimization}.

\subsubsection{Challenges of nonlinear design}
\label{sss:nonlinear}

In problems where dependence on the model parameters $\param$ is nonlinear, the Fisher information matrix $F$ will generally vary with $\param$. Since $\param$ is unknown, it is then unclear where to evaluate $F(\param, \design)$ and any of the associated design criteria.

A relatively crude approach is simply to choose some reference or
`best-guess' parameter value $\param_0$ and proceed; this is known
as `local' design, but it clearly ignores parameter uncertainty and
the impact of nonlinearity, and may have sharp dependence on the
choice of $\param_0$. One could iterate this approach, by estimating $\param$ after collecting data from a first local design, and then using the \rev{estimated} parameter value to produce a new local design, and so on \citep{Korkel_1999}.
A more principled alternative is a minimax formulation
\citep{Fedorov_1997,King_2000,Berger_2009}.
If $\phi(\param, \design)$ is any `local' design criterion (containing the parameter-dependent Fisher information matrix $F(\param, \design)$, for instance, $ \phi(\param, \design) = \log \det F^{-1}(\param,\design)$), then we seek
$$
\design^\ast \in  \argmin_{\design \in \Xi} \max_{\param \in \paramset} \phi(\param, \design).
$$
An interpretation of this objective is that it seeks the design with
best performance for the worst-case parameter value, i.e., the value
of $\param$ that is most challenging to estimate (in the sense of
$\phi$). This idea has been explored by \citet{Sun_2007} and \citet{Siade_2017},
among others, but generally leads to a rather difficult optimization problem. Typically $\phi$ is chosen to be a classical alphabetic optimality criterion (as in the example of D-optimality above), and thus the only `prior' information on $\param$ in these formulations is via the set $\paramset$.

Another natural alternative is to introduce a prior distribution
$\pdf_\Param$ for $\param$ and to average any local design criterion
$\phi(\param, \design)$ over this prior. This approach is widely
adopted, due in no small part to its computational tractability; see
\citet{Pronzato_1985} and \citet{Fedorov_1997}. If $\phi$ is solely based on the
Fisher information, however, then such a formulation is
only `pseudo-Bayesian' \citep{Atkinson_2007,Ryan_2016}.
Indeed, in some such works, the prior is used as a means
of handling the parameter dependence arising from nonlinearity but
then discarded for subsequent analysis. \citet{Ryan_2016} argue that
any `fully Bayesian' design criterion must be a functional of the
posterior distribution. Interestingly, however, \citet{Walker_2016}
shows that prior expectation of the trace of the Fisher information
matrix,
\begin{align*}
  \int \tr (F(\param, \design)) \pdf_{\rev{\Param}} (\param) \, \mathrm{d} \param ,
\end{align*}
has an \rev{information-theoretic} interpretation. \citet{Overstall_2022} and
\citet{Prangle_2023} both explore using this quantity as a design
criterion in nonlinear problems, and show that it has a
decision-theoretic justification as well.

As a step in the `fully Bayesian' direction, given a nonlinear model and a Gaussian prior $p_\Param$ with full rank covariance matrix $\Gamma_\Param$, one could seek, as proposed in \citet{Chaloner_1995},
\begin{align}
  \design^\ast \in  \arg \max_{\design \in \Xi}  \int    \log \det \left (F(\param, \design) + \Gamma_{\Param}^{-1} \right ) \pdf_\Param(\param) \,\mathrm{d} \param.
  \label{e:nonlinearheuristic}
\end{align}
This can be loosely interpreted as minimizing the average, over the prior predictive distribution of $Y$, of the log-determinant of the covariance of \rev{a Gaussian} approximation of each resulting posterior. Yet this interpretation is rather imprecise, as for a general nonlinear model, $F(\param', \design) + \Gamma_{\Param}^{-1}$ can be very far from the precision matrix of the posterior distribution that results from a realization of the data $y \sim \pdf_{Y \vert \param, \design}( \cdot \vert \param', \design)$. Criteria such as this are best viewed as approximations of an expected utility function arising from a more principled \rev{decision-theoretic} formulation, which we discuss next.

\subsection{Decision- and information-theoretic formulations}
\label{ss:decision_theoretic}

The decision-theoretic approach to OED was
formalized by \citet{Lindley_1956} (see also \citealt{stone1959application,Raiffa_1961}) and has two primary ingredients: a utility function $u$, chosen to reflect the purpose of the experiment; and the Bayesian principle of averaging over what is uncertain \citep{Berger_1985}. In this framework, any design criterion takes the form of an \textit{expected utility}, to be maximized with respect to $\design$:\footnote{Beginning in this section, we will drop subscripts from probability density functions when the arguments are explicit, letting these arguments make the choice of density clear.}
\begin{align}  
  U(\design) = \mathbb{E}_{Y, \Param|\design}[u(\design,Y,\Param)] =  \iint \pdf(y, \param \vert \design)  \, u(\design, y, \param)
  \,\mathrm{d} \param \, \mathrm{d}y.  \label{e:EU}
\end{align}
Here, the utility function $u(\design,y, \param)$ quantifies the value of an experiment performed with a design $\design$ that yields observations $y$, if the true parameter value is $\param$. Since the outcome of the experiment is not known when selecting $\design$, and since the parameter value is also uncertain, we average over the joint prior distribution of $Y$ and $\Param$. This process yields the expected utility $U(\design)$. Many choices of utility function $u$ have been proposed and explored in the literature. We review some of the possibilities below.

\subsubsection{Expected information gain in parameters}
\label{sss:EIGparams}

The influential paper of \citet{Lindley_1956} proposed using the expected gain in Shannon information, from prior to posterior, as an optimal experimental design criterion. It is evocative to think of this quantity in at least two ways, namely
\begin{align}
  U_{\text{KL}}(\design)
  &= \mathbb{E}_{Y|\design} \left [ \DKL ( \pdf_{\Param|Y,\design} \vert \vert \pdf_{\Param}) \right ] \label{e:EKL} \\[6pt]
  & = H(\Param) - H(\Param \vert Y, \design)  \label{e:entropydiff}
\end{align}
where $\DKL$ denotes the Kullback--Leibler (KL) divergence, or
relative entropy, from prior to posterior,
\begin{equation}
\DKL ( \pdf_{\Param|y,\design} \vert \vert  \pdf_{\Param} ) = 
  \int   \pdf(\param | y,\design) \log \frac{\pdf(\param | y, \design)} {\pdf( \param ) }  \,\mathrm{d} \param, 
\end{equation}
and the two terms in \eqref{e:entropydiff} are the entropy and conditional entropy, respectively:
\begin{align}
  H(\Param) & = - \int \pdf(\param)  \log \pdf(\param)  \, \mathrm{d} \param, \\
  H(\Param \vert Y, \design) & = - \iint \pdf (y , \param \vert \design) \log \pdf (\param \vert y, \design)   \,  \mathrm{d} \param \, \mathrm{d} y .
\end{align}
Equality of the expressions \eqref{e:EKL} and \eqref{e:entropydiff} is
easily verified. Note that $U_{\text{KL}}$ is always non-negative:
conditioning reduces entropy \textit{on average} (not necessarily for
each realization $Y=y$, but when averaging over values of $y$), with
zero entropy reduction $H(\Param) = H(\Param \vert Y, \design)$ if and
only if $\Param$ and $Y \vert \design$ are independent.
Similarly the KL divergence, whose expectation yields \eqref{e:EKL},
is always non-negative \rev{and reaches zero if and only if the two
  distributions being compared are identical} \citep{Cover_2006}. Some
intuition for maximizing this criterion is that the design $\design$
yielding data $Y$ that \textit{most} reduce Shannon entropy is, in
this \rev{information-theoretic} sense, the most informative. Another
intuition is that an optimal experiment should maximize the
\rev{`change'} (here, quantified by the KL divergence) from the prior to the posterior,
averaged over the prior predictive distribution of the data $Y$.

Lindley's original rationale for the design criterion $U_{\text{KL}}$ was not rooted in decision theory, but the criterion was later given a decision-theoretic justification by \citet{Bernardo_1979}; see also the discussion in \citet{Prangle_2023}. Bernardo frames the task of inference as a decision problem, where making a decision amounts to returning a probability density function for the parameters of interest $\Param$. He argues that the utility function for this decision should be a \emph{proper, local scoring rule} \citep{Gneiting_2007}, 
and that these desiderata in turn dictate that $u$ must specifically be a \emph{logarithmic} scoring rule. In the language of \eqref{e:EU} above, this means that one should choose
\begin{align}
u^{\text{score}}(\design, y,\param) = \log p(\param \vert y, \design) - \log p(\param). \label{e:uscore}
\end{align}
Substituting this utility into \eqref{e:EU} immediately yields \eqref{e:EKL} and \eqref{e:entropydiff}. 

Note also that choosing the utility to be the KL divergence from prior to posterior, which depends explicitly on $y$ and $\design$ but not on $\param$,
\begin{align}
u^{\text{div}}(\design, y, \param) = \DKL ( \pdf_{\Param|y,\design} \vert \vert  \pdf_{\Param} )  = u^{\text{div}}(\design, y) \label{e:udivergence}
\end{align}
and substituting this utility into \eqref{e:EU}, yields the same
expected utility $U_{\text{KL}}$ \eqref{e:EKL}. We also call
$U_{\text{KL}}$ the \textit{expected information gain} (EIG) in $\Param$, from
prior to posterior. It is useful for subsequent computations (see Section~\ref{s:estimation}) to write the EIG more explicitly as follows:
\begin{align}
  U_{\text{KL}}(\design) & =   \iint \pdf(y, \param \vert \design) \log \frac{\pdf(\param \vert  y, \design)}{\pdf( \param )} \,\mathrm{d} \param \, \mathrm{d}y \label{e:EKLpost} \\
                         &=   \iint \pdf(y, \param \vert \design) \log \frac{\pdf(y \vert  \param, \design)}{\pdf( y \vert \design )} \,\mathrm{d} \param \, \mathrm{d}y \label{e:EKLmarg} \\ 
                         &=   \iint \pdf(y, \param \vert \design) \log \frac{\pdf(y , \param \vert \design)}{\pdf( y \vert \design ) \pdf(\param)} \,\mathrm{d} \param \, \mathrm{d}y \label{e:EKLMI} \\ 
                         & \eqqcolon \mathcal{I}(Y ; \Param \vert\design). \nonumber
\end{align}
In all of these expressions, the joint prior predictive density of $Y$ and $\Param$ can be factored as $\pdf( y, \param \vert \design) = \pdf(y \vert \param, \design) \pdf( \param)$, i.e., as a product of likelihood and prior. Moving from \eqref{e:EKLpost} to \eqref{e:EKLmarg} is an application of Bayes' rule \eqref{eq:Bayes}. Both \eqref{e:EKLpost} and \eqref{e:EKLmarg} make clear that some kind of posterior calculation is necessary: the former involves the \emph{normalized} posterior density $\pdf(\param \vert  y, \design)$, while the latter involves the posterior normalizing constant $\pdf( y \vert \design )$. Moreover, these expressions must be evaluated for a range of $y$ values---i.e., for `all possible' posterior distributions---via the outer expectation. The last expression above, \eqref{e:EKLMI}, shows that EIG is equivalent to the \emph{mutual information} (MI) between the parameters and observations given the design, $\mathcal{I}(Y ; \Param \vert\design)$. Henceforth, we will use the terms EIG and MI interchangeably.

Expanding \eqref{e:EKLmarg} into two terms also shows that the EIG is a difference of entropies of the data, paralleling \eqref{e:entropydiff},
\begin{align}
U_{\text{KL}}(\design) & = H(Y \vert \design) - H(Y \vert \Param, \design).
\end{align}
For some statistical models, $H(Y \vert \Param, \design)$ is a
constant function of $\design$. One example is the case of a nonlinear
model with additive noise, $Y = G(\Param, \design) + \mathcal{E}$,
where $\mathcal{E}$ is independent of $\Param$ and its distribution
does not depend on $\design$; then the entropy $H(Y \vert \Param =
\param, \design) = H(\mathcal{E})$ and hence does not depend on
$\param$ or $\design$. Maximizing EIG then \textit{specializes} to
maximizing the entropy of the prior marginal distribution of $Y$. This
design strategy is called `maximum entropy sampling'; see
\citet{Shewry_1987} and \citet{Sebastiani_2000}.

The EIG objective has additional desirable properties. For one, it is invariant under bijective transformations of $\param$; this property follows from the invariance of the KL divergence to such transformations, and thus includes rescalings of the parameters as well as more complex reparameterizations. We also emphasize that there are \emph{no assumptions of linearity or Gaussianity} in the motivation for the EIG objective, and in any of its expressions above. 

In the linear-Gaussian case, however, maximizing EIG is equivalent to seeking a Bayesian D-optimal design \eqref{e:bayesdopt}. To see this, note that when $\Param$ and $Y \vert \design $ are jointly Gaussian (as in Section~\ref{s:bayesalphopt}), the EIG or MI can be written in closed form as
\begin{align}
  \mathcal{I}(Y ; \Param \vert\design)
  & = \frac12 \left ( \log \det \Gamma_{\Param}   - \log \det  \Gamma_{\Param \vert Y}(\design) \right )\label{e:LG_EIG1}\\
  & = \frac12 \left ( \log \det  \Gamma_Y(\design) - \log \det \Gamma_{Y \vert \Param}(\design) \right ),\label{e:LG_EIG2}
\end{align}
where $\Gamma_Y$ is the marginal (prior predictive) covariance of $Y$, $$\Gamma_Y(\design) = G(\design) \Gamma_\Param G(\design)^\top + \Gamma_{Y \vert \Param}(\design).$$
\revnew{From \eqref{e:LG_EIG1}}, it is apparent that maximizing EIG with respect to $\design$ is equivalent to minimizing the log-determinant of the posterior covariance. If, additionally, the observational error covariance is independent of the design, then $\Gamma_{Y \vert \Param}(\design) = \Gamma_{Y \vert \Param}$ and an equivalent goal, \revnew{via \eqref{e:LG_EIG2}}, is to maximize the log-determinant of the marginal covariance $\Gamma_Y$; this is a specific \rev{(linear-Gaussian)} case of the maximum entropy sampling described above.

It is interesting to note that the Bayesian D-optimality criterion for linear-Gaussian models can be derived from other utility functions, besides \eqref{e:uscore} and \eqref{e:udivergence}; see \citet[Section 2.2]{Chaloner_1995} for details.

\subsubsection{Other utility functions}
\label{sss:otherutilities}

Having defined an information- and decision-theoretic design criterion for inference of the model parameters $\Param$, it is natural to extend this construction to other goals.

Suppose that we are interested in only a subset of the model parameters. Partitioning $\Param$ as $\Param = (\Param_1, \Param_2)$, information gain in $\Param_1$ is captured by the KL divergence from its prior \textit{marginal} distribution to its posterior \textit{marginal} distribution:
\begin{align}
  u(\design, y) & = \DKL( p_{\Param_1 \vert y, \design} \vert \vert p_{\Param_1}),
  \label{e:subsetKL}
\end{align}
where
$$p(\param_1) = \int p(\param_1, \param_2) \, \mathrm{d} \param_2 \ \  \  \text{and} \ \ \  p(\param_1 \vert y, \design) = \int p(\param_1, \param_2 \vert y , \design) \, \mathrm{d} \param_2.$$
The outer expectation over $Y$ \revnew{in \eqref{e:EU}}, yielding the \emph{expected} information gain in $\Param_1$, must still account for uncertainty in both blocks of $\Param$. The optimal design criterion in this case becomes
\begin{align}
  U(\design) & = \iiint p(y \vert \param_1, \param_2, \design) p (\param_1, \param_2) \, \log \frac{p(\param_1 \vert y, \design)} {p(\param_1)} \, \mathrm{d} \param_1 \, \mathrm{d} \param_2 \, \mathrm{d} y \label{e:EKLfocusedpost} \\
             & = \iiint p(y \vert \param_1, \param_2, \design) p (\param_1, \param_2) \,  \log \frac{p(y \vert \param_1, \design)} {p(y \vert \design)} \, \mathrm{d} \param_1 \, \mathrm{d} \param_2 \, \mathrm{d} y.
               \label{e:EKLfocusedmarg}
\end{align}
The second expression is analogous to \eqref{e:EKLmarg} in that it uses densities for the data $Y$, but now even the numerator of the density ratio involves marginalization, as
\begin{align}
  p(y \vert \param_1, \design) &= \int p(y \vert \param_1, \param_2, \design) p(\param_2 \vert \param_1) \, \mathrm{d} \param_2 .
\end{align}
Compared to the EIG in $\Param$, evaluating this objective therefore requires an additional integration over $\Param_2$. Note also that \eqref{e:EKLfocusedpost} and \eqref{e:EKLfocusedmarg} are equivalent to the MI, $\mathcal{I}(Y ; \Param_1 \vert \design)$. In this formulation, $\Param_2$ could represent a variety of possible `nuisance' parameters in the statistical model, i.e., any parameters that are uncertain but simply not the modeler's immediate object of interest \rev{\citep{Feng_2019,alexanderian2021optimal}.} Special examples include the parameters of a discrepancy model \citep{Kennedy_2001} designed to capture model error, or the background medium in an inverse scattering problem \citep{borges2018reconstruction}.

A generalization of the preceding formulation is to consider the EIG in some (generally nonlinear) \textit{function} of the parameters $\Param$, $Z = \Psi(\Param)$, where $\Psi: \paramset \to \RR^q$ for some $q \leq p$.
This can be thought of as a `goal-oriented' objective (just like
Bayesian $\text{D}_{\text{A}}$-optimality in the linear-Gaussian case)
where the function $\Psi$ encodes the true quantity of interest.
Now we seek to maximize the expected KL divergence from the prior predictive distribution of $Z$ to its posterior predictive distribution:
\begin{align}
  U(\design) & = \mathbb{E}_{Y \vert \design} \left [ \DKL( \pdf_{Z \vert Y, \design} \vert \vert \pdf_{Z}) \right ] \nonumber \\
             & = \iint p(y, z \vert \design) \, \log  \frac{p(z \vert y, \design)} {p(z)} \, \mathrm{d}z \, \mathrm{d}y \label{e:goalpost} \\
             & = \iint p(y, z \vert \design) \, \log \frac{p(y \vert z, \design)} {p(y \vert \design)} \, \mathrm{d}z \, \mathrm{d}y  \label{e:goalmarg} \\
             & = \mathcal{I}(Y ; Z \vert \design). \nonumber
\end{align}
Notably, this objective is also the original object of interest in \citet{Bernardo_1979}. While it is straightforward to write down, it raises significant computational challenges, \textit{beyond} those associated with calculating EIG in the parameters alone. For generic $\Psi$, $p(\param)$, and $p(y \vert \param, \design)$, we do not have simple expressions for the prior density $p(z)$ or the posterior density $p(z \vert y, \design)$ of $Z$ (even up to a normalizing constant) appearing in \eqref{e:goalpost}. Nor do we have easy access to the marginal likelihood $p(y \vert z, \design)$ to instead evaluate \eqref{e:goalmarg}. Numerical approximations are needed, involving density estimation or approximate Bayesian computation; see Section~\ref{s:estimation}. Of course, in specific cases (such as linear $\Psi$ with Gaussian priors) some aspects of the expressions above become more tractable. We note also that if $\Psi$ is bijective, then EIG in $Z$ is identical to EIG in $\Param$, since (as noted earlier) the information gain objective is invariant under bijective transformations; otherwise the EIG in $Z$ is smaller \citep[Theorem~1]{Bernardo_1979}.

\begin{figure}[p]
  \centering
  \subfloat[EIG in $\Param$]{\includegraphics[width=0.75\linewidth]{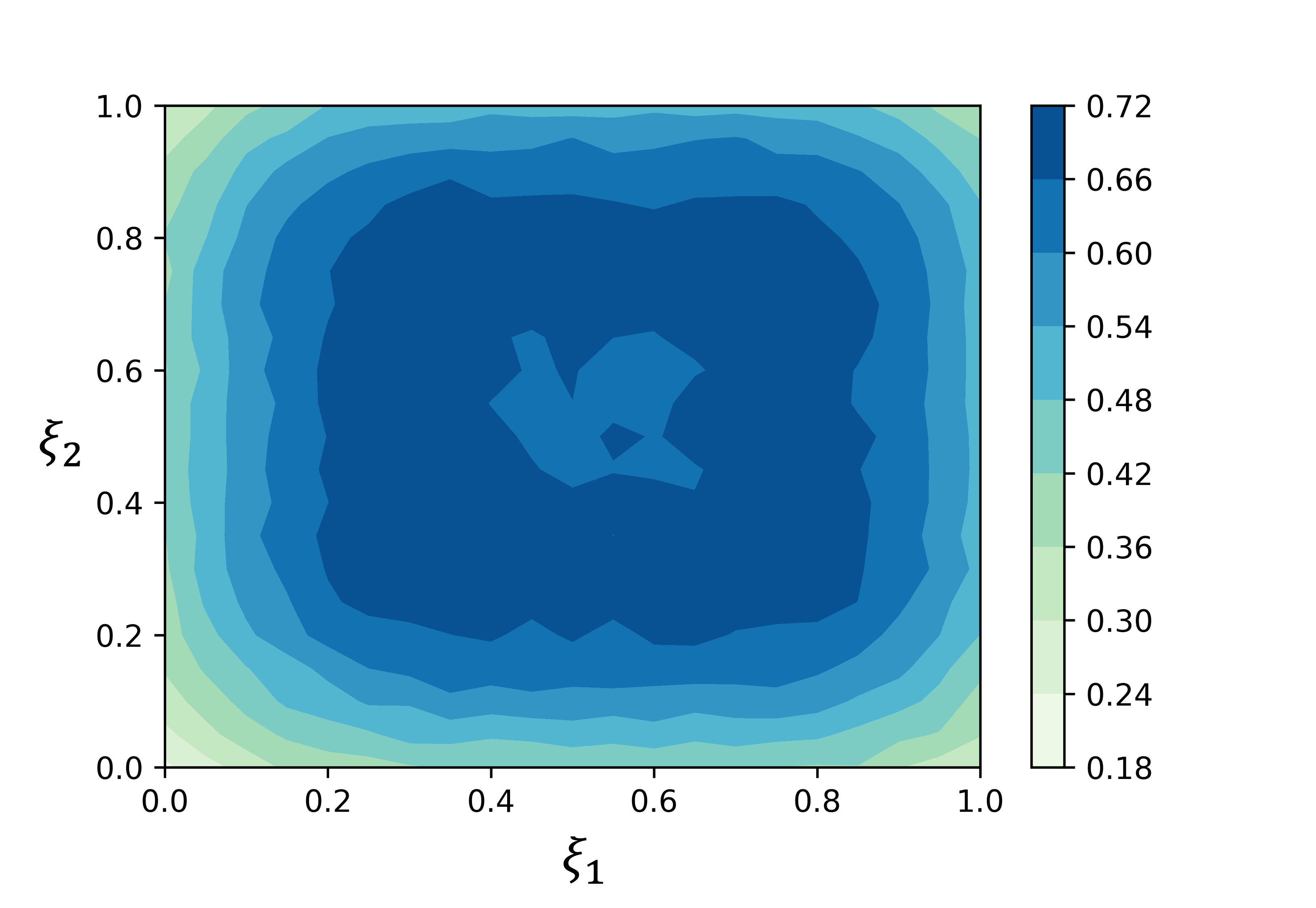}\label{f:KL_nonGOOED}}\\
  \subfloat[EIG in various $Z$]{\includegraphics[width=0.95\linewidth]{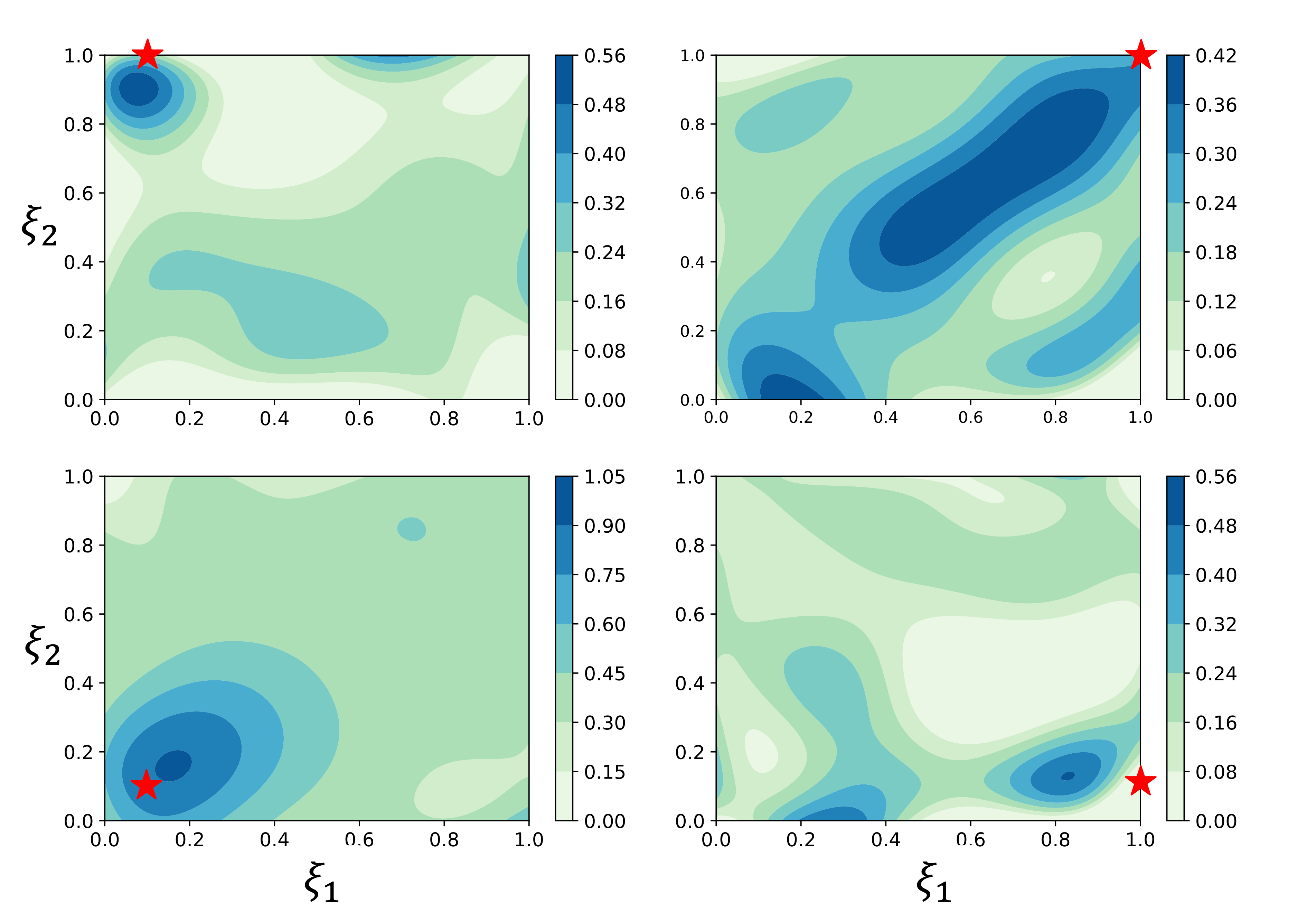}\label{f:KL_GOOED}}
  \caption{Optimal sensor placement in a time-dependent advection-diffusion problem. Each figure shows a map of expected information gain (EIG) in a chosen quantity, as a function of the sensor location $\design \in [0,1]^2$. Measurements are made at time $t_1 > 0$, and advection is towards the top right. (a) EIG in the unknown source location $\Param$. (b) EIG for different quantities of interest $Z$, where each $Z$ is the predicted concentration at some future time $t_2 > t_1$, at the location marked by the red star. The optimal designs, maximizing EIG in each case, differ significantly.}
  \label{f:GOOED_example}
\end{figure}

\vspace{0.5em}
The optimal design obtained by maximizing the EIG in \rev{some} $Z$ can differ drastically from the design maximizing \rev{EIG in $\Param$.}
Figure~\ref{f:GOOED_example} illustrates these contrasts for sensor
placement in a time-dependent advection-diffusion problem
\citep{Zhong_2024}. The parameter $\Param$ is the unknown source
location, endowed with a uniform prior on 
$[0,1]^2$. The source emits a scalar quantity that diffuses and is
advected towards the top-right of the $[0,1]^2$ domain, with the
advection velocity increasing linearly in time, from a value of zero
at $t=0$. The design entails placing a single sensor, with coordinates
$\design=(\design_1,\design_2) \in [0,1]^2$, that measures the
concentration of the scalar at time $t_1 >
0$. Figure~\ref{f:GOOED_example}\subref{f:KL_nonGOOED} shows a
map of EIG in $\Param$ as function of sensor location
(estimated numerically; see Section~\ref{s:estimation}). We see that
the optimal measurement location is not unique, but lies roughly 0.2
units of distance from the center of the domain; the slight asymmetry
is due to the direction of
advection. Figure~\ref{f:GOOED_example}\subref{f:KL_GOOED}, in
contrast, shows maps of EIG for four different
choices of $Z$; each $Z$ is the predicted concentration of the passive scalar  at a future time $t_2>t_1$ and at the specific location marked by a red star in each panel. We see that the optimal sensor locations, maximizing these goal-oriented EIG criteria, are markedly different from those in Figure~\ref{f:GOOED_example}\subref{f:KL_nonGOOED}.

\vspace{0.5em}
If the quantity of interest $Z$ is random given $\param$, e.g., if it is described by a conditional density $p(z \vert \param)$, then the formulation above still applies. Computations may actually be easier: the problem of estimating $p(z)$ or $p(z \vert y, \design)$ is \emph{smoothed} by the kernel $p(z \vert \param)$, and the restriction that $q \leq p$ is lifted as long as the conditional density $p(z \vert \param)$ on $\mathbb{R}^q$ exists. A special case is when $Z = Y\vert \design^+$, that is, we wish to maximize information gain in the model prediction for some given design $\design^+$. We call this future prediction $Y^+$, to distinguish it from the potential outcomes $Y$ of the experiment being currently designed. In this case, the utility $u$ can be set to
\begin{align}
    u(\design, y) & = \DKL( p_{Y^+ \vert y, \design, \design^+} \vert \vert p_{Y^+ | \design^+}),
  \end{align}
\vspace{0.5em}
which yields, as an expected utility, the expected information gain in $Y^+$:
  \begin{align}
    U(\design) & = \iint p(y, y^+ \vert \design, \design^+) \, \log \frac{p(y^+ \vert y, \design, \design^+)} {p(y^+ \vert \design^+)} \, \mathrm{d}y^+ \, \mathrm{d}y  \\
               & = \iiint p(y \vert \param, \design) p(y^+ \vert \param, \design^+) p(\param) \, \log \frac{p(y^+ \vert y, \design, \design^+)} {p(y^+ \vert \design^+)} \, \mathrm{d}y^+ \, \mathrm{d}y \, \mathrm{d} \param \\
    & = \mathcal{I}(Y ; Y^+ \vert \design, \design^+).
  \end{align}
The penultimate line above reflects the fact that $Y$ and $Y^+$ are conditionally independent given $\param$, and hence their joint prior predictive can be expanded and factored by introducing the parameter $\param$ explicitly.

A final \rev{information-theoretic} utility that we will consider arises
from problems of \emph{model discrimination} in the Bayesian setting
\citep{Myung_2009}. Suppose we have a countable set of models
$\mathcal{M}_m$, $m=1, 2, \ldots$, each with its own parameters,
$\param_m \in \paramset_m \subseteq \RR^{p_m}$, and its own prior on
parameters, $p(\param_m)$.
Suppose there is also a (discrete) prior distribution over the model
indicators $m$.
As a utility, we choose the relative entropy from this prior to the posterior distribution over model indicators,
\begin{align}
  u(\design, y) =  \sum_m P( m \vert y, \design) \log \frac{ P( m \vert y, \design)} { P( m ) }.
  \label{e:modelutility}
\end{align}
Following \eqref{e:EU}, we must take an expectation over the prior predictive distribution of $Y$ to obtain an expected utility. Now, however, because there are multiple possible models, the prior predictive distribution is itself a mixture of the prior predictives of each model:
\begin{align}
  p(y \vert \design) = \sum_m P( m ) p(y \vert m, \design) = \sum_m P( m ) \int_{\paramset_m}
  p(y \vert \param_m, \design)
  p(\param_m ) \, \mathrm{d} \param_m,
  \label{e:priorpredictivemodels}
\end{align}
where conditioning on $\param_m$ also implies conditioning on $m$ at
the same time.
The EIG in the model indicator $m$ follows by combining \eqref{e:modelutility} and \eqref{e:priorpredictivemodels}: $U(\design) = \int u(\design, y)  p(y \vert \design) \, \mathrm{d}y$.
As noted in \citet{Ryan_2016}, this design approach applies in the
$\mathcal{M}$-closed framework for model selection (see \citealt[Chapter~6]{Bernardo_2000});
that is, the true (data-generating) model is assumed to be within the
set of models considered, and one must assign a prior weight
$P(m)$ to the event that each model is true.
Examples of Bayesian OED
for model discrimination can be found in
\citet{Myung_2009,Cavagnaro_2010,McGree_2012,Drovandi_2014,Aggarwal_2016}, and \citet{Hainy_2022}.

While most of the preceding discussion has focused on \rev{information-theoretic} utilities, the decision-theoretic framework described at the start of Section~\ref{ss:decision_theoretic} is certainly not limited to utility functions of this kind. As described in \citet{Ryan_2016} and \citet{Chaloner_1995}, another natural utility is a quadratic loss, motivated by the desire to extract a point estimate of $\param$ from the posterior. For instance, let $\bar{\param}(y, \design) \coloneqq \mathbb{E}[\Param \vert y, \design]$ denote the posterior mean. Then we can write
\begin{align}
u(\design, y, \param) = - \left ( \param - \bar{\param}(y, \design) \right  )^\top B \left ( \param - \bar{\param}(y, \design) \right  ) \label{e:quadloss}
\end{align}
for some symmetric positive semi-definite
matrix $B \in \mathbb{R}^{p \times p}$. The expected utility is then
\begin{align}
U(\design) = - \iint \left ( \param - \bar{\param}(y, \design) \right  )^\top B \left ( \param - \bar{\param}(y, \design) \right  ) p(y, \param \vert \design) \, \mathrm{d} \param \, \mathrm{d} y,
\end{align}
which is the negative Bayes risk of the posterior mean under a \rev{\emph{weighted squared error}} loss. Maximizing the expected utility over $\design$ thus minimizes this risk. As noted in \citet{Chaloner_1995}, in the case of a linear-Gaussian model, this formulation reverts to Bayesian A-optimal design with weight matrix $B$, i.e., $\min_\design \tr \left ( B  \,  \Gamma_{\Param \vert Y}(\design)  \right )$.

Another family of \rev{non-information-theoretic} utilities involves scalar functionals of the posterior covariance matrix; these are not strictly motivated by point estimation, but rather can be seen as a more computationally tractable alternative to \rev{information-theoretic} utilities that require calculation of posterior normalizing constants. \citet{Ryan_2016} specifically suggest using as a utility the determinant of the posterior precision matrix:
\begin{align}
  u(\design, y) = \frac{1}{\det \left ( \Cov (\Param \vert y, \design) \right ) }
\end{align}
and then, as usual, averaging this quantity over the prior predictive
of $Y$ to obtain an expected utility:
\begin{align}
  U(\design) & = \int \left ( \det ( \Cov (\Param \vert y, \design) ) \right )^{-1} p(y \vert \design) \, \mathrm{d} y \\
                    & = \iint \left  ( \det ( \Cov (\Param \vert y, \design) ) \right )^{-1} p(y \vert \param, \design) p(\param) \, \mathrm{d} \param \, \mathrm{d} y.
\label{e:EUpostdprecision}
\end{align}
We emphasize that this criterion is intended for nonlinear/non-Gaussian problems, and thus calculation of the posterior covariance for different realizations of $Y$ is not a computationally trivial undertaking. It is instructive to compare this criterion to the similar but cruder heuristic \eqref{e:nonlinearheuristic}, which is motivated by a series of Gaussian approximations as described in \citet{Chaloner_1995}.

To close this section, we point the reader to a more general formalism
for what comprises a `valid' notion of information gain from a
statistical experiment, due to \citet{Ginebra_2007}. \rev{In this formalism,} an information
measure must satisfy a \emph{minimal} set of requirements: \rev{(i)} it is
real-valued; \rev{(ii)} it returns zero for a `totally non-informative experiment,'
where $Y$ is independent of $\Param$; and \rev{(iii)} it satisfies sufficiency
ordering \citep{Blackwell_1951,Blackwell_1953,LeCam_1964}.
The last requirement can be understood as follows. Let
$Y \vert \param, \design_1$ and $Y | \param,\design_2$ be the outcomes
of two different experiments, for the same parameter value $\param$,
and let $\eta$ be an independent random variable with fixed and known
distribution, introducing auxiliary randomness. If there exists a
function $W$ such that $W(Y,\eta) \vert \param,\design_1$ has the same
distribution as $Y|\param,\design_2$ for all $\param$, then the
experiment with design $\design_1$ is said to be `sufficient for' or
`always at least as informative as' the experiment with design
$\design_2$. That is to say, the data from $\design_1$ can generate
data from $\design_2$ with an additional randomization mechanism and
without knowing $\param$. In such a situation, $\xi_1$ is preferred.
This generalized notion of an information measure broadly encompasses
several commonly used objectives in OED, including mutual information.
We refer readers to \citet{Ginebra_2007} for an extended discussion,
including connections to likelihood ratio and posterior-to-prior ratio
statistics.

\subsection{Design criteria for infinite-dimensional problems}

Infinite-dimensional statistical models arise in the Bayesian approach to inverse problems \citep{Stuart_2010, StuartDashtihandbook,vanZanten} and, more broadly, in nonparametric estimation and nonparametric Bayesian procedures \citep{gine2021mathematical}. These problems can be understood as estimation or inference of \textit{functions}; in other words, the parameter $\param$ of the statistical model now belongs to a function space, rather than to $\mathbb{R}^p$ for finite $p$. Application domains of such models are vast, and we will not attempt to review them here. Instead we will focus on two classes of problems where the integration of OED with infinite-dimensional models has proved to be particularly fruitful.

\subsubsection{Inverse problems in the Bayesian setting}
\label{s:BIPs}

The infinite-dimensional setting is natural for inverse problems
involving partial differential equations, where the parameter to be
learned is typically an initial condition, a source term, a boundary
condition, or a heterogeneous coefficient---and thus a function of
space and/or time. An important research theme, at the intersection of
applied mathematics and statistics, has been to create statistical
formulations of inverse problems that are well-defined in the
infinite-dimensional setting \citep{Stuart_2010}. This is necessary,
for instance, to create consistent Bayesian models for inverse
problems---Bayesian models that have a well-defined limit as the
discretization of the underlying functions is refined
\citep{BuiThanh2013}. Another important practical result of these
efforts is algorithms with discretization-invariant (and hence
dimension-independent) performance, for example Markov chain Monte Carlo methods whose sampling efficiency does not deteriorate with grid refinement \citep{HairerStuart,Cotter_etal_2013,Cui_2016,rudolf2018generalization,PetraTOMS}.

OED for infinite-dimensional Bayesian inverse problems is well
explored in the setting of Gaussian priors, particularly for linear
inverse problems \citep{AlexanderianSISC2014}, and summarized in the
recent review of \citet{Alexanderian_2021}. To explain these
developments, we first briefly sketch the setting and refer the reader
to \citet{Stuart_2010} and \citet{StuartDashtihandbook} for full details and precise results. The parameter $\Param$ is modeled as a random variable taking values in an infinite-dimensional separable Hilbert space $(\mathcal{H}, \langle \cdot, \cdot \rangle_{\mathcal{H}})$. $\Param$ is further assumed to be Gaussian, which means that the real scalar-valued random variable $\langle v, \Param \rangle_{\mathcal{H}}$ is Gaussian for any $v \in \mathcal{H}$. The mean of $\Param$ can be defined as an element $\bar{\param}$ of $\mathcal{H}$ satisfying $\langle \bar{\param}, v \rangle_{\mathcal{H}} = \mathbb{E}[\langle \Theta, v \rangle_{\mathcal{H}} ]$ for all $v \in \mathcal{H}$. The covariance operator of $\Theta$ is the positive, self-adjoint, and compact linear operator $C: \mathcal{H} \to \mathcal{H}$ defined through $\langle v, Cw \rangle_{\mathcal{H}} = \mathbb{E} [ \langle \Param - \bar{\param}, v \rangle_{\mathcal{H}} \langle \Param - \bar{\param}, w \rangle_{\mathcal{H}} ]$ for all $v, w \in \mathcal{H}$. Because $\Theta$ takes values in $\mathcal{H}$, the trace of $C$ is finite; it is then said that $C$ is of \textit{trace class}. We can write this Gaussian (prior) measure of $\Param$ as $\mu = \mathcal{N}(\bar{\param}, C)$, and will assume that $C$ is strictly positive.

A common assumption on the statistical model for the observations $Y
\in \mathbb{R}^n$ is that they result from the action of a (possibly
nonlinear) `forward' operator $G_\design: \mathcal{H} \to \RR^n$
perturbed with additive Gaussian noise: $Y = G_\design(\Param) +
\mathcal{E}$, where $\mathcal{E} \sim \mathcal{N}(0, \Gamma_{Y \vert
  \Param})$ is independent of $\Param$. For any fixed $Y=y$, this in
turn defines a likelihood function that is proportional to
$$\mathcal{L}^y_\design: \theta \mapsto -\frac{1}{2} \exp \Bigl ( \bigl (y -
G_\design(\param) \bigr )^\top \Gamma_{Y \vert \Param}^{-1}  \bigl (y -
G_\design(\param)  \bigr ) \Bigr )  .$$
Under appropriate conditions on
$G_\design$ and the prior $\mu$, detailed in \citet{Stuart_2010}, the
posterior distribution of $\Param$, i.e., the distribution of $\Param$
given $Y=y$, denoted by $\mu^{y}_\design$, is well defined and
dominated by the prior measure, $\rev{ \mu^{y}_{\design} } \ll \mu$. One can then write $\mu^{y}_\design$ in terms of its Radon--Nikodym derivative with respect to $\mu$:
\begin{align}
\frac{\mathrm{d} \mu^{y}_\design}{ \mathrm{d} \mu}(\theta) \propto \mathcal{L}^y_\design(\theta).
\end{align}
The KL divergence from prior to posterior, $\DKL(\mu^y_{\design} \vert \vert \mu)$, can also be defined under these conditions.

With this background in hand, we can summarize several design criteria that have been proposed for infinite-dimensional Bayesian inverse problems. When the forward operator $G$ is \textit{linear}, the posterior measure $\mu^{y}_\design$ is again Gaussian, with a \rev{covariance operator $C_{\text{pos}}(\design)$}  that is independent of $y$. In this setting, \citet{AlexanderianSISC2014} propose minimizing the trace of the posterior covariance operator with respect to $\design$: $\min_{\design \in \Xi} \tr(C_{\text{pos}}(\design))$ . This is the infinite-dimensional version of Bayesian A-optimality; the objective is well defined because the posterior covariance $C_{\text{pos}} (\design)$ is also of trace class, under the conditions noted above. A marginalized version of infinite-dimensional Bayesian A-optimality, focused on the covariance of a subset of variables of interest, was used for design in \citet{alexanderian2021optimal}. This can be compared to L-optimality in the finite-dimensional setting.

The analog of Bayesian D-optimality is somewhat less straightforward,
as the eigenvalues of the posterior covariance operator
$C_{\text{pos}}(\design)$ accumulate at zero, and hence minimizing the
log-determinant of this operator is not meaningful
\citep{Alexanderian_2021}. Instead, \citet{Alexanderian_2016ba} use
the correspondence between D-optimality and maximizing EIG in linear
problems (see the discussion at the end of
Section~\ref{sss:EIGparams}) to derive an alternative objective for
Bayesian D-optimality in the linear setting. \rev{Specifically, they start with the EIG---taking advantage of the fact that the KL divergence from prior to posterior is well defined, under conditions summarized above. Specializing this quantity and its expectation over $Y$ to the linear-Gaussian setting, the objective thus obtained is}
$\frac12 \log \det (\text{Id} + \widetilde{H})$,
where $\widetilde{H}$ is the prior-preconditioned Hessian operator of the negative log-likelihood \citep[Theorem~1]{Alexanderian_2016ba}. This expression coincides with the log-determinant of the posterior covariance in the finite-dimensional case. Efficient ways to estimate this objective, leveraging low rank structure, are discussed in \citet{Alexanderian_Saibaba_2018}. We note also that  $\widetilde{H}$ is a central quantity in dimension reduction for Bayesian inverse problems, and will appear again in Section~\ref{s:estimation}.

For nonlinear forward operators, a full treatment of EIG in the
infinite-dimensional setting has not (to our knowledge) been used as
\rev{a} design criterion.
\rev{This may be largely due to computational tractability, though some theoretical gaps (e.g., checking that the mutual information between $Y$ and the infinite-dimensional $\Param$ is well defined; see \citealt{duncan1970calculation}) may remain.}
Instead, researchers have focused on simpler design criteria and their further approximations. For instance, \citet{Alexanderian_2021}, motivated by the finite-dimensional approach in \citet{haber2009numerical}, discusses design that minimizes the Bayes risk of the posterior mode $\hat{\param}(y)$ under the \rev{squared error} loss defined by the inner product on $\mathcal{H}$, i.e., $\| \cdot \|\rev{^2} \equiv \langle \cdot, \cdot \rangle_{\mathcal{H}}$. This is analogous to \eqref{e:quadloss} but using the posterior mode \rev{(also called the \textit{maximum a posteriori} (MAP) estimate \citep{dashti2013map})} $\hat{\param}(y)$ rather than the posterior mean, as the former is typically more computationally tractable.
Another \rev{closely related heuristic} %
suggested in \citet{Alexanderian_2021} is to minimize the trace of the posterior covariance operator, in expectation over the data $Y$:
\rev{\begin{align}
\min_{\design \in \Xi} \int_{\mathcal{H}} \int_{\mathbb{R}^n} \tr \left ( C_{\text{pos}}(y, \design) \right ) \mathcal{L}^y_\design(\param) \, \mathrm{d}y \, \mu(\mathrm{d} \param) . \label{e:tracemininfinite}
\end{align} }
This is \rev{essentially} the `Bayesian A-posterior precision' \rev{expected} utility described in \citet[Section~3.1.2]{Ryan_2016}. As the posterior covariance operator is \rev{difficult} to approximate in nonlinear inverse problems (not to mention \textit{many} posterior covariances, one for each realization of the data \rev{used to evaluate the integral in \eqref{e:tracemininfinite}}), \citet{Alexanderian_2016} instead propose replacing $\tr ( C_{\text{pos}}(y, \design) )$ above with the trace of the inverse Hessian of a Laplace approximation of the posterior at the MAP point $\hat{\param}(y)$. \rev{This approximation is reasonable when the posterior is `close' to Gaussian \citep{schillings2020convergence,helin2022non,spokoiny2023dimension}.}

Computing any of these design criteria in the setting of infinite-dimensional Bayesian inverse problems is a computationally challenging undertaking, due to the high discretization dimension used to represent the parameter $\param$ in practice, \rev{as well as the cost of forward operator evaluations}. Considerable computational ingenuity is required; an essential step towards \rev{mitigating the impact of high discretization dimension} is to take advantage of low-rank structure in the prior-preconditioned Hessian, and to exploit randomized numerical linear algebra methods for computing eigendecompositions, estimating the trace, and so on.

\subsubsection{Gaussian process regression}
\label{sss:GPR}

Gaussian process (GP) regression, also known as kriging, is a ubiquitous tool in %
\rev{spatial statistics, time series modeling,} machine learning,
engineering design, surrogate modeling, and countless other
applications. \citet{Rasmussen_2006} and \citet{gramacy2020surrogates} provide excellent expositions of both applications and \rev{some} theoretical foundations. Experimental design for GP regression has thus received considerable attention.

From the perspective of the previous section, GP regression can be
viewed as a linear Bayesian `inverse problem' on function space,
with a \rev{trivial} forward operator:  \rev{a selection operator}.
The underlying true function,
$\param^{\ast}: \mathcal{X} \to \RR$, for
$\mathcal{X} \subseteq \RR^d$, \rev{is observed directly through 
evaluation at a finite collection of points $x_i \in \mathcal{X}$, $i=1, \ldots, n$, perhaps with additive Gaussian noise,
e.g., $Y_i = \param^*(x_i) + \mathcal{E}_i$ with $\mathcal{E}_i\sim \mathcal{N}(0,\sigma^2)$.}
\rev{Here we let the prior model for the
true function be a Gaussian process $\Param$ on 
$\mathcal{X}$, with mean function $m(x)\coloneqq\EE[\Param(x)]$ and
positive semi-definite covariance \rev{function}
$c(x,x') \coloneqq \EE[(\Param(x)-m(x))(\Param(x')-m(x'))]$, which defines the prior covariance operator $C$ via $$(Cv)(x): x \mapsto \int_\mathcal{X} c(x, x') v(x') \, \mathrm{d}x' ,$$ for functions $v \in L^2(\mathcal{X})$.
Given a collection of observations $y_s \coloneqq (y_i )_{i=1}^n \in \mathbb{R}^n$, taken at corresponding covariate values $x_s \coloneqq (x_i)_{i=1}^n$, performing GP regression entails conditioning 
$\Param$ on these data. The posterior distribution, describing this conditioned process, remains Gaussian,
\begin{align}
  \Param|(x_s,y_s) \sim \mathcal{N}  (m_{\text{pos}}, C_{\text{pos}}  ),
\end{align}
with the posterior mean $m_{\text{pos}}$ and the posterior covariance function  $c_{\text{pos}}$ (yielding $C_{\text{pos}}$) expressible in closed form:}
$$
m_{\text{pos}}(x)=m(x)+\alpha^{\top}c(x_s,x) \ \  \text{and} \ \ 
c_{\text{pos}}(x,x' ; x_s)=c(x,x')-c(x_s,x)^{\top}R c(x_s,x'),
$$
where the matrix $R^{-1}\in\RR^{n\times n}$ has entries
$[R^{-1}]_{ij}=c(x_i,x_j)+\delta_{ij}\sigma^2$, the coefficient
vector $\alpha\in \RR^{n}$ has entries
$\alpha_i=R{[i,:]}(y_s-m(x_s))$, \rev{and $c(x_s, x)  \coloneqq  \left ( c(x_1, x), \ldots, c(x_n, x) \right ): \mathcal{X} \to \mathbb{R}^n$ \citep{Rasmussen_2006,gramacy2020surrogates}.}

Since the purpose of GP regression is generally to make predictions about $\param^{\ast}$ at unseen values of the covariates $x$, most design criteria involve the (posterior) \textit{predictive} variance, i.e., the variance of $\Param|(x_s,y_s)$. In this %
\rev{setting}, however, the `parameter' is the process \rev{$\Param$,} and hence the boundary between parameters and predictions is rather blurred. 

Another perspective on \rev{GP regression} follows intuitively from
finite-dimensional distributions of the process \rev{$\Param$,} which are
always multivariate Gaussian (both before and after conditioning on
the data).
For a finite number of sites $x_{\mathcal{S}} \coloneqq (x_i)_{i=1}^m \in \mathcal{X}$, $\Param(x_{\mathcal{S}}) \coloneqq (\Param(x_i) )_{i\in \mathcal{S}}$ is simply a multivariate normal random vector. We can observe some components of this vector, and we wish to use these observations to predict other components.

Maximum entropy sampling \citep{Shewry_1987} originates with this discretized perspective. Let $x_s \subset x_{\mathcal{S}}$ denote the $n < m$ \rev{distinct} locations selected for a candidate design and let $x_{s^c}  =x_{\mathcal{S}} \setminus x_s$ denote its complement. Then the chain rule for entropy yields
\begin{align}
H \left (\Param(x_{\mathcal{S}}) \right  ) = H \left ( \Param(x_s) \right ) + H \left ( \Param(x_{s^c}) \vert \Param(x_s) \right ) .
\label{e:entropysum}
\end{align}
Since the left-hand side of \eqref{e:entropysum} is fixed, minimizing entropy in the predictions at unobserved sites $x_{s^c}$ given the observations (the second term on the right) can be accomplished by maximizing the first term on the right. Thus finding an optimal design is cast as maximizing the entropy of the model predictions (the \rev{\emph{joint} entropy} of these predictions) at the observed locations, $H \left ( \Param(x_s) \right )$. (Recall that we also discussed maximum entropy sampling for general parametric statistical models in Section~\ref{sss:EIGparams}.) More explicitly, the optimization problem is typically posed with some cardinality constraint on the number of observations, e.g., $| x_s| \leq n$:
\begin{align}
\argmax_{x_s \subset {x_\mathcal{S}}, \ | x_s| \leq n} H \left ( \Param(x_s) \right ).
\end{align}
The objective above can be understood as a \textit{set function}, i.e., a function of all subsets of $x_\mathcal{S}$. In the present case, since $\Param(x_s)$ is a Gaussian vector, closed-form expressions for the entropy are immediately available. The problem is equivalent to finding the principal submatrix of $\Cov (\Theta(x_{\mathcal{S}}) )$ that has largest determinant. This problem is NP-hard, but many practical algorithms have been developed to tackle it \citep{Ko_1995}.

A crude approximation to maximum entropy sampling is to choose the elements of $x_s$ one at a time, in a greedy fashion: beginning with $x_s = \emptyset$ and $x_{s^c} = x_\mathcal{S}$, at each iteration select from $x_{s^c}$ the point with maximum predictive variance,\footnote{Recall that the entropy of a univariate Gaussian random variable is an increasing function of its variance.}
by ranking the diagonal elements of $\Cov ( \Param(x_{s^c}) \vert \Param(x_s) )$. Then add this point to $x_s$ and repeat. \rev{\citet{seo2000gaussian} and subsequent papers call this approach `active learning MacKay', after \citet{Mackay_1992}.} Its performance can be far from optimal, however, as the entropy objective is not submodular (see Section~\ref{s:optimization}).

An alternative design approach, advocated by \citet{Krause_2008} (see also \citet{caselton1984optimal}) is to maximize MI, rather than entropy. Specifically, the problem is posed as
\begin{align}
  \argmax_{x_s \subset x_\mathcal{S}, \ | x_s| \leq n} \mathcal{I} \left ( \Param(x_s) ;  \Param(x_{s^c}) \right ). 
\end{align}
Greedy approaches are typically applied to this problem, for reasons of computational tractability. Section~\ref{s:optimization} will discuss these algorithmic considerations in much more detail. Here, however, we will note that greedy approaches tend to work far better for MI maximization than for entropy maximization, as MI is \textit{submodular} \citep{Krause_2008,Nemhauser1978,Fisher_1978}. \citet[Section 4.1]{Krause_2008} provide some useful intuition contrasting greedy selection via MI and greedy selection via predictive entropy. A~key consideration in the setup above is to ensure also that the objective is \textit{monotone} increasing for $|x_s| \leq n$, which is required for optimization guarantees to hold. \citet{Krause_2008} shows that $\mathcal{I} \left ( \Param(x_s) ;  \Param(x_{s^c}) \right )$ is approximately monotone in this regime as long as the discretization of the underlying domain $\mathcal{X}$, via $x_\mathcal{S}$, is sufficiently fine. \rev{\citet{beck2016sequential} present improvements to greedy MI maximization tailored to computer model emulation.}

A rather different class of design approaches is more rooted in the continuous view of GPs, seeking the observation locations that minimize the resulting posterior predictive variance, \textit{integrated} over the domain of the process, $\mathcal{X}$. Letting $x_s \subset \mathcal{X}$ denote a \rev{finite collection of} observation locations (not necessarily chosen from a~\rev{finite} candidate set), the objective to be \rev{minimized, over feasible $x_s$,} can be written as \rev{
\begin{align}
  \int_{\mathcal{X}} c_{\text{pos}}(x,x ; x_s) \, \mathrm{d}x.
  \label{e:ivar}
\end{align}}
\rev{This} objective has appeared in many papers \citep{Sacks_1989,seo2000gaussian,Santner_2018, Gorodetsky_2016} and has been variously called the integrated mean-squared error (IMSE) criterion, \rev{the integrated mean-squared prediction error (IMSPE) criterion,} or the integrated variance (IVAR) criterion. With discrete candidate sets \rev{and a greedy one-point-at-a-time approach to constructing $x_s$ (see below), it is also called `active learning Cohn' (ALC), after \citet{Cohn_1996}. } In the language of classical design, minimizing \eqref{e:ivar} can be understood as a kind of (Bayesian) V-optimality, in that one minimizes the predictive variance integrated over a region.
\rev{We should also note that \eqref{e:ivar} is precisely $\tr ( C_{\text{pos}} )$, i.e., the trace of the posterior covariance operator, which is the infinite-dimensional notion of A-optimality discussed in Section~\ref{s:BIPs}.}
In practice, the integral \eqref{e:ivar} is approximated by a large
set of points chosen uniformly over $\mathcal{X}$, or perhaps
non-uniformly to reflect some desired weight. Both discrete selection
methods (see \citep{seo2000gaussian}) and methods that optimize over
the continuous coordinates of $n$ points $x_s =(x_1, \ldots, x_n)$
have been explored in the literature.  \rev{For the latter, see
  \citet{Sacks_1989} and \citet{Gorodetsky_2016}. Here, as with maximum entropy sampling, one can also optimize for all $n$ elements of $x_s$ simultaneously (a `full batch' design procedure), or proceed in a greedier but sub-optimal fashion: select a subset of design points to minimize \eqref{e:ivar}, `freeze' these points and update $c_{\text{pos}}$ accordingly, and repeat for the next subset. Batch approaches are more computationally demanding, but generally yield better performance; see demonstrations in \citet{Gorodetsky_2016}. We will discuss related optimization issues further in Section~\ref{s:optimization}. }

Finally, we mention several approaches that rely on spectral
decompositions of the GP, specifically the Karhunen--Lo\`{e}ve
\rev{representation of $\Param \sim \mathcal{N}( \rev{m}, C)$} \citep{Karhunen_1947,Loeve_1948}.
The idea is to find the leading eigenvalues and eigenfunctions
$(\lambda_i, \phi_i)$ of the prior covariance operator $C$, and to write the GP as
\begin{align}
\Param(x) = \rev{m(x)} + \rev{ \sum_{i=1}^\infty } \sqrt{\lambda_i} \phi_i(x) \zeta_i, 
\end{align}
where the \rev{scalar-valued random variables} $\zeta_i$ \rev{are standard Gaussian and mutually independent}.
If the eigensystem is truncated  to $r < \infty$ eigenpairs $(\lambda_i, \phi_i)_{i=1}^r$, then GP regression is
reduced to parametric regression with coefficients $(\zeta_1,\ldots,
\zeta_r)$. Then any standard Bayesian alphabetic optimality criterion
can be applied; \rev{see
  \citet{fedorov1997optimal,fedorov2007optimum}, and \citet{harari2014optimal}. See also \citet{spock2012spatial} for a related approach based on the polar spectral representation of $\Param$, assuming that the process is stationary and isotropic. 
The truncated eigendecomposition can also be used to approximate the integrated posterior variance objective \eqref{e:ivar}; see \citet{Fedorov_1996}.}

\rev{
One outstanding issue in model-based design for GP regression is that hyperparameters of the prior covariance function (controlling, e.g., the scale of the prior variance, correlation lengths, and smoothness) must often be learned from data as well. The methods discussed above all take the prior covariance as fixed, and thus ignore the `outer' hyperparameter learning process---as well as the impact that uncertainty in the hyperparameters has on the predictive distribution. In some applications, moreover, the main interest is not in prediction of an unknown function, but rather in learning the parameters $\gamma$ of the covariance function itself \citep{pardo1998maximum}; it is natural to expect that optimal designs for this purpose should differ from optimal design for prediction.
A common way of learning $\gamma$ is an `empirical Bayes' approach,
which provides a point estimate of the covariance parameters by
maximizing the \revnewnew{log-marginal} likelihood, \revnewnew{$\hat{\gamma} \in \argmax_{\gamma}
\log p( y_s \vert x_s,  \gamma)$, where}
\begin{align*}
  \revnewnew{
  \log p( y_s \vert x_s, \gamma) =   -\frac{1}{2} y_s^\top R_\gamma y_s - \frac{1}{2} \log \det R_\gamma^{-1} - \frac{n}{2} \log 2 \pi,
}
\end{align*}
\revnewnew{and the notation $R_\gamma$ emphasizes that $\gamma$ controls the matrix $R \in \mathbb{R}^{n \times n}$ defined earlier via $[R^{-1}]_{ij}=c(x_i,x_j)+\delta_{ij}\sigma^2$. Here, $\gamma$ may include parameters of the covariance function $c$ as well as the noise variance $\sigma^2$.}
Solving this optimization problem is
easier, computationally, than treating $\gamma$ in a fully Bayesian
way. That said, there are many papers and software packages
\citep{GPfullyBayessoftware} that do the latter,  endowing $\gamma$
with a prior distribution and inferring it jointly with $\Param$;
usually, this task requires \revnew{Markov chain Monte Carlo (MCMC)} or sequential Monte Carlo methods, as the joint distribution of $\Param$ and $\gamma$ is non-Gaussian. 
These fully Bayesian approaches are therefore computationally demanding---even more so in the setting of design. In principle, one could use the joint posterior distribution of $(\Param, \gamma)$ to find designs that maximize EIG in $\Param$, $\gamma$, or both, following the criteria developed in Sections~\ref{sss:EIGparams}--\ref{sss:otherutilities}.
We are not aware of methods that completely realize this approach. Instead, a variety of more practical computational schemes have been devised for experimental design in GP regression when the prior covariance parameters are uncertain.}

\rev{
\citet{zhu2005spatial} focus on design for parameters of the covariance function only, using the Fisher information matrix derived from the marginal likelihood. Since dependence on these parameters $\gamma$ is generally nonlinear, the authors propose using either `local' D-optimal design, a minimax approach, or a prior-averaged D-optimality criterion (cf.\ Section~\ref{sss:nonlinear}). It is interesting to note that the resulting designs involve points that are non-uniformly spaced over $\mathcal{X}$; intuitively, such point sets are useful for learning correlation lengths in the prior covariance function \citep{gramacy2020surrogates}. 
\citet{zhu2006spatial} then suggest a two-stage design process, where some fraction of the design points are chosen according to a criterion focused on estimation of the covariance parameters $\gamma$, while the rest are chosen to improve prediction of  $\Param$; uncertainty in the covariance parameters, given some asymptotic approximations, is accounted for in the design criterion for the latter.
\citet{spock2010spatial} cast design for prediction, using a spectral decomposition of the GP, within a minimax formulation over a compact set of parameterized covariance functions.
On the other hand, the local, sequential design scheme of
\citet{gramacy2015local} interleaves design for prediction with local
Fisher-infomation matrix-based design for the length-scale parameter
in $\gamma$. And simpler sequential schemes, where batches of standard
(e.g., IMSE) design for $\Param$ are interleaved with maximum
likelihood estimation of covariance parameters, are pursued in
\citet{harari2014optimal} and \citet{Gorodetsky_2016}. Further perspective on such sequential schemes is given in \citet[Section~6.2]{gramacy2020surrogates}.
Indeed, it is very natural to interleave updates of the covariance function parameters with actual observations, in a sequential design fashion. In the purely Gaussian case (i.e., with a fixed covariance function), the realized values of $y_s$ would have \emph{no} impact on subsequent designs, since the posterior covariance and entropy depend only on $x_s$, not on $y_s$. When parameters of the covariance function must \textit{also} be inferred, however, we are in the nonlinear design setting and hence there is value to feedback: $y_s$ informs the covariance parameters, and in turn these parameters reshape the predictive uncertainty of the Gaussian process for the next stage.}

\rev{We will discuss closed-loop sequential design much more {systematically} in Section~\ref{s:sequential}. Here we will mention just a few more instantiations in the setting of GP regression. \citet{riis2022bayesian} perform myopic sequential design using criteria based on the marginal posterior of $\Param$, where marginalization over the kernel hyperparameters is performed with MCMC samples. \citet{hoang2014nonmyopic} develop a non-myopic sequential design policy that can be understood as approximating a sequential variant of maximum entropy sampling; the authors argue that this policy naturally balances effort between informing covariance hyperparameters and directly reducing uncertainty in the prediction of $\Param$ itself.}

\rev{\subsection{Related problems and their distinctions}}
\label{ss:otherdesigns} 

To help orient the reader, here we discuss several classes of problems in the broader literature that have some conceptual overlap with optimal experimental design, but also some essential differences.

\paragraph{Space-filling and other non-model based designs.}
Space-filling designs, as the name suggests, spread design points
throughout the domain $\mathcal{X}$ so that one can reasonably assess
variations of a generic response or the parameters of an associated
statistical model.  In their simplest form, these methods do not
attempt to exploit the structure of a statistical model
for the response, or make any assumptions on such a model;
they are hence \emph{not} model-based designs, in contrast to the focus of this article.
Spread in the design space is achieved by formulating an optimization
problem, cases of which are primarily distinguished by whether the
focus is exclusively on the distance among the design points in
$\design$ (\textit{maximin distance} design), or the distance to all
points in the ground set $\mathcal{X}$ (\textit{minimax distance} design)
\citep{johnson1990minimax}. These notions have clear analogies to the
well-known problem of sphere packing \citep{zong2008sphere}.
Pure space-filling designs tend to  have poor projection properties,
failing to retain their optimality properties when viewed in
subspaces. This is undesirable in circumstances when the response is
insensitive to one or more of the design variables. In such cases,
Latin hypercube designs \citep{mckay1979comparison}, which ensure that
any of their projections along a single coordinate axis
yields a \textit{maximin distance} design, are a suitable alternative.
Space-filling properties in larger subspaces can be induced through orthogonal array extensions of Latin hypercube design \citep{Owen_1992,Tang_1993}. Further, `maximum projection' (MaxPro) designs have been developed to achieve space-filling properties on \emph{all} possible subsets of factors \citep{Joseph_2015,Joseph_2020}.

Other approaches using entropy maximization have also been suggested for space-filling design \citep{jourdan2010optimal}. Note that this is not akin to the maximum entropy sampling methods we previously discussed in Section~\ref{sss:GPR}. Here, `entropy' is that of the empirical distribution of design points on $\mathcal{X}$, and is used as a design criterion to be maximized. The core idea is to relate the space-filling quality of the design to the uniform distribution on $\mathcal{X}$, motivated by the fact that the uniform distribution has maximum entropy among all distributions with prescribed finite support.
  A similar idea is pursued through designs that seek to minimize
  the \textit{discrepancy} \citep{Niederreiter_1992} of a set of
  design points. This notion relates to the much broader topic of
  low-discrepancy sequences and quasi-Monte Carlo methods for
  integration \citep{caflisch1998monte,dick2013high}.
  We refer the reader also to \citet{pronzato2012design} and \citet{Santner_2018} for a
  more comprehensive discussion of space-filling designs targeting
  computer experiments.

  Other standard non-model-based design strategies include factorial designs,
  with blocking and fractional variants, as well as composite designs;
  \revarx{see \citet{wu2011experiments} and} \citet[Chapter~7]{Atkinson_2007} for more.

\paragraph{Active learning.} Active learning is a term originating in the computer science and machine learning communities, referring to a diverse array of algorithms for choosing which data to `label,' usually in a supervised learning setting \citep{dasgupta2011two}. In statistical terms, we can understand this setting as regression with real or discrete-valued outcomes $y_i$ (where the latter case is classification). In so-called `pool-based active learning,' there is a large pool of candidate covariates or feature values $\{x_i\}_i$, referred to as the unlabeled data \citep{schein2007active}. To label a chosen data point is to obtain its associated outcome, thus creating the pair $(x_{i^\ast}, y_{i^\ast})$ for some chosen index $i^\ast$. We can thus think of this problem as OED with a countable and even finite design space $\mathcal{X}$, corresponding to which indices should be chosen from the unlabeled pool.
Other learning scenarios might select covariates $x_i$ from an infinite set (e.g., with $\mathcal{X}$ now a region of $\mathbb{R}^d$); alternatively, one might be presented with a stream of successive $x_i$ and be required to choose, on-the-fly, whether to label the current value \citep{settles.tr09}.
In any of these cases, active learning usually refers to a \emph{sequential} version of the OED problem where data to be labeled are selected one at a time or in a batch, then labeled, then used to update a model, and then the process repeats. Most often, these iterations take a \emph{greedy} approach (see Section~\ref{s:sequential}).

An important point of differentiation among active learning methods is \emph{by what mechanism} this selection occurs. One class of selection methods, known as uncertainty sampling, selects the unlabeled data point(s) for which the model's current predictions are most uncertain. The notions of uncertainty used here vary widely, and can include many heuristics that do not have a statistical justification. This is an important distinction from OED. If the notion of uncertainty is tied to the posterior predictive distribution of a Bayesian model, however, then we can recover maximum entropy sampling, discussed in Section~\ref{sss:EIGparams} and~\ref{sss:GPR} above. Indeed, predictive entropy is commonly used to rank candidate points in active learning \citep{lewis1995sequential}.
Another widely used class of selection methods is `query-by-committee'
\citep{freund1997selective}, where disagreement among an ensemble of
models is used to rank candidate points, such that points with greater
disagreement are chosen. Yet another class of criteria ranks candidate
points by how much their label would reduce uncertainty in the
predictions of the model being trained. An example of the latter,
where uncertainty is captured by integrated variance, is the ALC
approach discussed in Section~\ref{sss:GPR}. Other selection methods
focus on more tailored prediction goals; for instance, \citet{blanchard2021output}
describe selection criteria favoring regions of the input space
that yield unusual output values, to help build regression models capable of predicting extreme events.

We should also emphasize that many active learning methods are not based on explicit design criteria, but rather on other heuristics \citep{settles.tr09}. Moreover, even criterion-based active learning methods usually focus on reducing uncertainty in predictions, rather than on improving estimation or inference of the \textit{parameters of a statistical model}. Again, this is an important distinction from many OED problems.

\paragraph{Bayesian optimization.} Bayesian optimization (BO)
\citep{movckus1975bayesian,jones1998efficient,wang2023recent} is
widely used in applications from engineering design to machine
learning, to name just a few. It is essentially a derivative-free
optimization method, used to maximize `black-box' (and often
computationally expensive) objective functions---that is, functions
which can be evaluated pointwise, but whose derivatives cannot be
directly evaluated. Gaussian process regression is a key ingredient of
modern BO. GP regression is used to build an approximation of the
objective function (via the mean of the GP) and an estimate of
uncertainty in the predicted value of this objective (via the variance
of the GP), and both are refined over the course of the optimizaton
iterations. A design-type question then arises in choosing where
(i.e., at what points in the input domain $\mathcal{X}$) to evaluate
the objective. In BO, this question is typically resolved by defining
a real-valued, easy-to-evaluate `acquisition function' over $\mathcal{X}$ and finding its maxima. A point at which the acquisition function is maximized is then taken to be the next evaluation point for the objective. The realized value of the objective at each iteration affects the acquisition function at the next stage, and hence the design process is sequential and adaptive. The acquisition process is generally formulated in a myopic way, though there are a few exceptions \citep{lam2017lookahead,wu2019practical}.

Many acquisition functions have been proposed in the literature,
beginning with \citet{jones1998efficient} and in the decades since;
these functions generally balance a notion of `exploration' (learning
the objective in unseen places, where the GP model has large
predictive variance) with `exploitation' (evaluating the objective at
points where it is expected to be larger than the current best value).
A prominent choice is the `expected improvement' function and its many
variants \citep{movckus1975bayesian,zhan2020expected}. Yet many other choices are
possible, with batch/parallel
\citep{chevalier2013fast,wang2020parallel} and even multi-fidelity
\citep{song2019general} schemes. BO is a large and active field which
we cannot hope to survey here; instead we point the reader to a few
recent reviews and tutorials \citep{shahriari2015taking,
  frazier2018tutorial,wang2023recent}. To set it in context relative
to OED, however, we emphasize that BO and OED are essentially
different: the goal of BO is to \emph{maximize} a function, not to predict the value of the function over all of $\mathcal{X}$ or even some \textit{a priori} chosen subset of $\mathcal{X}$. The resulting sets of evaluation points thus differ, both in their configuration and in their purpose, from the GP regression designs discussed in Section~\ref{sss:GPR}.

\paragraph{Data summarization.} Reducing or somehow `summarizing' large data sets is a problem of frequent practical interest, motivated by considerations of both storage and computation. The cost of most Bayesian inference algorithms, for example, scales at least linearly with the size of the data. Random subsampling of data is of course an option, but a more effective approach is to identify a small weighted subset of the data that is somehow representative of the full dataset. This is the notion of a \textit{coreset}, which originated in computational geometry and computer science \citep{Agarwal2005,Feldman2011}. It has since been formulated in a statistical setting, and notably the Bayesian setting \citep{huggins2016coresets,campbell2018bayesian,campbell2019automated,Campbell2019}, where the idea is to find weighted data subset of given size that least changes the likelihoood or the posterior distribution from its original full-data version. Ostensibly this problem seems similar to OED, but a crucial difference is that coresets are generally identified \textit{after} the data $Y$ are realized, and depend on the realized values of data. In optimal design, on the other hand, a design must be chosen \textit{before} $Y|\design$ are observed.

\section{Numerical approximation of design criteria}
\label{s:estimation}
Now we turn to one of the central computational questions of optimal design: how to approximate, numerically, the value of a chosen design criterion at any candidate design? This issue is not particularly vexing for standard alphabetic optimality criteria and linear models, where closed-form expressions are generally available. Evaluating these expressions can become costly when parameters $\Param$ are high- or infinite-dimensional, however, and we discuss dimension reduction methods relevant to this setting in Section~\ref{ss:dimred}.
The information-theoretic design criteria introduced in
Section~\ref{ss:decision_theoretic}, on the other hand---which have
become a mainstay of modern OED due to their flexibility and their applicability to complex nonlinear models---can be very challenging to evaluate, even when the parameter dimension is low. This section will discuss a variety of computational approaches for approximating such objectives, resting on nested Monte Carlo estimation (Section~\ref{ss:nmc}), approximation of the relevant densities within tractable families (Section~\ref{ss:densities}), and more general methods for constructing variational bounds (Section~\ref{ss:varbounds}).
Some of these methods differ with regard to which aspects of the underlying Bayesian model are assumed to be computationally accessible---i.e., is the problem in the `standard' setting where the likelihood function can be evaluated, or is it in the `implicit model' setting where likelihood evaluations and possibly prior density evaluations are unavailable? Throughout this section, we will comment on the applicability of the methods being discussed to either setting.

\subsection{Nested Monte Carlo estimators}
\label{ss:nmc}

For a generic expected utility design criterion, as formulated in
\eqref{e:EU}, a standard Monte Carlo estimator of the expectation
employs pairs of samples $(y^{(i)},\param^{(i)})$ drawn from the joint
prior of parameters and observations given the design $\design$,
$\pdf(y, \param \vert \design)$:
\begin{align}
  U(\design) = \mathbb{E}_{Y, \Param|\design}[u(\design,Y,\Param)]
  \approx \frac{1}{N}\sum_{i=1}^{N} u(\design, y^{(i)},
  \param^{(i)}).
  \label{e:general_EU_est}
\end{align}
These samples are typically obtained by first drawing
$\param^{(i)}\sim \pdf(\param)$ from the prior and then drawing
$y^{(i)}\sim \pdf(y|\param^{(i)},\design)$ from the conditional
density of the observations. The utility function $u$ must then be
evaluated at the samples generated:
$u(\design, y^{(i)}, \param^{(i)})$. This may not be easy to do. For
example, evaluating either of the utility functions $u^{\text{score}}$
\eqref{e:uscore} or $u^{\text{div}}$ \eqref{e:udivergence}, which
render $U$ equal to the expected information gain (EIG) in $\Param$,
requires evaluating a \textit{normalized} posterior density, where the
normalizing constant may change for each realization of $y$ and each
value of $\design$. One way forward is to estimate these normalizing
constants by \textit{another} Monte Carlo simulation, which gives
rise to \textit{nested Monte Carlo} (NMC) estimators.

As a canonical/representative case, we describe NMC approaches to estimating the EIG in parameters $\Param$, from prior to
posterior---i.e., the $U_{\text{KL}}$ defined in \eqref{e:EKL} and the
expressions thereafter. A widely used estimator, proposed in
\citet{ryan2003estimating}, employs the form of $U_{\text{KL}}$ given in
\eqref{e:EKLmarg}: we simply replace the posterior normalizing
constant $\pdf( y \vert \design )$ with a Monte Carlo estimate, as follows,
\begin{align}
  U_{\text{KL}}(\design) &=   \iint \pdf(y, \param \vert \design) \log \frac{\pdf(y \vert  \param, \design)}{\pdf( y \vert \design )} \,\mathrm{d} \param \, \mathrm{d}y \nonumber\\ 
  &=   \iint \pdf(y, \param \vert \design) \log \frac{\pdf(y \vert
    \param, \design)}{ \int \pdf( y \vert \tilde{\param}, \design )
    \revnew{\pdf(\tilde{\param}) } \, \revarx{\text{d}} \tilde{\param} } \,\mathrm{d} \param \, \mathrm{d}y  \nonumber \\
  &\approx \frac{1}{N}\sum_{i=1}^{N} \left( \log\pdf(y^{(i)} \vert
  \param^{(i)}, \design) - \log \left[\frac{1}{M} \sum_{j=1}^{M} \pdf(
    y^{(i)} \vert \tilde{\param}^{(i,j)}, \design )\right]\right) 
   \eqqcolon \widehat{U}_{\text{KL}}^{N,M}(\design). \label{e:DNMC}
\end{align}
Here the `outer loop' sample pairs $\{(y^{(i)},\param^{(i)})\}_{i=1}^N$
are drawn from $\pdf(y,\param\vert\design)$ as before, but we also
require an \emph{independent} collection of samples from the prior,
$$\{ \tilde{\param}^{(i,j)}\}_{i=1, j=1}^{i=N, j=M} \sim \pdf(\param),$$
amounting to $M$ independent samples for each outer-loop
iteration. Overall, evaluating $\widehat{U}_{\text{KL}}^{N,M}$
requires (i) the ability to sample from the prior, (ii) the ability to
sample from the statistical model for $Y \vert \design$, and (iii) the
ability to evaluate the likelihood function.

Properties of the estimator $\widehat{U}_{\text{KL}}^{N,M}$ have been
analyzed in \citet{ryan2003estimating,Beck_2018}, and \citet{rainforth2018nesting}. It is biased at
finite $M$, but asymptotically unbiased and consistent. More
precisely, the leading order terms of its bias and variance are given by
\begin{align}
  \mathbb{E}\left[\widehat{U}_{\text{KL}}^{N,M}(\design)\right] -  U_{\text{KL}}(\design) & =
          \frac{C_1(\design)}{M}  + \mathcal{O}\left ( \frac{1}{M^2} \right ),  \label{e:NMCbias} \\
  \Var \left [\widehat{U}_{\text{KL}}^{N,M}(\design) \right ] &=
                                                                \frac{C_2(\design)}{N} +  \frac{C_3(\design)}{NM} + \mathcal{O}\left ( \frac{1}{NM^2} \right ),
\label{e:NMCvar}
\end{align}
where $C_1, C_2, C_3$ are design-dependent constants. The constant $C_1$ is always positive (see \citealt[Proposition~1]{Beck_2018}), and thus the NMC estimator $\widehat{U}_{\text{KL}}^{N,M}$ is, to leading order, \textit{positively} biased. Intuitively, bias arises because the Monte Carlo estimator of $\pdf( y \vert \design )$ (which is unbiased) is transformed by a nonlinear function.

It is useful then to ask how to optimally allocate the sample sizes $M$ and $N$ to minimize the mean-square error $$\text{MSE} = \mathbb{E}  \big[  ( \widehat{U}_{\text{KL}}^{N,M} - U_{\text{KL}} )^2 \big]$$ for any given budget. The total number of samples drawn is $W = (M+1)N$; similarly, if cost lies in evaluating the statistical model $\pdf(y \vert \param, \design)$ for any new value of $\param$, then computing $\widehat{U}_{\text{KL}}^{N,M}$ incurs $W = (M+1)N$ evaluations. Letting $\alpha^2 = M/N$ denote the ratio of inner-to-outer loop sample sizes, one can show that the optimal value of this ratio for any given $W$ scales as $\alpha_\ast^2  = \mathcal{O}(W^{-1/3})$ \citep{Beck_2018,Feng_2019, rainforth2018nesting}. In other words, the ratio should decrease slowly as the computational budget increases. This scaling translates to setting  $M = \mathcal{O}( \sqrt{N})$, and an optimal convergence rate of $\text{MSE} = \mathcal{O}(W^{-2/3})$. Significantly, this is \emph{slower} than the standard Monte Carlo rate! Put another way, the computational effort required to achieve an $\text{MSE}$ of $\epsilon^2$ is $\mathcal{O}(\epsilon^{-3})$, rather than $\mathcal{O}(\epsilon^{-2})$.

Many improvements upon the `vanilla' NMC estimator \eqref{e:DNMC} have been proposed. A straightforward idea is to use importance sampling to estimate the evidence $p(y \vert \design)$ in the inner loop, i.e., to sample from some possibly $y$-dependent biasing distribution rather than the prior $\pdf_\Param$. To make this explicit, we write only the outer loop of \eqref{e:DNMC} and replace the log-evidence term with a plugin estimate,
\begin{align}
  U_{\text{KL}}(\design)  & \approx \widehat{U}_{\text{KL}}^{N,\text{is}}(\design) \coloneqq \frac{1}{N}\sum_{i=1}^{N}  \revnew {
                            \left (  \log\pdf(y^{(i)} \vert \param^{(i)}, \design) - \log \widehat{\pdf}( y^{(i)} \vert \design ) \right )}, \label{e:NMCis} 
\end{align}
where 
\begin{align}
  \widehat{\pdf}( y^{(i)} \vert \design) = \frac{1}{M} \sum_{j=1}^M \pdf(y^{(i)} \vert \tilde{\param}^{(i,j)}, \design ) w^{(i,j)}, \quad
  \tilde{\param}^{(i,j)} \stackrel{\text{iid}}{\sim}  q^{i,\design}, \quad  w^{(i,j)} = \frac{p(\tilde{\param}^{(i,j)})}{q^{i,\design}(\tilde{\param}^{(i,j)})}.
  \label{e:innerbiasing}
\end{align}
Here the superscripts on the density $q^{i,\design}$ of the biasing distribution emphasize that it can depend on the outer-loop index $i$, and thus on the point $y^{(i)}$ where the evidence is being evaluated. The biasing distribution generally depends on the design $\design$ as well. The question then becomes to how to choose this biasing distribution, for each summand of the outer loop.

One natural choice, proposed in \citet{Beck_2018}, is to use a Laplace approximation of the posterior distribution $\pdf(\param \vert y^{(i)}, \design)$ associated with each outer-loop sample. The Laplace approximation seeks the point of highest posterior density, $\param_{\text{map}}^i \revnew{\equiv \param_{\text{map}}(y^{(i)}, \design) } \in \argmax_\param \log \pdf(\param \vert y^{(i)}, \design)$, and builds a Gaussian approximation centered at this point, with covariance equal to the negative Hessian of $\log \pdf(\param \vert y^{(i)}, \design)$ \revnew{evaluated} at $\param_{\text{map}}^i$. \revnew{In general, we define}
\revnew{
  \begin{align*}
\Sigma_{\text{map}}(y, \design) =  \bigl ( \left . \nabla_{\param}^2 \log \pdf \left ( \param \vert y, \design \right )\right \vert_{\param = \param_{\text{map}}(y, \design)} \bigr )^{-1}.
\end{align*}
    }
\revnew{The biasing distribution suggested by \citet{Beck_2018} is then}
\begin{align}
\revnew{  q^{i,\design}= \mathcal{N} \bigl (\param_{\text{map}}^i, \Sigma_{\text{map}}(y^{(i)}, \design) \bigr ).}
  \label{e:laplacebias}
\end{align}
This choice aims to focus samples of the biasing distribution onto regions of higher posterior density, thus reducing the variance of the evidence estimate $\widehat{\pdf}( y^{(i)} \vert \design)$ and hence both the bias and the variance of estimates of $U_{\text{KL}}$, as demonstrated in \citet{Beck_2018}. As is typically the case with importance sampling, choosing a better biasing distribution reduces the magnitudes of the constants $C_1, C_2, C_3$ in \eqref{e:NMCbias}--\eqref{e:NMCvar}, but does not change the rates of convergence of the estimator with $M, N$.

\revnew{Computing a Laplace approximation efficiently, which must be done here $N$ times (once for each outer-loop sample), generally requires gradients of the log-posterior density, $\nabla_\param \log \pdf(\param \vert y^{(i)}, \design)$, and a good approximation of the Hessian of this log-density \cite{BuiThanh2013,schillings2016scaling}. We also note that, in general, $\param_{\text{map}}^i \neq \theta^{(i)}$; that is, the posterior mode does not coincide with the data-generating value of the parameter. \citet{englezou2022approximate} propose a simplification of Laplace-based importance sampling that instead centers the biasing distribution at the data-generating value of the parameter, drawn from the outer loop. Specifically, both the mean of the Gaussian and the position of the Hessian evaluation in \eqref{e:laplacebias} are set to $\param^{(i)}$, rather than $\param_{\text{map}}^i$. Doing so avoids the cost of numerical optimization to find $\param_{\text{map}}^i$, and may yield only a modest increase in the MSE of the estimator at finite sample sizes \citep{englezou2022approximate}. An (approximate) Hessian of the log-posterior, however, is still required.
We should also emphasize that choosing \revnew{a Gaussian approximation} as a biasing distribution can become unstable when the posterior is strongly non-Gaussian, and particularly if the posterior has heavy tails. Here other choices of biasing---for instance, using the mean and covariance matrix of the Laplace approximation to parameterize a heavier-tailed multivariate-$t$ distribution---could be more robust \cite{mcbook}.}

Another approach to importance sampling, developed in \citet{Feng_2019}, is derivative-free. It is a multiple importance sampling scheme that proceeds iteratively over the outer-loop index $i$: at any given $i$, past inner-loop samples $\{ ( \tilde{\param}^{(k,j)}) _{j=1}^M \}_{k< i}$ and their associated likelihood evaluations are used to create a \emph{mixture} biasing distribution $q^{i, \design}$ tailored to estimating the current evidence $p(y^{(i)} \vert \design)$. To make this process efficient and maximize reuse of information, the outer-loop iterations are ordered from largest to smallest prior density of $\param^{(i)}$, $p(\param^{(i)})$. Again, this approach can substantially decrease the bias and variance of EIG estimates relative to a vanilla NMC scheme.

A different way of accelerating nested Monte Carlo involves multilevel
formulations \citep{Giles2015}. \citet{goda2020multilevel}
introduce a multilevel Monte Carlo estimator of $U_{\text{KL}}$, where
the level controls the number of inner-loop samples. They also develop
an antithetic coupling for the inner-loop estimates that reduces
variance, and show that the overall construction reduces the
computational complexity required to achieve an MSE of $\epsilon^2$ to
$\mathcal{O}(\epsilon^{-2})$, improving on the
$\mathcal{O}(\epsilon^{-3})$ complexity of the standard NMC
estimator. In other words, we recover the optimal Monte Carlo
rate. \citet{goda2020multilevel} also show how importance
sampling---for instance, the Laplace approximation-based importance
scheme discussed above---can be incorporated within their multilevel
Monte Carlo estimator to reduce constants of the error
terms. \citet{beck2020multilevel} also introduce a multilevel scheme
for EIG, varying not only the number of inner-loop samples but also
the discretization/approximation of some underlying partial
differential equation (PDE) model used to define the likelihood.

All of the estimators discussed so far are consistent, that is, they
converge in probability to the true EIG as the relevant sample sizes
(e.g., $M$ and $N$, or their analogues in a multilevel scheme) are
sent to infinity. Other \revnew{schemes} proposed in the literature,
with the goal of improving computational efficiency, are not
consistent. For instance, \citet{Long_2013} propose replacing the
utility $u^{\text{div}}(\design, y)$ \eqref{e:udivergence}, whose
expectation yields the EIG, with \revnew{an approximation of} the
\revnew{Kullback--Leibler (KL)} divergence from the prior to a {Laplace approximation} $\mathcal{N} \bigl ( \param_{\text{map}}(y, \design),  \Sigma_{\text{map}}(y, \design) \bigr )$ of the posterior $p(\param \vert y, \design)$. 
\revnew{Their construction introduces a series of additional approximations---replacing the true MAP estimate $\param_{\text{map}}(y, \design)$ with the data-generating value of the parameter $\param$, using the Gauss--Newton approximation of the Hessian at this point rather than at the MAP (the inverse of which yields a `covariance' matrix $\widetilde{\Sigma}(\param, \design)$ that does not depend on $y$), and taking further asymptotic approximations of the relevant integrals---to obtain an approximation of EIG that involves only integration over the prior,}
\begin{align}
\revnew{\text{EIG}(\design) \approx  \int \left (  -\frac12 \log \det \widetilde{\Sigma}(\param, \design)  - \frac{p}{2} \left ( \log 2 \pi + 1 \right ) - \log p(\param)  \right ) p(\param) \, \revarx{\text{d}} \param . }
  \label{e:Laplaceinfogain}
\end{align}
\revnew{As a design criterion, we note that this EIG approximation is similar in structure to \eqref{e:nonlinearheuristic}.} 
The error of this approximation \revnew{can be related} to the Gaussianity of the
posterior \emph{and} the number of independent repeated trials
$N_{\text{tr}}$ of the experiments specified by $\design$, and is bounded by $\mathcal{O}(1/N_{\text{tr}})$ in probability \citep{Long_2013}. Intuitively, repeated trials cause the posterior to concentrate around its mode, and this concentration controls the error of integral approximations leading to \eqref{e:Laplaceinfogain}. In practice, for small $N_{\text{tr}}$ and a non-Gaussian problem, this error can be large.
\revnew{An alternative and perhaps more straightforward way of using the Laplace approximation is proposed by \citet{overstall2018approach}; here the approach is simply to replace the estimate of the log-evidence term $\log p(y^{(i)} \vert \design)$ in \eqref{e:NMCis} with the log-evidence of the standard Laplace approximation of the posterior $p(\param \vert y^{(i)})$. The explicit likelihood term and the outer Monte Carlo sum in \eqref{e:NMCis} are unchanged. The bias of this EIG approximation increases from zero as the posterior departs from Gaussianity, but compared to the scheme in \citet{Long_2013}, it does not rely \textit{directly} on posterior concentration.} 

A different family of approximations follows by replacing computationally expensive aspects of statistical model $p(y \vert \param, \design)$ with a computationally cheaper `surrogate.' For instance, suppose that the data arise from a nonlinear forward \revnew{operator} $G_\design: \mathbb{R}^p \to \mathbb{R}^n$ perturbed with additive noise $\mathcal{E}$, i.e., $Y = G_\design(\param) + \mathcal{E}$. This setup corresponds to a discretization of the Bayesian inverse problems discussed in Section~\ref{s:BIPs}. Here, evaluating the function $G_\design$ often involves solving a set of PDEs or integral equations, and it is natural to replace this solution with a cheaper approximation $\widetilde{G}_\design \approx G_\design$, which in turn induces an approximation $\widetilde{p}(y \vert \param, \design)$ of the likelihood and an approximation $\widetilde{p}(\param \vert y, \design)$ of the posterior distribution via \eqref{eq:Bayes}. Inserting any of these approximations into \eqref{e:entropydiff} and equivalent expressions thus yields an {approximate} EIG.
By the same token, using evaluations of $\widetilde{p}(y \vert \param, \design)$ and samples $\tilde{y}^{(i)}$ drawn from this approximate model in any of the NMC estimators discussed above will yield consistent estimates of this \textit{approximate} EIG. 

\citet{huan2013simulation} proposed such a procedure, using polynomial approximations $\widetilde{G}_\design$ of $G_\design$ built via sparse quadrature. But a wide variety of other approximation schemes and formats are possible: direct function approximation, whether via polynomials, Gaussian processes, or neural networks \citep{herrmann2020deep}; but also reduced-order models \citep{benner2015survey} or even coarser numerical discretizations of the model giving rise to the parameter-to-observable map $G_\design$.
The impact of such approximations on the posterior distribution is by now reasonably well understood, especially when one can build a family of approximations $\widetilde{G}_\design^\ell$, indexed by $\ell$, that converge (in some appropriate sense) to $G_\design$ as $\ell \to \infty$ \citep{marzouk2009stochastic, Stuart_2010,stuart2018posterior,sprungk2020local}. The impact of such approximations on the EIG has only recently been analyzed, however. \citet{duong2023stability} show that the difference between the EIG and its approximation is controlled by the prior expectation of likelihood perturbations under the KL divergence.
As a consequence, in the setting of Gaussian likelihoods that we sketched here, \citet[Theorem~4.4]{duong2023stability} show that closeness of $\widetilde{G}_\design^\ell(\param)$ to $G_\design(\param)$ in a prior-weighted $L^2$ sense, uniformly over designs $\design \in \Xi$, guarantees uniform control over the error in the approximate EIG $U^\ell_{\text{KL}}(\design)$ and convergence of the maximizers of $U^\ell_{\text{KL}}$ as $\ell \to \infty$.

\begin{figure}
  \centering
  \includegraphics[width=0.75\linewidth]{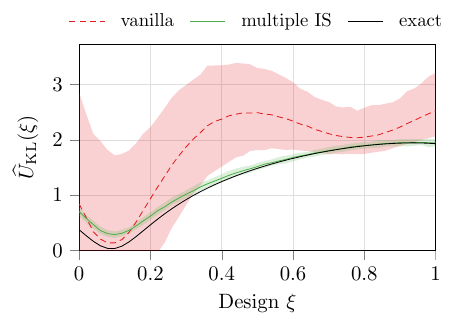}
  \caption{Estimated EIG as a function of a scalar design parameter
    $\xi$ for a linear-Gaussian model, using vanilla \revnew{NMC}
    (red) or an improved multiple importance sampling scheme (green),
    compared to the true EIG (black). Shaded areas represent the
    interval containing 95\% of 2000 independent estimates of EIG at
    each $\xi$; red dashed and solid green lines are the means of
    these estimates. Figure adapted from \citet{Feng_2019}.}
  \label{f:bias}
\end{figure}

We close this section with a caution. Even the consistent \revnew{NMC} estimators of $U_{\text{KL}}(\design)$ discussed so far are
biased at finite \revnew{inner loop} sample sizes. This bias can be significant but, more
importantly, will vary with $\xi$ in general. Figure~\ref{f:bias},
adapted from \citet{Feng_2019}, illustrates this phenomenon for a
four-dimensional linear-Gaussian problem (thus allowing comparison
with exact solutions). EIG in a subset of the model parameters
\eqref{e:EKLfocusedpost}--\eqref{e:EKLfocusedmarg} is estimated using
an adaptation of the vanilla NMC estimator
$\widehat{U}_{\text{KL}}^{N,M}$ \eqref{e:DNMC} (in red), and with the
multiple importance sampling scheme developed in \citet{Feng_2019} (in
green). Sample sizes are \emph{fixed} for all $\xi$, and the true EIG
is shown in black. We see that the bias of the vanilla estimator
actually obscures the location of the true maximum. The bias is
largest where posterior concentration is maximized, as this is where
the prior-weighted estimates of the evidence have greatest
variance. The multiple importance scheme fares better, but there is
generally no guarantee regarding stability of the maxima for any
finite sample size.
Care is thus needed to adjust the approximation of EIG in conjunction
with the optimization procedure. We will revisit this issue in
Section~\ref{s:optimization}.

\subsection{Mutual information bounds via density approximations}
\label{ss:densities}

As anticipated in the discussion following \eqref{e:EKLpost}--\eqref{e:EKLmarg}, one of the core computational tasks of the NMC estimators discussed in Section~\ref{ss:nmc} is to estimate the posterior normalizing constant $p(y \vert \design)$ across many different values of $y$. The vanilla NMC approach does this entirely independently for each value of $y$, as does the Laplace-based importance sampling method. (The adaptive importance sampling scheme of \citet{Feng_2019}, on the other hand, could be said to `borrow' information from other values of $y$ to create a local biasing distribution.) A rather different way of approaching this problem is to approximate the marginal density $p(y \vert \design)$, or similarly, the \textit{normalized} posterior density $p(\param \vert y, \design)$, directly, e.g., in some parametric family of densities.

Let us recall \eqref{e:EKLpost}--\eqref{e:EKLMI} in a more concise form:
\begin{align}
 \mathcal{I}(Y ; \Param \vert\design) & = \mathbb{E}_{Y, \Param \vert \design} \left [ \log \frac{p(Y \vert \Param, \design)}{p(Y \vert \design)} \right ] \label{e:Imarg} \\  & = \mathbb{E}_{Y, \Param \vert \design} \left [ \log \frac{p(\Param \vert Y, \design)}{p(\Param)} \right ]. \label{e:Ipost}
\end{align}
In either \eqref{e:Imarg} or \eqref{e:Ipost}, one could in principle
seek to approximate the density in the numerator, the density in the
denominator, or both. Suppose that we replace $p(y \vert \design)$ in
\eqref{e:Imarg} with some approximating probability density function
$q^{\text{mar}}(y \vert \design)$ (where the superscript stands for
`marginal'). Then, as noted in
\citet{barber2004algorithm,poole2019variational}, and \citet{Foster_2019},
\begin{align}
  \mathcal{I}(Y ; \Param \vert\design)
  & = \mathbb{E}_{Y, \Param \vert \design} \left [ \log \frac{p(Y \vert \Param, \design) \, q^{\text{mar}}(Y \vert \design)}{q^{\text{mar}}(Y \vert \design) \, p(Y \vert \design)} \right ]  \nonumber \\
  & = \mathbb{E}_{Y, \Param \vert \design} \left [ \log \frac{p(Y \vert \Param, \design)}{q^{\text{mar}}(Y \vert \design)} \right ] - \DKL( p_{Y \vert \design} \vert \vert q^{\text{mar}}_{Y \vert \design} ) \nonumber \\
  & \leq \mathbb{E}_{Y, \Param \vert \design} \left [ \log \frac{p(Y \vert \Param, \design)} {q^{\text{mar}}(Y \vert \design)} \right ] = \mathbb{E}_{\Param} \left [ \DKL( p_{Y \vert \Param, \design} \vert \vert q^{\text{mar}}_{Y \vert \design} ) \right ] \label{e:upperbound},
\end{align}
where the inequality follows from the non-negativity of the KL divergence. Hence, for \textit{any} approximation $q^{\text{mar}}(y \vert \design)$ of the marginal density of $Y \vert \design$, \eqref{e:upperbound}  is an {upper bound} on the mutual information (EIG).

Similarly, if we replace $p(\param \vert y, \design)$ in \eqref{e:Ipost} with some approximating probability density function $q^{\text{pos}}(\param \vert y, \design)$ (where the superscript denotes `posterior'), we obtain
\begin{align}
  \mathcal{I}(Y ; \Param \vert\design)
  & = \mathbb{E}_{Y, \Param \vert \design} \left [ \log \frac{ q^{\text{pos}}(\Param \vert Y, \design) \, p(\Param \vert Y, \design)}{ p(\Param) \, q^{\text{pos}}(\Param \vert Y, \design) } \right ] \nonumber \\
  &= \mathbb{E}_{Y, \Param \vert \design} \left [ \log \frac{q^{\text{pos}}(\Param \vert Y, \design) }{ p(\Param) } \right ] + \mathbb{E}_{Y \vert \design} \left [ \DKL( p_{\Param \vert Y, \design} \vert \vert q^{\text{pos}}_{\Param \vert Y,\design} )\right ]\nonumber \\
  & \geq    \mathbb{E}_{Y, \Param \vert \design} \left [ \log \frac{q^{\text{pos}}(\Param \vert Y, \design) }{ p(\Param) } \right ] = \mathbb{E}_{Y, \Param \vert \design}[ \log q^{\text{pos}}(\Param \vert Y, \design) ] + H(\Param)          .\label{e:BAlower}
\end{align}
Hence, for any approximation $q^{\text{pos}}(\param \vert y, \design)$ of the posterior density, \eqref{e:BAlower} is a \textit{lower bound} on the mutual information (EIG). This bound is sometime called the Barber--Agakov bound, after \citet{barber2004algorithm}. Note that all expectations above, and specifically in \eqref{e:upperbound} and \eqref{e:BAlower}, are with respect to the \textit{true} distribution $p_{Y, \Param \vert \design}$.

Evaluating the upper bound \eqref{e:upperbound} requires the ability to evaluate the likelihood function, and thus it does not apply to the implicit model setting where the likelihood is intractable. Evaluating the lower bound, on the other hand, only requires access to the prior density (or the differential entropy of the prior) and the ability to find a tractable approximation $q^{\text{pos}}(\param \vert y, \design)$.

\citet{Foster_2019} were the first to suggest using these mutual
information bounds in OED, and in practice selected the approximations
$q$ from simple parametric families of densities (e.g., Gaussian,
uniform) that were tailored to the design problem at hand.  Once such a
family $\mathcal{Q}$ is specified, the best member of the
family---i.e., the density yielding the closest approximation of the EIG---can
be found by tightening the bound. Specifically, for the marginal approximation, we seek (for any given design $\design$)
\begin{align}
  q^{\text{mar},\ast} \in \argmax_{q \in \mathcal{Q}} \mathbb{E}_{Y \vert \design} \left [ \log q(Y \vert \design) \right ] ,
  \label{e:optimizemarg}
\end{align}
which minimizes the upper bound \eqref{e:upperbound}, while for the posterior approximation we seek
\begin{align}
  q^{\text{pos},\ast} \in \argmax_{q \in \mathcal{Q}} \mathbb{E}_{Y, \Param \vert \design} \left [ \log q(\Param \vert Y, \design) \right ],
  \label{e:optimizepost}
\end{align}
which maximizes the lower bound \eqref{e:BAlower}. In practice, the expectations in \eqref{e:optimizemarg} or \eqref{e:optimizepost} are approximated using samples from the model; for example, \eqref{e:optimizemarg} becomes
\begin{align}
  \hat{q}^{\text{mar}} \in \argmax_{q \in \mathcal{Q}} \sum_{i=1}^M \log q(y^{(i)} \vert \design)  , \quad y^{(i)} \sim \pdf(y \vert \design),
  \label{e:MLEmarg}
\end{align}
and analogously for \eqref{e:optimizepost}. Inspecting
\eqref{e:MLEmarg}, it is apparent that identifying a member of the
variational family in this way is none other than \emph{maximum
  likelihood estimation} of either the marginal  density $p(y\vert \design)$ or conditional  density $p(\param \vert y, \design)$.\footnote{`Maximum likelihood' here refers to the density estimation problem immediately at hand, i.e., estimating  the best $q \in \mathcal{Q}$ given samples, and should \revnew{not} be confused with \revnew{the idea of estimating $\param$ by maximizing} $\param \mapsto \pdf(y \vert \param, \design)$.}

With this link in mind, we can immediately generalize the machinery
used to construct good density approximations. One powerful approach
for estimating both joint and conditional densities rests on
transportation of measure
\citep{villani2009optimal,Marzouk_2016,spantini2018inference}. Given
some generic target distribution $\pi$ on $\mathbb{R}^d$, the idea
behind transport methods is to find an invertible transformation $S:
\mathbb{R}^d \to \mathbb{R}^d$ that pushes forward $\pi$ to a simple
reference distribution $\rho$ on $\mathbb{R}^d$ whose density can be easily evaluated. Abusing notation (by not distinguishing measures from densities), we can write this as $S_\sharp \pi = \rho$, which means that $\rho(A) = \pi \left ( S^{-1}(A) \right )$ for any $\rho$-measurable set $A$.
Crucially, for any diffeomorphism $\tilde{S}$ on $\mathbb{R}^d$, the distribution $\tilde{S}^\sharp \rho \coloneqq \tilde{S}^{-1}_\sharp \rho $, called the \textit{pullback} of $\rho$ under $\tilde{S}$, has a closed-form expression for its density:
\begin{align}
\tilde{S}^\sharp \rho = (\rho \circ \tilde{S}) \, \det \nabla \tilde{S},
\end{align}
which is guaranteed to be positive and to integrate to one. Density estimation can thus be recast as the problem of finding a map $\tilde{S}$ in some suitable class such that $\tilde{S}^\sharp \rho$ is `close' to $\pi$ \citep{wang2022minimax}.

In practice, these models can be quite expressive. Any measure absolutely continuous with respect to the Lebesgue measure on $\mathbb{R}^d$ can be represented as the pullback, under some map, of a Gaussian $\rho$ on $\mathbb{R}^d$; in fact, for any pair of equivalent continuous measures $(\rho, \pi)$, there exist infinitely many transport maps $S$ that achieve $S^\sharp \rho = \pi$.
Normalizing flows \citep{kobyzev2020normalizing,papamakarios2021normalizing} are a special case of this formulation, i.e., a particular class of transport map parameterizations that guarantee invertibility, differentiability, and easy evaluation of the Jacobian determinant $\det \nabla S$. 
But many other representations are useful. Monotone triangular maps
\citep{bogachev2005triangular}, for instance, can represent arbitrary
absolutely continuous distributions and be parameterized in a way that
endows the maximum likelihood estimation problems
\eqref{e:optimizemarg}--\eqref{e:optimizepost} with optimization
guarantees; see \citet{baptista2023representation} for
details. Continuous optimal transport maps can also be estimated by first solving a discrete optimal transport problem (i.e., between empirical measures) and smoothing the result \citep{manole2021plugin,pooladian2021entropic} or by parameterizing a differentiable convex potential \citep{huang2020convex}.

In the context of OED, it is useful to employ
\revnew{\emph{block-triangular} maps} \citep{baptista2020conditional}
\revnew{(a class which includes strictly triangular maps but infinitely many other choices)}, as they naturally capture conditional densities.
Consider the following block arrangement for a map $S:\mathbb{R}^{n + p} \to \mathbb{R}^{n+p}$:
\begin{align}
\label{e:blockmap}
    S(y, \param) =  \begin{bmatrix*}[l]
        S^Y(y) \\
        S^\Param(y, \param)
    \end{bmatrix*},
\end{align}
where $S^Y: \mathbb{R}^n \to \mathbb{R}^n$ and $S^\Param:
\mathbb{R}^{n+p} \to \mathbb{R}^p$. \revnew{Let the reference
  distribution $\rho$ on $\mathbb{R}^{n+p}$ factor as $\rho = \rho_n
  \otimes \rho_p$, where the subscripts denote the dimension of the
  factors. (The standard Gaussian distribution on $\mathbb{R}^{n+p}$
  naturally factorizes in this way, but any distribution with
  appropriate block-independence would suffice.) Then,} as shown in
\citet{baptista2020conditional} and \citet{Marzouk_2016}, if $S_\sharp \pdf_{Y, \Param} = \revnew{\rho}$, then $S^Y_\sharp \pdf_Y = \rho_n$ and \revnewnew{$\bigl (\param \mapsto S^\Param(y, \param) \bigr )_\sharp p_{\Param \vert Y = y} = \rho_p$} for any $y \in \supp p_Y$. Thus the two component functions of the map capture the evidence ($Y$-marginal) and the posterior, respectively. Note also that \eqref{e:blockmap} can easily be extended to depend on the design, that is,
\begin{align}
    S(y, \param; \design) =  \begin{bmatrix*}[l]
        S^Y(\design, y) \\
        S^\Param(\design, y, \param)
      \end{bmatrix*}.
  \label{e:blockmapdesign}
\end{align}

Suppose now that we obtain a maximum likelihood estimate $\widehat{S}$ of \revnew{a map of the form \eqref{e:blockmapdesign}}, given samples from $p(y, \param \vert \design)$ and some tractable class of candidate block-triangular maps $\mathcal{S}$, \revnew{for instance} as described in \citet{baptista2023representation}:
\revnew{\begin{align}
\widehat{S}( \, \cdot \, \, ; \design) \in \arg \max_{S \in \mathcal{S}} \sum_{i=1}^M \log S^\sharp \rho (y^{(i)}, \param^{(i)}).
\label{e:mapMLE}
\end{align}}
The map $\widehat{S}$ yields a plug-in estimate of the desired densities, \revnew{via its two component functions:}
\begin{align*}
  \hat{q}^{\text{mar}}(y \vert \design) & = \rho_n\bigl (\widehat{S}^Y(\design, y) \bigr ) \det \nabla_y \widehat{S}^Y(\design, y), \\
  \hat{q}^{\text{pos}}(\param \vert y, \design) & = \rho_p \bigl (\widehat{S}^\Param(\design, y, \param) \bigr ) \det \nabla_\param \widehat{S}^\Param(\design, y, \param).
\end{align*}
(For a statistical convergence analysis of transport-based density \revnew{estimators, in a general} nonparametric setting, see \citealt{wang2022minimax}.)
In fact, since \revnew{the reference $\rho$} is a product distribution, \revnew{\eqref{e:mapMLE} splits into two separate optimization problems such that} the component functions $S^Y$ and $S^\Param$ can be estimated separately, and hence only the component needed for the desired variational bound needs to be learned. The resulting density estimates $\hat{q}^{\text{mar}}$ or $\hat{q}^{\text{pos}}$ can then be substituted into \eqref{e:upperbound} or \eqref{e:BAlower}, respectively. The transport approach effectively defines the approximating class of densities $\mathcal{Q}$ in \eqref{e:optimizemarg} or \eqref{e:optimizepost} as the set of all densities that can be expressed as $S^\sharp \rho$ for $S \in \mathcal{S}$.

\revnew{Several recent instantiations of this transport approach in
  OED have parameterized the maps as normalizing flows; see
  \citet{kennamer2023design,Chen2024}, and \citet{Dong2024}. In particular, the lower component function $S^\Param$ can be represented as a \emph{conditional} normalizing flow, which is essentially a structured invertible function of $\param$ that is parameterized by $y$. Another canonical choice of $S^\Param$ is a conditional Brenier map \citep{carlier2016}, which can be understood as a family of $L^2$-optimal transport maps from $p_{\Param | Y = y}$ to $\rho_p$, parameterized by $y$. The essential requirements are to respect the overall block structure of \eqref{e:blockmap} in an invertible, differentiable map that pushes forward $p_{Y, \Param}$ to a block-independent reference distribution.}

\begin{figure}
  \centering
  \includegraphics[width=0.8\linewidth]{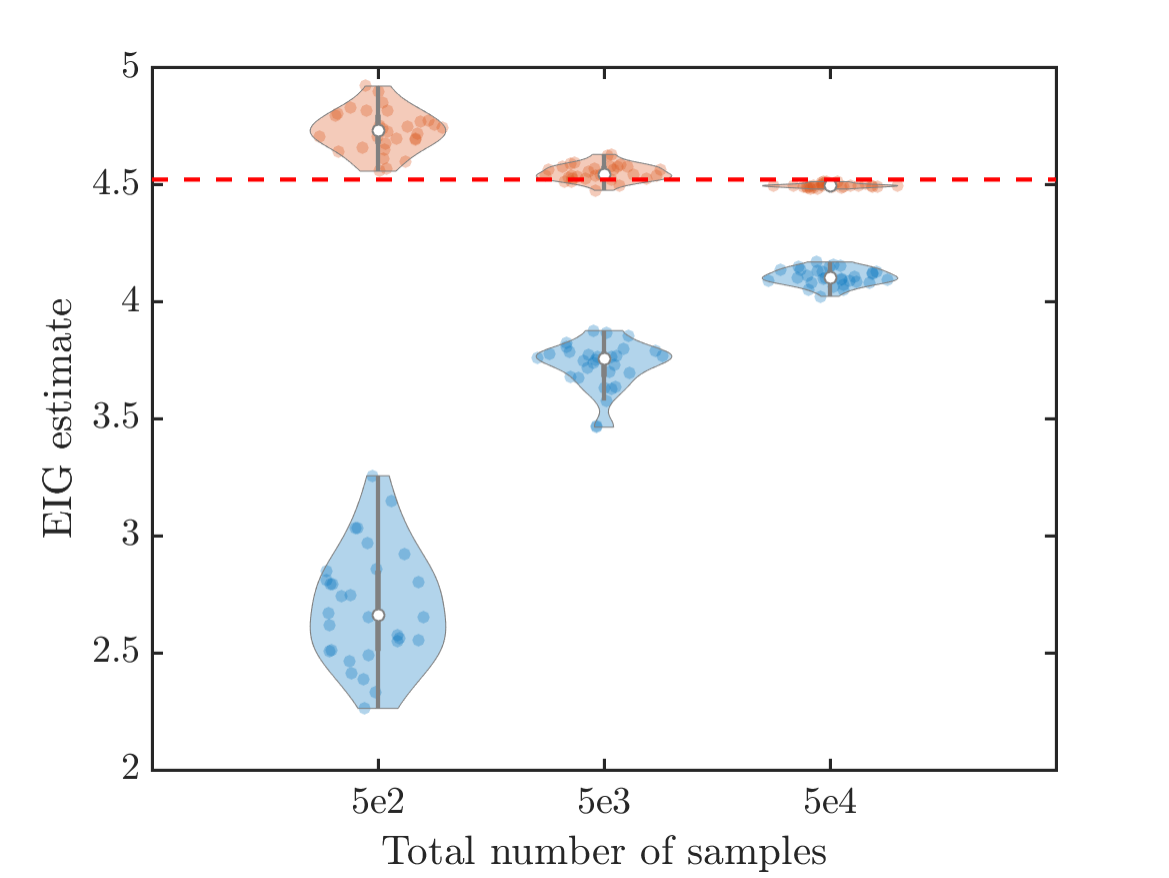}
  \caption{Variational upper (orange) and lower (blue) bounds on the
    \revnew{EIG} in a nonlinear design problem, compared to a biased
    estimate obtained via \revnew{NMC} (dashed red line). See the
    discussion in Section~\ref{ss:densities}. Figure adapted from \citet{fengyi2024forthcoming}.}
  \label{fig:densitysandwich}
\end{figure}

Figure~\ref{fig:densitysandwich}, adapted from \citet{fengyi2024forthcoming}, shows an application of transport-based density estimators to the nonlinear M\"ossbauer spectroscopy example described in \citet[Section~4.2]{Feng_2019}, for a fixed design $\xi$. The orange violin plots and circles show repeated independent estimates of the upper bound \eqref{e:upperbound}, while the blue violin plots and circles illustrate repeated independent estimates of the lower bound \eqref{e:BAlower}. These estimates are produced by learning the appropriate transport map component, $S^Y$ or $S^\Param$, via the adaptive semi-parametric procedure described in \citet{baptista2023representation}, which naturally enlarges the family of maps being considered as the sample size available for estimation increases. \revnew{Here the maps $S$ are strictly triangular, and hence approximations of the Knothe--Rosenblatt rearrangement; see  \citet{rosenblatt1952remarks,knothe1957contributions} and \citet[Section 2.3]{santambrogio2015optimal}.} The horizontal axis shows the total number of independent samples drawn from the model, comprising two batches: one to estimate the map and a separate batch to estimate the outer expectations $\mathbb{E}_{Y, \Param \vert \design}$ in \eqref{e:upperbound} and \eqref{e:BAlower}. (Asymptotically optimal allocations of this sample budget are discussed in \citet{fengyi2024forthcoming}.) The upper and lower bounds `sandwich' the true EIG as the sample size increases.

It is useful to understand the source and nature of the randomness in
these results. Each estimate of a transport map yields a density
estimate $\hat{q}$,
and plugging \textit{any} such estimate into \eqref{e:upperbound} and \eqref{e:BAlower} yields a guaranteed upper or lower bound, as appropriate. Approximating the expectations in \eqref{e:upperbound} and \eqref{e:BAlower} with samples, however, yields an unbiased \textit{estimate} of the bound. There is no guarantee that each realization of the estimator will be above or below the true EIG, though the figure suggests that no realizations cross this threshold at the sample sizes considered here. Overall, however, fluctuation in the estimates reflects randomness in both $\hat{q}$ and in  the Monte Carlo approximation of the outer expectation. The red dashed line in Figure~\ref{fig:densitysandwich} is an NMC estimate of the EIG obtained with $1.58 \times 10^6$ samples, which is two to four orders of magnitude greater than the number of samples used in the variational estimators. The NMC estimate appears to be positively biased, even at this large sample size.

As noted earlier, estimating the posterior density to construct a
lower bound for EIG is well suited to implicit models, since \revnew{\eqref{e:BAlower} does
not require the ability to evaluate} $p(y \vert \param, \design)$. But the flexibility of using such density estimates also applies to other information-based design criteria discussed in Section~\ref{sss:otherutilities}. For instance, the expected marginal information gain objective in \eqref{e:EKLfocusedpost} can be bounded from below by estimating $p(\param_1 \vert y , \design)$, i.e., by replacing this marginal posterior density with some $q(\param_1 \vert y, \design)$. \citet{baptista2022bayesian} uses this approach to assess the information value of different observables in a phase-field model (a nonlinear PDE), where high-dimensional random initial conditions take the role of the `nuisance parameters' $\Param_2$. If the prior marginal density $p(\param_1)$ is not available in closed form, it can also be estimated from samples, with the caveat that if $p(\param_1 \vert y , \design)$ and $p(\param_1)$ are \textit{both} replaced with approximations, the resulting EIG approximation  will be neither an upper nor a lower bound for the true marginal EIG in general \citep{Foster_2019}.

In the case of goal-oriented OED \eqref{e:goalpost}--\eqref{e:goalmarg}, the need for learning both the numerator and denominator in the associated density ratios is more acute: for a generic $\Psi$, none of the densities in \eqref{e:goalpost} or \eqref{e:goalmarg} may be readily available. But as long as we can simulate from $p(z, y \vert \design)$---which can be done by drawing $\theta^{(i)} \sim p(\theta)$, drawing $y^{(i)} \sim p(y \vert \theta^{(i)}, \design)$, and evaluating $z^{(i)} = \Psi(\theta^{(i)})$---the density estimation techniques discussed above will apply. If we choose to use transport in this setting, then two block-triangular maps will be relevant:
\begin{align}
    S_1(y, z; \design) =  \begin{bmatrix*}[l]
        S_1^{Y}(\design, y) \\[6pt]
        S_1^{Z}(\design, y, z)
      \end{bmatrix*} \quad  \text{and} \quad
      S_2(z, y; \design) =  \begin{bmatrix*}[l]
        S_2^{Z}(\design, z) \\[6pt]
        S_2^{Y}(\design, z, y)
      \end{bmatrix*}.
\end{align}
Approximating the densities in \eqref{e:goalpost} would use the lower component of $S_1$ and the upper component of $S_2$; conversely, approximating the densities in \eqref{e:goalmarg} would use the upper component of $S_1$ and the lower component of $S_2$.

In closing, we note that transport approximations of the densities relevant to an EIG calculation can be built by means other than maximum likelihood estimation. For instance, \citet{koval2024tractable} first prescribe some density over the continuous real-valued design variables $\xi$, and then build triangular transport maps directly from unnormalized evaluations of the joint density $p(y, \param \vert \design) p(\design)$, using a functional tensor-train format \citep{cui2023scalable}. This construction method does not correspond to tightening a variational bound, in contrast to maximizing the likelihood, but the error in EIG due to the map approximation can nonetheless be analyzed.

Finally we note that another use for the variational approximations discussed in this subsection---specifically, the approximation $q^{\text{pos}}(\param \vert y, \design)$---is as a biasing distribution in NMC estimation of EIG. This idea, proposed in \citet{Foster_2019}, fits directly into \eqref{e:innerbiasing} by setting $q^{i,\design}(\param) = q^{\text{pos}}(\param \vert y^{(i)}, \design)$ and recovers the consistency guarantees of NMC.

\subsection{More general variational bounds for mutual information}
\label{ss:varbounds}

So far, we have discussed bounds on mutual information that follow from parameterizing a density---or a transport map from which a density estimate can be derived. These bounds are variational in the sense that they can be tightened by solving an optimization problem over some class of densities or transport maps. For the bounds discussed so far, if the class of densities or transports is sufficiently rich, in principle there is a member of the class that corresponds to the true density and the bound will attain the true mutual information.

But there are many other variational bounds for mutual information, based on different approximation formats and more general classes of functions (i.e., not just densities or invertible transport maps). Such bounds are in fact an important topic in information theory, with many applications in machine learning. And several have found recent use in OED. \citet{poole2019variational} provide a unifying framework for understanding many variational bounds, which inspires some of the discussion below.

To provide some context, we first recall a classical nonparametric
estimator of mutual information (which is not variational).
The $k$-nearest-neighbor
estimator of \citet{Kraskov2004} (KSG) takes joint sample pairs
$(y^{(i)},\param^{(i)}) \sim \pdf_{Y, \Param}$ as input, and uses the
statistics of nearest-neighbor distances in both the joint space and
in the marginal directions to construct an estimator. Its construction
is related to $k$-nearest-neighbor estimators of the Shannon entropy
\citep{Kozachenko1987}. The KSG estimator has been used for OED
\citep{Terejanu2012}, but is known to scale poorly with
dimension. Rigorous statistical analyses of this estimator, e.g.,
proofs of consistency for fixed $k$ and bounds on the rate of
convergence of the MSE with sample size $N$, have appeared relatively
recently \citep{Gao2018}. For instance,
\citet[Corollary~2]{Gao2018}
establish an upper bound on the MSE
that is, up to polylogarithmic factors,
$\mathcal{O}(N^{-2/(n+p)})$, where $n$ is the dimension of $Y$ and $p$ is the dimension of $X$. They suggest that this convergence rate cannot be refined even if the densities of interest are assumed to be  H\"older-smooth.

Nonparametric estimators of the entropy (rather than the mutual
information) have been used to construct lower bounds for mutual
information in \citet{ao2020approximate}. Here the idea is to write $\mathcal{I}(Y ; \Param \vert \design) = H(Y \vert \design) - H(Y \vert \Param, \design)$ and then to seek an upper bound for the conditional entropy $H(Y \vert \Param, \design)$. The authors develop an upper bound that requires only (unconditional) entropy estimation, performed via a $k$-nearest-neighbors approach \citep{Kozachenko1987}.

\subsubsection{Variational lower bounds and approximate density ratios}
Now we turn our discussion back to variational bounds, parameterized by \textit{learnable functions}.
Dropping \revnew{our} earlier requirement that the function be a properly normalized density \revnew{or a smooth invertible transport map}, \citet{poole2019variational} develop the following bound, which holds for any function $f: \mathbb{R}^p \times \mathbb{R}^n \to \mathbb{R}$ and any positive function $a: \mathbb{R}^n \to \mathbb{R}_{>0}$:
\begin{align}
  \mathcal{I}(Y ; \Param) & \geq \mathbb{E}_{\Param, Y} \left [ f(\Param, Y) \right ] - \mathbb{E}_{Y} \left [ \frac{\mathbb{E}_{\Param}\left [e^{f(\Param, Y)} \right ]}{a(Y)}   + \log a(Y) - 1 \right ] \nonumber \\
                          & \eqqcolon \mathcal{L}^{\text{TUBA}}(f,a) \, .
                            \label{e:TUBA}
\end{align}
Here `TUBA' denotes `tractable unnormalized Barber--Agakov (BA),' for
reasons we will explain shortly. The function $f$ is known as the
`critic.' This bound is tightened by maximizing simultaneously over
$f$ and $a$. Since the mutual information will generally depend on the
design $\design$, the critic and the function $a$ can both depend on
the design as well, but here we have temporarily dropped dependence on
$\design$ to lighten notation.\footnote{For the remainder of this
section, we will keep notation lighter by not explicitly writing
dependence of the mutual information, the critic, the likelihood, and
other quantities on $\design$, with the understanding that everything
is appropriately conditioned on $\design$---since we are considering
the \revnew{mutual information} at a particular design.}
We can link this bound to the
BA bound in \eqref{e:BAlower}, which uses variational approximation over a class of normalized densities, by writing
\begin{align}
q(\param \vert y) = \frac{p(\param)}{Z(y)} e^{f(\param, y)}
\end{align}
and
\begin{align}
  \label{eq:defZ}
Z(y) = \mathbb{E}_{\Param} [\exp f(\Param, y) ].
\end{align}
Hence $f$ parameterizes an approximation ${q(\param \vert y)}/
{p(\param)}$ of the true posterior-to-prior density ratio
${p(\param \vert y)}/ {p(\param)}$,
and $Z$ is the $y$-dependent normalizing constant. The bound
$\mathcal{L}^{\text{TUBA}}(f,a)$ is then tight when
\revnew{
  $$
f(\param, y) = \log p(y \vert \param) + c(y) \quad \text{and} \quad a(y) = Z(y),$$
where $c$ is any function of $y$.} One specific case is when $c=0$, and hence $f(\param, y) = \log p(y \vert \param)$ and $a(y) = p(y)$. 

\citet{poole2019variational} point out that the lower bound for
\revnew{mutual information} introduced earlier by \citet{Nguyen2010}, often called the NWJ bound, is a special case of \eqref{e:TUBA} that follows from setting $a = e$:
\begin{align}
  \mathcal{I}(Y ; \Param) & \geq  \mathbb{E}_{\Param, Y} \left [ f(\Param, Y) \right ]  - \frac{1}{e} \, \mathbb{E}_Y \left [Z(Y) \right ] \eqqcolon \mathcal{L}^{\text{NWJ}}(f) .
                            \label{e:NWJ}
\end{align}
In this case, the unique optimal critic is $$f^\ast(\param, y) = 1 + \log \frac{p(\param \vert y)}{p(\param)}.$$

Unbiased estimates of these lower bounds, for any $f$ and $a$, can be constructed by simple Monte Carlo estimation of the expectations therein. Similarly, one can produce unbiased estimates of the gradient of each bound with respect to $f$ and $a$ (or some parameterization thereof). The modern approach is to parameterize $f$ and $a$ with deep neural networks, and to maximize over the network parameters, with the hope that the resulting class of functions is rich enough to yield a good approximation of the optimal critic. 
\revarx{The statistical properties of these nonlinear M-estimators are not yet fully understood, though there are some recent results. For instance, \citet{sreekumar2022neural} develop non-asymptotic error bounds for neural estimators of statistical divergences, using shallow neural network parameterizations of $f$ (i.e., with a single hidden layer) and under strong smoothness assumptions, capturing the impact of sample size and neural network size.}
\revarx{Statistical convergence results have also been developed for the NWJ estimator in the more classical setting}
where the critic is constrained to lie in a reproducing kernel Hilbert space
and parameterized with kernels \citep{Nguyen2010}.

A more general observation is that while these \revarx{variational bounds} can in principle become tight,
their estimators exhibit high variance---especially when the true value
of the mutual information is large; see \citet{McAllester2020}
for a discussion.

For completeness, we should note that the classical Donsker--Varadhan
bound \citep{Donsker1983}, a variational lower bound on KL divergence, can be adapted to mutual information to yield
\begin{align}
  \mathcal{I}(Y ; \Param) & \geq  \mathbb{E}_{\Param, Y} \left [ f(\Param, Y) \right ]  - \log  \mathbb{E}_Y \left [Z(Y) \right ] \eqqcolon \mathcal{L}^{\text{DV}}(f).
\end{align}
When the critic $f$ is parameterized by a neural \revnew{network}, this is the `mutual information neural estimation' (MINE) approach of \citet{Belghazi2018}.   The challenge in using this bound is that the second term on the right
yields a \textit{nested} expectation, \revarx{following the definition of $Z$ in \eqref{eq:defZ}}.
Hence, in practice, when estimating the bound with Monte Carlo as in
\citet{Belghazi2018}, we revert to the NMC setting of
Section~\ref{ss:nmc}, where estimates are biased at finite \revnew{inner loop} sample
sizes.  Then, as emphasized in \citet{poole2019variational}, we will obtain (to leading order) a positively biased estimator of a lower bound. The mean of this estimator is neither an upper nor a lower bound on the desired $\mathcal{I}(Y ; \Param)$.

\subsubsection{Multi-sample lower bounds}
\label{sss:multi_sample}
A different class of lower bounds are called `multi-sample' in that they involve not only an expectation over the joint distribution of $\Param, Y \sim \pdf_{\Param, Y}$ but simultaneously another expectation over $\Param_{2:M} \sim \prod_{j=2}^M r_{\Param}$, where $r_{\Param}$ could be different from the marginal $p_\Param$. 

The simplest among these is the so-called `prior contrastive
estimator' (PCE) proposed in \citet{Foster2020}:
\begin{align}
  \mathcal{I}(Y ; \Param) & \geq \mathbb{E}_{Y_1 \vert \Param_1} \mathbb{E}_{\Param_1} \mathbb{E}_ {\Param_{2:M}} \left [ \log \frac{p(Y_1 \vert \Param_1)} {\frac{1}{M} \sum_{j=1}^M p(Y_1 \vert \Param_{j})} \right ] \eqqcolon \mathcal{L}^{\text{PCE}}(M) ,
                            \label{e:PCE}
\end{align}
where the expectation is over $\Param_1, Y_1 \sim \pdf_{\Param, Y}$
and $\Param_{2:M} \stackrel{\text{iid}}{\sim}
\pdf_{\Param}$.  Keep
in mind that to \textit{estimate} the right-hand side of
\eqref{e:PCE}, one must draw many independent and identically
distributed (i.i.d.) realizations of $(Y_1, \Param_1,
\Param_{2:M})$. Suppose that we draw $N$ such (outer-loop)
realizations, with sample index $i$. The result is very much like the
NMC estimator \eqref{e:DNMC},
with one crucial modification: each outer loop sample of the parameters, $\Param_1^{(i)}$, is used once more, within the inner-loop estimate of the evidence for the corresponding $Y_1^{(i)}$. Doing so guarantees that $\mathcal{L}^{\text{PCE}}(M)$ is a lower bound of the mutual information for any $M > 1$; moreover, like NMC,  the bound becomes tight as $M \to \infty$ \citep[Theorem~1]{Foster2020}. Intuitively, recycling the outer-loop parameter sample makes the denominator of \eqref{e:PCE} over-estimate the evidence and hence under-estimate the mutual information. The samples $\Param_{2:M}^{(i)}$ are called contrastive samples, in that, unlike the original sample $\Param_1^{(i)}$, they are not responsible for $Y_1^{(i)}$.

Just as the variance and bias of the NMC estimators in Section~\ref{ss:nmc} were improved by importance sampling, a $y$-dependent, properly normalized, biasing distribution $q(\param \vert y)$ can be used within a contrastive estimator as well. (The biasing distribution will generally depend on $\design$ too, but recall that we are not explicitly writing dependence on $\design$ in this section, to keep notation simple.) This leads to the idea of an `adaptive contrastive estimator' (ACE) proposed by \citet{Foster2020},
\begin{align}
                            \label{e:ACE}
  \mathcal{I}(Y ; \Param) & \geq \mathbb{E}_{Y_1, \Param_1} \mathbb{E}_ {\Param_{2:M}} \left [ \log \frac{p(Y_1 \vert \Param_1)} %
                            {\frac{1}{M} \sum_{j=1}^M \frac{ p(Y_1 \vert \Param_{j}) p(\Param_{j}) } {q(\Param \vert Y_1 )} } \right ] \eqqcolon \mathcal{L}^{\text{ACE}}(M, q),
\end{align}
where now the expectation is over $\Param_1, Y_1 \sim \pdf_{\Param, Y}$ and $\Param_{2:M} \stackrel{\text{iid}}{\sim} q_{\Param \vert Y_1}$. The optimal biasing distribution in this case is $q(\param \vert y) = p(\param \vert y)$, which leads to zero-variance estimates of the evidence $p(Y_1)$. Seeking this $q$ is exactly the idea described at the end of Section~\ref{ss:densities}---combining a variational approximation of the posterior density with nested Monte Carlo---with only the addition of a contrastive recycling of the outer-loop $\Param_1$. With the optimal choice of $q$, this bound is tight for any finite $M$; otherwise, for sub-optimal $q$, it becomes tight as $M\to \infty$. 

Both \eqref{e:PCE} and \eqref{e:ACE} require the ability to evaluate the likelihood, and thus in contrast with the variational bounds \eqref{e:BAlower}, \eqref{e:TUBA},  and \eqref{e:NWJ}, they are not suited to implicit models. A further modification can make \eqref{e:PCE} likelihood-free. It introduces a critic function $f$ as follows:
\begin{align}
  \mathcal{I}(Y ; \Param) & \geq \mathbb{E}_{Y_1, \Param_1} \mathbb{E}_ {\Param_{2:M}} \left [ \log \frac{\exp f(\Param_1, Y_1)} {\frac{1}{M} \sum_{j=1}^M \exp f(\Param_j, Y_1)} \right ]  \eqqcolon \mathcal{L}^{\text{NCE}}(M,f) ,
                            \label{e:NCE}  
\end{align}
where the expectation is again over $\Param_1, Y_1 \sim \pdf_{\Param, Y}$ and $\Param_{2:M} \stackrel{\text{iid}}{\sim} \pdf_{\Param}$. This is essentially the `information noise-contrastive estimation' (InfoNCE) approach proposed in \citet{VandenOord2018}. $ \mathcal{L}^{\text{NCE}}(M,f)$ is a lower bound of the mutual information for all $f$ and $M$ \citep{poole2019variational}. An amalgam of \eqref{e:ACE} and \eqref{e:NCE}, incorporating both a biasing distribution $q$ and a critic $f$, is called `likelihood-free ACE' in \citet{Foster2020}. We note that both $\mathcal{L}^{\text{PCE}}(M)$ and $\mathcal{L}^{\text{NCE}}(M,f)$ are bounded above by $\log M $---no matter how the critic is chosen in the latter. Hence if $\mathcal{I}( Y ; \Param) > \log M $, these contrastive bounds cannot become tight; caution is therefore needed when the mutual information is high or the contrastive sample size $M$ is small. Note also that the $\log M$ limit does not apply to \eqref{e:ACE} and to other variants that use importance sampling, because in principle there exists a biasing distribution that yields a zero-variance estimate of the evidence, and in such a case even $M=1$ is sufficient.

An alternative way of estimating the PCE and InfoNCE bounds given only a single set of i.i.d.\ sample pairs $(Y_i, \Param_i)_{i=1}^M \sim p_{Y, \Param}$ is to rotate the role of the `data-generating sample'
  and the `contrastive samples' through the set. We can take
  the $i$th pair as the `data-generating sample' and use the remaining
  $M-1$ samples as `contrastive samples' to obtain
  $$\log \frac{\exp
    f(\Param_i, Y_i)} {\frac{1}{M} \sum_{j=1}^M \exp f(\Param_j,
    Y_i)}.$$
  Repeating this for $i=1,\ldots,M$ so that each pair
has the opportunity to be the `data-generating sample,'
  and then taking the average, we obtain the estimator,
\begin{align}
  \frac{1}{M} \sum_{i=1}^M \log \frac{\exp f(\Param_i, Y_i)} {\frac{1}{M} \sum_{j=1}^M \exp f(\Param_j, Y_i)} .
  \label{e:NCEestimator}
\end{align}
As shown in \citet{poole2019variational}, the expectation of this estimator is a lower bound of the true mutual information, for any critic $f$:
\begin{align}
  \mathcal{I}(Y ; \Param) & \geq \mathbb{E}_{(Y_i, \Param_i)_{i=1}^M}     \left [ \frac{1}{M} \sum_{i=1}^M \log \frac{\exp f(\Param_i, Y_i)} {\frac{1}{M} \sum_{j=1}^M \exp f(\Param_j, Y_i)} \right ].
                            \label{e:NCEreuse}
\end{align}
The expectation is also bounded above by $\log M$, in the same way as \eqref{e:PCE} and \eqref{e:NCE} above. Inequality \eqref{e:NCEreuse} is in fact the form of InfoNCE bound originally proposed in \citet{VandenOord2018}. It is in essence the same as \eqref{e:NCE} but using correlated, rather than independent, terms to estimate the expectation of $$\log \Bigl ( \exp f(\Param_1, Y_1) \Big / \frac{1}{M} \sum_{j=1}^M \exp f(\Param_j, Y_1) \Bigr ).$$

If the conditional density $\pdf(y \vert \param)$ is tractable, then we can replace $f$ with an optimal critic, $f(\param, y) = \log p(y \vert \param)$, such that \eqref{e:NCEreuse} becomes
\begin{align}
  \mathcal{I}(Y ; \Param) & \geq \mathbb{E}_{(Y_i, \Param_i)_{i=1}^M}  \left [ \frac{1}{M} \sum_{i=1}^M \log \frac{ p(Y_i \vert \Param_i)} {\frac{1}{M} \sum_{j=1}^M p(Y_i \vert \Param_j) } \right ],
\end{align}
which can be understood as a `sample re-use' analogue of the PCE bound \eqref{e:PCE} above. Interestingly, the corresponding estimator (i.e., \eqref{e:NCEestimator} with the log-likelihood substituting for $f$) was used in \citet{huan2013simulation}, before it was formally understood to be a lower bound in expectation. There, the rationale was to lower the cost of nested Monte Carlo by reducing the number of expensive model evaluations from $\mathcal{O}(NM)$ to $\mathcal{O}(M)$, in a setting where evaluations of the likelihood $p(y \vert \param)$ are costly for each new value of $\param$, but cheap for new values of $y$ given a fixed $\param$.

\subsubsection{Deploying variational bounds} 
\label{sss:deploy}

The relative merits of the variational bounds discussed above are very
much a current subject of research, involving both empirical
comparisons and some theory \citep{Czyz2023,Song2020,letizia2023variational}.
In general, the TUBA, NWJ, and DV bounds can become tight with increasingly good choices of critic, but their Monte Carlo estimates generally have much higher variance than estimates of multi-sample bounds such as PCE and InfoNCE. The latter bounds, on the other hand, are hampered by the fact they cannot become tight for arbitrarily large mutual information when using finite contrastive sample sizes. We also emphasize that new variants of these variational bounds continue to be proposed; we have attempted to cover the key constructions relevant to OED, but do not claim to be comprehensive. 
We refer to \citet{Song2020} and \citet{Czyz2023} for empirical evaluations of the mutual information estimators (and estimators of mutual information bounds) discussed here, along with many others. Important settings for these evaluations include low versus high values of mutual information, heavy versus light tails, various forms of non-Gaussianity, high dimensionality, sparse interactions, and so forth. As emphasized by \citet{Czyz2023}, evaluations limited to the Gaussian setting, while convenient in that exact analytical results are available for comparison, are not fully representative.

Deploying the bounds discussed in Sections~\ref{ss:densities} and
\ref{ss:varbounds} requires optimization. Specifically, it is useful
to have an \emph{unbiased estimate} of the gradient of each bound with
respect to parameters representing the critic $f$ (and possibly $a$)
or the density approximation $q$, along with an unbiased estimate of
the gradient of the bound with respect to the design $\design$, for
any given critic or density approximation. Explicit forms of these
gradient estimates, for several bounds above, are detailed in
\citet{Kleinegesse2021a}. Given these gradients, one can \emph{simultaneously} maximize the bound with respect to both the variational parameters and the design, e.g., via stochastic gradient ascent. We will return to this idea in Section~\ref{ss:contopt}.

\subsection{Low-dimensional structure}
\label{ss:dimred}

So far, we have discussed a variety of methods that, essentially,
approximate the densities or density ratios appearing in expressions
for the mutual information (expected information gain). A different
and complementary class of approximation methods involve
\emph{dimension reduction}. Here we will discuss methods for Bayesian
OED that exploit possible low dimensionality in the \emph{update} from
prior to posterior; such methods rest on the idea that the posterior
is well-approximated by a distribution that departs from the prior
only on a subspace of dimension $r \ll p$. At the same time, we will
consider dimension reduction for the observation vector $y \in
\mathbb{R}^n$;
the central idea is that conditioning on $y$ could be replaced by conditioning on the \emph{projection} of $y$ onto a subspace of dimension $s \ll n$.

Both of these notions of low dimensionality appear quite often in Bayesian inverse problems, where the data are related to the parameters $\param$ by the action of a forward operator that is somehow smoothing, e.g., in that it suppresses information about finer scales \citep{Cui2014a}. Low dimensionality more generally results from limited informativeness of the data in certain directions of $\paramset$ \revnew{and/or} redundancy among the components of $y$.
\revnew{There is additional potential for dimension reduction in \emph{goal-oriented} problems, as in Bayesian $\text{D}_{\text{A}}$-optimality and \eqref{e:goalpost}--\eqref{e:goalmarg}, where the experimenter wishes to learn about a specific quantity of interest $Z = \Psi(\Param)$.}
Identifying and exploiting this low dimensionality can lead to much more computationally efficient approximations of the mutual information, as we shall discuss below.

\subsubsection{Linear models}
\label{ss:lineardimred}

We first consider dimension reduction, of both parameters and data, in the context of finite-dimensional Bayesian linear-Gaussian models. \revarx{We begin with design for parameter inference and then consider goal-oriented design.}

\paragraph{Optimal low-rank approximations for parameter inference.}
\revarx{Recall the setup of the Bayesian linear-Gaussian model from the start of Section~\ref{s:bayesalphopt}. The mutual information between parameters and observations} in this setting can already be written in closed form as a ratio of log-determinants, as in \eqref{e:LG_EIG1} and \eqref{e:LG_EIG2}. It is nonetheless useful to express this quantity in a way that reveals the structure and \emph{intrinsic dimensionality} of the linear-Gaussian model.

The central objects in such a construction are {generalized eigenvalue
  problems} involving combinations of the prior covariance matrix
$\Gamma_\Param$; the noise covariance $\Gamma_{Y \vert \Param}$; the
posterior covariance $\Gamma_{\Param \vert Y}$ \eqref{e:LG_post_cov};
the negative Hessian of the log-likelihood, $H \coloneqq G^\top
\Gamma_{Y \vert \Param}^{-1} G$; and the marginal covariance of the data, $\Gamma_Y = G \Gamma_{\Param} G^\top + \Gamma_{Y \vert \Param}$. Several different generalized eigenvalue problems involving these matrices can be written, all closely related via simple transformations. Perhaps the most direct is the symmetric definite generalized eigenproblem
\begin{align}
  H w_j = \lambda_j \Gamma_{\Param}^{-1} w_j \, ,
  \label{e:gevoriginal}
\end{align}
where, since $H \succeq 0$ and $\Gamma^{-1}_{\Param} \succ 0$, we have $\lambda_j \geq 0$. This eigenproblem can be understood as balancing informativeness of the likelihood with informativeness of the prior, via the Rayleigh ratio $w^\top H w / w^\top \Gamma_{\Param}^{-1} w$: an eigendirection $w_j \in \mathbb{R}^p$ associated with $\lambda_j > 1$ is more constrained by the likelihood than by the prior, and thus more important to the update from prior to posterior. See \citet{Spantini2015} for thorough interpretations and demonstrations of this eigenstructure.
Closely related to \eqref{e:gevoriginal} are two other $p \times p$ parameter-space matrix pencils: $(\Gamma_{\Param} - \Gamma_{\Param \vert Y}, \Gamma_{\Param \vert Y})$ and $(\Gamma_{\Param}, \Gamma_{\Param \vert Y})$. The first has eigenvalue-eigenvector pairs $(\lambda_j, v_j)$, where $v_j = \Gamma_{\Param}^{-1} w_j$, and the second has eigenvalue-eigenvector pairs $(1+\lambda_j, v_j)$. See \citet[Proposition 10]{jagalur2021batch} for a proof of this equivalence. The leading eigendirections $v_j$ (ordered from largest to smallest $\lambda_j$) are the directions along which the ratio of posterior to prior variance is minimized; see \citet[Section 3.1]{Spantini2015}.

In the linear-Gaussian setting, there is a duality between spectral
structure in the parameter space and in the data space. Consider the
$n \times n$ data-space matrix pencils $(\Gamma_Y - \Gamma_{Y \vert
  \Param}, \Gamma_{Y \vert \Param})$ and $(\Gamma_Y, \Gamma_{Y \vert
  \Param})$. \citet[Proposition 10]{jagalur2021batch} show that $(\Gamma_Y
- \Gamma_{Y \vert \Param}, \Gamma_{Y \vert \Param})$  and
$(\Gamma_{\Param} - \Gamma_{\Param \vert Y}, \Gamma_{\Param \vert Y})$
have identical non-trivial generalized eigenvalues, $\lambda_j > 0$,
and that $(\Gamma_Y, \Gamma_{Y \vert \Param})$ and $(\Gamma_{\Param},
\Gamma_{\Param \vert Y})$ have identical eigenvalues that are strictly
greater than one, i.e., $1 + \lambda_j > 1$. The eigenvectors $u_j$ of
the two data-space pencils are identical, with $w_j =
\frac{1}{\alpha_j}\Gamma_{\Param} G^\top u_j$ for all $j$ with
$\lambda_j > 0$, where $\alpha_j$ is a scaling parameter that can be set
to $\alpha_j = \sqrt{\lambda_j}$ to obtain $\langle U_i, U_j
\rangle_{\Gamma_{Y \vert \Param}} = \delta_{ij}$, given $\langle W_i, W_j \rangle_{\Gamma_{\Param}^{-1}} = \delta_{ij}$.

\revarx{A convenient way of simultaneously capturing spectral structure in the parameter space and the data space is through the `whitened' forward operator $\Gamma_{Y \vert \Param}^{-1/2} G \Gamma_{\Param}^{1/2}$. The singular value decomposition of this matrix is
  $$\Gamma_{Y \vert \Param}^{-1/2} G \Gamma_{\Param}^{1/2} = \sum_{j \geq 1}
  \sqrt{\lambda_j} a_j b_j^\top,$$
  where the left singular vectors are related to the preceding data-space eigendirections via $a_j = \Gamma_{Y \vert \Param}^{1/2} u_j$ and the right singular vectors similarly are $b_j =  \Gamma_{\Param}^{-1/2} w_j$. See also \citet[Remark 4]{Spantini2015}.
  The whitened forward operator thus summarizes much of the essential structure of the linear-Gaussian Bayesian inverse problem and appears in many contexts. In the case of uncorrelated observational errors $\Gamma_{Y \vert \Param} = \sigma^2 I$, for example, the adjoint of the whitened forward operator, $\sigma^{-1} \Gamma_{\Param}^{1/2} G^\top$, is the central object in \textit{column subset selection} approaches to finding D-optimal designs \citep{eswar2024bayesian}. We will comment further on these methods in Section 4.}

\citet[Theorem 2.3]{Spantini2015} consider optimal approximations of the
posterior covariance $\Gamma_{\Param \vert Y}$ that take the form of
\textit{low-rank updates} of the prior covariance. Optimality is cast
in terms of minimizing a class of loss functions $\ell(M)$ over the manifold of all symmetric positive definite (SPD) matrices $M$; this class includes the geodesic distance from $M$ to $\Gamma_{\Param \vert Y}$ on SPD manifold, and the KL divergence or Hellinger distance from $\normdist(\mu,M)$ to $\normdist(\mu, \Gamma_{\Param \vert Y})$, where $\mu \in \mathbb{R}^p$ is an arbitrary (common) mean vector. The approximations sought are of the form $\widetilde{\Gamma}^r_{\Param \vert Y} \in \{\Gamma_{\Param} - K K^T \succ 0, \ \text{rank}(K) \leq r \}$, and the optimal approximation for any $r \leq n$, simultaneously minimizing all loss functions in the class, is given by
$$KK^\top = \sum_{i=1}^r \frac{\lambda_i}{1 + \lambda_i} w_i w_i^\top;$$
hence it follows from the leading eigenpairs of the parameter-space pencil $(H, \Gamma_{\Param}^{-1})$. The decay of the eigenvalue sequence
$(\lambda_j)_{j \geq 1}$, common to all of the matrix pencils above, captures the intrinsic dimensionality of the Bayesian model.

These optimal approximation results are related to the approximation of a more central object of interest in OED: the mutual information. Using the fact that the generalized eigenvectors $v_j$ simultaneously diagonalize $\Gamma_{\Param}$ and $\Gamma_{\Param \vert Y}$, or that the generalized eigenvectors $u_j$ simultaneously diagonalize $\Gamma_Y$ and $\Gamma_{Y \vert \Param}$, the mutual information between parameters $\Param$ and data $Y$ \eqref{e:LG_EIG1}--\eqref{e:LG_EIG2} can be written as
\begin{align}
\mathcal{I}(Y ; \Param) = \frac12 \sum_{j} \log (1 + \lambda_j) \, .
\label{e:spectralMI}
\end{align}
This expression for mutual information has appeared in many places in
recent literature, e.g., \citet{Alexanderian_2016ba} and \citet{giraldi2018optimal}. We note also that $\mathcal{I}(Y ; \Param)$ can be written using the squared canonical correlation scores \citep{Bach2002} between $Y$ and $\Param$; these scores follow from generalized eigenvalue problems that are slightly different from those discussed above, but closely related.
Truncating the sum in \eqref{e:spectralMI} after $r < \min(n,p)$ terms yields a low-rank approximation of the mutual information. From a computational perspective, since the MI is dominated by the largest generalized eigenvalues (though the $\log$ function slows this decay), problems with quickly decaying spectra and hence low intrinsic dimension are easier to approximate: one must compute only the leading eigenvalues of \textit{any} of the matrix pencils described above. 

Truncating \eqref{e:spectralMI} in this way in fact corresponds to the mutual information obtained from an \textit{optimal} $r$-dimensional projection of the data $Y$, of the parameters $\Param$, or of both simultaneously. Specifically, \citet{giraldi2018optimal} show that the leading eigenvectors $U_{1:r} = [u_1 \cdots u_r]$ of the data-space matrix pencils define low-dimensional projections of the data that are optimal at any given dimension $r \leq n$, in the sense of maximizing $\mathcal{I}(A_r^\top Y ; \Param)$ over matrices $A \in \mathbb{R}^{n \times r}$. The resulting mutual information is $$\mathcal{I}(U_{1:r}^\top Y; \Param) = \frac12 \sum_{j=1}^r \log(1 + \lambda_j).$$  Similarly, replacing $\Param$ with the $r$-dimensional projection $V_{1:r}^\top \Param$ yields $$\mathcal{I}(Y; V_{1:r}^\top\Param) = \frac12 \sum_{j=1}^r \log(1 + \lambda_j).$$ Going further, one can show that the same, optimal value of mutual information is also achieved by  $$\mathcal{I}(U_{1:r}^\top Y; V_{1:r}^\top\Param) = \frac12 \sum_{j=1}^r \log(1 + \lambda_j).$$

Note that this truncated mutual information is equivalent to the mutual information that would be obtained if the forward \revnew{operator} $G$ in \eqref{eq:linearGaussianmodel} were replaced by certain `projected' versions. Define
$$\mathbb{P}_{\text{obs}} \coloneqq \Gamma_{Y \vert \Param} U_{1:r} U_{1:r}^\top  \quad  \text{and} \quad  \mathbb{P}_{\text{param}} \coloneqq \Gamma_{\Param} V_{1:r} V_{1:r}^\top = W_{1:r} W_{1:r}^\top \Gamma_{\Param}^{-1}.$$
Then one can show that
\begin{align}
  \mathbb{P}_{\text{obs}} G = G \mathbb{P}_{\text{param}} = \mathbb{P}_{\text{obs}} G \mathbb{P}_{\text{param}} ,
\end{align}
and that these projected forward \revnew{operators} achieve the mutual information $$\frac12 \sum_{j=1}^r \log (1 + \lambda_j).$$ We emphasize, however, that the equivalence of these three projected models, and more broadly the strict duality between parameter and observation reduction discussed above, is specific to the linear-Gaussian setting.

The expression \eqref{e:spectralMI} naturally appears in infinite-dimensional formulations of linear Bayesian inverse problems, as discussed in Section~\ref{s:BIPs}. Indeed, finding the leading  eigenpairs of \eqref{e:gevoriginal} is equivalent to constructing a low-rank approximation of the prior-preconditioned Hessian \revarx{operator} $\widetilde{H}$, as discussed in \citet{Alexanderian_2016ba}. To see this intuitively in finite dimensions, note that $\widetilde{H} = \Gamma_{\Param}^{1/2} H \Gamma_{\Param}^{1/2} $, and that its (simple, not generalized) eigenvalue-eigenvector pairs are $(\lambda_j, \Gamma_{\Param}^{-1/2} w_j )$, where $(\lambda_j, w_j)$ are the eigenpairs of $(H, \Gamma_{\Param}^{-1})$. The algorithms proposed in \citet{Alexanderian_Saibaba_2018} (see also \citealt{Ghattas2021}) compute these low-rank approximations efficiently in \revnew{a discretization-invariant manner},
\revnew{when} the forward operator $G$ is described by partial differential equations. 

\paragraph{Optimal low-rank approximations for goal-oriented design.}
\revnew{Now we turn to the \emph{goal-oriented} linear setting. Our interest here lies not in learning the parameters $\Param$ \textit{per se}, but rather in informing a quantity of interest $Z = A^\top \Param$, where $A \in \mathbb{R}^{p \times s}$ has full column rank $s < p$. Maximizing the mutual information $\mathcal{I}(Y ; Z) $ is equivalent to maximizing the Bayesian $\text{D}_{\text{A}}$-optimality criterion mentioned in Section~\ref{s:bayesalphopt}, and is a linear special case of \eqref{e:goalpost}--\eqref{e:goalmarg}. Now, however, there is further potential for dimension reduction, stemming from the interaction of $A$ with other elements of the problem. Intuitively, the update from the prior predictive distribution to posterior predictive distribution of $Z$ (and hence the mutual information) should be well captured by directions in the parameter space $\mathbb{R}^p$ that are simultaneously informed by the data (relative to the prior) \emph{and} influential on $Z$. We can expect the latter consideration to help `screen out' parameter directions that are dampened by $A^\top$. (Consider, as a limiting example, elements of the kernel of $A^\top$.)} 

\revnew{This idea is formalized by \citet{spantini2017goal}. Beginning with the linear-Gaussian model $Y = G \Param + \mathcal{E}$ \eqref{eq:linearGaussianmodel}, a Gaussian prior distribution $\Param \sim \normdist(0, \Gamma_\Param)$ (zero mean for simplicity), and the specification of $Z = A^\top \Param$, \citet[Lemma 2.2]{spantini2017goal} introduce an \emph{equivalent} linear-Gaussian model that directly relates $Y$ and $Z$,
\begin{align}
Y = G A_\dagger Z + \Delta,
  \label{e:goalorientedlinear}
\end{align}
where $A_\dagger \coloneqq  \Gamma_\Param A \Gamma_Z^{-1}$, $\Gamma_Z
\coloneqq A^\top \Gamma_{\Param} A$, $Z \sim \normdist(0, \Gamma_Z)$
is the prior induced on $Z$, and
\begin{align*}
  \Delta \sim \normdist(0, \Gamma_\Delta) \quad   \text{with} \quad  \Gamma_\Delta  \coloneqq \Gamma_{Y \vert \Param} + G(\Gamma_\Param - \Gamma_\Param  A \Gamma_Z^{-1} A^\top \Gamma_\Param ) G^\top.
  \end{align*}
Crucially, these choices render the `effective noise' $\Delta$ \emph{independent} of $Z$. Now the low-rank approximation methods discussed earlier for the $Y$--$\Param$ system can be applied directly to the $Y$--$Z$ system above.}

\revnew{For example, as a goal-oriented counterpart to \eqref{e:gevoriginal}, one can compute the leading eigenpairs of the matrix pencil $( A^\top_\dagger G^\top \Gamma_{\Delta}^{-1} G A_\dagger , \Gamma_Z^{-1})$. Letting $(\tilde{\lambda}_j)_{j \geq 1}$ denote the generalized eigenvalues of this pencil, arranged in descending order, we can then write the mutual information as $\mathcal{I}(Y ; Z) =  \frac{1}{2} \sum_j \log (1 + \tilde{\lambda}_j)$ and truncate this sum as desired to obtain a low-rank approximation; see \citet{wu2023offline} for more details and a first application of this approach in OED. As demonstrated in \citet{spantini2017goal}, $(\tilde{\lambda}_j)$ can decay much more quickly than the spectrum $(\lambda_j)$  of \eqref{e:gevoriginal}, yielding a more efficient approximation than would be obtained by first approximating the posterior covariance matrix of $\Param$ and then applying $A$. 
Moreover, analogously to the discussion above, one can obtain optimal \emph{goal-oriented} projections of the data using the leading eigenvectors of $(\Gamma_Y, \Gamma_\Delta)$. Each of these eigenproblems optimally balances all the ingredients of the goal-oriented inference problem: the structure of the forward operator, the scale and structure of the observational noise covariance and prior covariance, and the ultimate prediction goals.}

\medskip

In closing, we should point out that (as in Section~\ref{ss:varbounds}) we have not explicitly noted dependence of the quantities above on the design $\design$. Recall, however, that our generic linear-Gaussian model \eqref{eq:linearGaussianmodel} \revnew{can be written more explicitly as} $Y \vert \Param, \design \sim \normdist( G(\design) \Param, \Gamma_{Y \vert \Param}(\design) )$. Hence all of the eigenvalues, eigenspaces, and projectors described above may depend on $\design$ via $G$ and $\Gamma_{Y \vert \Param}$. The details of this dependence are problem-dependent, but it is of interest to understand how to exploit any smoothness or other regularity that may be present---for instance, by differentiating the eigenvalues $\lambda_j$ with respect to $\design$. To our knowledge, algorithmic approaches in this vein are in their infancy.

\subsubsection{Nonlinear models}
\label{ss:nonlineardimred}

Extending the low-dimensional structure discussed in
Section~\ref{ss:lineardimred} to nonlinear models is considerably more
complex, and a subject of ongoing research. The tools we will
highlight here originate in dimension reduction for Bayesian inverse
problems and Bayesian statistical models more generally:
likelihood-informed subspaces \citep{Cui2014a,Cui2022}, certified
dimension reduction \citep{Zahm2022,li2023principal},  and related efforts. Our perspective is similar to that taken in the linear case: find the subspace of the parameter space where the posterior differs `most' from the prior, which is equivalent to finding the subspace that is best informed by the data; and find the subspace of the data that is most informative of the parameters.

We focus here on the approximation of mutual information, highlighted
in \citet{Baptista2022}, which makes the intuitive notions just stated more precise. In the nonlinear case, it is generally difficult to find \emph{optimal} low-dimensional projections. Instead, one can develop gradient-based upper bounds on the error induced by projecting $Y$ and $\Param$ onto lower-dimension subspaces, i.e., find upper bounds on the difference
$$\mathcal{I}(Y ; \Param) - \mathcal{I}(U_{1:s}^\top Y; V_{1:r}^\top \Param)$$
for some matrices $U_{1:s} \in  \mathbb{R}^{n \times s}$ and $V_{1:r} \in \mathbb{R}^{p \times r}$, and then minimize these bounds over $U$ and $V$. (Note that the matrices $U$ and $V$ discussed in this section are generally different from the $U$ and $V$ found in Section~\ref{ss:lineardimred}.)

The key assumption underlying this analysis is that the joint
distribution $p_{Y, \Param}$ must satisfy a \emph{subspace logarithmic
Sobolev inequality}; see \citet[Definition 2]{Baptista2022} or
\citet[Theorem 2.10]{Zahm2022}. Letting $Z = (Y, \Param)$ be a random variable taking values in $\mathbb{R}^{n+p}$, and letting $W \in \mathbb{R}^{(n+p) \times (n+p)}$
be a unitary matrix, the essence of this assumption is that any
conditional distribution $p(Z_t \vert Z_\perp = z_\perp)$ defined by
the decomposition $W = [W_t \ W_\perp]$, with $W_t \in
\mathbb{R}^{(n+p) \times t}$, $Z_t = W_t^\top Z$, and $Z_\perp =
W_\perp^\top Z$, satisfies a logarithmic Sobolev inequality with
constant $C \left ( p(Z_t \vert Z_\perp = z_\perp) \right )$ bounded
above by some $\overline{C} < \infty$, for all $W$, $t$, and
$z_\perp$. As shown in \citet{Zahm2022}, a \textit{sufficient}
condition for a distribution to satisfy the subspace logarithmic
Sobolev inequality is that it has convex support and that its
log-density is a bounded perturbation of a strongly convex function;
see \citet[Examples~2.5--2.9]{Zahm2022} and
\citet[Examples~1--2]{Baptista2022}. This allows, for example, Gaussian mixtures (and hence certain multi-modal distributions with strictly positive densities) but not distributions with exponential tails.

Now define the diagnostic matrices
\begin{align}
  H_\Param & =\iint (\nabla_\param \nabla_y \log \pdf_{Y \vert \Param}(y \vert \param) )^\top
  (\nabla_\param \nabla_y \log \pdf_{Y \vert \Param}(y \vert \param) ) \pdf_{Y, \Param}(y, \param) \, \mathrm{d} \param \, \mathrm{d} y, \\
  H_Y & =  \iint (\nabla_\param \nabla_y \log \pdf_{Y \vert \Param}(y \vert \param) )
  (\nabla_\param \nabla_y \log \pdf_{Y \vert \Param}(y \vert \param) )^\top \pdf_{Y, \Param}(y, \param) \, \mathrm{d} \param \, \mathrm{d} y ,
\end{align}
where $\nabla_\param \nabla_y \log \pdf_{Y \vert \Param} \in \mathbb{R}^{n \times p}$. 
We then have the following theorem, adapted from \citet{Baptista2022}.
\begin{theorem}
  Let $p_{Y, \Param}$ satisfy a subspace logarithmic Sobolev inequality with constant $\overline{C} < \infty$. Then, for any unitary matrices $V = [V_{1:r} \ V_{r+1:p}] \in \mathbb{R}^{p \times p}$ and $U = [U_{1:s} \ U_{s+1:n}] \in \mathbb{R}^{n \times n}$, we have
  \begin{align}
\mathcal{I}(Y ; \Param) - \mathcal{I}(U_{1:s}^\top Y; V_{1:r}^\top \Param)
    & \leq \overline{C}^2 \Bigl (  \tr \bigl ( V_{r+1:p}^\top H_\Param V_{r+1:p}  \bigr )    + \tr \bigl (U_{s+1:n}^\top H_Y U_{s+1:n} \bigr )  \Bigr )\,.
      \label{e:LSIbound}
  \end{align}
  \label{t:LSItheorem}
\end{theorem}

Crucially, the right-hand side of \eqref{e:LSIbound} can be minimized for any dimensions $r \leq p$ and $s \leq n$ by choosing $V_{1:r}$ to be the leading eigenvectors of $H_\Param$ and $U_{1:s}$ to be the leading eigenvectors of $H_Y$. This choice yields the bound
\begin{align}
  \mathcal{I}(Y ; \Param) - \mathcal{I}(U_{1:s}^\top Y; V_{1:r}^\top \Param) & \leq  \overline{C}^2
                                                                               \left ( \sum_{i=r+1}^p \lambda_i(H_\param)  + \sum_{i=s+1}^n \lambda_i(H_Y) \right ) .
                                                                               \label{e:LSIeigs}
\end{align}
Fast decay of the eigenvalues $\lambda_i$ of the two diagnostic matrices thus provides more opportunity for dimension reduction, with controlled error.

Computationally, this dimension reduction approach can be viewed as a first step towards approximating a mutual information design \revnew{objective:} first, find low-dimensional projections of the data \revnew{and} parameters, $Y_s = U_{1:s}^\top Y$ and $\Param_r  = V_{1:r}^\top \Param$, for some $r$ and $s$, by finding the dominant eigenspaces of $H_Y$ and $H_\Param$. (Note that, unlike in the linear-Gaussian case, the spectra of these two matrices are generally different.) Then, use any of the techniques discussed in Sections~\ref{ss:nmc}--\ref{ss:varbounds} to estimate or bound $\mathcal{I}(Y_s; \Param_r)$. By the data-processing inequality (or the positivity of the right-hand side of \eqref{e:LSIbound}) we always have $\mathcal{I}(Y_s; \Param_r) \leq \mathcal{I}(Y; \Param)$.
\revnew{For example, \citet{fengyi2024forthcoming} use the transport-based density estimators of Section~\ref{ss:densities} to bound the projected mutual information $\mathcal{I}(Y_s; \Param_r)$. The density estimation problems to be solved are thus of reduced dimension, and involve estimating transport maps that act on lower-dimensional spaces. Alternatively, \citet{wu2023large} apply $V_{1:r}$ for parameter dimension reduction as proposed above, but use a simpler dimension reduction scheme for the data (principal component analysis based on the marginal covariance $\Cov(Y)$); both reductions are a prelude to constructing a reduced-dimensional surrogate $\widetilde{G}$ for the forward operator in a Bayesian inverse problem (see Section~\ref{ss:nmc}), and then using this surrogate within a standard nested Monte Carlo estimator of the mutual information.}

\revnew{In Theorem~\ref{t:LSItheorem}}, for simplicity, we restricted the matrices $U$ and $V$ to be unitary. This constraint can effectively be relaxed by preconditioning the problem. Note that with any change of variables $\overline{\Param} = A \Param$ and $\overline{Y} = B Y$,  the left-hand side of \eqref{e:LSIbound} does not change, but the right-hand side does---via the diagnostic matrices and the log-Sobolev constant. It is unclear how to choose an optimal change of variables, as the log-Sobolev constant is in general very hard to compute, but one heuristic suggested in \citet[Section 4]{Baptista2022}, for the case of a Gaussian prior and Gaussian additive noise, is to `whiten' the problem so that $\Gamma_{\overline{\Param}}
$ and $\Gamma_{\overline{Y} \vert \overline{\Param}}$ become identity matrices. In the case of a linear-Gaussian model, this change of variables leads the proposed method to recover the optimal subspaces of Section~\ref{ss:lineardimred}. We also note that the change of variables need not be linear; any bijective transformation will leave the mutual information terms on the left-hand side of \eqref{e:LSIbound} unchanged. Thus, in principle one could compose a nonlinear change of variables with linear dimension reduction to effectively achieve nonlinear dimension reduction!

Another remark is that the bounds in Theorem~\ref{t:LSItheorem} are
generally not tight. One can intuitively see this in the
linear-Gaussian setting where, after the linear preconditioning
described above, we have $\lambda_i = \lambda_i(H_\Param) =
\lambda_i(H_Y)$. Then we can compare \eqref{e:spectralMI}, where the
exact truncation error involves \textit{logarithms} of trailing
eigenvalues $\log(1 + \lambda_i)$, with \eqref{e:LSIeigs}, which sums
the trailing eigenvalues directly. These quantities are not the same,
and the difference is analyzed in \citet[Section 4.3]{Baptista2022}; see
also \citet[Section 2.3]{Zahm2022}. There is reason to believe that
recent dimensional improvements to log-Sobolev inequalities
\citep{eskenazis2024intrinsic} could help tighten the bounds used here.

\section{Design optimization methods}
\label{s:optimization}
Now we will discuss methods for optimizing (generally, maximizing) OED criteria over possible designs $\design \in \Xi$. Different representations of candidate designs and choices of design criterion can yield very different classes of optimization problems. This section will therefore review a broad array of optimization approaches, ranging from algorithms for combinatorial optimization (including continuous relaxations thereof) to gradient-based algorithms for intrinsically continuous problems.
Section~\ref{ss:lineardesign} focuses on exact designs for linear models, as linearity lends the problem special structure; here, we discuss various discrete algorithms, as well as convex continuous relaxations that are followed by rounding. Section~\ref{ss:discrete_opt} discusses exact designs for more general nonlinear problems, with design criteria treated as set functions. Section~\ref{ss:contopt} turns to designs that are parameterized by real variables (e.g., sensor positions or times, dosages), and discusses how a variety of derivative-free or gradient-based optimization approaches interact with estimators or bounds for decision-theoretic design criteria.

\subsection{Optimization methods for linear design}
\label{ss:lineardesign}

We state a  prototypical linear design problem, by recalling the linear regression model with features $f: \mathcal{X} \to \mathbb{R}^p$ and uncorrelated observational errors. If the design $\design$ is supported on $n$ equally weighted points $x_i \in \mathcal{X} \in \mathbb{R}^{d}$, $i=1, \ldots, n$, then the Fisher information matrix of the model is given by \eqref{e:FIregression}; as pointed out earlier, this expression follows from \eqref{eq:fisherinfolinear} by setting 
\[ G = \left [ f(x_1)^\top;   f(x_2)^\top; \ldots; f(x_n)^\top \right ] \in \mathbb{R}^{n \times   p} \]
and $\Gamma_{Y | \param} = \sigma^2 I_n$. The classical design objectives discussed in Section~\ref{s:lineardesign} can be viewed as functions of the Fisher information matrix.

We will use the index set $\mathcal{V} \coloneqq \{1, \ldots, m\}$ to represent the support of the design space. It is instructive to think of $\mathcal{V}$ as a specific instance of $\mathcal{X}$ in the setting of a discrete design domain. In the simplest setting, the task of finding an optimal \textit{exact design} amounts to selecting a set $\mathcal{S} \subseteq \mathcal{V}$ of specified cardinality $n$ that maximizes the chosen objective, i.e., any of the design criteria for linear models given in Section~\ref{s:lineardesign}. More complex cases can involve constraints on the overall budget and variable costs associated with each individual design instance. 

Strictly speaking, the set $\mathcal{S}$  is better characterized as a multiset, where elements can be duplicated. In a conventional set, elements are typically assumed to be different from one another. The notion of a multiset is more general, and accommodates the `with repetition' scenario where a given design point can be chosen multiple times. (Repeated experiments at the same design point are not necessarily redundant, since they may yield different outcomes due to statistically independent noise.)  The contrasting scenario is `without repetition,' where each point is selected at most once; this is applicable, for instance, in the setting of deterministic computer experiments.

\subsubsection{Local search and exchange algorithms}
\label{sss:exchange_alg}
Algorithms guided by local search and exchange heuristics are non-sequential in nature. (We discuss sequential algorithms later in Section~\ref{sss:sequential_alg}.)
In the non-sequential setting, we start with an initial design of the required size, and at each iteration attempt to improve the quality of the design by deleting, adding, or exchanging points as guided by certain rules.  The non-sequential nature of these algorithms implies that the resulting solution sets are non-nested, meaning that optimal design set of cardinality $n$ may look quite different from the optimal set of cardinality $n-1$ or $n+1$.

In the context of exchange algorithms, the approach of \citet{fedorov1972theory} has been quite influential. Fedorov's exchange method starts with an arbitrary initial set $\mathcal{S}_{0}$ of $n$ candidate points, and at each iteration $t$ attempts to improve the objective by exchanging one of the points,  $$\mathcal{S}_{t} \leftarrow   \{j\} \cup \mathcal{S}_{t-1}  \setminus \{i\},$$  where $i \in \mathcal{S}_{t-1}$ and $j \in \mathcal{V} \setminus \mathcal{S}_{t-1}$. The search at each iteration is exhaustive: the improvement resulting from every possible exchange is calculated, and the best exchange is selected. The process is continued as long as an exchange improves the design objective. 

Fedorov's original work focused exclusively on D-optimal design, although its applicability to other design criteria has also been demonstrated.
While the exchange algorithm is relatively simple, it is accompanied by a sizable computational overhead due to the exhaustive search at each step. \citet{cook1980comparison}  proposed a modified Fedorov exchange procedure, where they consider each design point in turn, perhaps in random order, carrying out any beneficial exchange as soon as it is discovered.  \citet{johnson1983some} suggested further improvements by focusing the search on fewer points from the current design set, specifically those that have the lowest predictive variance.

The underlying ideas of the exchange algorithm can be extended to the
setting where design points are selected from a continuous compact
space, without an underlying candidate set
\citep{meyer1995coordinate}. A preliminary step here is to specify the
expected utility of changing one factor or coordinate at a time, while
holding the others constant. This step can be made more efficient by
using Gaussian process emulators, as demonstrated by
\citet{Overstall2017}. These ideas in principle are also applicable for nonlinear design problems.

Although the original Fedorov exchange algorithm was proposed several decades ago, rigorous analyses establishing \textit{approximation guarantees} have only recently emerged.
Here, an approximation guarantee is a provable bound on the ratio between the value of the objective function evaluated at the solution returned by the algorithm, and the optimal solution. It is thus a bound on the worst-case performance of the algorithm (in a relative sense). This is a typical way to assess the performance of an algorithm that produces approximate solutions to optimization problems, in particular NP-hard problems \citep{hochba1997approximation,vazirani2001approximation}.

\citet{madan2019combinatorial} establish approximation guarantees for local search algorithms for D-optimal design and A-optimal design; their results for the case of A-optimality are restricted to the less general `with repetition' setting. They show that the algorithms are asymptotically optimal when $n/p$ is large, and that in the case of A-optimal design there could be arbitrarily bad local optima.
They also propose approximate local search algorithms, where exchanges are made only when the objective improves by a factor of $1 + \delta$, leading to improved running times with a slight degradation in the approximation guarantees.  \citet{lau2022local} improve upon these results using novel analysis tools, and provide approximation  guarantees for D/A/E-optimal designs. In  particular they show that Fedorov’s exchange method for A-optimal design works well as long as there exists an near-optimal solution with a well-conditioned design matrix.

\subsubsection{Continuous convex relaxation approaches}
\label{sss:convex_relaxation_alg}

The combinatorial linear design problem is strictly speaking an integer program, whose exact solution is often intractable. One elegant solution approach is to perform a continuous relaxation of the design variables into a convex program that can be solved using established techniques \citep{boyd2004convex,ben2001lectures}.  A solution to the original combinatorial problem is then obtained by appropriately rounding the convex solution to an integer solution. Various polynomial-time (and possibly randomized) rounding algorithms, which convert the convex solution to an integer solution for the combinatorial problem, have been proposed.

Consider the linear design problem, where $\phi$ denotes an operator that acts on the Fisher information matrix to form the design criterion (e.g., \revnew{$\phi = - \tr$} for A-optimality, \revnew{$\phi  = -\log \det$} for D-optimality).
The relaxed continuous optimization problem takes the form
\begin{align}
  \argmin_{\xi = (\xi_1,\ldots,\xi_n)}
  \phi\left( \sum_{i=1}^m \xi_i f(x_i)
  f(x_i)^\top \right) \quad \text{subject to} \quad \norm{\xi}_1 \leq n.
  \label{e:convexopt}
\end{align}
Written in this way, the traditional optimality criteria discussed in
Section \ref{sss:classical} are all convex functions of the Fisher
information matrix. The convex constraint $\norm{\xi}_1 \leq n$
ensures that only $n$ points are selected. If we further specify $0
\leq \xi_i \leq 1$, then we constrain each point to be selected at
most once, corresponding to the without-replacement scenario. More
complex constraints, e.g., encoding varied costs associated with
selecting each point and a ceiling on the cumulative cost (i.e., a
knapsack constraint), can easily be added to  \eqref{e:convexopt}.
The fractional solution weights from this convex program have natural analogues to the notion of designs as probability measures, specifically to \textit{continuous designs}, as discussed in Section \ref{sss:design_measure}. However the notion of a continuous design arose organically from the work of early experimental design researchers, independently of these modern continuous relaxation formulations.

There are a number of convex programming relaxation approaches for the D/A/E-optimal design \revnew{problems}, differing from each other in how the rounding algorithm provides integer solutions. Even though these design criteria share attributes such as convexity, they behave differently in terms of approximability.
\citet{allen2017near} connect experimental design to matrix sparsification \citep{spielman2008graph,batson2009twice} and use regret minimization methods \citep{allen2015spectral} to obtain %
approximate designs for popular optimality criteria.
\citet{singh2020approximation} \revnew{devise} an  %
approximation algorithm for D-optimal design where the rounding strategy \revnew{uses} approximate positively correlated distributions.
\citet{nikolov2022proportional} \revnew{develop} an %
approximation algorithm for D/A-optimal design using proportional volume sampling; interestingly, they also show that the same approach will not work for E-optimal design.
\citet{lau2022local} modify the iterative randomized rounding algorithm based on the regret minimization framework of \citet{lau2020spectral} \revnew{for} D/A/E-optimal design problems with knapsack constraints.
Approximation algorithms using more involved relaxation techniques have enabled D-optimal design under partition \citep{nikolov2016maximizing} and matroid \citep{madan2020maximizing} constraints.

\subsubsection{Sequential algorithms \revarx{for finding optimal designs}}
\label{sss:sequential_alg}
\revarx{Sequential algorithms} arrive at a solution set either by the gradual addition of candidate points to a smaller design set, or by the sequential deletion of candidate points from a larger design set. These are called `forward' and `backward' procedures, respectively \citep{Atkinson_2007}. Typically the greedy heuristic guides the selection of candidate points at each step. In some cases, designs obtained using sequential procedures can be further improved using the non-sequential techniques outlined in Section \ref{sss:exchange_alg}.  For the case of D/A-optimal designs, \citet{madan2019combinatorial} analyze the sequential forward procedure, and provide %
approximation  guarantees that retain a specificity to the initialized set. The dependence on the initial set is rather undesirable, which can be ameliorated by leveraging distributed computing resources and running the algorithms with different initializations. Sequential design algorithms and greedy heuristics have been more generally studied using the language of set functions for combinatorial problems; we will discuss these methods more deeply in Section \ref{ss:discrete_opt}.

\subsection{Combinatorial approaches for discrete design variables}
\label{ss:discrete_opt}

In this section we formulate experimental design problems as optimization of set functions. The latter topic has been studied extensively \citep{Wolsey_Nemhauser_book,lovasz_book,Schrijver_book,Papadimitriou_Steiglitz_book}, and has a rich mathematical structure. We discuss some fundamental properties of set functions and their relevance to popular design criteria, including criteria for nonlinear design. We also discuss algorithms for their optimization.

\subsubsection{Background}

Previously we defined the candidate index set as $\mathcal{V}
\coloneqq \lbrace 1, \ldots, m \rbrace, \ m \in\mathbb{Z}_{>0}$. Let
its power set (i.e., the set of all subsets) be denoted by $2^{\mathcal{V}}$.  An important property of a certain class of set functions is submodularity:  any real-valued set function $\psi : 2^{\mathcal{V}} \rightarrow \mathbb{R}$ such that $\psi(\emptyset)=0 $  is submodular \citep{fujishige2005,Bach2013} if and only if, for all subsets $\mathcal{A},\mathcal{B} \subseteq \mathcal{V}$, we have
\begin{equation} \label{eqn:submodularity}
\psi(\mathcal{A}) + \psi(\mathcal{B}) \geq  \psi\lr{\mathcal{A} \cup \mathcal{B}} +  \psi\lr{\mathcal{A} \cap \mathcal{B}}.
\end{equation}
A function is \emph{supermodular} if its negation is submodular, and it is \emph{modular} if it is both supermodular and submodular. 
Many objective functions that arise in experimental design are naturally submodular. Consider, for instance, sensor placement where each sensor has a certain coverage area, and our interest is in optimizing the collective locations of a fixed number of sensors. There is a natural  \emph{diminishing returns} property that accompanies such objectives, which is characteristic of submodular functions.
An alternative but equivalent \citep{fujishige2005,Bach2013} definition of submodularity using first-order differences highlights this \emph{diminishing returns} property 
and is often easier to demonstrate in practice: the set function $\psi$ is submodular if and only if, for all $\mathcal{A},\mathcal{B} \subseteq  \mathcal{V}$ and $\nu \in \mathcal{V}$
such that $\mathcal{A} \subseteq \mathcal{B}$ and $\nu \notin \mathcal{B}$, we have
\begin{displaymath}
  \psi(\mathcal{A} \cup \lbrace \nu \rbrace) - \psi(\mathcal{A}) \geq \psi(\mathcal{B} \cup \lbrace \nu \rbrace) - \psi(\mathcal{B}).
\end{displaymath}
The term $\rho_{\nu}(\mathcal{A}) \coloneqq  \psi(\mathcal{A} \cup \lbrace \nu \rbrace) - \psi(\mathcal{A})$ is the incremental gain associated with the element $\nu$ when added to the set $\mathcal{A}$. It refers to the amount of change in the objective when the individual item $\nu$ is added to an existing pool of items in set $\mathcal{A}$. The definition is applicable by extension when we add a set $\mathcal{B}$ to set $\mathcal{A}$.

In certain cases when the function is not strictly submodular, it is helpful to understand how much it deviates from submodularity, or in other circumstances to quantify how close a set function is to being modular. Such characterizations also prove useful in analyzing the performance of many optimization algorithms.
The notion of curvature introduced by \citet{Conforti_Cornuejols_1984} in the context of non-negative set-functions is a bound on the intrinsic value of an item against its value in conjunction with all items in the candidate set. Formally, curvature is defined as the scalar $c$:
\[
c \coloneqq \max_{\nu \in \mathcal{V}} \frac{\rho_{\nu}(\emptyset) - \rho_{\nu}(\mathcal{V}\setminus \{ \nu \}) }{\rho_{\nu}(\emptyset)} = 1 - \min_{\nu \in \mathcal{V}} \frac{\rho_{\nu}(\mathcal{V}\setminus \{ \nu \}) }{\rho_{\nu}(\emptyset)}.
\]
The classical notion of curvature  \citep{Conforti_Cornuejols_1984} measures how close a submodular set function is to being modular, while the notion of generalized curvature \citep{Bian_etal_2017}  measures how close a set function---not necessarily submodular---is to being supermodular.
The submodularity ratio introduced by \citet{Das_Kempe_2011} for a general non-negative set function is a lower bound on the ratio of the sum of incremental gains associated with elements in a set, to the incremental gain associated with the set itself. Intuitively it captures `how close' to submodularity is the  function in question. Formally it is defined as the scalar $\gamma$:
\[
\gamma \coloneqq  \min_{\mathcal{B} \subseteq \mathcal{V}, \mathcal{A} \cap \mathcal{B} = \emptyset} \frac{ \sum_{\nu \in \mathcal{A}} \rho_{\nu}(\mathcal{B})}{\rho_{\mathcal{A}}(\mathcal{B})}.
\]
In \citet{Bian_etal_2017} the submodularity ratio and the generalized curvature together quantify how close a set function is to being modular. 

The counterpart of the submodularity ratio termed the supermodularity ratio  \citep{Tzoumas_etal_2017_SupModRatio,Bogunovic_etal_2018_SupModRatio,Karaca_etal_2018_SupModRatio,jagalur2021batch}  has also proved useful in analyzing many algorithms. It is defined as the scalar $\eta$:
\[
\revnew{\eta \coloneqq} \min_{\mathcal{B} \subseteq \mathcal{V},  \mathcal{A} \cap \mathcal{B} = \emptyset} \frac{\rho_{\mathcal{A}}(\mathcal{B})}{ \sum_{\nu \in \mathcal{A}} \rho_{\nu}(\mathcal{B})}.
\]
In \citet{jagalur2021batch}, the product of supermodularity and
submodularity ratios measures deviation from modularity. We will
revisit these notions and their implications for performance
guarantees in Section \ref{sss:greedy}. We now discuss set function attributes of widely used experimental design criteria.

\subsubsection{Set function attributes of design criteria}
\label{sss:set_function_design_criteria}

In Section \ref{sss:classical} we considered several classical alphabetic optimality design criteria. The D-optimality criterion defined using the log determinant is submodular. Using strictly linear algebraic techniques, it can be shown that the log determinant of a principal submatrix is a submodular function with respect to the indices defining the
submatrices \citep{Gantmacher_Krein_1960,Kotelyanskii_1950,Fan_1967,Fan_1968,Kelmans_1983,Johnson_1985}. Since the log determinant evaluations may be non-positive, however, most existing algorithms for submodular maximization are not directly applicable.
The A-optimality criterion in general is not submodular, as shown in \citet{Krause_2008}. We can, however, qualify its non-submodular nature by bounding the  submodularity ratio $\gamma$  and curvature $c$  for the Bayesian A-optimal design objective \citep{Bian_etal_2017}.

In Section \ref{ss:decision_theoretic}, we introduced a variety of
information theoretic design criteria. Specifically, we defined the
expected information gain \revnew{(EIG)} in parameters and showed that it is equivalent to the mutual information  between the parameters and the observations given the design, $\mathcal{I}(Y;\Theta|\xi)$. Suppose that selecting a design corresponds to choosing individual components of the $\mathbb{R}^m$-valued random variable $Y$. Then we can recast the optimization problem as finding a selection operator $\mathcal{P} \in \mathbb{R}^{m \times n}$,  $n < m$,
$\mathcal{P} \coloneqq  [ e_{i_1},  \ldots, e_{i_{n}}   ]$, where $e_{i_j}$ are distinct canonical unit vectors in $\mathbb{R}^m$:
\begin{equation}
        \mathcal{P}_{\text{opt}} = \argmax_{\mathcal{P} \in  \mathbb{R}^{m \times n}}  \mathcal{I} \revnew{ \lr{ \mathcal{P}^{\top} Y ; \Theta }. }  
\end{equation}
Given a desired number of observations $n < m$, we seek a selection operator $\mathcal{P}_{\text{opt}} $  such that the mutual information between the inference parameter $\Theta$ and the selected observations $Y_{\mathcal{P}_{\text{opt}}}\coloneqq\mathcal{P}_{\text{opt}}^{\top}Y$ is maximized. \citet{jagalur2021batch} show that the mutual information between parameters $\Theta$ and subsets of data $Y_{\mathcal{P}}$ is a submodular function if the observations are conditionally independent. Interestingly, this property holds even when the underlying joint distribution is non-Gaussian. In the setting of Bayesian inverse problems with additive noise, correlated noise renders the mutual information design objective non-submodular. When the inverse problem is linear, it is possible to quantify the non-submodularity of the information-theoretic objective by bounding the submodularity ratio $\gamma$ and supermodularity ratio $\eta$. It was shown in \citet{jagalur2021batch}  that those parameters can be both lower-bounded by ${\log \zeta_{\text{min}}}/{\log \zeta_{\text{max}}}$, where $\zeta$ is any generalized eigenvalue of the definite pair $\lr{\Gamma_Y, \Gamma_{Y|\Theta}}$.

\subsubsection{Greedy algorithms for cardinality-constrained designs}
\label{sss:greedy}
The case of cardinality-constrained optimization commonly arises in \revnew{OED}, with the prototypical example being sensor placement.  When the design objective is monotone and non-negative, the greedy heuristic of successively picking the candidate corresponding to the highest incremental gain performs well despite its simplicity.
\citet{Nemhauser1978} were the first to analyze the greedy heuristic
for the class of submodular functions, and showed that the algorithm has a constant factor $1-1/e$ approximation guarantee. 
\citet{Nemhauser1978b} further showed that the approximation guarantee cannot be improved in general by any other polynomial-time algorithm.

In the above paragraph, the term `constant factor'  refers to the approximation guarantee not depending on the particular instance of the function and only requiring the function to be submodular. By incorporating parameters that capture more specific attributes of the function, however, more expressive results can be obtained and offer useful insights into performance of the greedy heuristic in different scenarios. For instance, \citet{Conforti_Cornuejols_1984} proved the more refined guarantee $\frac{1}{c}(1-e^{-c})$ for submodular functions\revnew{,} where $c \in [0,1]$ is the curvature. When the function is known to have a small curvature, the improved performance of the greedy heuristic is easily explained.
Now suppose the  function is not submodular but has a submodularity
ratio $\gamma \in [0,1]$; \citet{Das_Kempe_2011} proved that the
greedy heuristic has a $1-e^{- \gamma}$ approximation
guarantee. \citet{Bian_etal_2017} improved upon that result by
incorporating the notion of a generalized curvature $\alpha \in [0,1]$
to obtain the factor $\frac{1}{\alpha}(1-e^{-\alpha \gamma})$. As we
alluded to in Section~\ref{sss:set_function_design_criteria}, these parameters can be concretely bounded for many experimental design objectives.

The greedy algorithm has a $\mathcal{O}(mn)$ complexity, where $m$ is the size of the candidate set and $n$ is the desired cardinality.  Closely related variants of the greedy heuristic can be better choices depending on the context and needs.
\citet{Robertazzi_Schwartz_1989} explored an accelerated version wherein the computed incremental gains are stored and exploited in \revarxtwo{subsequent steps}, possibly reducing the overall number of function evaluations. This is much like the modified Fedorov algorithm we discussed previously in Section~\ref{sss:exchange_alg}.
To reduce the complexity further and make it independent of the cardinality constraint,  \citet{lazier_than_lazy_greedy} analyze a randomized version of the greedy heuristic, termed stochastic greedy. This algorithm achieves, in expectation, a ($1-1/e -\epsilon$) approximation guarantee relative to the optimum solution. The number of function evaluations does not depend on the cardinality constraint, but linearly on the size of the candidate set, thus reducing the complexity substantially.

To benefit from modern HPC platforms, \citet{Mirzasoleiman_etal_2013} proposed a two-stage parallelized version which reduces the number of function evaluations per parallel process.  The approximation guarantee for the algorithm, however, in general depends on the size of the candidate set and cardinality constraint, which can only be overcome in special cases.
\citet{jagalur2021batch} analyzed a batch version of stochastic and distributed greedy algorithms with applicability to non-submodular objectives.  The heuristic involved choosing multiple candidates in each step but relying solely on the incremental gains associated with individual candidates. This reduces the computational overhead but with reduced approximation guarantees.

\subsubsection{Beyond greedy: algorithms for more general design problems}
As mentioned earlier in passing, many experimental design problems involve constraints more complex than simple cardinality bounds.
Consider, for instance, sensor placement with non-uniform costs $c(\nu)$ associated with placing the sensors. Here the goal could be to maximize the \revnew{design objective} $\psi(\cdot) $ while ensuring that the cumulative cost is within a budget $\sum_{\nu \in \mathcal{S}} c(\nu) \leq b$:
\[
\max_{\mathcal{S} \subset \mathcal{V}} \psi(\mathcal{S}) \quad
\text{subject to} \quad \sum_{\nu \in \mathcal{S}} c(\nu) \leq b.
\]
This is the well known knapsack problem, where the goal is to maximize the set function \revnew{$\psi$} subject to a non-negative modular constraint.
A natural modification of the the greedy algorithm \revnew{in this setting is
to pick the element maximizing the benefit-to-cost ratio, at each
step. Surprisingly, for} the case when \revnew{$\psi$} is submodular, it can be shown that either the output of the standard greedy algorithm or the index set returned by the cost-benefit greedy algorithm will be within ${(1-1/e)}/{2}$ of the optimal solution \citep{leskovec2007cost}.
A stronger result ($1-1/e$) is possible via a more computationally involved algorithm that exhaustively enumerates all subsets of size 3, and augments them using the cost-benefit greedy algorithm \citep{sviridenko2004note}.

Although we exclusively focused on non-negative monotone functions in
Section~\ref{sss:greedy},  several design criteria can be
non-monotone. In Section \ref{sss:GPR} we discussed an information-theoretic design objective arising in sensor placement that is symmetric submodular but strictly speaking non-monotone: $ \mathcal{I} \left ( \Param(x_s) ;  \Param(x_{s^c}) \right ) $. For the maximization of such submodular objectives with cardinality constraints, \citet{buchbinder2014submodular} have analyzed discrete random greedy and continuous double greedy algorithms. We refer the interested readers to this work for more details on \revnew{such} methods.

An important tool in the field of submodular optimization is the use of \textit{extensions}, particularly the multilinear extension \citep{Vondrak_2008} in the context of maximization.
The multilinear extension of the set function $\psi$ is the function $\Psi: [0,1]^{m} \rightarrow \revnew{\mathbb{R}}$ defined as
\[
\Psi(\mathbf{\xi}) = \sum_{\mathcal{S} \subseteq \mathcal{V}} \psi(\mathcal{S}) \prod_{i \in \mathcal{S}} \xi_i \prod_{i \in \mathcal{V} \setminus \mathcal{S}} (1-\xi_i).
\]
An intuitive interpretation of this extension is to think of the original set function as defined over the corners of the hypercube $\{0,1\}^{m}$, while the extension is valid over the entire unit cube $ [0,1]^{m} $. 
$\Psi(\xi)$ is the expected value of $\psi$ over sets, where for any set $\mathcal{S}$, each element $i$ is included independently with probability $\xi_i$, and not included with  probability $1-\xi_i$. 
If we write $\mathcal{S} \sim \mathbf{\xi}$ to indicate that $\mathcal{S}$ is the random subset sampled according to $\mathbf{\xi}$, then the multilinear extension is simply $\Psi(\mathbf{\xi})  = \mathbb{E}_{\mathcal{S} \sim \mathbf{\xi}} [\psi(\mathcal{S})]$.  
The multilinear extension \citep{Vondrak_2008} is quite different from
the Lov\'{a}sz extension \citep{Lovasz1983}; the latter maps any
discrete submodular function to its continuous convex counterpart
while the former maps it to its continuous concave counterpart. Note that these continuous extensions are quite different from the convex relaxation formulations we described in Section~\ref{sss:convex_relaxation_alg}!
Many of the more general approaches to maximizing submodular functions under a wide class of constraints rely on the multilinear extension. The approach here involves first approximately solving the problem
\[ \max \Psi(\mathbf{\xi}) \quad \text{subject to} \quad  \mathbf{\xi} \in \mathcal{X} \subseteq  [0,1]^m \]
and then rounding the continuous solution to obtain a near-optimal set \citep{Vondrak_2011,Chekuri_Vondrak_Zenklusen_2014,Vondrak_Chekuri_Zenklusen_2011,Calinescu_Chekuri_Pal_Vondrak_2011,Sviridenko_etal_2017}. We refer the interested reader to the cited works for details of the algorithms. For a broader survey on submodular maximization, see \citet{Krause_Golovin_2014}.

\subsection{Optimizing real-valued design variables}
\label{ss:contopt}

In this section we consider OED problems where the design is \revnew{naturally} parameterized by real-valued coordinates. These coordinates could be, for example, the spatial locations of a finite collection of sensors, times at which measurements should be taken, the values of certain experimental controls (pressure, temperature), and so on.
In Section~\ref{sss:design_measure} we formalized this
continuous-parameter case by letting these coordinates be elements of
a set $\mathcal{X} \subseteq \mathbb{R}^d$ and considering the design
$\xi$ to be a probability measure on $\mathcal{X}$. An exact $n$-point
design can then be understood as a mixture of Dirac measures, $\xi =
\frac{1}{n} \sum_{i=1}^n \delta_{x_i}$, where each $x_i \in
\mathcal{X}$. Here we will focus on continuous optimization methods
for such exact designs, and assume the number of support points $n$ to
be prescribed. The design measure $\xi$ is thus entirely parameterized
by $(x_1, \ldots, x_n) \in \bigtimes_{i=1}^n \mathcal{X} \subseteq
\mathbb{R}^{D}$, where $D = nd$. To keep notation intuitive in this setting, we will \textit{elide} the notions of the design measure and its parameterization and simply write $\xi = (x_1, \ldots, x_n)$. In other words, throughout this section we will consider $\xi$ to be a vector in $\mathcal{X}^{n} = \bigtimes_{i=1}^n \mathcal{X} = \Xi$. This way, the design remains exact, and gradients with respect to $\xi$ are straightforward to define.

Of course, one could discretize $\mathcal{X}^{n}$ and employ approaches from Section~\ref{ss:discrete_opt} to find a design within this discretized space. This approach may be convenient when dealing with complicated constraints on the set of feasible designs or when the number of desired experiments $n$ is not fixed (as in the case of knapsack constraints). For any na\"{i}ve discretization of $\mathcal{X}^{n}$, however, the number of candidate designs would grow exponentially with $nd$. Instead, we will focus here on optimization algorithms that make use of a continuous (Euclidean) parameterization of $\xi$.

When the design criterion can be evaluated exactly (i.e., in closed form) for any given $\xi$, the problem of optimizing over a real-valued $\xi$ in some sense becomes standard---though of course it inherits the natural difficulties associated with how the parameterized design enters the objective, or with the geometry of the feasible domain $\mathcal{X}$ (e.g., nonlinear constraints, non-convexity). These issues are generally quite problem dependent, however, and are not specific to OED.

The decision-theoretic OED objectives formulated in Section~\ref{ss:decision_theoretic}, on the other hand, do raise certain cross-cutting challenges for optimization. The expected utility $U$, which is the objective function to be maximized, typically must be estimated using Monte Carlo techniques. The optimal design problem can thus be written as
\begin{align}
  \design^{\ast} \in \arg \max_{\design\in\Xi} \EE_{W \vert \design} \big[\widehat{U}(\design,W) \big], \label{e:cont_opt}
\end{align}
where $\widehat{U}$ is an estimator of $U$, and the random variable $W$ is the source of stochasticity in this estimator.
For example, the Monte Carlo estimator of a general expected utility $U$ given in \eqref{e:general_EU_est},
\begin{align}
  \widehat{U}(\design,W)=\frac{1}{N}\sum_{i=1}^{N} u(\design, Y^{(i)}, \Param^{(i)}),
  \label{e:Uhatdef}
\end{align} %
with $\Param^{(i)}\sim \pdf(\param)$ and
$Y^{(i)} \sim \pdf(y|\param^{(i)},\design)$, has $W=(Y^{(i)},\Param^{(i)})_{i=1}^{N}$.
If, at any given design $\design$, all we can compute are estimates $\widehat{U}(\design,\revnew{W})$ for different realizations of $\revnew{W} \sim \pdf(W|\design)$, the optimization problem \eqref{e:cont_opt} is essentially a \emph{stochastic approximation} (SA) problem.

Implicit in writing \eqref{e:cont_opt} is the assumption that we are
content to maximize \revnew{$\mathbb{E}[\widehat{U}]$: that is,}
either $\widehat{U}(\xi, W)$ is an unbiased estimator of the desired
utility $U(\xi)$ at any $\xi$ {or} the bias
\revnew{$\mathbb{E}[\widehat{U}] - U$} is small and acceptable to the experimenter. In \revnew{Sections~\ref{sss:unbiased} and \ref{sss:simult}} we will discuss additional techniques that can be deployed when this is not the case, as in the case of mutual information maximization with biased nested Monte Carlo estimators or variational bounds. \revnew{Starting below, however, we will use the notation  $\Ubar(\design) \coloneqq \EE_{W \vert \design} [\widehat{U}(\design,W) ]$ for the objective in \eqref{e:cont_opt} to make the possible presence of bias clear.}

Many nonlinear optimization methods can be applied to \eqref{e:cont_opt}, and we do not attempt to survey this vast literature here. Instead, we will briefly recall a number of derivative-free and gradient-based approaches that have been used in the context of OED with continuous design variables. We call a method `derivative-free' if it only requires evaluations of $\widehat{U}$ (even if these evaluations are used to estimate derivatives) and we call a method `gradient-based' if it requires additional derivative information as an input, e.g., unbiased estimates of the gradient of \revnew{$\Ubar$ with respect to $\design$}. In both cases, we assume that the objective \revnew{$\Ubar$} is differentiable, but we make no assumptions on convexity or other structure. \revnew{We also note that estimating $\nabla_\design \Ubar$ may, in many cases, require evaluating derivatives of the log-likelihood or an underlying simulation model with respect to $\design$, and that this task may not be straightforward for PDE-based forward operators or intractable likelihoods.}

\subsubsection{Derivative-free methods}
\label{sss:dfo}

\citet{Larson2019} provide a comprehensive survey of derivative-free optimization methods, focusing on local optimization, with methods for solving stochastic problems of the form \eqref{e:cont_opt} specifically discussed in \citet[Section 6]{Larson2019}.  %
A key desideratum for algorithms for this setting is that they perform
well with \textit{noisy} objective evaluations. Below we briefly point
out several classes of derivative-free optimization methods that do
exactly this, and that are therefore useful for OED. We make no attempt to be comprehensive, and instead refer the reader to the preceding survey for more information.

\paragraph{Bayesian optimization.}

We introduced Bayesian optimization (BO) in
Section~\ref{ss:otherdesigns} to elucidate its differences from OED
for Gaussian processes. But BO can be quite useful as a tool
\textit{within} OED---specifically, as a means of solving
\eqref{e:cont_opt} when $\xi$ is of moderate dimension. BO
\citep{movckus1975bayesian,jones1998efficient,wang2023recent} is
essentially a derivative-free algorithm for global optimization, well
suited to `black-box' objective functions that are expensive to
evaluate and potentially noisy. BO uses Gaussian process regression to
construct and refine a model for the objective \revnew{(in the case of OED, the function $\Ubar$);}
\revnew{this regression}
naturally handles noisy \revnew{pointwise evaluations} and smooths the underlying
function estimate. Both the level of noise and the smoothness of the
reproducing kernel Hilbert space (RKHS) containing this estimate (the posterior mean of the Gaussian
process) can be adjusted by learning hyperparameters of the covariance
function of the Gaussian process. Examples of the application of BO to
OED include, among others, works by
\citet{Weaver2016,Overstall2017,Xu2020}, and \citet{Zhong_2024}. For more
information on BO, we refer to the surveys by
\citet{shahriari2015taking} and \citet{frazier2018tutorial}.

\paragraph{Stochastic approximation methods based on finite differences.}

A prototypical deterministic optimization method is gradient ascent:
\begin{align}
  \design_{k+1} = \design_k + \alpha_k g(\design_k),
  \label{e:grad_ascent}
\end{align}
where $g(\design_k) \coloneqq \nabla_\design \revnew{\Ubar(\design_k)}$ is the
gradient of the objective $\revnew{\Ubar(\design)} = \mathbb{E}_{W \vert
  \design}[\widehat{U}(\design, W)]$ evaluated at $\design_k$, and $\{
\alpha_k\}$ is a sequence of scalar step sizes with $\alpha_k > 0$. 
The Kiefer--Wolfowitz algorithm \citep{Kiefer1952} estimates \revnew{$g(\design_k)$} by applying centered differences to unbiased estimates of \revnew{$\Ubar$}. Specifically, for the $i$th component of $g$, where \revarx{$i=1, \ldots, D$,} we have
\begin{align}
  g_i(\design_k) \approx \frac{\widehat{U}(\design_k+ e_i\Delta_k , w_i^{+}) -
    \widehat{U}(\design_k- e_i \Delta_k, w_i^{-}) }{2\Delta_k},
  \label{e:kw}
\end{align}
where $w_i^+$ and $w_i^-$ are independent draws from $p_{W \vert
  \design_k}$, $e_i$ is the unit vector in coordinate $i$, and the
scalar $\Delta_k > 0$ is a difference parameter. Hence we need to
evaluate the estimator $\widehat{U}$ $2D$ times to compute each
gradient estimate. Under appropriate conditions on the step size
sequence $\{ \alpha_k\}$  and difference sequence $\{ \Delta_k \}$,
the objective function \revnew{$\Ubar$} and the noise $W$, Kiefer--Wolfowitz
iterations converge to a first-order critical point of \revnew{$\Ubar$} almost
surely; see, e.g., \citet{Blum1954} and \citet{Bhatnagar2013}.
Simultaneous perturbation stochastic approximation (SPSA) \citep{Spall1998,Spall1998a} is similar in form to Kiefer--Wolfowitz, except that it always uses a single centered difference at each iteration, the direction of which is chosen from a particular probability distribution. SPSA thus uses only \emph{two} independent realizations of $\widehat{U}$ at each step $k$, independent of the dimension of $\xi$.
An intuitive justification for SPSA is that error in the gradient produced by restricting the direction of the finite differences `averages out' over a large number of iterations~\cite{Spall1998a}.

\paragraph{Model-based methods.}
Model-based methods for derivative-free optimization use pointwise
evaluations of the objective function (and constraints, if applicable)
to form local approximations (called models) that can be analyzed to
produce the next optimization step. There is an extensive literature
on these methods, with many effective algorithms; for thorough
reviews, see \citet{Larson2019} and \citet{Conn2009}. Local models can take the form of low-order polynomials or radial basis function approximations, constructed via interpolation or regression on carefully chosen point sets. Updating of these models is often set within a trust region framework that carefully manages the point sets, the trust region radius, and the acceptance of each optimization step. Rigorous convergence guarantees have been developed for most prevalent algorithms \citep{Larson2019}.

Model-based derivative-free optimization has been applied in settings where only noisy estimates $\widehat{U}$ of the objective are available. The models fit by these methods (e.g., via local regression) are stochastic, and hence certain modifications to the algorithms and certainly to their convergence analyses are required, as explained in \citet[Section~6.3]{Larson2019}. On the algorithmic side, one modification that has been proposed is to introduce adaptive schemes for Monte Carlo sampling that balance noise in $\widehat{U}$ with other errors \citep{Shashaani2018}.

\paragraph{Direct search methods.}

Direct search methods \citep{Torczon1997,Larson2019} are also good choices for \eqref{e:cont_opt}. In general, these methods do not explicitly build local approximations of the objective or attempt to approximate its gradient, but rather compute optimization steps based on the \textit{relative ordering} of the evaluated function values.  A classical pattern search approach is the Nelder--Mead nonlinear simplex \citep{Nelder1965}, which has been explored for OED by \citet{huan2013simulation}. Modifications to Nelder--Mead for strongly noisy objectives include re-sampling schemes to account for the impact of noise on rankings \citep{Chang2012}. Other direct search methods, such as generalized pattern search methods \citep{Audet2002,Audet2004}, have variants for stochastic objectives as well \citep{Sriver2009}.

\subsubsection{Gradient-based methods}
\label{sss:cont_opt_gradient}

\paragraph{Robbins--Monro.}

The Robbins--Monro (RM) algorithm \citep{Robbins1951}---the progenitor
of modern stochastic gradient descent algorithms---follows
\eqref{e:grad_ascent} but requires access to an unbiased estimator
$\hg$ of the gradient \revnew{$\nabla_\design \Ubar$. That is, we need a random $\hg$ satisfying}
$$
\revnew {\EE[\hg(\design)] = g(\design) \coloneqq \nabla_\design \Ubar(\design). }
$$
\revnew{One way of constructing such an estimator is to reparameterize $\widehat{U}$ such that the distribution over which we take the expectation is functionally independent of the design variable $\design$. Such a reparameterization is always feasible \citep{mohamed2020monte}. For example, in \eqref{e:Uhatdef} 
where $W=(Y^{(i)},\Param^{(i)})_{i=1}^N$, we initially have $\EE_{W|\design}=\EE_{Y^{(1:N)}|\param^{(1:N)},\design} \, \EE_{\Param^{(1:N)}}$, but it is easy to replace the design-dependent observations $Y$ in this expectation with random variables that do not depend on $\xi$. If the observations are described as $Y = G_\design(\theta) + \mathcal{E}$, for instance, with an observation noise $\mathcal{E}$ whose distribution is independent of the design, this goal is immediately achieved: we can replace the original expectation with $\EE_{\mathcal{E}^{(1:N)}} \, \EE_{\Param^{(1:N)}}$ and reparameterize the expected utility estimator accordingly.
More fundamentally, however, the source of randomness in any statistical model is always a transformation of some \emph{independent} random input.}
\revnew{Call this random input $\check{W} \equiv \left ( \check{W}^{(i)} \right )_{i=1}^N \sim \pdf_{\check{W}}$. (In the example we just raised, $\check{W}^{(i)} = (\mathcal{E}^{(i)}, \Param^{(i)})$.) After reparameterization, the estimator \eqref{e:Uhatdef} of the expected utility is rewritten as
\begin{align}
  \check{U}(\design, \check{W}) = \frac{1}{N} \sum_{i=1}^N \check{u}\left ( \design,  \check{W}^{(i)} \right ) ,
\label{e:Ucheck}
\end{align}
and satisfies $\mathbb{E}_{\check{W}}[ \check{U}(\design, \check{W})]
= \EE_{W \vert \design} [\widehat{U}(\design,W) ] =
\Ubar(\design)$.
Since $\check{W}$ is independent of $\xi$, we can then write
$$\nabla_\design \Ubar(\xi) = \nabla_\design \mathbb{E}_{\check{W}}[
\check{U}(\design, \check{W})] = \mathbb{E}_{\check{W}} [
\nabla_\design \check{U}(\design, \check{W})] = \mathbb{E}_{\check{W}}
[ \hat{g}(\design, \check{W}) ],$$
where
\begin{align}
 \hat{g}(\design, \check{W}) \coloneqq \nabla_\design \check{U}(\design, \check{W})  =  \frac{1}{N} \sum_{i=1}^N  \nabla_\design \check{u}\left ( \design,  \check{W}^{(i)} \right )
\label{e:unbiasedgradubar}
\end{align}
is hence an unbiased estimator of the gradient.
}

Under appropriate conditions on the objective, the noise, and the step size
sequence $\{ \alpha_k \}$, it can be shown that RM converges almost \revnew{surely to} the global optimum \revnew{of $\Ubar$} in convex problems \citep{Kushner2003}
or to the first-order critical set in certain non-convex problems
\citep{mertikopoulos2020almost}. Many improvements on the original RM algorithm
have been devised and widely deployed, including Polyak--Ruppert
averaging \citep{Ruppert1988,Polyak1992}, which can improve robustness
and convergence rates by averaging iterates along the optimization
path, and Nesterov acceleration \citep{Nesterov1983}.  RM-type
algorithms have seen extensive use in OED, for example by
\citet{Huan2014,Foster_2019,Carlon2020,Foster2020,Foster_2021,Kleinegesse2020,Ivanova_2021,Shen_2021,Zhang2021},
and \citet{Shen_2023}.

\paragraph{Sample-average approximation.}

Sample-average approximation (SAA) (also referred to as the retrospective method or the sample-path method) \citep{Shapiro1991,Healy1991,Gurkan1994,Kleywegt2002} takes a different strategy: it seeks to reduce the stochastic objective to a deterministic one by fixing the randomness throughout the entire optimization process.
\revnew{This again requires reparameterizing the distribution over which we take the expectation (i.e., $\EE_{W|\design}$) to be independent of the design variable $\design$, as in the Monte Carlo estimator $\check{U}$ of \eqref{e:Ucheck}. The SAA problem then becomes}
\begin{align}
  \design^{\ast}_{\text{SAA}} \in \arg \max_{\design\in\Xi} \check{U}(\design, \check{W}=w_s), \label{e:SAA}
\end{align}
where $w_s$ is a realization of $\check{W}$ that is \emph{fixed} across $\Xi$. SAA can also be viewed as an application of common random numbers, where holding the sample of $\check{W}$ fixed essentially correlates the estimates of \revnew{$\Ubar$} across the design space and yields a smoother objective surface. \revnew{(For further discussion of the relationships between common random numbers, stochastic finite difference approximations to the gradient as in \eqref{e:kw}, and stochastic gradient estimates as in \eqref{e:unbiasedgradubar}, see \citet[Chapter VII.2]{Asmussen2007}.)}
The optimization problem \eqref{e:SAA} is \emph{deterministic}, and hence any standard deterministic optimization algorithm can be adopted.
For instance, \citet{Huan2014} applied SAA to a nonlinear OED problem with an EIG objective, and used the Broyden--Fletcher--Goldfarb--Shanno (BFGS) quasi-Newton scheme \citep[Chapter~6]{Nocedal2006} to find optimal designs. Stochastic bounds on the optimality gap \revnew{$\Ubar(\design^{\ast})-\Ubar(\design^{\ast}_{\text{SAA}})$} can also be
obtained~\cite{Norkin1998, Mak1999}.

\subsubsection{Unbiased \revnew{EIG} gradient estimators}
\label{sss:unbiased}

\revnew{As discussed near the start of Section~\ref{ss:contopt}, the algorithms described thus far focus on the objective $\Ubar$; in other words, they assume that we have an unbiased estimator of the desired objective at any $\design$ (or, practically speaking, that the bias is `small enough' to be ignored). Section~\ref{sss:cont_opt_gradient} then describes how to obtain an unbiased estimator of the gradient of $\Ubar$. But we have not yet addressed what to do when $\Ubar$ departs from the true design objective $U$.} 

For the mutual information (EIG) objectives discussed throughout Section~\ref{s:estimation}, \revnew{we do not have unbiased estimators; this is due to the presence of nested expectations and hence nested Monte Carlo estimators, or the use of density or density-ratio estimators within a nonlinear function (i.e., the logarithm).  Yet} directly applying any of the optimization methods that we have just discussed \revnew{to a} biased estimator of EIG can lead to arbitrarily large departures from the true maximizer (recall the example of Figure~\ref{f:bias}), as the objectives at hand may have multiple local maxima over $\Xi$ and the bias itself may vary significantly as a function of $\design$. Similarly, stochastic gradient-based methods will converge to critical points specified by the biased gradient, i.e., points where \revnew{$\nabla_{\design} \Ubar(\design) = 0$}, which again differ from points where \revnew{$\nabla_\design U (\design) = 0$} if we cannot enforce \revnew{$\nabla_{\design} \Ubar(\design)  = \nabla_{\design}U(\design)$} for all $\design$.

To remedy this situation for gradient-based optimization methods, \citet{Goda2022} propose an unbiased estimator of the gradient of the EIG \eqref{e:EKL} with respect to $\design$, i.e., of $\nabla_\design U_{\text{KL}}(\design)$.  Their approach builds on the multilevel nested Monte Carlo estimator of \citet{goda2020multilevel} by combining it with the random truncation approach of \citet{Rhee2015}, which generally aims to `de-bias' a consistent sequence of estimators. \citet{Goda2022} use the Rhee and Glynn scheme to randomly truncate the infinite telescoping sum in the inner loop of nested Monte Carlo, in a way that guarantees unbiasedness of the overall gradient estimator, allowing it to satisfy the requirements of the RM algorithm.

\revnew{A different estimator of  $\nabla_\design U_{\text{KL}}(\design)$ is proposed by \citet{ao2024estimating}. Here, the authors first use the reparameterization trick to express the observations $Y$ as a function of the parameters $\Param$ and additional design-independent random variables $\mathcal{E}$, i.e., as $Y = h(\Param, \mathcal{E}, \design)$, and then apply the following identity to rewrite the gradient of the log-evidence term in \eqref{e:Imarg},
\begin{align*}
  \nabla_\design \log p_{Y \vert \design} (y^{(i)} \vert \design ) = - \mathbb{E}_{\Param \vert y^{(i)}, \design} \bigl [ \nabla_\design \log p_{Y \vert \Param, \design} \bigl ( h(\param^{(i)}, \mathcal{E}^{(i)}, \design) \, \vert \,  \Param, \design \bigr )  \bigr ] ,
\end{align*}
where $y^{(i)} = h(\param^{(i)}, \mathcal{E}^{(i)}, \design)$ is any
realization of $Y$. MCMC sampling from the
posterior $p(\param \vert  y^{(i)}, \design)$ yields an estimate of
$\nabla_\design \log p(y^{(i)} \vert \design )$, which is then embedded in an outer loop of standard Monte Carlo
sampling over the joint distribution of $\Param$ and
$\mathcal{E}$. (One must also estimate the gradient of the expected
log-likelihood term in \eqref{e:Imarg}, but this is straightforward to
do in an unbiased way using the reparameterization trick.)
Insofar as the inner posterior samples are exact, the resulting nested
estimator of $\nabla_\design U_{\text{KL}}(\design)$ is unbiased. Of
course, finite-length MCMC chains produce biased estimates of the associated expectations, but the authors demonstrate empirically that this bias can be made negligible. Moreover, other posterior sampling schemes could be used instead.}

\subsubsection{Simultaneous bound tightening and design optimization}
\label{sss:simult}

Focusing again on design objectives that involve EIG in parameters,
predictions, or some other quantity of interest, an important theme of
Sections~\ref{ss:densities} and \ref{ss:varbounds} was the
construction of variational lower bounds for the EIG, i.e., functions
$\mathcal{L}(f, \design) \leq \text{EIG}(\design)$. Most of these
lower bounds (with some exceptions, e.g., $\mathcal{L}^{\text{PCE}}$
\eqref{e:PCE}) are parameterized by learnable functions $f$---which
could take the form of probability densities, transport maps, or more
generic `critic' functions.  Moreover, most of these learnable bounds
can become tight for an appropriate choice of $f$ (again with some
exceptions, e.g., certain contrastive multi-sample bounds in
Section~\ref{sss:multi_sample}). In these cases, it is then natural to
maximimize \emph{simultaneously} over $f$ and $\design$, as proposed
in \citet{Foster2020} and \citet{Kleinegesse2020,Kleinegesse2021a}:
\begin{align}
  \design^\ast \in \argmax_{\design \in \Xi} \max_f \mathcal{L}(f, \design) \, .
  \label{e:simultaneousmax}
\end{align}
This approach simultaneously seeks to tighten the lower bound and to find a good design. The maximizer should in principle be a maximizer of the true EIG.

Recall that all of the variational bounds $\mathcal{L}$ were expressed
as expectations. To apply a stochastic gradient technique to
\eqref{e:simultaneousmax}, unbiased estimates of gradients
$\nabla_\design \mathcal{L}(f, \design)$ and $\nabla_f \mathcal{L}(f,
\design)$ are required. As noted in Section~\ref{sss:deploy}, these
estimates are generally available through simple Monte Carlo
estimation, without the need for nested Monte Carlo or other
complexities. Simultaneous maximization \eqref{e:simultaneousmax} via
stochastic gradient techniques has been used in the setting of batch
OED by \citet{Foster2020,Kleinegesse2020,Kleinegesse2021a,Zhang2021}
and \citet{Chen2024},
and in the setting of sequential experimental design by \citet{Ivanova_2021} and \citet{Shen_2023a}, which we will discuss \revnew{in the next section}. 

\revnew{A related approach to leveraging bounds on EIG is proposed by \citet{zheng2020sequential}. Here, the authors take advantage of the ability to obtain \emph{both} lower and upper bounds (in expectation) for the EIG---e.g., the prior-contrastive estimator \eqref{e:PCE} as the lower bound and the standard NMC estimator \eqref{e:DNMC} as the upper bound. Both of these bounds can be refined and made tight by increasing the inner-loop sample size $M$. \citet{zheng2020sequential} thus develop an adaptive refinement strategy, {in which} the bounds are tightened over the course of optimization, that achieves regret-style guarantees.}

\section{Sequential optimal experimental design}
\label{s:sequential}
Sequential experimental design is concerned with the planning of
multiple experiments that are conducted in a sequence, where the
\emph{results} of previous experiments in the sequence can inform the
design of subsequent experiments. This is the crucial difference
between sequential design and the `batch' \rev{(also called
  `static') design} approaches that were the focus of previous
sections. Here we will focus on a Bayesian approach to sequential
experimental design.

One straightforward \rev{sequential design} procedure is to apply
the batch \revnew{optimal experimental design (OED)} framework and methods from
Sections~\ref{s:criteria}--\ref{s:optimization} to one experiment, or
subset of experiments, at a time: optimally choose the next design,
perform that experiment, update the prior to the posterior based on
the outcome of the experiment, and repeat the process for the
next. Such a procedure is called \emph{greedy} or \emph{myopic},
because it does not take into account future experiments when finding
the (immediate) next experiment; it involves no `lookahead.'  Yet
greedy design is conceptually simple to implement, especially if
computational tools for batch design already exist, and it is flexible
for situations where the total number of desired experiments is
unknown. As a result, a large body of sequential experimental design
research has been based on some form of greedy design \citep{Box_1992,
  Dror_2008, Cavagnaro_2010, Solonen_2012, Drovandi_2013,
  Drovandi_2014, Kim_2014,Hainy_2016,Kleinegesse_2021}.

Of course, to improve coordination among the experiments, one could
design \emph{all} the experiments simultaneously and thus revert to
\rev{an overall {batch} design}. But doing so would forgo the
opportunity to adapt to new observations; in other words, it would
allow for coordination among the experiments, but \emph{no
feedback}. A hallmark of sequential experimental design is precisely
the idea of feedback.

\revarxtwo{We note that the greedy design approach described just now, and more
generally the sequential design approaches to be detailed in this section, differ from
the sequential \textit{algorithms}  discussed in Sections~\ref{ss:lineardesign} and \ref{ss:discrete_opt} for solving batch design problems.}
In particular, the latter are not for sequential experimental design, as they do not involve
observing the results of experiments between design decisions, and
thus do not incorporate any feedback. Rather, they still seek a
static design for a single batch of experiments, but break the search into
stages to control the dimension of the design space and avoid
combinatorial scaling of computational complexity.

In the remainder of this section, we will present \rev{a sequential optimal}
experimental design formulation that includes
\rev{both} lookahead and feedback; \rev{see}, e.g., \citet{Muller_2007},
\citet[VII.G]{vonToussaint_2011}, and \rev{\citet[Chapter~3]{Huan_2015}}. We refer to this formulation
as \emph{sequential OED} (sOED).
sOED generalizes both greedy and batch design \citep[Section
  2.3]{Shen_2023}. The key ideas of sOED are that (i) each design
should be selected while taking into consideration \emph{all remaining
experiments} to be performed, and (ii) designs should be given in the
form of \emph{functions}, called policies, that adaptively specify the
next experiment in the sequence given the current state of
information. We will formalize these ideas using the framework of
Markov decision \revnew{processes,} to be presented shortly.

As we shall see, the numerical solution of the sOED problem is rather
challenging, and there have been relatively few attempts to solve it
in great generality. For example, \citet{Carlin_1998,Gautier_2000,Pronzato_2002,
  Brockwell_2003, Christen_2003, Murphy_2003,
  Wathen_2006,Muller_2022}, and \citet{Tec_2023} have all made advances largely
limited to discrete settings, or did not employ a Bayesian framework
with information-theoretic design criteria. In this section, we will
survey recent progress towards realizing fully Bayesian sOED, including methods
that make use of dynamic programming \cite{Huan_2015,Huan_2016},
reinforcement learning \cite{Blau_2022,Shen_2023}, and information
bounds \cite{Foster_2021,Ivanova_2021,Shen_2023a}.

\subsection{Background}

We focus on the design of a finite number of experiments,
indexed
by
$k=0,1,\ldots,N-1$. We assume $N$ is known and fixed.  The Bayesian
update for the $k$th experiment then becomes
\begin{align}
\revarx{    \pdf(\param|y_k,\design_k,\info_k) =
    \frac{\pdf(y_k|\param,\design_k,\info_k)\,
      \pdf(\param| \kern 0.05em \info_k)}{\pdf(y_k|\design_k,\info_k)},
    }
    \label{e:seq_Bayes}
\end{align}
where $\design_k$ and $y_k$ are, respectively, the design and the value
of the observation realized in the $k$th experiment, and
\revarx{$\info_k \coloneqq \left[ \design_0,y_0,\dots,\design_{k-1},y_{k-1}
\right]$ is a realization of the random `background information' sequence $\Info_k \coloneqq \left[ \design_0,Y_0,\dots,\design_{k-1},Y_{k-1}
    \right]$}
composed of the
history of designs and observations from all experiments preceding
the current one, \revarx{with} $\Info_0=\emptyset$.  
The prior density \revarx{$\pdf(\param| \kern 0.05em \info_k)$} represents the
\rev{state of knowledge} about \rev{the uncertain parameters} $\Param$ before the $k$th experiment, and the
posterior \rev{density} \revarx{$\pdf(\param|\rev{y_k,\design_k},\info_k)$} represents the
updated \rev{state of knowledge} after having observed the outcome
\rev{of} the $k$th
experiment. Equation \eqref{e:seq_Bayes} uses the simplification
\revarx{$\pdf(\param|\design_k,\info_k)=\pdf(\param| \kern 0.05em \info_{k})$}, since the prior
density does not depend on the pending choice of design.  The
posterior after the $k$th experiment
\revarx{$\pdf(\param|y_k,\design_k,\info_k)=\pdf(\param| \kern 0.05em \info_{k+1})$} then
becomes the prior for the $(k+1)$th experiment, and
\eqref{e:seq_Bayes} can be \rev{applied recursively.}
In \eqref{e:seq_Bayes} we present the general setting where the
density of the current data \revarx{$\pdf(y_k|\param,\design_k,\info_k)$} may depend on the design
and observations from previous experiments, i.e., on \revarx{$\info_k$}. In
many situations, however, $Y_k$ is conditionally
independent of the other observations in the sequence given $\param$
and $\design_k$, in which case its density simplifies to
\revarx{$\pdf(y_k|\param,\design_k,\info_k)=\pdf(y_k|\param,\design_k)$}.

\rev{We note that the experiments in the sequence do not need be of
  the same type, as long as they share the same parameters
  $\Param$. The spaces of possible designs $\Xi_k \ni \design_k$ and observations
  $\mathcal{Y}_k \ni y_k$ can differ from one experiment to the next,
  and even change dimension. Similarly, the conditional density of the
  data in \eqref{e:seq_Bayes} is,
  in more explicit notation, \revarx{$p_{Y_k|\Param,\design_k,\Info_k}(y_k|\param,\xi_k,\info_k)$} and thus has a $k$-dependence as well.
  For example, in learning the properties of a fluid, a first 
  experiment might entail selecting the shape of an obstacle to place
  in the flow and observing the velocity in its wake,
 while the next experiment might involve choosing where on
 the surface of this obstacle to place a pressure sensor.}

\subsection{Formulation as a Markov decision process} 
\label{ss:formulationMDP}

Sequential experimental design can be modeled through a \rev{Markov
  decision process (MDP)} \revarx{specified} by
a tuple $\bigl ( \CS, \rev{\{ \CA_k \}_k}, s_0, \{ r_k(\cdot) \}_k,
\{T_k(\cdot) \}_k \bigr )$. 
\revarx{$\CS$ is the state space of the Markov process, where the
  state $S_k$ is a random element with values in $\CS$ and realizations of this state are  denoted by  $s_k \in \CS$;}
\revarx{the action spaces $\CA_k$ comprise} possible actions (\rev{which here are designs})
$\design_k\in \rev{\CA_k}$;\footnote{\rev{$\CA_k \equiv \Xi_k$, i.e., spaces of
    candidate designs indexed by $k$, but we use $\CA_k$ in this
    section to be consistent with the usual notation in the MDP literature.}}
$s_0$ is an initial state; \revarx{$r_k(s_k,\design_k,y_k)$ are} scalar-valued reward \rev{functions} 
that evaluate the instantaneous \rev{reward} when taking action $\design_k$ and observing $y_k$ at
state $s_k$; and \rev{$T_k(\mathcal{S}' \vert
  s_k, \design_k)$  are state transition \rev{kernels}
that evaluate the probability of transitioning to any set of states
$\mathcal{S}' \subseteq \mathcal{S}$ at stage $k+1$ having taken} action $\design_k$ at state $s_k$.
In the context of experimental design, the action being taken is the selection of a
design; thus we use the terms `action' and `design'
interchangeably.

\medskip

\noindent\textit{State.}
The state \rev{of the system} before the
$k$th experiment is described by
\revarx{$S_k= \bigl ( S_{k}^{b},S_{k}^{p} \bigr )$},
\rev{a quantity} that summarizes all information deemed relevant \rev{to} future
design decisions.
We split \revarx{$S_k$} into a `belief state' \revarx{$S_{k}^{b}$},
representing the state of knowledge/uncertainty in \revarx{the parameters} $\Param$, and a
`physical state' \revarx{$S_{k}^{p}$} comprising any other deterministic
variables that may be relevant \rev{to the design process.} Since
$\Param$ is not directly observed and \rev{can only be} inferred from
observations $Y_k$, this setup can be viewed as a partially observed
MDP (POMDP) for $\Param$ or a belief MDP \revarx{with (belief) state $S_k$}
\citep{Kaelbling_1998}. 

\revarx{A realization of} the belief state, \revarx{$s_k^b$,} is simply the posterior distribution \rev{of $\Param$}
given \revarx{a set of past} experimental designs and realized observations,
\revarx{$\info_k$}. For \rev{$\Param$ taking values on $\RR^p$}, \revarx{$s_k^b$} is thus a probability distribution on
$\RR^p$. Numerically, it can be \rev{represented by, for example, a
  probability density function approximation, or a set of weighted particles, or
  by} tracking \revarx{$\info_k$} directly.
Tracking \revarx{$\info_k$} \rev{is easiest to implement since it does not require
additional calculations translating  $(\design_i)_{i < k}$ and
  $(y_i)_{i < k}$ to another representation, but the dimension of \revarx{$\info_k$}
grows with $k$---though it is
bounded for finite $N$.} In general, the \rev{set of possible
posterior distributions} that can
be realized is uncountably infinite,\footnote{Unless both $\design_k$
and $Y_k$ are discrete.} and hence \rev{this setting}
differs from a discrete or finite-state system.

\rev{Maintaining only a belief state, in the form of the posterior, does
  not suffice to preserve the Markov property of the system if the likelihood 
  depends on the history of past experiments \revarx{$\info_k$}, as presented
  in \eqref{e:seq_Bayes}. This can be fixed by introducing a
  physical state.}
With regard to the \revarx{$\info_k$}-dependence in \revarx{$\pdf(y_k|\param,\design_k,\info_k)$}, the
physical state essentially extracts and tracks relevant features
from \revarx{$\info_k$} that allow the likelihood to be evaluated or the
observations $Y_k$ simulated.  An example of the physical
state would be the known position of a mobile sensor platform, which
might evolve from one stage of the design sequence to the next. Note
that if
$\Info_k$ is adopted as the belief state, then information about the physical state is already
contained in $\Info_k$, even if only implicitly.

\medskip

\noindent\textit{Action (design) and policy.}
Sequential experimental design is adaptive in nature. Whereas a batch
design problem seeks a \rev{single} design $\design$ (see
Section~\ref{s:criteria}), sequential design now looks for a strategy,
called a \emph{policy}, describing \emph{how to choose} the design
depending on the \rev{current state}.
The policy \rev{is a collection of functions}
$\policy = \{\mu_k : \CS \to \rev{\CA_k}, k=0,\ldots,N-1\}$, where the policy
function $\mu_k$ returns the design for the $k$th experiment given the
current state, $\design_k=\mu_k(s_k)$. In general, $\mu_k$ differs from
experiment to experiment for finite \revnew{$N$.}\footnote{\rev{In contrast,}
  in the infinite-horizon setting, \rev{the policy must be stationary,
    i.e., independent of $k$}. Thus only a single $\mu$ \rev{needs to
    be found, making infinite-horizon problems generally} \revnew{easier to tackle.}}
\revnew{Some} intuition for this fact follows by considering that even
  when starting from the same belief and physical state, a `good' design can be
  quite different depending on how many experiments \rev{remain} in
the overall sequence. We focus on deterministic policies in this
\rev{formulation}, although stochastic policies \rev{that evaluate the
  probability of choosing different candidate designs can also be
  adopted.}

\medskip

\noindent\textit{State transition dynamics.}  When an experiment is performed,
the state changes according to a transition kernel
\rev{$T_k(\mathcal{S}' \vert
  s_k, \design_k)$
describing the probability of
transitioning from the current state $s_k$, having chosen design
$\design_k$ and observed the outcome of resulting experiment, to
any set of states
  at stage $k+1$, $\mathcal{S}' \subseteq \mathcal{S}$.}
This kernel is generally intractable to evaluate, but \rev{it can
  instead be \emph{simulated}, by sampling from the prior predictive
  distribution of \revarx{$Y_k$} given the design $\design_k$ and then applying
  Bayes' rule \eqref{e:seq_Bayes}. Of course, both of these steps make
  use of the statistical model for \revarx{$Y_k$}. We denote the latter transition dynamics by
$s_{k+1}=\CF_k(s_k,\design_k,y_k)$; the function $\CF_k$ encapsulates
the transition from prior to posterior, given values of $\design_k$
and the realized data $y_k$, following \eqref{e:seq_Bayes}.}

The physical state, if present, evolves according to \rev{a model for} the relevant physical process.

\medskip

\noindent\textit{Reward (utility).}
Here $r_k(s_k,\design_k,y_k)$ denotes the \rev{real-valued}
reward immediately obtained after performing the $k$th experiment,
\rev{which} may depend on the state, design, and observation values.  For
example, the reward can reflect the cost of carrying out the $k$th
experiment \revnew{and/or} the information gained \rev{in}
$\Param$ \rev{as a result}. \rev{On the other hand}, $r_N(s_N)$ denotes the
terminal reward, which \rev{reflects} any rewards that can only
be \rev{quantified} after all experiments are completed.
The choice of reward \rev{functions $(r_k)_{k=0}^N$ is} quite flexible and can be based on the
wide range \rev{of optimal design utilities} discussed in Section~\ref{ss:decision_theoretic}. In the
next section we will highlight the \rev{reward functions}
corresponding to $u^{\text{div}}$ \eqref{e:udivergence}, \rev{which
  reflects} information gain in \rev{the} model parameter\rev{s} $\Param$.

\begin{figure}[tb]
  \centering
  \includegraphics[width=1.0\linewidth]{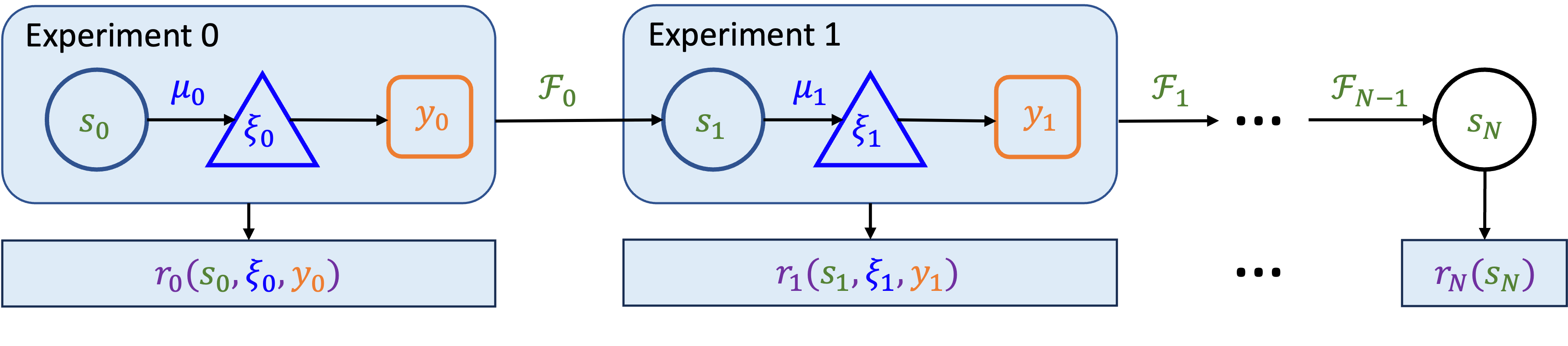}
  \caption{In the MDP progression of sequential experimental design,
    we start with an initial state $s_0$ (i.e., initial prior and
    physical state), evaluate the policy function at the state
    $\mu_0(s_0)$ to obtain the design $\design_0$ for experiment 0,
    conduct the experiment to obtain its outcome $y_0$ and immediate
    reward $r_0(s_0,\design_0,y_0)$, update the state to the new state
    via the transition dynamics $s_1=\CF_0(s_0,\design_0,y_0)$
    (i.e., updated posterior and physical state), and repeat for the
    next experiments. Once the last experiment $N-1$ is completed, the
    terminal state $s_N=\CF_{N-1}(s_{N-1},\design_{N-1},y_{N-1})$ can
    be computed along with the corresponding terminal reward
    $r_N(s_N)$. \revnew{Figure adapted from \citet{Shen_2023}.}}
  \label{f:MDP_diagram}
\end{figure}

The overall \rev{progression of an} MDP for the sequential design of $N$
experiments is depicted in Figure~\ref{f:MDP_diagram}\revnew{, adapted from \citet{Shen_2023}}.

\subsection{Problem statement}

The sOED problem entails finding a design policy
$\policy^{\ast}$ that maximizes the expected utility
$U(\policy)$:
\begin{align}
    \policy^{\ast} \in
  \rev{\argmax_{\policy=\{\mu_0,\ldots,\mu_{N-1}\}}} & \ \ 
    \left\{ U(\policy) \coloneqq
    \EE_{Y_{0:N-1}|\policy,s_0}\left[\sum_{k=0}^{N-1} \revarx{r_k(S_k,\design_k,Y_k)+r_N(S_N)}\right]\right\}
        \label{e:sOED}\\
    \revnew{\text{subject to}}\hspace{0.4em}
    & \ \ \ \revarx{\design_k = \mu_k(S_k) \in \CA_k,}  \nonumber\\
    & \ \ \   \revarx{S_{k+1}=\CF_k(S_k,\design_k,Y_k)\revnew{,}}
    \quad \text{for}\,\, k=0,\dots,N-1. \nonumber
\end{align}
Here \rev{$Y_{0:N-1}=Y_0,Y_1,\ldots,Y_{N-1}$. The} initial state $s_0$
is assumed known; if \rev{it is} not known, then another expectation can be taken
over \revarx{the random initial state $S_0$}. Optionally, a discount \rev{factor
$\gamma\in (0,1]$} \rev{can
  multiply} the \rev{reward at each successive stage (hence yielding
  terms $\gamma^k
  r_k$) to} 
artificially
reduce the \rev{value} of rewards obtained \rev{further in} the
future. The expected utility here is also known as the \emph{expected
  return} (or expected total reward) in MDP terminology; we will use
these terms interchangeably. 

\rev{In the field of reinforcement learning \citep{Kaelbling1996,Sutton2018}, problem} \eqref{e:sOED}
corresponds to a model-based planning problem, described by a finite-horizon belief-state MDP \rev{with} continuous action and observation
spaces. Embedded in each transition \rev{$\CF_k$ is a step of Bayesian
  inference, which can be quite expensive to perform, especially in
  nonlinear and non-Gaussian settings with computationally intensive
  likelihoods.}

\subsection{Solution approaches}
\label{ss:solution_approaches}

In general, problem \eqref{e:sOED} cannot be solved analytically. Different
numerical strategies \rev{must} be adopted to find an approximate solution.
Before we highlight \rev{representative} approaches from recent literature, we
introduce two types of value function that are central to the
MDP formulation, \rev{as many solution strategies are built on approximating
these functions.}

The \emph{action-value \rev{functions}} (or \emph{Q-\rev{functions}})
corresponding to a policy $\policy$ \rev{are}
\begin{align}
Q_k^{\policy}(s_k,\design_k)&=\EE_{Y_{k:(N-1)}|\policy,s_k,\design_k}\left[r_k(s_k,\design_k,\revarx{Y_k}) + \sum_{t=k+1}^{N-1} r_t\revarx{(S_t,\mu_{t}(S_t),Y_t)} + r_N(\revarx{S_N})\right]
\label{e:Q1}\\
&=\EE_{Y_k|\revarx{\pi},s_k,\design_k} \left[ r_k(s_k,\design_k,\revarx{Y_k}) + Q^{\policy}_{k+1}(\revarx{S_{k+1},\mu_{k+1}(S_{k+1})})\right],
\label{e:Q2}
\end{align}
for $k=0,\ldots,N-1$ and subject to
\revarx{$S_{t+1}=\CF_t(S_t,\design_t,Y_t)$}, \revarx{with $Q_{N}^{\policy}(s_N) = r_N(s_N)$}.  \rev{The value of the Q-function
  $Q_k^{\policy}(s_k,\design_k)$} is the \rev{expected remaining
  cumulative reward (i.e., the expected sum of all remaining rewards)}
for performing the $k$th experiment \rev{at} design $\design_k$
\rev{from} state $s_k$ and thereafter following policy
$\policy$.

The \emph{state-value \rev{functions}} (or \emph{V-\rev{functions}})
corresponding to a policy $\policy$ \rev{are} 
\begin{align}
V_k^{\policy}(s_k)&=\EE_{Y_{k:(N-1)}|\policy,s_k}\left[\revarx{r_k(s_k,\mu_k(s_k),Y_k)}+\sum_{t=k+1}^{N-1} r_t\revarx{(S_t,\mu_{t}(S_t),Y_t)} + r_N(\revarx{S_N})\right]
\label{e:V1}\\
 &= \EE_{Y_{k}|\policy,s_k} \left[ r_k\revarx{(s_k,\mu_{k}(s_k),Y_k)} + V^{\policy}_{k+1}(\revarx{S_{k+1}}) \right], \label{e:V2}
\end{align}
\revarx{for $k=0,\ldots,N-1$ and subject to
$S_{t+1}=\CF_t(S_t,\design_t,Y_t)$, with $V_N^{\policy}(s_N) = r_N(s_N)$}. \rev{The value of the V-function $V_k^{\policy}(s_k)$} is the \rev{expected
  remaining cumulative reward}
starting \rev{from a given} state $s_k$ and following policy $\policy$ for all
remaining experiments.  The expected utility in \eqref{e:sOED} can be
succinctly written as $U(\policy)=V_0^{\policy}(s_0)$.

\rev{The V-function} and Q-function are related to each other via
\begin{align}
V_k^{\policy}(s_k)=Q_k^{\policy}(s_k,\mu_{k}(s_k)),
\end{align}
and thus often only one of the two value functions needs to be
\rev{solved for}.
We note that both value functions can also be expressed in \rev{the}
recursive forms \eqref{e:Q2} and \eqref{e:V2}.  When the policy is the
optimal policy $\policy^{\ast}$ from \eqref{e:sOED}, the recursive
relations become the well-known \emph{Bellman optimality equations}:
\begin{align}
V_k^{\ast}(s_k) &= \max_{\design_k\in \CA_k} \EE_{Y_{k}|\design_k,s_k} \left[
  r_k(s_k,\design_{k},\revarx{Y_k}) + V^{\ast}_{k+1}(\revarx{S_{k+1})}
  \right],  \label{e:VB1} \\ %
 V_N^{\ast}(s_N) &= r_N(s_N), \label{e:VB2}
\end{align}
\revarx{for $k=0, \ldots, N-1$. In this form, the optimal policy
  $\policy^{\ast}$ is expressed implicitly, since the designs selected by
  $\policy^{\ast}$ are maximizers of \eqref{e:VB1}.}

Furthermore, if the reward terms are chosen to \rev{create the
  sequential analogue of $u^{\text{div}}$ in
  \eqref{e:udivergence}---i.e., to capture the total
expected information gain (EIG) in the model parameter $\Param$ from
\emph{all} experiments---}two natural formulations arise. A \emph{terminal
formulation} places in the terminal reward a single \revnew{Kullback--Leibler (KL)} divergence from
the initial prior to the final posterior:
\begin{align}
  r_k(s_k, \design_k, y_k) &= 0, \quad
  k=0,\ldots,N-1, \label{e:terminal1}\\ r_N(s_N) &= \DKL\left(
  \pdf_{\Param| \kern 0.05em \revarx{\info_N}} \vert \vert \pdf_{\Param} \right).
  \label{e:terminal2}
\end{align}
An \emph{incremental formulation} instead captures all
incremental KL divergence terms from each intermediate experiment's
prior to its corresponding posterior:
\begin{align}
  r_k(s_k, \design_k, y_k) &= \DKL\left(
  \pdf_{\Param| \kern 0.05em \revarx{\info_{k+1}}} \vert\vert \pdf_{\Param| \kern 0.05em \revarx{\info_k}} \right),
  \quad k=0,\ldots,N-1,
  \label{e:incremental1}\\
  r_N(s_N) &=
  0. \label{e:incremental2}
\end{align}
As pointed out in Theorem 1 of \citet{Shen_2021,Shen_2023} and
Theorem 1 of \citet{Foster_2021}, the expected utility
$U_T(\cdot)$ produced by substituting \eqref{e:terminal1}--\eqref{e:terminal2} into \eqref{e:sOED}, and the
expected utility $U_I(\cdot)$ produced by substituting
\eqref{e:incremental1}--\eqref{e:incremental2} into \eqref{e:sOED}\revnew{,} are
\emph{equal}: $U_T(\policy)=U_I(\policy)$, for any given policy
$\policy$. 
We can also show this equality using the chain rule of mutual
information. Starting from the incremental formulation, each immediate
reward's contribution to \eqref{e:sOED} is 
\begin{align}
  &
  \EE_{Y_{0:N-1}|\policy,s_0}\left[\DKL\left( \pdf_{\Param| \kern 0.05em \Info_{k+1}}
    \vert\vert \pdf_{\Param| \kern 0.05em \Info_k} \right)\right] \nonumber\\
  &=
  \EE_{Y_{0:k-1}|\policy,s_0} \, \EE_{Y_{k}|Y_{0:k-1},\policy,s_0}\left[\EE_{\Param| \kern 0.05em \Info_{k+1}}\left [\log\frac{\pdf(\revarx{\Param}| \kern 0.05em \Info_{k+1})}{\pdf(\revarx{\Param}|{\Info_k})}\right ] \right] \nonumber\\
  &=
  \EE_{Y_{0:k-1}|\policy,s_0} \, \EE_{\Param,Y_{k}|\design_k,\Info_k}\left [\log\frac{\pdf(\revarx{\Param,Y_k}|\design_k,{\Info_{k}})}{\pdf(\revarx{\Param}|{\Info_k})\pdf(\revarx{Y_k}|\design_k,{\Info_k})}\right ] \nonumber\\
  &=
  \EE_{Y_{0:k-1}|\design_{0:k-1}}\left[\DKL\left(\pdf_{\Param,Y_k|\design_k,\Info_{k}}
      \vert \vert \pdf_{\Param| \kern 0.05em \Info_k} \otimes \pdf_{Y_k|\design_k,\Info_k}\right) \right] \nonumber\\[6pt]
   &= \mathcal{I}(\Param;Y_k|\design_k,{\Info_{k}}),
     \label{e:incrementalCMI}
\end{align}
where all $\design_k$ follow from the given policy $\policy$. In the
\revnew{first} equality above, the expectation over \revnew{$Y_{k+1:N-1}$} collapses since the
immediate reward does not depend on these observations. \revnew{In the second
  equality, we use} the fact that 
$$\revnew{\EE_{Y_{k}|Y_{0:k-1},\policy,s_0}=
  \EE_{Y_{k}|Y_{0:k-1},\design_{0:k}}}=\EE_{Y_{k}|\design_k,\Info_k},$$
\revarx{where $\design_k$ follows from the policy $\policy$.}
Summing the conditional mutual information terms \eqref{e:incrementalCMI} according to \eqref{e:sOED} yields
\begin{align}
  U_I(\policy) = \sum_{k=0}^{N-1}
  \mathcal{I}(\Param;Y_k|\design_k,{\Info_{k}}) =
  \mathcal{I}(\Param;Y_{0:N-1}|\design_{0:N-1}) = U_T(\policy),
\end{align}
via the chain rule of mutual information.

We note that the incremental reward functions can be augmented with additional
terms, e.g.,
rewards reflecting the costs of
candidate designs, without affecting the correspondence of these two
ways of expressing total \revnew{EIG}.

Computationally, the terminal formulation
requires only a single KL divergence estimate \rev{per
trajectory, at its
terminal point. (A `trajectory' is a realization of the sequence
  of designs and observations, also known as an
  `episode' in MDP terminology\revnew{.)}} 
\rev{In contrast, the incremental} formulation needs many more intermediate
KL divergence calculations, which can be quite costly. On the other hand, the
terminal formulation is a case of \emph{delayed reward}, where the
feedback from the reward occurs only at the completion of all experiments, which can
make learning of the intermediate value functions more
difficult. 
Lastly, we note that a greedy design strategy requires calculating
\revnew{\emph{all}} intermediate posteriors and KL divergence terms, which is
  computationally much more expensive than the sOED terminal
formulation.

\vspace{1em} %
\subsubsection{Approximate dynamic programming (ADP-sOED)}
 
While approximate dynamic programming (ADP) can refer to a range of
computational techniques for finding an optimal policy
\citep{Powell2011}, we refer here to the approach introduced in
\citet{Huan_2015} and \citet{Huan_2016}, \rev{which centers on the idea of
  numerically approximating the optimal V-functions $V_k^{\ast}(s_k)$
  that satisfy the Bellman optimality equations \eqref{e:VB1} and \eqref{e:VB2},
  using some parametrized $\tilde{V}_k^{\ast}(s_k)$. We
  call this method ADP-sOED.}

In their work, the belief state $s_k^b$ is represented by discretizing
the posterior density function on an adaptive tensor-product grid, which 
\rev{is expanded/shrunk and refined/coarsened} based on 
local density values. Alternatively, \citet{Huan_2015} also
\rev{explores} representing all possible posterior distributions by constructing a triangular transport map
\citep{ElMoselhy_2012,Marzouk_2016} jointly on $(\design_k, Y_k,
\Param)$, where $\design_k$ is sampled from a probability distribution
with full support on $\CA_k$ and $(Y_k, \Param)$ are drawn from the
corresponding prior predictives at stage $k$. See more discussion of
such constructions in Section~\ref{ss:densities}.
In either case, a linear architecture
$\tilde{V}_k^{\ast}\revarx{(s_k)} =\sum_{i=1}^m \beta_{k,i}\psi_{k,i}(s_k)$ is used
to \rev{approximate the optimal V-functions}, where features
$\psi_{k,i}(s_k)$ are selected to be polynomial \revarx{functions} of \revnew{the} physical state and
of the posterior moments.

Once the representations of $s_k^b$ and
$\tilde{V}_k^{\ast}$ are chosen, a procedure known as approximate
value iteration (also referred to as backward induction, especially for
finite-horizon settings)
is used to build the approximate V-functions. The main
steps of the overall algorithm are summarized as follows.
\begin{enumerate}
\item \textit{Trajectory simulation.} Generate trajectories induced by the current
  approximate V-functions $\tilde{V}_k^{\ast}$, by choosing
  the design $ \tilde{\design}_k^{\ast}$ at each stage via
  \begin{align}
    \tilde{\design}_k^{\ast} \in \arg \max_{\design_k\in \CA_k}
    \EE_{Y_{k}|\design_k,s_k} \left[
      r_k\left(s_k,\design_{k},\revarx{Y_k}\right) +
      \tilde{V}^{\ast}_{k+1}\left(\CF_{k}(s_{k},\design_{k},\revarx{Y_k})\right)\right]. \label{e:ADP_sample}
  \end{align}
  To realize the corresponding state trajectories
  $(s^{(i)}_k)_{k=0}^N$ sequentially, \rev{where $i$ indexes 
    trajectories, } \revnew{an observation $y_k^{(i)}$ is simulated} at
  each stage using the statistical model \revnew{$\pdf\big(y_k \vert
    \param^{(i)}, \tilde{\design}_k^{\ast,(i)}, \revarx{\info_k^{(i)}}\big)$,} 
    the current design, and a
  realization $\param^{(i)}$ of
  $\Param$ generated
  from the prior \rev{at $k=0$} and then fixed for that entire trajectory; in other
  words, $\param^{(i)}$ serves as the true value of $\Param$ for that trajectory. Belief states are updated
  using the grid or transport representations noted above.
  An exploration policy, \rev{which simply generates random designs,}
  can be used if $\tilde{V}_{\revarx{k+1}}^{\ast}$   is not yet
  available, and otherwise may provide supplementary trajectory
  samples.

\item \textit{Approximate value iteration (backward induction).}
  Start from the final stage $k=N-1$ and evaluate \eqref{e:VB1} at the sample
  states $s_{N-1}^{(i)}$ generated in step 1, 
  \begin{align}
    \tilde{V}_{\text{tr}}^{(i)}=\max_{\design_k\in \CA_k} \EE_{Y_{k}|\design_k,\revarx{s_k^{(i)}}} \left[
      r_k\big(s_k^{(i)},\design_{k},\revarx{Y_k}\big) +
      \tilde{V}^{\ast}_{k+1} \big(\CF_{k}\big(s_{k}^{(i)},\design_{k},\revarx{Y_k}\big)\big)\right],
  \end{align}
  and then use these evaluations as training points to update the approximate V-\rev{functions}
  $\tilde{V}^{\ast}_{k}$ via linear regression:
  \begin{align}
    \big\{s_k^{(i)},\tilde{V}_{\text{tr}}^{(i)}\big\} \rightarrow
    \tilde{V}^{\ast}_{k}(s_{k}).
    \label{e:ADP_V_eval}
  \end{align}
  Once $\tilde{V}^{\ast}_{k}$ is updated, repeat the same process
  stepping backwards from $k=N-2, N-3,\ldots$ to $k=0$, thus 
  completing the update for all of the V-function approximations.

\item \textit{Refinement.} Optionally repeat steps 1--2 to improve the
  pool of trajectory samples and hence (state, value) pairs for
  training, using the newly updated V-function approximations. Once iterations are
  terminated, a final set of functions $\tilde{V}^{\ast}_{k}$ is returned, which
  can be used to evaluate approximate optimal design actions
  through \eqref{e:ADP_sample}.
\end{enumerate}
The approximation structure in \rev{\eqref{e:ADP_sample} is} known as
\emph{one-step lookahead} due to its \rev{invocation} of an approximation
function after one step of dynamic programming\footnote{For example,
two-step lookahead \rev{\citep[p.~304]{Bertsekas_2005}} would take the
form of \begin{align*} \tilde{\design}_k^{\ast} &\in \arg
  \max_{\design_k\in \CA_k} \EE_{Y_{k}|\design_k,s_k} \Bigg[
    r_k\left(s_k,\design_{k},\revarx{Y_k}\right) + \max_{\design_{k+1}\in \CA_{k+1}}
    \EE_{Y_{k+1}|\design_{k+1},\revarx{S_{k+1}}} \Big[
      r_{k+1}\left(\revarx{S_{k+1}},\design_{k+1},\revarx{Y_{k+1}}\right) \\ &\qquad
      \qquad
      +\check{V}^{\ast}_{k+2}\left(\CF_{k+1}(\revarx{S_{k+1}},\design_{k+1},\revarx{Y_{k+1}})\right)\Big]\Bigg]\end{align*}
for some approximate V-function $\check{V}^{\ast}_{k+2}$ two steps
ahead.} \rev{\citep{Bertsekas_2005}}. This is not to be
confused with a greedy or myopic design \rev{strategy that is based on
  truncating the problem \emph{horizon}}; doing so would not
incorporate value from any future experiments \rev{beyond that
  truncated horizon}.

It is important to note that both \eqref{e:ADP_sample} and
\eqref{e:ADP_V_eval} require solving a stochastic approximation
problem (e.g., \rev{using the} Robbins--Monro algorithm
\citep{Robbins1951}), since the expectation therein \rev{is
  typically} estimated using Monte Carlo. In fact, since the policy is
only implicitly represented by the V-functions, \rev{using} the
final policy still involves solving \eqref{e:ADP_sample}. As a result,
\rev{ADP-sOED is} quite computationally expensive.

\subsubsection{\rev{Actor-critic policy} gradient (PG-sOED)}
\label{sss:pgsoed}

In response to the heavy computations required by \rev{ADP-sOED, rooted
  in approximating the optimal V-functions in lieu of the policy},
faster \rev{methods} have been developed by explicitly representing
the policy and extracting its gradient, \rev{generally known as policy
  gradient (PG) approaches \citep{Sutton_2000}}. One such \rev{method}
is the \rev{PG}-based sOED (PG-sOED) introduced in \citet{Shen_2023}
(and its earlier version, \citealt{Shen_2021}), which makes
  use of actor-critic techniques \citep{Konda_1999,
    Peters_2008}---specifically, actor-critic techniques
  for deterministic \rev{policies with} deep neural network
  parameterizations \citep{Silver_2014,Lillicrap_2015,Mnih_2015}.

\rev{Suppose that each policy function $\mu_k$ is given a parametric
  representation $\mu_{k,w_k}$ with parameters $w_k$. Collecting
  these functions in}
$$\revnew{ \policy_{w} \coloneqq \{\mu_{0,w_0}, \mu_{1,w_1},\ldots,\mu_{N-1,w_{N-1}}\}\quad
\text{with}\quad w \coloneqq \{w_0,w_1,\ldots,w_{N-1}\},}$$
the sOED problem searching within the new
parameterized policy space now entails solving
$$\max_{w} U(\policy_w).$$
An expression for the policy
gradient can be \rev{derived \citep[Theorem~2]{Shen_2023}:}
\begin{align}
  \nabla_{w} U(\policy_{w}) = \sum_{k=0}^{N-1}
  \EE_{Y_k|\policy_w,s_0}\left[\nabla_{w} \mu_{k,w_k}(\revarx{S_k})
    \nabla_{\design_k}Q_{k}^{\policy_{w}}(\revarx{S_k},\design_k)\right] \label{e:PG}
\end{align}
with $\design_{k}=\mu_{k,w_k}(\revarx{S_k})$ \revarx{and $S_{k+1}=\mathcal{F}_k(S_k,\xi_k,Y_k)$}. 
The \rev{appearance of} $\nabla_{\design_k}Q_{k}^{\policy_{w}}$ in
\eqref{e:PG} further motivates a parameterization of the Q-functions,
similarly denoted by
$$\revnew{Q^{\policy_{w}}_{\eta} \coloneqq \big\{Q^{\policy_{w}}_{0,\eta_0},
Q^{\policy_{w}}_{1,\eta_1},\ldots,Q^{\policy_{w}}_{N,\eta_{N}}\big\} \quad
\text{with}
\quad \eta \coloneqq \{\eta_0,\eta_1,\ldots,\eta_{N}\}.}$$ \rev{S}imultaneously
learning the parameterized policy (the \textit{actor} $\policy_{w}$) and
the Q-\rev{functions} (the \textit{critic} $Q^{\policy_{w}}_{\eta}$) makes this
an \emph{actor-critic} \rev{method}.
Overall, access to the policy gradient opens the door to a wide
range of gradient-based optimization methods to iteratively improve
the policy $\policy_{w}$ \textit{en route} to maximizing $U(\policy_{w})$.

\rev{To this end, \citet{Shen_2023} use the policy gradient to develop
  several numerical methods for solving \eqref{e:sOED}. In what
  follows, the background information sequence $\Info_k$
  is always used to
  represent the state, as discussed in
  Section~\ref{ss:formulationMDP}.} First, a Monte Carlo estimator
of \eqref{e:PG} is formed via
\begin{align}
    \label{e:PG_MC}
    \nabla_w U(\policy_{w}) \approx \frac{1}{M} \sum_{i=1}^M
    \sum_{k=0}^{N-1} \nabla_w \mu_{k,w_k}\big(s^{(i)}_k\big)
    \nabla_{\design_k} Q^{\policy_w}_k\big(s^{(i)}_k,\design^{(i)}_k\big)
\end{align}
with $\design^{(i)}_k=\mu_{k,w_k}\big(s^{(i)}_k\big)$ computed from the
current policy. \rev{Here} for the $i$th trajectory,
a `true' data-generating $\param^{(i)}$ is drawn from the
prior $s_{0}^b$ and used to generate all subsequent $y_k^{(i)}$ via
the models $\pdf\big(y_k|\param^{(i)},\design_k^{(i)},\revarx{\info_k^{(i)}}\big)$ for
that entire trajectory. 

\begin{figure}[htb]
  \centering
  \includegraphics[width=1.0\linewidth]{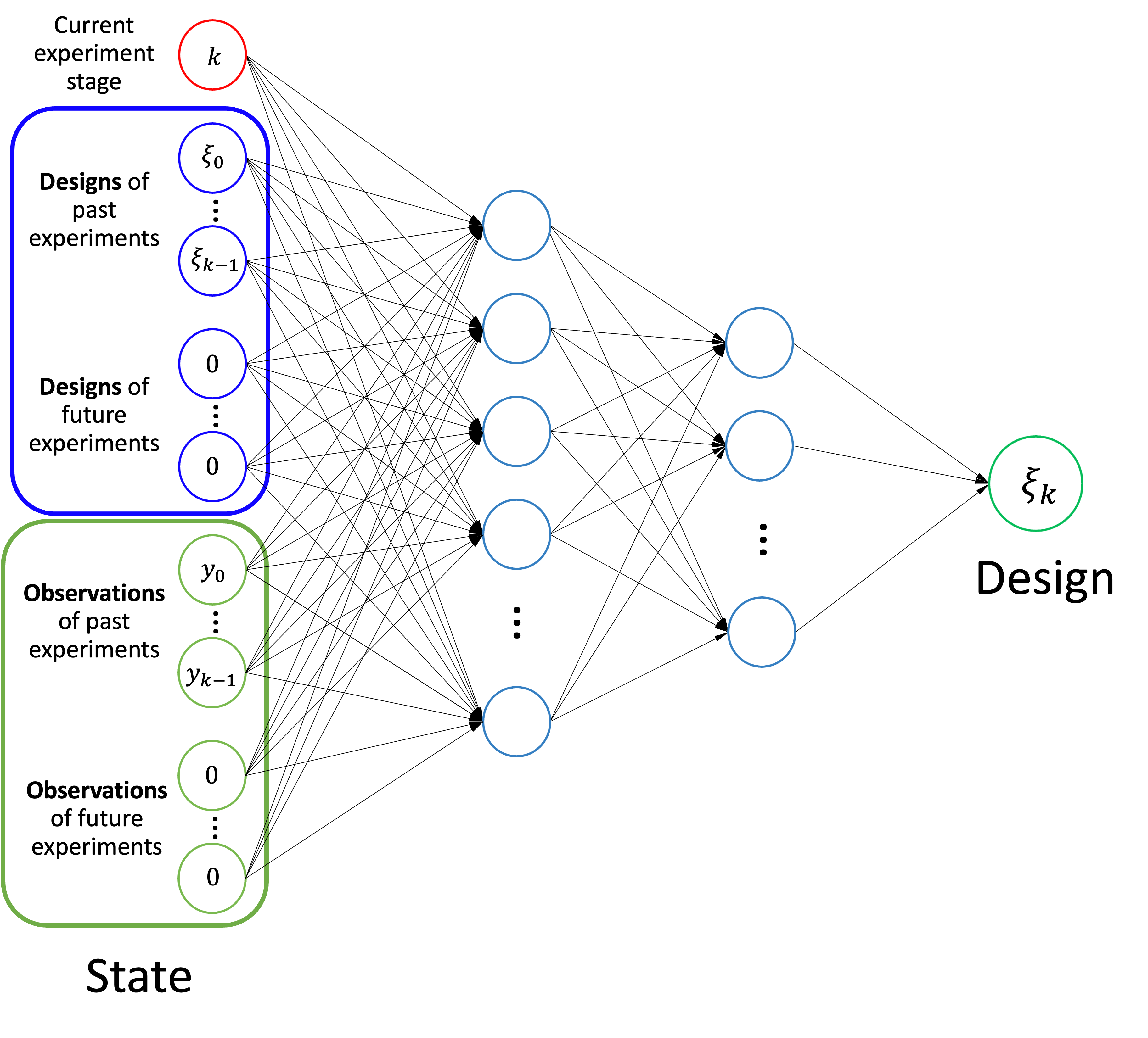}
  \caption{A policy is a mapping from state to design. In this DNN
    representation of the policy, its input entails the current
    experiment stage, and designs of past experiments and their resulting
    observations. The designs and observations of future experiments
    that have not yet taken place are padded with zeros.}
  \label{f:policy_network}
\end{figure}

Second, \rev{deep neural networks (DNNs)} are used to parameterize
\rev{both the policy and the Q-functions, following the ideas of deep
  Q-networks (DQN) \citep{Mnih_2015} and deep deterministic policy
  gradient (DDPG) \citep{Lillicrap_2015}}.
\rev{The policy  $\policy_w$, and hence all $\mu_{k,w_k}$ for
  $k=0,\ldots,N-1$, are represented by} a single \rev{DNN called the
  policy network}. \rev{The} consolidated policy network has an input
layer that takes in the state, specifically in the form of the
information sequence $\revarx{\info_k}$ and the current experimental stage $k$, as depicted in
Figure~\ref{f:policy_network}.
The stage $k$ can be represented either directly as an integer or via
its one-hot encoding \revarx{(i.e., a unit vector in $\mathbb{R}^N$)}.
Designs $\design_t$ and observations $y_t$ for future
experiments that have not yet taken place ($t\geq k$) are padded with
zeros. The output layer returns $\design_k$. The gradient of such a 
\rev{DNN}-based policy, which is needed to evaluate \eqref{e:PG_MC}, can be
computed efficiently via back-propagation. Note that the policy
network is \rev{therefore} not trained in a supervised learning manner, but
\rev{rather} by improving $w$ \textit{en route} to maximizing $U(\policy_w)$ in solving
\eqref{e:sOED}.

\rev{Similarly, the consolidated Q-function $Q^{\policy_{w}}_{\eta}$
  is represented by a separate, single DNN called the Q-network}.  The
Q-network is trained in a supervised learning manner from the Monte
Carlo \rev{trajectory samples} by minimizing a \rev{quadratic} loss derived from \eqref{e:Q2}:
\begin{align}
    \label{e:Qnet_loss}
    \ell(\eta) = \frac{1}{M} \sum_{i=1}^M \sum_{k=0}^{N-1}
    \left[ Q^{\policy_w}_{k,\eta_k}\big(s^{(i)}_k,\design^{(i)}_k\big) -
      \left(r_k\big(s^{(i)}_k,\design^{(i)}_k,y^{(i)}_k\big) +
      Q^{\policy_w}_{k+1}\big(s^{(i)}_{k+1},\design^{(i)}_{k+1}\big)\right)
      \right]^2
\end{align}
with $\design^{(i)}_k = \mu_{k,w_k}\big(s^{(i)}_k\big)$ computed from the
current policy.
Note that the last term in \eqref{e:Qnet_loss}, \revnew{$Q^{\policy_w}_{k+1}\big(s^{(i)}_{k+1},\design^{(i)}_{k+1}\big)$,} is
the \emph{true} Q-function as defined in \eqref{e:Q1}
and thus
does not depend on $\eta$. The true Q-function is usually not
available, however, and therefore this term is typically replaced by the Q-function {approximation} at the current iteration, \revnew{$Q^{\policy_w}_{k+1,\eta_{k+1}}\big(s^{(i)}_{k+1},\design^{(i)}_{k+1}\big)$,}
but with its contribution to the $\eta$-gradient of the loss in
\eqref{e:Qnet_loss} ignored.
\revnew{The trained Q-network $Q^{\policy_{w}}_{\eta}$ from
  \eqref{e:Qnet_loss} is then used to
  replace the true Q-function in evaluations of \eqref{e:PG_MC}. 
  Notably, this PG computation does not require evaluating gradients of the
   likelihood (or the underlying simulation model) with respect to $\design_k$, since the gradient operator
  now only needs to act on the Q-network. }

\rev{Third}, the information-based immediate \revnew{and/or} terminal rewards
$r_k$ and $r_N$, which require evaluation of the KL divergence based
on the belief state, are \rev{calculated by directly approximating
  integrals involving the relevant unnormalized densities, via
  numerical quadrature.}
This technique,
however, is only practical for low-dimensional $\revarx{\Param}$ (e.g., $p
\leq 4$).

Assembling these numerical methods, an optimal policy \rev{is} sought
using gradient-based optimization such as stochastic gradient
ascent. The main steps of the overall algorithm are summarized as
follows.
\begin{enumerate}
\item \textit{Trajectory simulation.} Generate trajectories. \rev{For
  each trajectory, first draw} $\param$ from the prior, then for each
  of the $k=0,\dots,\revnew{N-1}$ experiments in the trajectory,  sequentially \rev{compute}
  $\design_k$ from the current policy and draw $y_k$ from its
  \rev{statistical model}. Calculate the \rev{associated} trajectory of
  states; for example, if using $\Info_k$ as the state, simply store
  $\revarx{\info_N}$, from
  which all $\revarx{\info_{k< N}}$ can be easily extracted. Compute the corresponding
  immediate and terminal rewards $r_k$ \revnew{and $r_N$}.

\item \textit{Value function update.} Update the Q-network by finding an
  $\eta$ that minimizes the loss in \eqref{e:Qnet_loss}.

\item \textit{Policy update.} Estimate $\nabla_wU(w)$ through
  \eqref{e:PG_MC} \revnew{but using the Q-network,} and update the policy network through, for example,
  gradient ascent $w \revarx{\leftarrow} w + \alpha \nabla_wU(w)$, where $\alpha$ is the
  learning rate.
\item \textit{Refinement.} Repeat steps 1--3 to improve the trajectory
  sample pool with the newly updated policy \rev{network} and
  Q-\rev{network}. Once terminated, a final policy network is
  returned.
\end{enumerate}

\rev{Because of the many complex approximations in these algorithms, it is
useful to validate them in simple settings where an optimal policy can
be derived analytically. To this end, we show how ADP-sOED and PG-sOED
perform on a two-experiment linear-Gaussian benchmark problem
\citep{Huan_2016,Shen_2023}. The model} is a
simplified version of \eqref{eq:linearGaussianmodel}:
\begin{align}
    Y_k = \design_k \rev{\Param} + \rev{\mathcal{E}_k}
\end{align}
with $\rev{\mathcal{E}_k} \sim \mathcal{N}(0,1^2)$ and no physical
state.  The benchmark \rev{entails} $N=2$ experiments, with prior
$\Param\sim\mathcal{N}(0,3^2)$ and \rev{designs constrained to lie in $d_k \in
[0.1,3]$. The conjugate prior ensures that all subsequent posteriors
are Gaussian,} following \eqref{e:LG_post_cov} and
\eqref{e:LG_post_mean}.  The rewards are set to
\begin{align}
    r_k(s_k,\design_k,y_k) &= 0, \quad k=0,1, \label{e:LG_immediate}\\
    r_N(s_N) &= \DKL\left( \pdf_{\Param| \kern 0.05em \revarx{\info_N}} \vert\vert \pdf_{\Param}
    \right) - 2\left( \log{\sigma_N^2} - \log{2}
    \right)^2, \label{e:LG_terminal} 
\end{align}
where $\sigma_N^2$ is the variance of the final posterior, and the
additive penalty in $r_N$ is purposefully inserted to make the problem
more challenging. A derivation of the  resulting optimal policies can
be found in \citet[Appendix B]{Huan_2015}, with $U(\policy^\ast)\approx0.783$.

\rev{In this particular linear-Gaussian problem, sOED is equivalent to
  batch OED.  Upon substituting the terminal reward
  \eqref{e:LG_terminal} into the sOED objective \eqref{e:sOED} and
  taking an expectation over $Y_0, Y_1 \vert \policy$, the objective
  depends explicitly \emph{only} on the final posterior variance.  The
  portion of the objective following from the KL divergence term in
  \eqref{e:LG_terminal} only involves the posterior variance, as shown
  by \eqref{e:LG_EIG1}--\eqref{e:LG_EIG2}, and the portion resulting
  from the penalty term in \eqref{e:LG_terminal} only involves the
  posterior variance by construction. While this variance depends on
  the designs chosen by the policy, in the linear-Gaussian setting it
  is independent of the realized values of $Y_{0:N-1}$, $y_{0:N-1}$. Consequently,
  feedback or adaptation in response to $y_{0:N-1}$ would not affect
  the expected utility, and sOED and batch OED therefore coincide. See
  \citet[Appendix B]{Huan_2015} for a detailed derivation of this
  equivalence.}

\begin{figure}[tb]
  \centering
  \subfloat[ADP-sOED]{\includegraphics[width=0.49\linewidth]{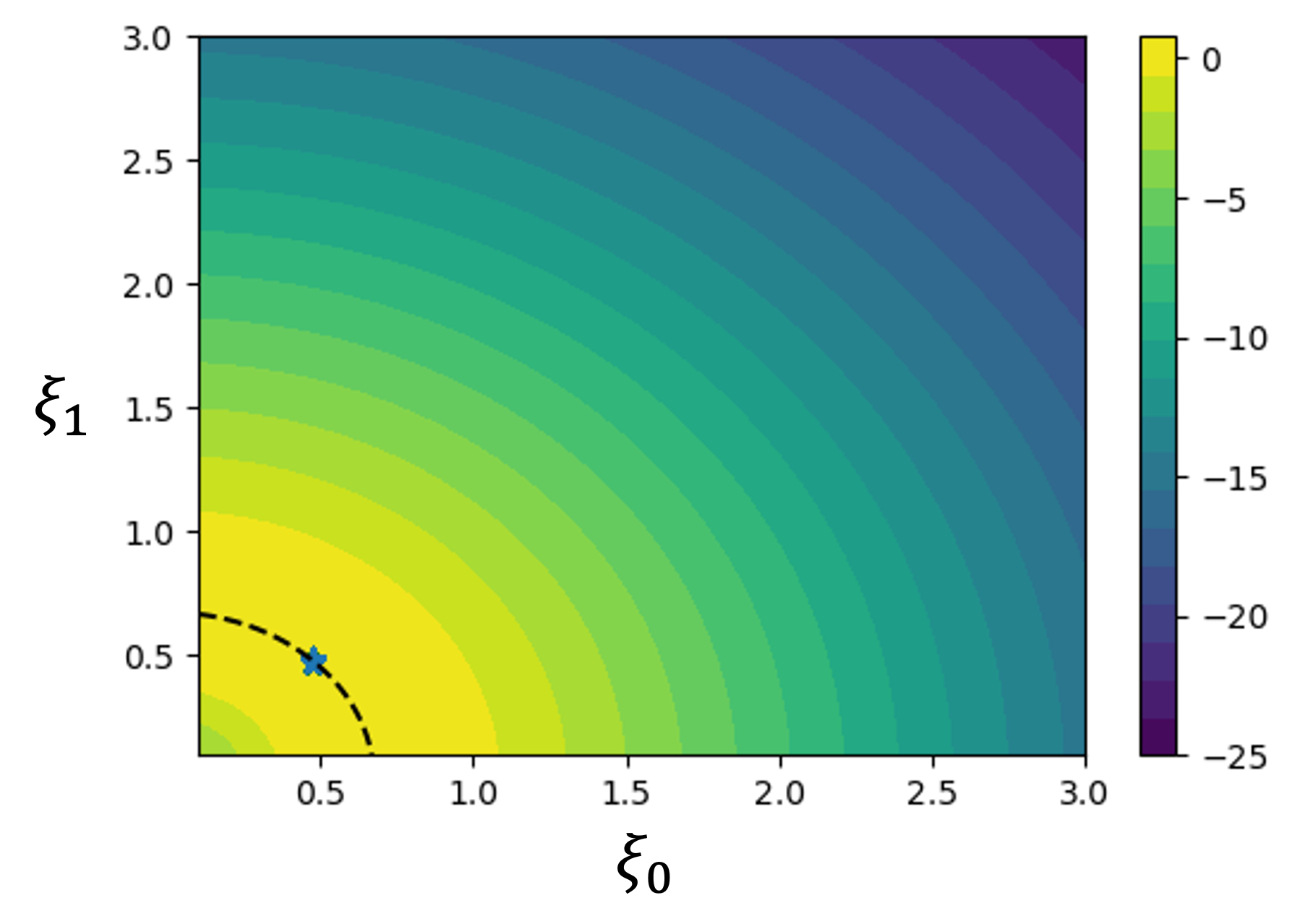}\label{f:sOED_linear_Gaussian_ADP}}
  \subfloat[PG-sOED]{\includegraphics[width=0.49\linewidth]{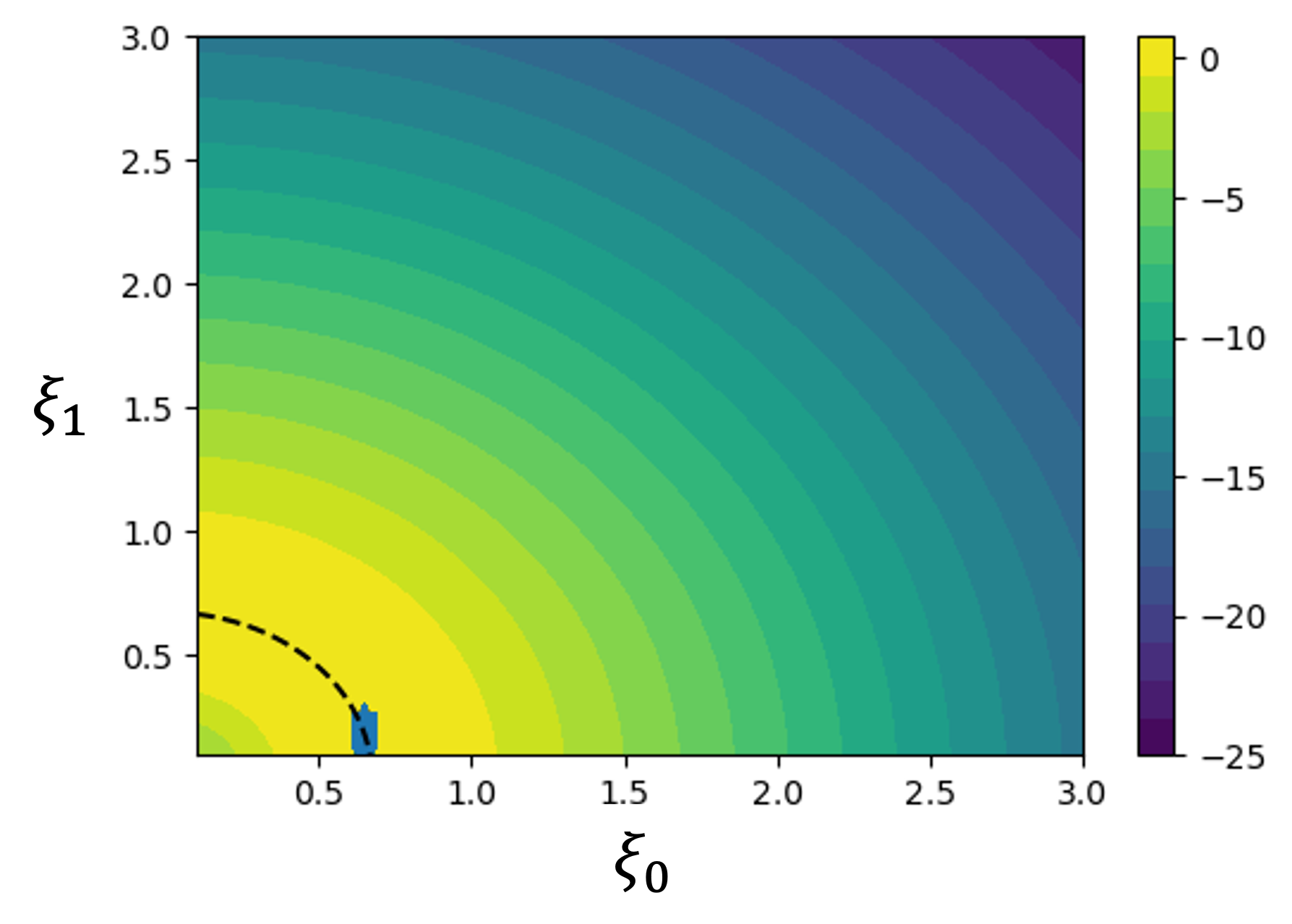}\label{f:sOED_linear_Gaussian_PG}}
  \caption{Expected utility contours for the linear-Gaussian benchmark
    with the dashed curve showing the set of optimal designs. The
    cluster of points in each plot indicates the designs selected by
    (a) ADP-sOED and (b) PG-sOED policies under repeated trials. Both
    algorithms arrived at the optimal set and achieved expected
    utility values consistent with the analytic optimal \revnew{policy:
    $U(\policy^{\ast}_{\text{ADP}})\approx 0.775$,
    $U(\policy^{\ast}_{\text{PG}})\approx 0.775$,
    $U(\policy^\ast)\approx 0.783$}.}
  \label{f:sOED_linear_Gaussian}
\end{figure}

\revnew{Figure~\ref{f:sOED_linear_Gaussian}
  illustrates}
 the expected utility for different design choices via
  colored contours, with the dashed curve illustrating the set
  of optimal designs. \revnew{The cluster of points in each plot indicates} the
  designs selected by the ADP-sOED and PG-sOED policies under repeated
  trials. Both algorithms arrive at the optimal set and
  achieve expected utility values consistent with the analytic
  optimal policy: $U(\policy^{\ast}_{\text{ADP}}) \approx 0.775$,
  $U(\policy^{\ast}_{\text{PG}}) \approx 0.775$,
  $U(\policy^\ast) \approx 0.783$.
  The computational times reported in Table~\ref{t:ADPvsPG}\revnew{,
    adapted from \citet{Shen_2023},} are
  obtained using a single 2.6 GHz CPU on a MacBook Pro laptop.  The
  timing values reflect 30 gradient ascent updates for PG-sOED in the
  training stage, and one policy update (the minimum needed) for
  ADP-sOED.
  PG-sOED is orders of magnitude faster than ADP-sOED, especially in
  testing times, making it suitable for
  applications that have real-time requirements. This drastic
  difference is due to ADP-sOED being a value-based (critic-only)
  approach wherein the policy (actor) is not explicitly represented,
  such that evaluating the policy requires solving a (stochastic)
  optimization problem.
In contrast, applying PG-sOED  requires only a single forward pass through
the policy network, without any additional optimization runs or forward model
evaluations.

\begin{table}[bt]
    \centering
    \caption{Comparison of computational costs between ADP-sOED and
      PG-sOED for the linear-Gaussian benchmark. \revnew{Data adapted from \citet{Shen_2023}.}}
    \label{t:ADPvsPG}
    \begin{tabular}{l| r r |r}
    \hline\hline
    & {Training time (s)} & {Forward model evaluations} & {Testing time (s)} \\
    \hline
    ADP-sOED & 837 %
    & $5.3 \times 10^8$ %
    & 24,396 %
    \\
    PG-sOED & 24 %
    & $3.1 \times 10^6$ %
    & 4 %
    \\
    \hline
    \end{tabular}
\end{table}

\begin{figure}[tb]
  \centering
  \includegraphics[width=0.99\linewidth]{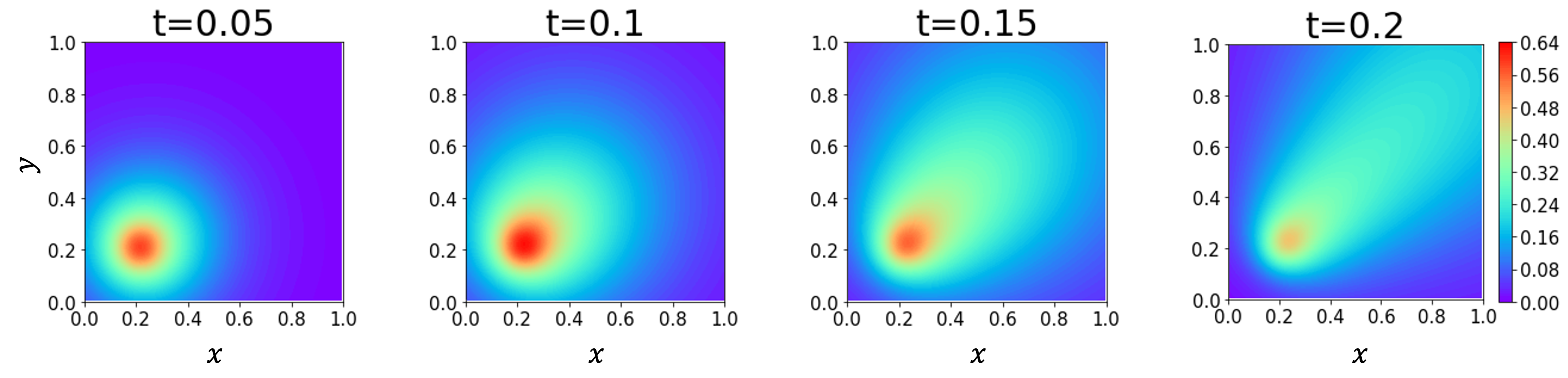}
  \caption{An example time-evolution of a convection-diffusion
    field. The contours show the concentration of the plume. \revnew{Figure
    adapted from \citet{Shen_2023}.}}
  \label{f:Source_example}
\end{figure}

\begin{figure}[tb]
  \centering
  \includegraphics[width=0.99\linewidth]{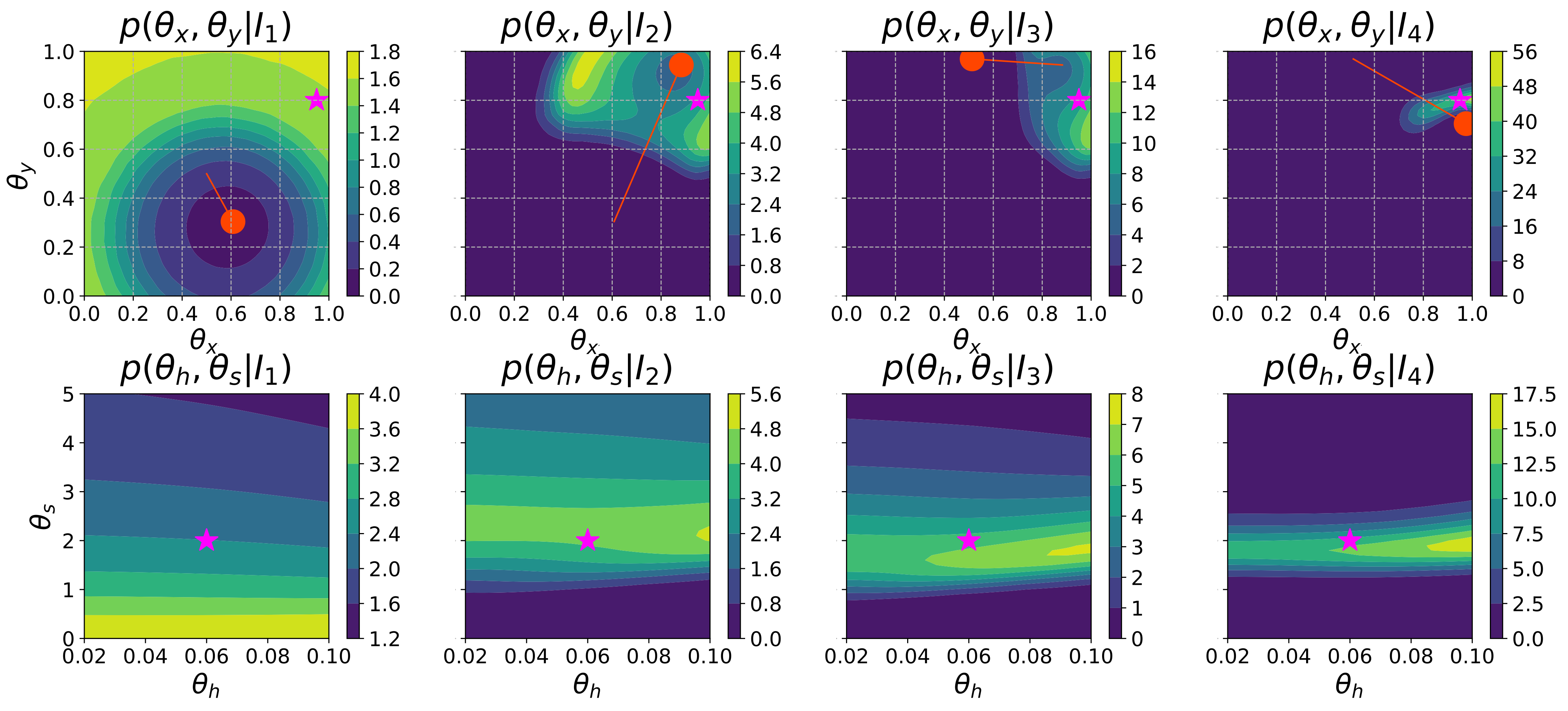}
  \caption{Sequence of marginal posterior densities (unknown source
    locations $\revarx{\Param_x,\Param_y}$ on the first row, unknown source width
    and strength $\revarx{\Param_h,\Param_s}$ on the second row) from an example
    trajectory instance using PG-sOED. The purple star represents the
    true data-generating $\param$ value, the red dot represents the
    physical state (\revarx{sensor} location), and the red line
    segment \revarx{depicts the design, which is the displacement of the sensor
    from its previous location}. \revnew{Figure
    adapted from \citet{Shen_2023}.}}
  \label{f:Source3_contour}
\end{figure}

\rev{Figures~\ref{f:Source_example}--\ref{f:Source3_contour}\revnew{, both
  adapted from \citet{Shen_2023},} illustrate the
application of PG-sOED to a problem of mobile sensor guidance in a
two-dimensional advection-diffusion \revnew{partial differential equation}. The location of the center of
a source term (e.g., emitting some contaminant) in the
advection-diffusion problem is uncertain, along with the strength and
radius of the source. Figure~\ref{f:Source_example} shows an example
of the contaminant plume concentration evolving in time, with the
advection velocity pointing up and to the right.
The design variables
are the displacements of the mobile sensor from one stage to the next. 
The inference goal is to learn the unknown source location, source
strength, and radius of the source, from noisy measurements of
contaminant concentration at successive times. All four of these
parameters are endowed with uniform priors, and the velocity field
advecting the contaminant field is assumed known. The rewards for the
sOED problem include the joint information gain (KL divergence from
the prior to the posterior after four experiments) in all four
parameters, instituted in the terminal formulation manner,
along with a negative stagewise reward corresponding
to a quadratic penalty on sensor movement from one stage to the
next. Figure~\ref{f:Source3_contour} illustrates an application of the
resulting PG-sOED \revnew{policy for} one particular realization of the
experimental sequence. The top row shows marginal posterior densities
of the source location $\revarx{(\Param_x, \Param_y)}$, while the bottom row
shows marginals of the source radius $\revarx{\Param_h}$ and source strength
$\revarx{\Param_s}$. The movements of the sensor, i.e., the chosen designs, are
visualized in the top row (red dots and \revnew{lines) along} with the true
source location (fixed purple star).}

\begin{figure}[htbp]
  \centering
  \subfloat[Batch]{\includegraphics[width=0.45\linewidth]{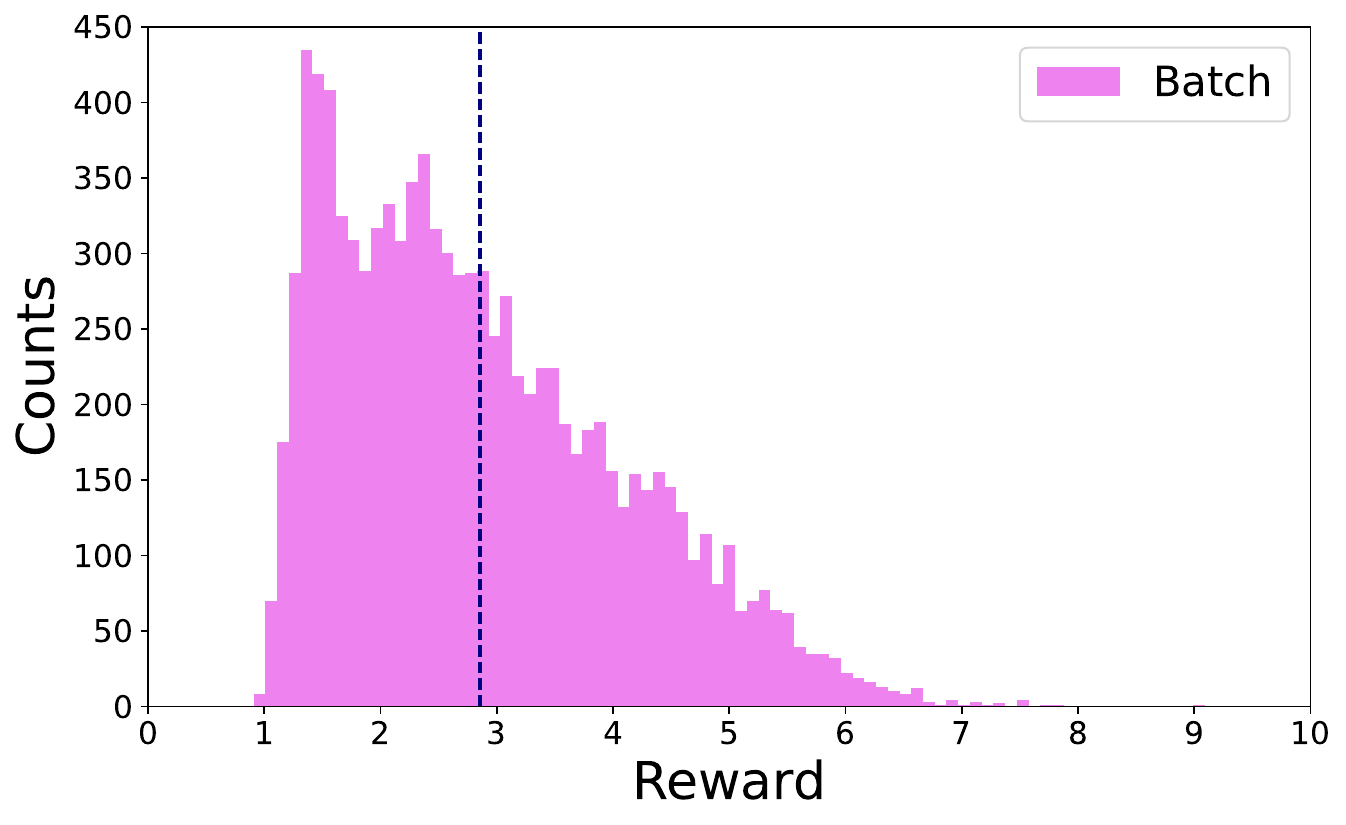}}
  \subfloat[Greedy]{\includegraphics[width=0.45\linewidth]{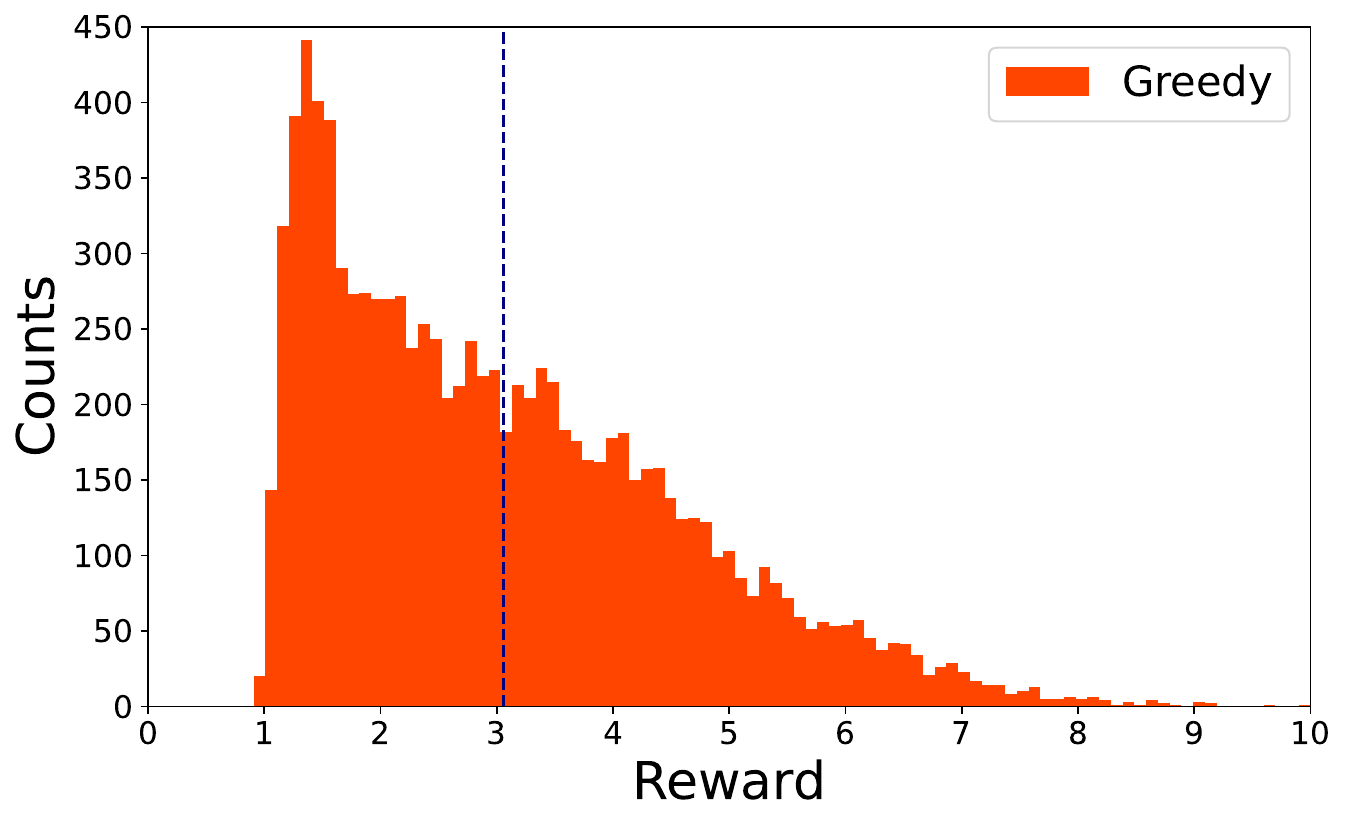}}\\
  \subfloat[PG-sOED]{\includegraphics[width=0.45\linewidth]{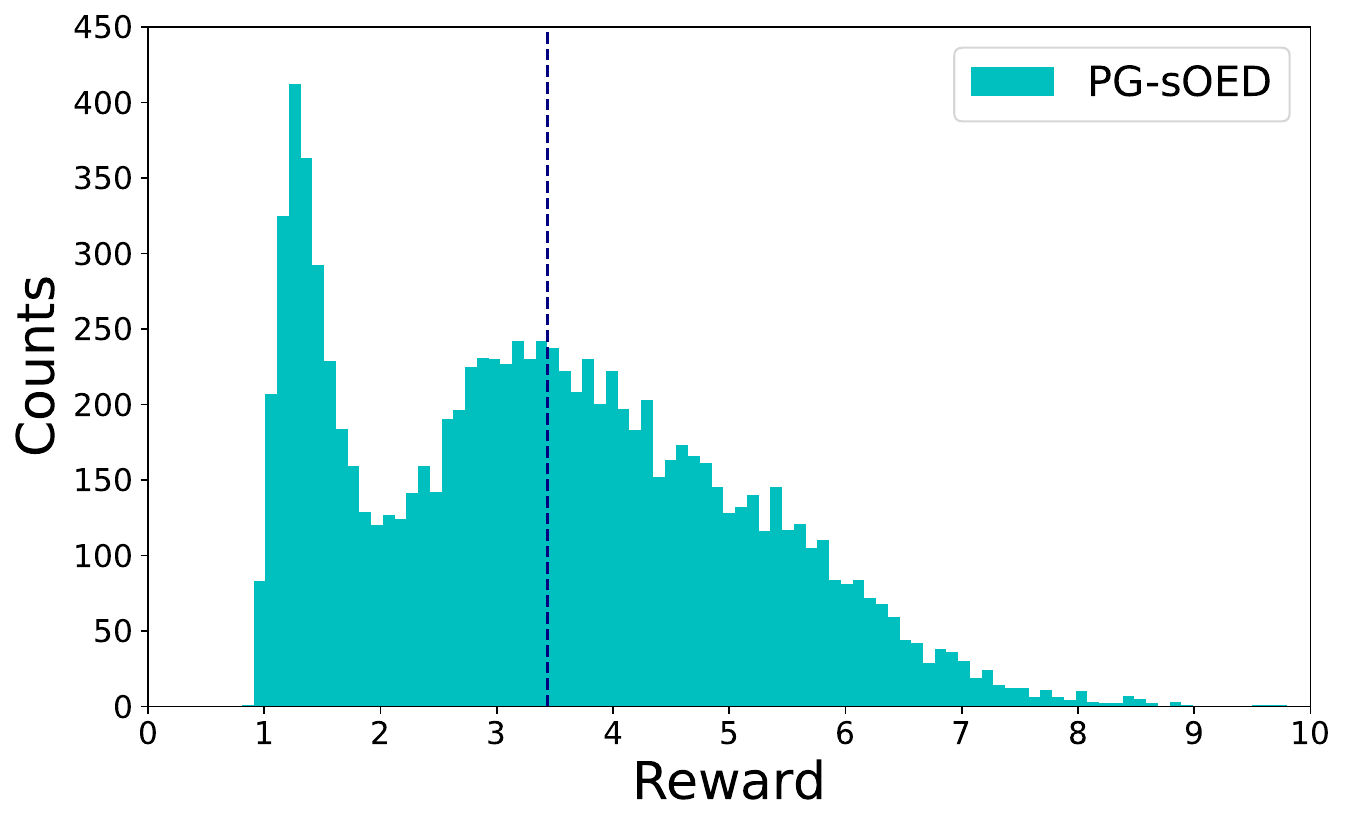}}
  \caption{Histograms of total rewards from $10^4$ test trajectories
    generated using batch, greedy, and PG-sOED policies, with
    respective expected total reward values (indicated by vertical
    lines) \revnew{$U(\policy^{\ast}_{\text{batch}})\approx 2.856$,
$U(\policy^{\ast}_{\text{greedy}}) \approx 3.057$, and
    $U(\policy_{\text{PG}}) \approx 3.435$}. \revnew{Figure adapted from \citet{Shen_2023}.}}
  \label{f:conv_diff_hist}
\end{figure}

\rev{Recall that the objective \eqref{e:sOED} being maximized involves an expectation over
$\Param$ and over all values of $Y_{0:N-1}$ realized under the
policy. While Figure~\ref{f:Source3_contour} shows a single trajectory
of experiments, a policy is designed to work well, on average,
over all possible trajectories. Thus, the most comprehensive way of
assessing the effectiveness of a policy is to study the reward it
produces over many realized trajectories.
Figure~\ref{f:conv_diff_hist}\revnew{, adapted from \citet{Shen_2023},} presents histograms of the total rewards
obtained from $10^4$ trajectories using the batch, greedy, and PG-sOED
policies. Their expected values, indicated by the vertical black lines, are respectively
\revnew{$U(\policy^{\ast}_{\text{batch}})\approx 2.856$,
$U(\policy^{\ast}_{\text{greedy}})\approx 3.057$, and
$U(\policy_{\text{PG}})\approx 3.435$,} with 
\revnew{PG-sOED achieving} the highest expected reward.
In this example, greedy design tends to `chase after' the
most recent estimate of the source location. Batch design can plan
ahead and take advantage of knowing where the plume will advect in
future experiments, but is unable to adapt to new measurements. PG-sOED
can do both.}

\subsubsection{Methods that leverage information bounds}
\label{sss:sOED_bounds}

\rev{Another prominent series of sequential experimental design
  algorithms, developed around the same time as those discussed in
  Section~\ref{sss:pgsoed}, leverage various information bounds.
  Here we briefly summarize these methods.}

\paragraph{DAD and related methods.} \citet{Foster_2021} introduced a technique they named
deep adaptive design (DAD).
Like PG-sOED, DAD performs the bulk of its computation offline,
constructing a \rev{DNN}-based policy that can quickly produce the next design online as data are
realized. Unlike PG-sOED, however, DAD does not use an actor-critic
approach. For a discussion of the relative merits of actor-critic
algorithms and more direct \revnew{PG} approaches, see
\citet{Konda_1999} and \citet[Chapter~13]{Sutton2018}.
The expected utility $U(\policy)$ in DAD is the \revnew{total EIG in
  the parameters $\Param$} from prior to
posterior after $N$ experiments, which we call
$U_{\text{KL}}(\policy)$ and is equivalent to the mutual information
$\mathcal{I}(Y_{0:N-1};\Param\vert \policy)$. 
DAD \textit{directly} targets maximization of the EIG,
by first expressing it in terms of log-density ratio of the
observations $Y_{0:N-1}$, that is,
\begin{align}
  U_{\text{KL}}(\policy)=
  \mathcal{I}(Y_{0:N-1};\Param\vert \policy)
  &=
  \EE_{Y_{0:N-1}|\policy,s_0}\left[D_{\text{KL}}(\pdf_{\Param| \kern 0.05em \Info_N} ||
    \pdf_{\Param})\right]\nonumber\\
  &=
  \EE_{Y_{0:N-1}|\Param,\policy,s_0}\EE_{\Param}\left[\log\frac{\pdf(Y_{0:N-1}|\Param,\design_{0:N-1})}{\pdf(Y_{0:N-1}|
      \design_{0:N-1})}\right],
\end{align} 
where $\design_{k}=\mu_k(\revarx{S_k})$
follows the given policy
$\policy$.
A sequential version of the prior contrastive estimator (PCE)
(cf.~\eqref{e:PCE}), called the sequential PCE (sPCE), is then
introduced, \revnew{taking the form:}
\begin{align}
  \mathcal{I}(Y_{0:N-1};\Param\vert \policy)  &\geq \EE_{Y_{0:N-1}|\policy, \Param_1} \EE_{\Param_1} \EE_{\Param_{2:M}} \left[\log
    \frac{\pdf(Y_{0:N-1}|\Param_1,\design_{0:N-1})}{\frac{1}{M}\sum_{j=1}^M
      \pdf(Y_{0:N-1}|\Param_{j},\design_{0:N-1})} \right] \nonumber\\
  &\eqqcolon
  \mathcal{L}^{\text{sPCE}}(\policy;M),  \label{e:sPCE}
\end{align}
where $\Param_1$ is the data-generating parameter for
$\design_{0:N-1},\revarx{Y_{0:N-1}}$
in the equation above with
$\design_{k}=\mu_k(\revarx{S_k})$ following the given policy $\policy$; the
subscripts for $Y$ and $\design$ refer to the experiment (stage)
index, while those for $\Param$ refer to the multi-sample index of sPCE as
introduced in Section~\ref{sss:multi_sample}.
The key here is that the value of the \emph{data-generating} $\Param_1$
is included in the Monte Carlo estimate
of the evidence in the denominator; including this value ensures that
\eqref{e:sPCE} is a lower bound to $\mathcal{I}(Y_{0:N-1};\Param\vert \policy)$ for any
$\policy$, which becomes tight as $M \to \infty$
\citep[Theorem~2]{Foster_2021}. (When $\Param_1$ is excluded from the
evidence estimate, \eqref{e:sPCE} 
reverts to the standard nested Monte Carlo estimator of EIG.)
Approximating the expectations in sPCE with Monte Carlo sampling from
the prior and from the statistical models for successive
$(Y_k)_{k=0}^{N-1}$ yields in the end a \textit{negatively biased
  Monte Carlo estimator} of $\mathcal{I}(Y_{0:N-1};\Param\vert \policy)$.

When the policy $\policy$ is parameterized as $\policy_w$, DAD seeks a
policy that satisfies
$$\max_{w} \mathcal{L}^{\text{sPCE}}(\policy_w; M).$$
The 
gradient of the
sPCE bound \rev{can be} obtained by moving the \rev{gradient} operator $\nabla_w$
inside the expectations. This may require first reparameterizing the
stochasticity in the observations to be independent of the policy.
(For instance, with a statistical model $Y_k=G_k(\Param,\design_k)+
\mathcal{E}_k $, an expectation over $Y_k$ can be rewritten as an
expectation over $\mathcal{E}_k$; if the distribution of
$\mathcal{E}_k$ is functionally independent of the design $\design_k$,
then the goal is already achieved. More generally, if the distribution of $\mathcal{E}_k$ does depend on $\design_k$, then the expectation can always be rewritten in terms of some other random variable whose distribution is independent of $\design_k$ and hence of $\policy$.)
The sPCE gradient is then
\begin{align}
  \rev{\nabla_{w}\revnew{\mathcal{L}^{\text{sPCE}}}(\policy_w;M) = \EE_{\mathcal{E}_{1:N}}\EE_{\Param_{1:M}} \left[\nabla_{w}\left(\log
    \frac{\pdf(Y_{0:N-1}|\Param_1,\design_{0:N-1})}{\frac{1}{M}\sum_{j=1}^M
      \pdf(Y_{0:N-1}|\Param_{j},\design_{0:N-1})} \right)\right]}
\end{align}
with $\design_{k}=\mu_{k,w_k}(\revarx{S_k})$ following the given policy $\policy_w$.
With this gradient in hand, one can use any gradient-based
optimization scheme, such as stochastic gradient ascent, to find a
parameterized policy that maximizes the estimated lower bound. With
regard to how to choose the parameterization $\policy_w$,
\citet{Foster_2021} recommend a `pooling-emitter'
\revnew{DNN} architecture, motivated by the permutation
invariance of the EIG under certain conditions (e.g., if all
the observations in the sequence are conditionally independent given
the parameters, and described by the same parametric statistical model).

\citet{Blau_2022} build on the sPCE bound from DAD, but seek a
stochastic policy via the randomized ensembled double Q-learning
approach of \citet{Chen_2021}, which is an actor-critic method. They
demonstrate applicability of this approach to discrete design spaces.

\paragraph{iDAD.} All of the sequential design methods discussed so
far require the ability to explicitly evaluate the likelihood (i.e.,
to evaluate the \rev{probability {densities}
  $\pdf(y_k|\param,\design_k,\revarx{\info_k})$}).
In certain problems, however, the likelihood may be only implicitly
defined and otherwise intractable to compute; yet sampling \revarx{$Y_k$}  from
the model may still be possible. To accommodate such 
models in sOED, a \rev{variation of} DAD, called implicit deep
adaptive design (iDAD), has been developed by \citet{Ivanova_2021}.

iDAD introduces several likelihood-free lower bounds.  The
main idea behind these bounds can be understood as
approximating the posterior-to-prior or likelihood-to-evidence density
ratio through a parameterized `critic' function, denoted here by
$f : \paramset \times \Xi_{0:N-1}\times \mathcal{Y}_{0:N-1}
\rightarrow \RR$.
Note that this notion of a critic \emph{differs} from the `critic'
in the actor-critic methods of Section~\ref{sss:pgsoed} (where the
role of the critic was played by the Q-function).
In these lower bound expressions, any choice of $f$ yields a tractable lower bound for the EIG;
thus $f$ can act as the `knob' to maximize the lower bound. One
\rev{example is} the sequential design version of the NWJ bound (\revnew{cf.}~\eqref{e:NWJ})
\begin{align}
\mathcal{I}(Y_{0:N-1};\Param\vert \policy) &\geq 
  \EE_{\Param}\EE_{Y_{0:N-1}|\Param,\policy} \left[f(\Param,\Info_N)\right] -
  \frac{1}{e} \EE_{Y_{0:N-1}|\policy}\EE_{\Param} \left[\exp f(
    \Param,\Info_N)\right] \nonumber\\
  &\eqqcolon \revnew{\mathcal{L}^{\text{sNWJ}}}(\policy; f), \label{e:iDAD_NWJ}
\end{align}
while another is the InfoNCE bound (cf.~\eqref{e:NCE})
\begin{align}
  \mathcal{I}(Y_{0:N-1};\Param\vert \policy) &\geq  
  \EE_{\Param_1}\EE_{Y_{0:N-1}|\Param_1,\policy}\EE_{\Param_{2:M}} \left[\log
    \frac{\exp f(\Param_0,\Info_N)}{\frac{1}{M}\sum_{j=1}^M
      \exp f(\Param_{j},\Info_N)} \right] \nonumber\\
  &\eqqcolon \revnew{\mathcal{L}^{\text{sNCE}}}(\policy;M,f).
\end{align}
Both bounds can be tight for an optimal selection of the critic
function $f$
and as $M\rightarrow \infty$ \citep{Ivanova_2021}.  \rev{Importantly,
  these bounds no longer involve the densities $\pdf(y_k\vert
  \param,\design_k, \revarx{\info_k})$.}  Upon parameterizing the policy $\policy$ as 
$\policy_w$ and critic $f$ as $f_{\phi}$, a policy
can be found by maximizing (tightening) the lower bound simultaneously
over $w$ and $\phi$. For example, in the NWJ case:
$$\max_{w,\phi} \revnew{\mathcal{L}^{\text{sNWJ}}}(\policy_w;f_{\phi}).$$
\rev{Similar} to DAD, the gradient of the lower bound with respect to both the policy and
critic parameters can be obtained by bringing the \rev{gradient} inside
the expectation upon reparameterizing the observations,
allowing for stochastic gradient ascent updates to tighten the
bound. We note that while iDAD does not require explicit likelihoods
and relies only on a simulator of $Y_k$, it does require access to
derivatives of the output of this simulator with respect to the design
$\design_k$, and with respect to all previous experiments' designs and
outcomes, $\revarx{\info_k}$.

\paragraph{vsOED.}

The variational sequential OED (vsOED) method in \citet{Shen_2023a}
derives a lower bound for the EIG by employing variational
approximations of the relevant posterior distributions
and then substituting these approximations in `one-point'
\revarx{reward functions that replace the KL divergence rewards, which we shall describe below.}
The framework
generalizes the objective functions used in previous approaches
in that it simultaneously accommodates multiple models, nuisance
parameters, predictive quantities of interest, and implicit
likelihoods.

\rev{Let $\mathcal{M}_m$ be a countable set of models indexed by $m=1,
  2, \ldots$, where each model has its own parameters \revarx{$\Param_m \in \paramset_m
  \subseteq \RR^{p_m}$} and predictive \revnew{quantity of interest
    (QoI)}
  \revarx{$Z_m=\Psi_m(\Param_m)$,} with $\Psi_m : \paramset_m \rightarrow
  \RR^{q_m}$ for some $q_m\leq p_m$.
  Suppose also that we have a prior distribution $P_M(m)$ over the model
  indicator $M$, a prior $\pdf_{\Param_m}(\param_m \vert m)$ for the parameters $\Param_m$ of each model,
  and a prior $\pdf_{Z_m}(z_m)$ for each \revnew{model's QoI $Z_m$} induced by $\pdf_{\Param_m}$
  and $\Psi_m$.
  A combined reward function 
  that includes a weighted combination of information gains in these
  random variables can then be formed.}
For example, the incremental formulation in
\eqref{e:incremental1} and \eqref{e:incremental2} becomes
\begin{align}
    r_k(s_k, \design_k, y_k) &= \alpha_M \DKL\left( P_{M| \kern 0.05em \revarx{\info_{k+1}}}
    \vert\vert P_{M| \kern 0.05em \revarx{\info_k}}
    \,\right) \nonumber\\
    &\quad + \EE_{M| \kern 0.05em \revarx{\info_{k+1}}} \revnew{\Big[} \alpha_\Param \DKL\left(
      \pdf_{\Param_m| \kern 0.05em \revarx{\info_{k+1}}} \vert\vert\pdf_{\Param_m| \kern 0.05em \revarx{\info_k}} \right) \nonumber\\
       &\hspace{3.5em} \revnew{+\alpha_Z
      \DKL\left( \pdf_{Z_m| \kern 0.05em \revarx{\info_{k+1}}}\vert\vert\pdf_{Z_m| \kern 0.05em \revarx{\info_k}} \right)
      \Big],} \quad k=0,\ldots,N-1,\label{e:vsOED_incremental1}
    \\
    r_N(s_N) &= 0, \label{e:vsOED_incremental2}
\end{align}
where $\alpha_M \in [0,1]$ (for the model \rev{indicator}), $\alpha_\Param \in
[0,1]$ (for \revnew{the} model parameters), and $\alpha_Z \in [0,1]$ (for the predictive
QoIs) are the weights of KL contributions from these variables.
The terminal formulation can be constructed in a similar manner. 

To compute the sOED objective in \eqref{e:sOED} with the
rewards \eqref{e:vsOED_incremental1} and \eqref{e:vsOED_incremental2},
an expectation needs to be taken over $Y_{0:N-1} \vert \policy,
s_0$.
\revarx{A natural approach entails generating samples of trajectories.
Consider the following procedure for trajectory generation. First, a
model indicator and corresponding parameter value are drawn from
the priors, $$m_0 \sim P_M, \quad \param_{m,0} \sim
p\big(\param_{m,0} \vert m_0\big)\revnew{,}$$
\revnew{with a corresponding QoI sample} $$z_{m,0}
= \Psi_{m_0}\big(\param_{m,0}\big).$$
\revnew{Then, $m_0$ and $\param_{m,0}$} generate \revnew{a
  trajectory} $\revarx{\info_N}$ using the policy $\policy$ and the statistical
models for each $Y_k$.}
\revarx{A key insight is that, \revnew{for} any such trajectory $\info_N$, we
  know exactly the `oracle' values of the model indicator and model
  \revnew{parameters} that generated the trajectory. New
  `one-point' reward functions (which avoid the integration involved
  in computing the incremental KL divergence) can be constructed using these
  oracle values:}
\revarx{
\begin{align}
  \tilde{r}_k(s_k,\design_k,y_k,m_0,\param_{m,0},z_{m,0})
  &= \alpha_M \log 
  \frac{P(m_0| \kern 0.05em \info_{k+1})}{P(m_0| \kern 0.05em \info_k)} + \alpha_{\Param} \log
  \frac{\pdf(\param_{m,0}| \kern 0.05em \info_{k+1})}{\pdf(\param_{m,0}| \kern 0.05em \info_k) }
    \nonumber\\
  & \hspace{1em}+ \alpha_Z \log \frac{\pdf(z_{m,0}| \kern 0.05em \info_{k+1})}{\pdf(z_{m,0}| \kern 0.05em \info_k)},
    \hspace{0.75em} k=0,\ldots,N-1,
 \label{e:one_point_incremental1}
  \\
  \tilde{r}_N(s_N,m_0,\param_{m,0},z_{m,0}) &= 0, \label{e:one_point_incremental2}
\end{align}
where the arguments $m_0,\param_{m,0},z_{m,0}$ are constrained to be
the oracle values that correspond to the trajectory
$\info_{k+1} = (\info_k,
  \design_k, y_k)$ (as represented by $s_k,\design_k,y_k$ in this notation)}.
\revarx{Computing the  expected utility $\tilde{U}_{\text{KL}}$ produced by the rewards $(\tilde{r}_k)_{k=0}^N$ thus requires taking the joint expectation over these random elements:
\begin{align}
    \label{eq:one_point_expected_utility}
    \tilde{U}_{\text{KL}}(\pi) &= \EE_{M_{0},\Param_{m,0} | s_0}
    \Bigg[\EE_{Y_{0:N-1}|\pi,s_0,M_{0},\Param_{m,0}}\Bigg[\sum_{k=0}^{N-1}\tilde{r}_k(S_k,\design_k,Y_k,M_0,\Param_{m,0},Z_{m,0})
\nonumber\\
        & \hspace{15em}+
        \tilde{r}_N(S_N,M_0,\Param_{m,0},Z_{m,0})\Bigg]\Bigg],
\end{align}
where we have factored the joint expectation so that the inner
conditional expectation reflects the generation of a trajectory
$\info_N$  from a single value of $m_0$ and  $\param_0$. (Since
$Z_{m,0}$ is a deterministic function of $\Param_{m,0}$ and $M_0$, we
drop it from the expectations.  If $Z_{m,0}$ had additional independent randomness, it could be included explicitly in the outer expectation,
but it is not needed for the inner conditional expectation since the
trajectory  $\info_N$ does not depend on $z_{m,0}$.)}
\revarx{Letting $U_{\text{KL}}$
denote
the expected
utility produced by the original rewards $(r_k)_{k=0}^N$ 
in
\eqref{e:vsOED_incremental1} and \eqref{e:vsOED_incremental2},}
\citet[Theorem~2]{Shen_2023a} show that
$\tilde{U}_{\text{KL}}(\policy) = U_{\text{KL}}(\policy)$ for any policy $\policy$.  In
other words, using the one-point \revarx{rewards} does not alter the
\revarx{expected utility objective of the} sOED
problem. \revarx{Note, however, that the realized utility (i.e., total reward)
  for any \textit{specific} trajectory naturally changes when using
  $\tilde{r}_k$ instead of $r_k$, and consequently the overall distribution of the utility
  (and its higher-order moments and quantiles)
  will be different, even though the expectation is preserved}.

\revarx{Evaluating the probability terms in
  \eqref{e:one_point_incremental1} remains a
  challenge.} 
To make the computation tractable, each of the
probability terms \rev{is} approximated within a parameterized family of
distributions---taking a variational inference approach---leading to
the following \rev{\emph{variational}} one-point \revarx{reward functions}:
\revarx{
\begin{align}
  \tilde{r}_{k,\phi}(s_k,\design_k,y_k,m_0,\param_{m,0},z_{m,0})
  &= \alpha_M \log 
  \frac{q(m_0| \kern 0.05em \revarx{\info_{k+1}};\phi_M)}{q(m_0| \kern 0.05em \revarx{\info_k};\phi_M)}
    \nonumber\\
&\hspace{-9.5em}+ \alpha_\Param \log
  \frac{q(\param_{m,0}| \kern 0.05em \revarx{\info_{k+1}};\phi_{\Param_m})}{q(\param_{m,0}| \kern 0.05em \revarx{\info_k};\phi_{\Param_m}) }
  + \alpha_Z \log \frac{q(z_{m,0}| \kern 0.05em \revarx{\info_{k+1}};\phi_{Z_m})}{q(z_{m,0}| \kern 0.05em \revarx{\info_k};\phi_{Z_m})},
  \quad k=0,\ldots,N-1,
 \label{e:vs_one_point_incremental1}
  \\
  \tilde{r}_{N,\phi}(s_N,m_0,\param_{m,0},z_{m,0})
  &= 0, \label{e:vs_one_point_incremental2}
\end{align}
with} $\phi=\{\phi_M, \phi_{\Param_m}, \phi_{Z_m}\}$ being the
variational parameters describing the approximations $q$ above. We use the notation $q_{\phi}$ to
represent succinctly the complete set of approximating distributions. Letting $\mathcal{L}^{\text{vsOED}}(\policy;q_{\phi})$ denote the
expected utility function obtained when using 
\eqref{e:vs_one_point_incremental1} and
\eqref{e:vs_one_point_incremental2} \revarx{and taking expectation in
  the same manner as \eqref{eq:one_point_expected_utility}},
\revnew{\citet[Theorem~3]{Shen_2023a} show}  %
that $\mathcal{L}^{\text{vsOED}}(\policy;q_{\phi})\leq \tilde{U}_{\text{KL}}(\policy) = U_{\text{KL}}(\policy)$ for any
policy $\policy$.  That is, adopting the variational approximation
forms a lower bound to the \revarx{original} expected utility
\revarx{(i.e., the EIG)}.

To solve the problem, we can use
an actor-critic approach as in PG-sOED, parameterizing the policy $\policy$ as
$\policy_{w}$ and the corresponding Q-function (critic) $Q^{\policy_w}$ as $Q_{\eta}^{\policy_w}$.  A policy can then be found by maximizing (tightening) the lower
bound simultaneously over $w$ and $\phi$: $$\max_{w,\phi} \mathcal{L}^{\text{vsOED}}(\policy_w;q_{\phi}).$$
(Recall that $\eta$ is not directly optimized, but updated via \eqref{e:Qnet_loss}.) 
This can be achieved by gradient ascent updates, where gradients with respect to $w$ and $\phi$ are found following similar steps to those used to obtain \eqref{e:PG}.  \rev{Notably, with the introduction
  of the approximating distributions $q_{\phi}$, evaluation of
  the reward no longer requires explicit likelihoods and thus the method can be
  used for implicit models. %
  \revnew{Unlike DAD and iDAD, vsOED does
    not require access to gradients of the log-likelihood or of an underlying simulator,
    due to the use of a Q-network in the same way as PG-sOED.}
}
 
\begin{figure}[tb]
  \centering
  \includegraphics[width=0.69\linewidth]{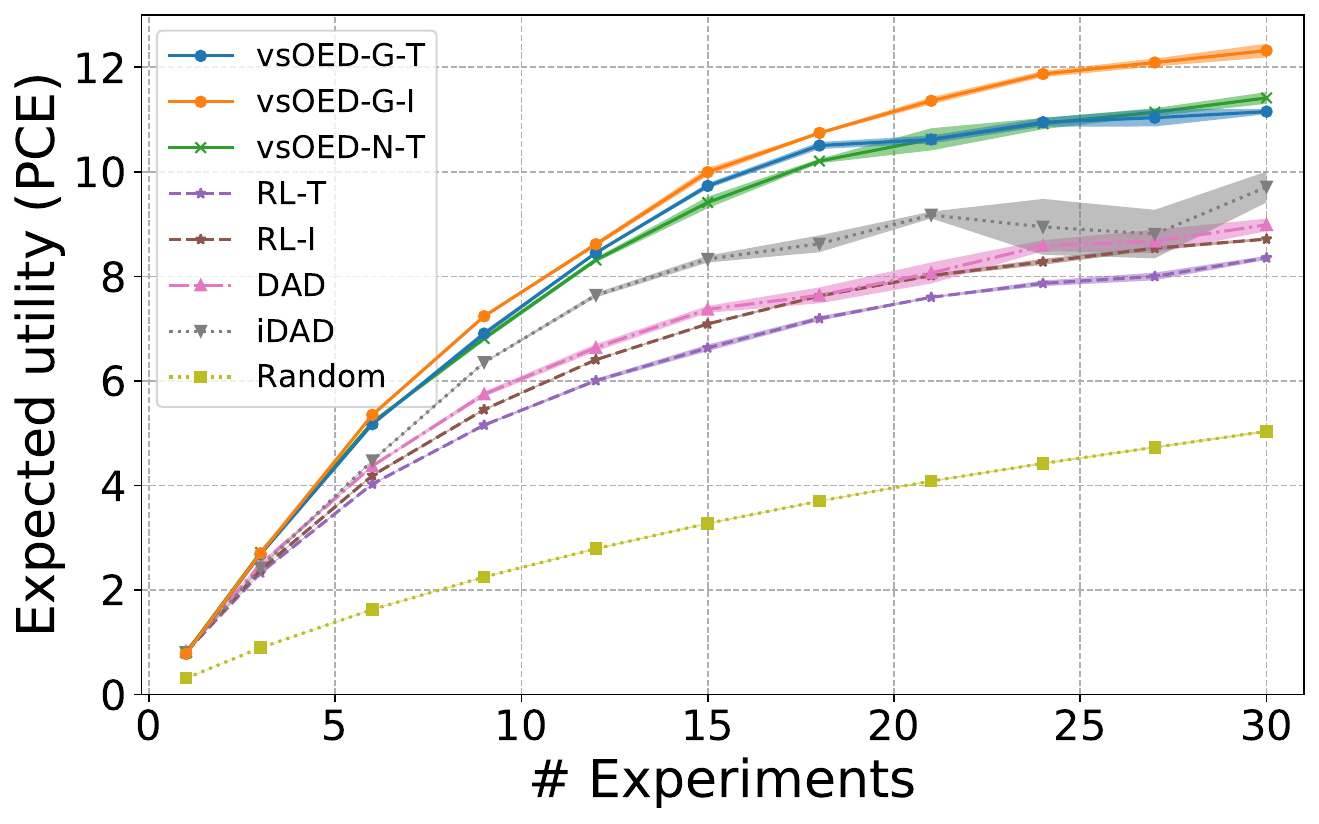}
  \caption{Expected utility lower bounds achieved by various
    sequential experimental design policies for a source location
    problem. vsOED-G and vsOED-N denote the vsOED method using
    Gaussian mixture models and normalizing flows, respectively, as variational
    distributions $q$. Suffixes -T and -I indicate whether the algorithm uses the
    terminal or incremental \revnew{formulation}. Figure adapted from \citet{Shen_2023a}.}
  \label{f:source_uni_poi_EU}
\end{figure}

Test problems in \citet{Shen_2023a} 
illustrate how vsOED can handle multiple models, 
predictive QoIs, and implicit likelihoods. For example,
Figure~\ref{f:source_uni_poi_EU}, adapted from that paper, compares
the EIG policies obtained via
several sequential experimental design approaches on a source location
problem introduced \rev{in~\citet{Foster_2021}, for} a range of
design horizons $N$. \rev{Results are obtained by fixing the total
  number of trajectory samples available to each algorithm. The 
  policies produced by each algorithm are then re-evaluated to assess their achieved EIG using a common
  estimator, namely the PCE lower bound estimator.}

\subsection{Open questions in sequential optimal design}

While the approaches discussed in this section represent promising
recent advances in sequential OED, there is a need for further
development in many important directions. We briefly mention a few
below.

One rather immediate need is for {more efficient state
  representations}---in particular, representations of the belief
state $\revarx{S_k^{b}}$ (the evolving posterior distribution) that achieve the
natural `compression' induced by Bayesian logic.
Many of the approaches described in this section simply
track $\revarx{\Info_k}$, which has the advantages of being trivial to update
(by appending the newly arrived $\design_k$ and $\revarx{Y_k}$) and free of
approximation error. Moreover, $\revarx{\Info_k}$ contains all the information
necessary to compute any intermediate and final physical states
(though they may not be accessible until evaluated via the transition dynamics for the physical state).
A key disadvantage of using $\revarx{\Info_k}$, however, is that its dimension
grows linearly with $k$, i.e., the number of experiments completed.
A resulting numerical difficulty can already be seen in the policy
network architectures used for PG-sOED (Figure~\ref{f:policy_network})
and related methods, where the state input layer needs to accommodate a variable-length
$\revarx{\Info_k}$. This is crudely handled by creating a layer of maximum
possible size and zero-padding the inapplicable entries.  The
efficiency of such a padding technique degrades as the design
horizon $N$ increases (i.e., there are many more stages where a large
fraction of the inputs are set to zero). 
Additionally, an $\Info_k$
of unbounded size introduces conceptual and
technical difficulties in the infinite-horizon setting.

In essence, tracking $\revarx{\Info_k}$---the trivial sufficient statistics of the
posterior---delays engaging with the actual
process of Bayesian inference and the computation of
posteriors.
$\revarx{\Info_k}$ is therefore not as `compressed' as the posterior, and the correspondence of
$\revarx{\Info_k}$ values to posterior distributions is generally non-injective (multiple
$\revarx{\Info_k}$ may lead to the same posterior). In other words, $\revarx{\Info_k}$ carries
additional, unnecessary information beyond that which is needed to
perform the Bayesian update
\citep{Jaynes_2003,Cox_1946a}.
Using state representations that directly capture the posterior---and
that simultaneously allow numerical accuracy, scalability to
high-dimensional parameter spaces, and easy online updating---would
therefore be quite desirable. Inspiration should arise from other
research areas that must maintain efficient online posterior
representations, such as sequential Bayesian inference, data
assimilation, and streaming variational inference.

More broadly, more efficient and scalable computational methods are
needed to realize sOED with computationally intensive models,
high-dimensional parameters and data, and complex design
objectives. Indeed, the development of sOED is rather new compared to
batch OED, and its demonstrations have thus far been confined to
rather simple formulations and models. An important goal is to expand
sOED's capabilities so that it can tackle problems of the complexity
now attained in batch design studies. To this end, accurate
calculations of the posterior and the associated information
measures---which occur \textit{repeatedly} in sOED---will need to be
pushed to higher dimensions.  More expressive function representations
and advanced reinforcement learning techniques will be needed to
improve the \textit{sample efficiency} of constructing a policy.

Along these lines, there is an enormous need for numerical analysis
and approximation theory in these settings. From an approximation
perspective, little is known about properties of \textit{optimal}
policies and value functions for sOED, outside of perhaps the simplest
(e.g., linear-Gaussian) cases. How smooth are these functions? Do they
enjoy anisotropic dependence on their inputs?  Are they well
approximated by low-rank tensors, or ridge functions, or in some other
format that is particularly tractable? What neural network
architectures are hence most suitable? Understanding these questions
is necessary to characterize the optimality gaps emerging from
current \textit{ad hoc} choices of representation and network
architecture. This understanding will also help to create more tailored and effective policy
and Q-function approximations, ideally endowed with error bounds
and performance guarantees.
At the same time, convergence guarantees and rates for the
reinforcement learning techniques used here, namely policy gradient
and actor-critic methods, should be strengthened for the case of
sequential experimental design in a Bayesian setting.

There is also a need to explore more sophisticated sequential design
formulations: optimal stopping of sequential experiments, interleaving
of batch and sequential designs, and policies that are robust to
horizon changes.

Finally, as sOED is generally more computationally demanding than
simpler batch and greedy designs, it would be extremely valuable to
develop ways of choosing the appropriate design strategy \textit{a
  priori}. Here we would advocate for inexpensive ways of estimating
or bounding the \textit{benefit of feedback} and the \textit{benefit
  of lookahead}, before actually solving the sOED problem, and then
balancing these potential upsides with their computational costs.

\section{Outlook}
\label{s:outlook}

The preceding sections have presented a broad overview of the current
state of the art in optimal experimental design (OED). Research continues in all of the threads we have discussed: (i) devising design criteria that are sufficiently expressive to capture diverse experimental goals, with complex models; (ii) estimating these criteria in efficient and structure-exploiting ways, in high dimensions \revnew{and/or} in the presence of strong non-Gaussianity; (iii) maximizing a chosen design criterion given different parameterizations of feasible designs, whether continuous or discrete, and developing guarantees for the associated optimization algorithms; and (iv) advancing formulations and algorithms for optimal sequential design. In this final section, we will highlight some broader questions and issues that have not yet been discussed. Many of these issues remain rather open-ended, and thus we believe they comprise fertile ground for ongoing research.

\subsection{Model misspecification}

\revnew{OED} is intrinsically model-based: all of the methods we have discussed use a statistical model, perhaps augmented with prior information, to \textit{predict} the outcomes of experiments at candidate designs and to \textit{assess} how these outcomes might improve knowledge of model parameters, reduce uncertainty in model-based predictions, and so on. In the general decision-theoretic terms of Section~\ref{ss:decision_theoretic}, models are needed to define the utility function $u$ and to specify the distribution $\pdf_{Y, \Param}$ over which we take expectations to yield the expected utility $U$. 

This reliance on models gives OED great power, but raises the question of what happens when models are, inevitably, misspecified---by which we mean that our statistical model $\mathcal{M}$ \eqref{eq:statmodel} of the data-generating process, for any given design, does not adequately capture how the data are generated in the actual experiment. In other words, $\mathcal{M}$ might not contain the distribution that generated the observed data.

A natural way to address misspecification is simply to augment the model, so that it can more closely capture the data-generating distribution.
One line of work, developed in the setting of Bayesian inverse
problems, is the Bayesian approximation error approach of
\citet{Kaipio2013, Kaipio2007}, and \citet{alexanderian2022optimal},
which augments the noise model---for example, in the case of additive Gaussian noise, enlarging the covariance matrix of this noise---to account for error in the forward \revnew{operator}.
Another approach \revnew{is to explicitly} introduce new `nuisance'
parameters: rather than considering only $p(y \vert \param, \design)$,
one could introduce a richer model $p(y \vert \param, \eta, \design)$,
where $\eta \in \bm{H}$ are additional parameters that capture
previously un-modeled phenomena, such that there exists some
$(\theta^\ast, \eta^\ast) \in \paramset \times \bm{H}$ for which $p(y
\vert \param^{\ast}, \eta^{\ast}, \design)$ matches the
data-generating distribution for \emph{all} designs $\design \in
\Xi$. This viewpoint underlies several recent efforts, e.g.,
\citet{alexanderian2022optimal} and \citet{Sargsyan2015}.
Here $\eta$ could represent variability or
imprecision in the placement or timing of observations, background
stochasticity or uncertain initial conditions, the impact of
unresolved scales, etc.
In a fully Bayesian setting, the design formulation discussed following \eqref{e:subsetKL} is relevant: the utility function can be chosen to depend on the posterior (and prior) marginal distributions of $\param$, with the impact of the additional uncertain parameters $\eta$ handled via integration over the prior $p(\eta)$. There are two equivalent ways of writing the posterior marginal of $\param$:
\begin{align}
  p(\param \vert y, \design) & = \int p( \param, \eta \vert y, \design) \, \mathrm{d} \eta, \\  \text{where}  \quad 
  p( \param, \eta \vert y, \design) & \propto p(y \vert \param, \eta, \design) p(\param \vert \eta) p(\eta) , \nonumber
\end{align}
and
\begin{align}
  p(\param \vert y, \design)  & \propto p(y \vert \param, \design)
  p(\param), \\  \text{where } \quad 
  p(y \vert \param, \design)  & = \int p(y \vert \param, \eta, \design) p(\eta \vert \param ) \,  \mathrm{d} \eta . \nonumber
\end{align}
In the first, we perform inference for both $\param$ and $\eta$, then
focus attention on the posterior marginal of $\param$. In the second,
we first create a marginal likelihood for $\param$ and then perform
inference with this integrated quantity.

{Although these viewpoints of
\textit{inference} are equivalent,} there are two distinct ways of modeling subsequent \textit{posterior predictions}. From a standard hierarchical Bayesian perspective, the distribution of predictions $Y^+$ at a design $\design^+$ (which may or may not be equal to $\design$) follows from the \emph{joint} posterior of $\param$ and $\eta$:
\begin{align}
  \pdf_1(y^+ \vert \design^+, y, \design ) = \iint \pdf(y^+ \vert \param, \eta, \design^+)\, \pdf(\param, \eta \vert y, \design)\, \mathrm{d}\param \, \mathrm{d} \eta.
  \label{e:predstandard}
\end{align}
In this setting, both $\param$ and $\eta$ were uncertain before observing $y$, but since they both influenced the observed value of $y$, conditioning on this observation updates one's knowledge of both parameters. Crucially, the unknown value of the nuisance parameter $\eta$ that influences subsequent predictions is assumed to be the same as the one that affected $y$ (just as $\param$ is common to both stages).
A different model for the predictions $Y^+ \vert \design ^+$, however, is that they arise from an independent, \emph{newly realized} $\eta$ (here called $\eta^+$, for emphasis) that has \emph{not} been conditioned on previous observations:
\begin{align}
  \pdf_2(y^+ \vert \design^+, y, \design ) = \iint \pdf(y^+ \vert \param, \eta^+, \design^+)\, \pdf(\param \vert y, \design)\, \pdf(\eta^+) \,  \mathrm{d}\param \, \mathrm{d} \eta^+.
  \label{e:predweird}
\end{align}
Embedded inadequacy models
\citep{Sargsyan2015,Sargsyan2019,Morrison2018}, which view uncertainty
in $\eta^+$ as somehow irreducible, can be understood according to
this prediction model \eqref{e:predweird}. This model is also somewhat related
to the viewpoint in \citet{koval2020optimal} and \citet{alexanderian2022optimal}.
We emphasize that, in this setting, the posterior \textit{marginal} $p(\param \vert y, \design)$ is the meaningful representation of uncertainty in the parameters $\param$ resulting from an experiment, even if $\eta$ is assumed irreducible; therefore it appears in the integrand of \eqref{e:predweird}. It is not meaningful to simply average the conditional posterior $p(\param \vert \eta, y, \design)$, or some functional thereof, over $p(\eta)$.

In any case, whether the goal is parameter inference, or prediction following \eqref{e:predstandard} or \eqref{e:predweird}, the ability to perform OED with implicit models as discussed throughout Section~\ref{s:estimation} is quite useful. In particular, the posterior marginal $p(\param \vert y, \design)$ or the marginal likelihood of $\param$, $p(y \vert \param, \design)$ are likely part of the utility function $u$; these densities are generally intractable and thus need to be estimated from samples. If the utility involves comparing prior and posterior predictions of $Y^+$, then again the densities \eqref{e:predstandard} and \eqref{e:predweird} are likely intractable.

The approaches just discussed, however, are at best a partial solution
to the problem of model misspecification, as they essentially rely on
the modeler being able to create a `better' statistical model for
the data. In many situations, doing so may not be feasible. Approaches
to this more challenging (and general) situation are very much a
subject of ongoing research. They are, at least in spirit, related to
earlier work on \textit{robust Bayesian analysis}, which, in the words
of \citet{Berger_1994}, ``studies \ldots the sensitivity of Bayesian
answers to uncertain inputs.''
Research on robust Bayesian methods
first flourished several decades ago, but much less has been done to
address the sensitivity of Bayesian OED. Some relevant analysis is
found in
\citet{duong2023stability},
which considers perturbations to the likelihood $p(y \vert \param,
\design)$ (in the sense of \revnew{the Kullback--Leibler} divergence) and elucidates the rate at
which the associated mutual information and its maximizers converge as
the likelihood perturbations \revarx{become} smaller. In a rather different (and
parametric) approach, \citet{Chowdhary2023} consider Bayesian linear
inverse problems in infinite dimensions and show how to compute
derivatives of the mutual information/\revnew{expected information gain} with respect to a finite set of `auxiliary' model parameters, i.e., parameters that are held fixed during the inference procedure.
\citet{Attia2023} propose a robust Bayesian A-optimal design formulation, again for linear (or linearized) models, where parametric families of prior and noise covariance matrices are specified and the design criterion is maximized for the worst-case element of these families. 

\revarx{\citet{Go2022} instead take a more general nonparametric approach to misspecification, proposing a robust OED formulation rooted in \textit{distributionally robust optimization} \citep{rahimian2019distributionally,kuhn2019wasserstein}.}
The robustness considered by \citet{Go2022} is with respect to the choice of prior $p_{\Param}$: the authors introduce an ambiguity set specified by the \revnew{Kullback--Leibler} divergence, $\mathcal{Q}_\epsilon = \{ q : \DKL(q \vert \vert p_{\Param}) \leq \epsilon \}$, and seek the design that maximizes the \emph{worst-case} mutual information, over priors drawn from $\mathcal{Q}_\epsilon$: 
\begin{align}
  \design^\ast = \arg \max_{\design \in \Xi} \inf_{q_{\Param} \in \mathcal{Q}_\epsilon} \mathbb{E}_{Y \vert \design} \left [ \DKL( p^{\revarx{q}}_{\Param \vert Y, \design} \vert \vert q_\Param) \right],
\end{align}
where the expectation is over $Y \vert \design \sim \int p(y \vert
\param, \design) \, q_\Param(\param) \, \mathrm{d} \param$ \revarx{and $p^{q}_{\Param \vert Y, \design}$ denotes the posterior derived from the prior $q_\Param$, i.e., $p^q _{\Param \vert Y, \design} (\param \vert y, \design) \propto p_{Y \vert \Param, \design}(y \vert \param, \design) q_\Param(\param) $.} This problem is not directly tractable, so \citet{Go2022} propose an approximation that is well-behaved for sufficiently small $\epsilon$.

We suggest that there is ample opportunity for further work at this
intersection of distributional robustness and OED. For instance, it is
important to consider robustness to other aspects of the joint
distribution $p_{Y, \Param \vert \design}$, especially the likelihood
\citep{zhang2022wasserstein} and the forward \revnew{operator} therein. It would
also be natural to consider ambiguity sets based on other divergences
or distances, e.g., Wasserstein distances. And it remains to understand
how to set the radius $\epsilon$ of any such ambiguity set, and to relate this value to other information one might have about the nature of the model misspecification.

\revarx{Also relevant to model misspecification, \citet{overstall2022bayesian} propose an extension
  of Bayesian decision-theoretic design that enables targeted
  explorations of certain aspects of model choice. Specifically, they formulate the design problem using \emph{two} models, modifying the decision-theoretic formulation of Section~\ref{ss:decision_theoretic}: the utility $u$ is defined according to one model, called the `fitted' model, but the distribution of $Y, \Param \vert \design$ yielding the outer expectation in \eqref{e:EU} (which transforms the utility $u(\design, y, \param)$ to an expected utility $U(\design)$) is given by an alternative model, called the `designer' model. \citet{overstall2022bayesian} suggest several potential justifications for this formulation. For instance, computing the utility $u$ might be intractable with the model that best represents the data-generating process, but tractable with a simpler alternative model; yet it could still be beneficial to use the more accurate model for the outer expectation. Alternatively, the fitted model might be quite complex and have weak prior information; in this case, computing the outer expectation with a more simplistic but informative model could yield more decisive designs. A related formulation appears in \citet{catanach2023metrics}.}

\revarx{More broadly}, we note that the goal of `robustifying' OED is very much
related to recent efforts to formulate robust notions of
\emph{inference} under model misspecification
(see \citealt{Kleijn2012,Bochkina2019}), and can benefit from advances in
this direction. These two threads are inextricably linked: a `robustified' OED might do a better job of producing data, but this data must then be interpreted through a model. Reverting solely to the misspecified model for this second phase makes little sense.
A multitude of interesting approaches to inference under model misspecification have been proposed in recent years, e.g., coarsened
inference \citep{Miller2019}; power posteriors
\citep{Grunwald2017}; reweighing data to reduce the effect of
outliers, data contamination, and other forms of misspecification
\citep{dewaskar2023robustifying};  averaging posteriors over bootstrapped datasets
\citep{Huggins2023,Pompe2021};
modularized inference \citep{Carmona2020,Jacob2017}; and many
others. It would be of interest to understand how these methods could
both mitigate the impact of model misspecification on OED procedures
and help to better process the resulting data.

It may also be natural to depart from reliance solely on optimality
criteria and to re-introduce some flavor of \textit{randomness} or \textit{space filling} to designs,
as a means of hedging against misspecification of the models that
produced the design criterion. This approach may be of particular interest in
the sequential OED setting, where model error could be assessed and to some extent quantified at each stage of experimentation, in a way that informs subsequent rounds of design.

\subsection{Risk-aware design criteria}
\label{ss:risk_OED}

The design formulations described in Section~\ref{ss:decision_theoretic} and addressed throughout this paper focused primarily on the \emph{expectation} of a utility function $u(\xi,Y,\Param)$, where randomness in the utility is induced by the randomness in $Y$ and $\Param$.
Yet the notion of \emph{risk} (see \citet{Royset_2022} for a recent review), aimed at quantifying  `hazard' or undesirable outcomes, is also relevant to OED. For instance, an experimenter might be interested in characterizing---and controlling---the probability \revnew{and/or} severity of an experimental outcome with very low utility (e.g., low information gain). Doing so requires moving beyond the expected value.

Popular risk measures include mean-plus-deviation and
mean-plus-variance \citep{Markowitz_1952}, quantiles and
superquantiles (also called the value-at-risk and the conditional
value-at-risk (CVaR), respectively), worst-case risk, entropic risk,
and many more; see \citet{Rockafellar_2013} and
\citet[Chapter~6]{Shapiro_2021}.  Desiderata for candidate risk measures
include the notion of coherence \citep{Artzner_1999}, with
consideration for robustness, elicitability, and backtesting
\citep{He_2022}.  A `risk quadrangle' system, characterizing the
relationships among measures of risk, regret, deviation, and error,
has also been proposed \citep{Rockafellar_2013, Rockafellar_2015}.

To our knowledge, risk measures have not been widely applied in the general setting of (nonlinear) design with generic $y$- \revnew{and} $\param$-dependent utility functions. Some initial work in this direction appears in \citet[Chapters~4, 5]{Shen_2023b}, which explores using a
variance-penalized expected utility (i.e., a mean-plus-variance risk)
for nonlinear OED. In the batch setting, the objective becomes
\begin{align}
  U(\design) = \EE_{Y,\Param|\design}[u(\design,Y,\Param)] - \lambda
    \Var_{Y,\Param|\design} \left[ u(\design,Y,\Param) \right],  \label{e:vp_OED}
\end{align}
where $\lambda >0$ is a scaling parameter reflecting the experimenter's degree of
risk-aversion (larger $\lambda$) or tolerance. Estimating this objective involves estimating the second moment of the utility, $\mathbb{E}_{Y,\Param|\design}[u^2(\design,Y,\Param)]$. When $u$ is chosen as in \eqref{e:uscore} or \eqref{e:udivergence}, i.e., to reflect information gain in the parameters $\Param$, \citet{Shen_2023b} constructs a consistent nested Monte Carlo estimator of this term and hence of the overall objective \eqref{e:vp_OED}, and characterizes its bias and variance to leading order. \citet{Shen_2023b} also demonstrates this variance-penalized utility for sequential OED.

A different form of risk-aware OED has been proposed by
\citet{Kouri_2022}, who address the setting of classical
(non-Bayesian) linear design for regression, where risk is now
associated with the distribution of the variance of the predicted
response, $v_\design(x) \coloneqq f^\top(x) F^{-1}(\design) f(x)$,
over \textit{inputs/covariates} $x \in \mathcal{X}$, given a
probability measure $\mu$ on $\mathcal{X}$ \revnew{and Fisher
  information matrix $F$}. (Recall our notation from
Section~\ref{sss:classical}.) In other words, the predictive variance
$v_\design(X)$ becomes random once the choice of input is treated as
random, $X \sim \mu$.
This formulation can be understood as a nonlinear interpolation between $G$-optimality \eqref{e:Gopt} (choosing $\design$ to minimize the maximum over $x \in \mathcal{X}$, and hence the worst case, of $v_\design(x)$) and I-optimality (minimizing the average over $X \sim \mu$ of the predictive variance, $\mathbb{E}_\mu [ v_{\design}(X) ]$). Specifically, \citet{Kouri_2022} propose  an \emph{R-optimality} criterion that involves minimizing the CVaR (also called the average value-at-risk) of $v_\design(X)$ at confidence level $\beta$:
\begin{align*}
  \design^\ast \in \argmin_{\design \in \Xi} \text{CVaR}_{\beta}\left( v_\design(X) \right )
\end{align*}
where, for any random variable $Z$ and $\beta \in [0,1]$,
\begin{align*}
  \text{CVaR}_{\beta}(Z) \coloneqq \frac{1}{1-\beta}\int_{\beta}^{1}
  q_{\alpha}(Z)\, \revarx{\text{d}} \alpha,
\end{align*}
and $q_{\alpha}(Z) \coloneqq \inf\{t\in \RR \,\vert\, \PP[X \leq t]
\geq \alpha\}$ is the upper $\alpha$-quantile of $Z$ \citep{Rockafellar_2002}. Thus the R-optimality
criterion seeks to mitigate the risk of large prediction variances by
minimizing the average of the \textit{upper tail} of predictive
variances, arising over the domain $\mathcal{X}$. \citet{Kouri_2022}
show that the R-optimality criterion satisfies standard properties of
classical design criteria, e.g., that it is an `information
function' in the sense of \citet[Chapter~5.8]{Pukelsheim_2006} on the set of
positive semi-definite matrices $F$, and also show how to compute its
gradient (or subgradient) to facilitate optimization over designs
$\design$.  In a nonlinear extension, the authors handle the
$\param$-dependence of the information matrix $F$
by averaging the R-optimality criterion over samples of $\param$.

\subsection{OED in practice}

This article has focused on mathematical/statistical formulations of the optimal design problem, and on computational methods for producing designs according to these formulations. We have not emphasized specific applications, but nonetheless it remains crucial to appreciate the interplay between the reality of practical applications, ways of formulating an OED problem, and the many \emph{modeling} choices therein. These modeling choices certainly may have computational implications.

For instance, what if the experiments we perform have a nonzero chance of failing: a computational simulation could abort or fail to converge, or a laboratory experiment could be interrupted or cancelled, for a variety of reasons. If the probability of failure is known, or can be parameterized and learned, it can become part of the statistical model of the experiment and hence the utility function, with failed experiments returning zero or negative utility. Similarly, if the designs realized in practice might not precisely match the intended design, this mismatch too can become part of the model; one could put $\xi' = \xi + \eta$, for some random variable $\eta$, or one could convolve the utility with a kernel $q(\xi' \vert \xi)$.

Sequential OED, as always, raises more complex possibilities. The non-myopic formulations discussed in Section~\ref{s:sequential} involve assessing information gain from future experiments, but what if the environment changes before these experiments can be executed? The experimental horizon might be cut short, the space of feasible designs $\Xi_k$ at some future stage might change due to supply or personnel constraints, or the timing of planned experiments might have to be altered. Robust policies able to hedge against these possibilities would be valuable.

\revnew{Another important question for OED, whether batch or
  sequential, is how to `validate' the designs that are produced. We
  can interpret this question in many ways, but an initial
  version is to ask how to verify that the designs produced by an OED
  procedure are in fact optimal. One possible approach, following the
  definition of expected utility in \eqref{e:EU}, is to generate prior
  samples of $\Param$ and $Y$ at different values of $\design$ and to
  scrutinize the results of estimation or inference. For example, when
  maximizing the expected information gain \eqref{e:EKL}, we can check
  if on average (over the $Y$ samples),
  the posteriors at the optimal design have lower entropy than those
  at other designs. Similarly, we could evaluate the Bayes risk of
  point estimators at different designs. Yet such checks are ultimately
  \emph{internal}---i.e., based only on simulation---and only reveal potential inconsistencies or errors in calculating and maximizing the intended expected utility.}

\revnew{Notions of \emph{external} validation, involving conducting
  the actual experiments being designed, are conceptually much more
  challenging. One difficulty arises from the fact that any experiment
  necessarily produces $Y$ values, and hence realized information
  gains, based on nature's `true' data-generating distribution. In the
  well-specified case, this distribution corresponds to \emph{one}
  value of $\Param$; otherwise it corresponds to \emph{no} value of
  $\Param$. While one could perform repeated experiments for a given
  design, or for a variety of feasible designs, it
  is generally not possible or meaningful to perform real experiments
  for `other' values of $\Param$ or other
  data-generating distributions. In other words, we cannot sample from the prior. 
  The real-world setting thus differs fundamentally from the OED methodology itself. In a Bayesian OED procedure, we introduce a prior distribution over $\Param$ in order to reflect our belief or lack of knowledge about the parameters, \emph{not} some intrinsic variability in the data-generating process. And the way in which we define optimality of the design requires the specification of this prior.
  At the same time, this situation is perhaps as it should be: when a decision (here the choice of design) is made under uncertainty, the correct decision is one conditioned only on the information available at the time of decision-making. Correctness should not be judged by how things actually turned out to be.}

\revnew{A} broader practical challenge here is to encode the many complexities of real-world experimentation into the elements of an OED formulation that we have discussed throughout this article. For instance, in human subject experiments, the space of feasible designs $\Xi$ must adhere to institutional review board requirements, and to other regulatory and ethical standards. Laboratory protocols for biological experiments, intended to ensure safety and minimize the potential for contamination, should similarly be reflected in $\Xi$. In Bayesian design, the choice of prior is crucial, and in some settings creating a suitable prior might involve techniques of  elicitation \citep{OHagan_2006a}. Even the design criterion could be refined to reflect specialized, expert knowledge. For instance, the information-based objectives we have discussed could be shaped or augmented using past examples of how experiments have been selected; here, an interesting approach could involve \emph{inverse} reinforcement learning \citep{Ng_2000}, which seeks to estimate unknown rewards from data describing how an agent selected actions (experiments) under varying conditions.
The ability to tailor and constrain elements of the OED problem in these ways are not only of interest mathematically: they might help improve the overall trust that practitioners place in OED methodologies, and stimulate the adoption of OED in new applications.

\bigskip

\noindent\textit{Acknowledgements}

\noindent We are grateful to all of our students, postdocs, and other
collaborators---too many to list by name---who have shaped our
thinking on optimal experimental design and related subjects, and
who contributed to the work described here. 
XH acknowledges support from the US Department of Energy (DOE), Office of
Science, Office of Advanced Scientific Computing Research (ASCR), under award
numbers DE-SC0021397 and DE-SC0021398.
JJ and YMM acknowledge support from DOE ASCR under award number
DE-SC0023188. 
JJ further acknowledges that this work was performed under the auspices of the DOE by Lawrence Livermore National Laboratory under contract DE-AC52-07NA27344, 
Release number LLNL-JRNL-861483.
YMM further acknowledges support from DOE ASCR awards
DE-SC0023187 and DE-SC0021226, from the Air Force Office of Scientific
Research under award FA9550-20-1-0397, and from the Office of Naval
Research under award N00014-20-1-2595.
\revarxtwo{We thank the editors of \textit{Acta Numerica} for their trust and
patience, and we thank our families for their considerable patience
and support throughout the writing of this article.}

\addcontentsline{toc}{section}{References}
\bibliography{xhuan,jj-bibfile,ymarz,refsnew}

\begin{thebibliography}{xx}

\bibitem[Agarwal {\em et~al.}(2005)Agarwal, Har-Peled and
  Varadarajan]{Agarwal2005}
P.~K. Agarwal, S.~Har-Peled and K.~R. Varadarajan  (2005), Geometric
  approximation via coresets, {\em Combinatorial and Computational Geometry}
  {\bf 52}, 1--30.

\bibitem[Aggarwal {\em et~al.}(2016)Aggarwal, Demkowicz and
  Marzouk]{Aggarwal_2016}
R.~Aggarwal, M.~J. Demkowicz and Y.~M. Marzouk  (2016), Information-driven
  experimental design in materials science, in {\em Information Science for
  Materials Discovery and Design} (T.~Lookman, F.~Alexander and K.~Rajan, eds),
  Springer, pp.~13--44.

\bibitem[Alexanderian(2021)Alexanderian]{Alexanderian_2021}
A.~Alexanderian  (2021), {Optimal experimental design for infinite-dimensional
  Bayesian inverse problems governed by PDEs: A review}, {\em Inverse Problems}
  {\bf 37}(4), 043001.

\bibitem[Alexanderian and Saibaba(2018)Alexanderian and
  Saibaba]{Alexanderian_Saibaba_2018}
A.~Alexanderian and A.~K. Saibaba  (2018), {Efficient D-optimal design of
  experiments for infinite-dimensional Bayesian linear inverse problems}, {\em
  SIAM Journal on Scientific Computing} {\bf 40}(5), A2956--A2985.

\bibitem[Alexanderian {\em et~al.}(2016{\em a})Alexanderian, Gloor and
  Ghattas]{Alexanderian_2016ba}
A.~Alexanderian, P.~J. Gloor and O.~Ghattas  (2016{\em a}), {On Bayesian A- and
  D-optimal experimental designs in infinite dimensions}, {\em Bayesian
  Analysis} {\bf 11}(3), 671--695.

\bibitem[Alexanderian {\em et~al.}(2022)Alexanderian, Nicholson and
  Petra]{alexanderian2022optimal}
A.~Alexanderian, R.~Nicholson and N.~Petra  (2022), {Optimal design of
  large-scale nonlinear Bayesian inverse problems under model uncertainty}.
\newblock Available at
  \href{https://arxiv.org/abs/2211.03952}{arXiv:2211.03952}.

\bibitem[Alexanderian {\em et~al.}(2014)Alexanderian, Petra, Stadler and
  Ghattas]{AlexanderianSISC2014}
A.~Alexanderian, N.~Petra, G.~Stadler and O.~Ghattas  (2014), {A-optimal design
  of experiments for infinite-dimensional Bayesian linear inverse problems with
  regularized $\ell_0$-sparsification}, {\em SIAM Journal on Scientific
  Computing} {\bf 36}(5), A2122--A2148.

\bibitem[Alexanderian {\em et~al.}(2016{\em b})Alexanderian, Petra, Stadler and
  Ghattas]{Alexanderian_2016}
A.~Alexanderian, N.~Petra, G.~Stadler and O.~Ghattas  (2016{\em b}), {A fast
  and scalable method for A-optimal design of experiments for
  infinite-dimensional Bayesian nonlinear inverse problems}, {\em SIAM Journal
  on Scientific Computing} {\bf 38}(1), A243--A272.

\bibitem[Alexanderian {\em et~al.}(2021)Alexanderian, Petra, Stadler and
  Sunseri]{alexanderian2021optimal}
A.~Alexanderian, N.~Petra, G.~Stadler and I.~Sunseri  (2021), {Optimal design
  of large-scale Bayesian linear inverse problems under reducible model
  uncertainty: Good to know what you don't know}, {\em SIAM/ASA Journal on
  Uncertainty Quantification} {\bf 9}(1), 163--184.

\bibitem[Allen-Zhu {\em et~al.}(2017)Allen-Zhu, Li, Singh and
  Wang]{allen2017near}
Z.~Allen-Zhu, Y.~Li, A.~Singh and Y.~Wang  (2017), {Near-optimal design of
  experiments via regret minimization}, in {\em Proceedings of the 34th
  International Conference on Machine Learning (ICML 2017)}, Vol.~70 of
  Proceedings of Machine Learning Research, PMLR, pp.~126--135.

\bibitem[Allen-Zhu {\em et~al.}(2015)Allen-Zhu, Liao and
  Orecchia]{allen2015spectral}
Z.~Allen-Zhu, Z.~Liao and L.~Orecchia  (2015), {Spectral sparsification and
  regret minimization beyond matrix multiplicative updates}, in {\em
  Proceedings of the 47th Annual ACM Symposium on Theory of Computing (STOC
  2015)}, ACM, pp.~237--245.

\bibitem[Ao and Li(2020)Ao and Li]{ao2020approximate}
Z.~Ao and J.~Li  (2020), {An approximate KLD based experimental design for
  models with intractable likelihoods}, in {\em Proceedings of the 23rd
  International Conference on Artificial Intelligence and Statistics}, Vol. 108
  of Proceedings of Mchine Learning Research, PMLR, pp.~3241--3251.

\bibitem[Ao and Li(2024)Ao and Li]{ao2024estimating}
Z.~Ao and J.~Li  (2024), On estimating the gradient of the expected information
  gain in {B}ayesian experimental design, in {\em Proceedings of the 38th AAAI
  Conference on Artificial Intelligence} (M.~Wooldridge, J.~Dy and
  S.~Natarajan, eds), AAAI Press, pp.~20311--20319.

\bibitem[Artzner {\em et~al.}(1999)Artzner, Delbaen, Eber and
  Heath]{Artzner_1999}
P.~Artzner, F.~Delbaen, J.~Eber and D.~Heath  (1999), {Coherent measures of
  risk}, {\em Mathematical Finance} {\bf 9}(3), 203--228.

\bibitem[Asmussen and Glynn(2007)Asmussen and Glynn]{Asmussen2007}
S.~Asmussen and P.~W. Glynn  (2007), {\em {Stochastic Simulation: Algorithms
  and Analysis}}, Springer.

\bibitem[Atkinson {\em et~al.}(2007)Atkinson, Donev and Tobias]{Atkinson_2007}
A.~C. Atkinson, A.~N. Donev and R.~D. Tobias  (2007), {\em {Optimum
  Experimental Designs, with SAS}}, Oxford University Press.

\bibitem[Attia {\em et~al.}(2018)Attia, Alexanderian and Saibaba]{Attia2018}
A.~Attia, A.~Alexanderian and A.~K. Saibaba  (2018), {Goal-oriented optimal
  design of experiments for large-scale Bayesian linear inverse problems}, {\em
  Inverse Problems} {\bf 34}(9), aad210.

\bibitem[Attia {\em et~al.}(2023)Attia, Leyffer and Munson]{Attia2023}
A.~Attia, S.~Leyffer and T.~Munson  (2023), {Robust A-optimal experimental
  design for Bayesian inverse problems}.
\newblock Available at
  \href{https://arxiv.org/abs/2305.03855}{arXiv:2305.03855}.

\bibitem[Atwood(1969)Atwood]{Atwood_1969}
C.~L. Atwood  (1969), {Optimal and efficient designs of experiments}, {\em The
  Annals of Mathematical Statistics} {\bf 40}(5), 1570--1602.

\bibitem[Audet(2004)Audet]{Audet2004}
C.~Audet  (2004), {Convergence results for generalized pattern search
  algorithms are tight}, {\em Optimization and Engineering} {\bf 5}(2),
  101--122.

\bibitem[Audet and Dennis(2002)Audet and Dennis]{Audet2002}
C.~Audet and J.~E. Dennis  (2002), {Analysis of generalized pattern searches},
  {\em SIAM Journal on Optimization} {\bf 13}(3), 889--903.

\bibitem[Bach(2013)Bach]{Bach2013}
F.~Bach  (2013), {Learning with submodular functions: {A} convex optimization
  perspective}, {\em Foundations and Trends in Machine Learning} {\bf 6}(2--3),
  145--373.

\bibitem[Bach and Jordan(2002)Bach and Jordan]{Bach2002}
F.~R. Bach and M.~I. Jordan  (2002), {Kernel independent component analysis},
  {\em Journal of Machine Learning Research} {\bf 3}(1), 1--48.

\bibitem[Baptista {\em et~al.}(2024{\em a})Baptista {\em
  et~al.}]{baptista2022bayesian}
R.~Baptista, L.~Cao, J.~Chen, O.~Ghattas, F.~Li, Y.~M. Marzouk and J.~T. Oden
  (2024{\em a}), {Bayesian model calibration for block copolymer self-assembly:
  Likelihood-free inference and expected information gain computation via
  measure transport}, {\em Journal of Computational Physics} {\bf 503}, 112844.

\bibitem[Baptista {\em et~al.}(2024{\em b})Baptista, Hosseini, Kovachki and
  Marzouk]{baptista2020conditional}
R.~Baptista, B.~Hosseini, N.~B. Kovachki and Y.~Marzouk  (2024{\em b}),
  {Conditional sampling with monotone GANs: From generative models to
  likelihood-free inference}, {\em SIAM/ASA Journal on Uncertainty
  Quantification}.
\newblock Available at
  \href{https://arxiv.org/abs/2006.06755}{arXiv:2006.06755}.

\bibitem[Baptista {\em et~al.}(2022)Baptista, Marzouk and Zahm]{Baptista2022}
R.~Baptista, Y.~Marzouk and O.~Zahm  (2022), {Gradient-based data and parameter
  dimension reduction for Bayesian models: An information theoretic
  perspective}.
\newblock Available at
  \href{https://arxiv.org/abs/2207.08670}{arXiv:2207.08670}.

\bibitem[Baptista {\em et~al.}(2023)Baptista, Marzouk and
  Zahm]{baptista2023representation}
R.~Baptista, Y.~Marzouk and O.~Zahm  (2023), {On the representation and
  learning of monotone triangular transport maps}, {\em Foundations of
  Computational Mathematics}.
\newblock Available at
  \href{https://doi.org/10.1007/s10208-023-09630-x}{doi:10.1007/s10208-023-09630-x}.

\bibitem[Barber and Agakov(2003)Barber and Agakov]{barber2004algorithm}
D.~Barber and F.~Agakov  (2003), {The {IM} algorithm: A variational approach to
  information maximization}, in {\em Advances in Neural Information Processing
  Systems 16}, MIT Press, pp.~201--208.

\bibitem[Batson {\em et~al.}(2009)Batson, Spielman and
  Srivastava]{batson2009twice}
J.~D. Batson, D.~A. Spielman and N.~Srivastava  (2009), Twice-{R}amanujan
  sparsifiers, in {\em Proceedings of the 41st Annual ACM Symposium on Theory
  of Computing (STOC 2009)}, ACM, pp.~255--262.

\bibitem[Beck and Guillas(2016)Beck and Guillas]{beck2016sequential}
J.~Beck and S.~Guillas  (2016), {Sequential design with mutual information for
  computer experiments (MICE): Emulation of a tsunami model}, {\em SIAM/ASA
  Journal on Uncertainty Quantification} {\bf 4}(1), 739--766.

\bibitem[Beck {\em et~al.}(2020)Beck, Dia, Espath and
  Tempone]{beck2020multilevel}
J.~Beck, B.~M. Dia, L.~Espath and R.~Tempone  (2020), {Multilevel double loop
  Monte Carlo and stochastic collocation methods with importance sampling for
  Bayesian optimal experimental design}, {\em International Journal for
  Numerical Methods in Engineering} {\bf 121}(15), 3482--3503.

\bibitem[Beck {\em et~al.}(2018)Beck {\em et~al.}]{Beck_2018}
J.~Beck, B.~M. Dia, L.~F. Espath, Q.~Long and R.~Tempone  (2018), {Fast
  Bayesian experimental design: Laplace-based importance sampling for the
  expected information gain}, {\em Computer Methods in Applied Mechanics and
  Engineering} {\bf 334}, 523--553.

\bibitem[Belghazi {\em et~al.}(2018)Belghazi {\em et~al.}]{Belghazi2018}
M.~I. Belghazi, A.~Baratin, S.~Rajeswar, S.~Ozair, Y.~Bengio, A.~Courville and
  R.~D. Hjelm  (2018), {Mutual information neural estimation}, in {\em
  Proceedings of the 35th International Conference on Machine Learning (ICML
  2018)}, Vol.~80 of Proceedings of Machine Learning Research, PMLR,
  pp.~531--540.

\bibitem[Ben-Tal and Nemirovski(2001)Ben-Tal and Nemirovski]{ben2001lectures}
A.~Ben-Tal and A.~Nemirovski  (2001), {\em Lectures on Modern Convex
  Optimization}, SIAM.

\bibitem[Benner {\em et~al.}(2015)Benner, Gugercin and
  Willcox]{benner2015survey}
P.~Benner, S.~Gugercin and K.~Willcox  (2015), {A Survey of projection-based
  model reduction methods for parametric dynamical systems}, {\em SIAM Review}
  {\bf 57}(4), 483--531.

\bibitem[Berger(1985)Berger]{Berger_1985}
J.~O. Berger  (1985), {\em {Statistical Decision Theory and Bayesian
  Analysis}}, Springer Series in Statistics, Springer.

\bibitem[Berger(1994)Berger]{Berger_1994}
J.~O. Berger  (1994), {An overview of robust Bayesian analysis (with
  discussion)}, {\em Test} {\bf 3}(1), 5--124.

\bibitem[Berger and Wong(2009)Berger and Wong]{Berger_2009}
M.~P.~F. Berger and W.~K. Wong  (2009), {\em {An Introduction to Optimal
  Designs for Social and Biomedical Research}}, Wiley.

\bibitem[Bernardo(1979)Bernardo]{Bernardo_1979}
J.~M. Bernardo  (1979), {Expected information as expected utility}, {\em The
  Annals of Statistics} {\bf 7}(3), 686--690.

\bibitem[Bernardo and Smith(2000)Bernardo and Smith]{Bernardo_2000}
J.~M. Bernardo and A.~F.~M. Smith  (2000), {\em {Bayesian Theory}}, Wiley.

\bibitem[Berry {\em et~al.}(2010)Berry, Carlin, Lee and
  M{\"{u}}ller]{Berry_2010}
S.~M. Berry, B.~P. Carlin, J.~J. Lee and P.~M{\"{u}}ller  (2010), {\em
  {Bayesian Adaptive Methods for Clinical Trials}}, Chapman \& Hall/CRC.

\bibitem[Bertsekas(2005)Bertsekas]{Bertsekas_2005}
D.~P. Bertsekas  (2005), {\em {Dynamic Programming and Optimal Control}},
  Vol.~1, Athena Scientific.

\bibitem[Bhatnagar {\em et~al.}(2013)Bhatnagar, Prasad and
  Prashanth]{Bhatnagar2013}
S.~Bhatnagar, H.~L. Prasad and L.~A. Prashanth  (2013), {\em {Stochastic
  Recursive Algorithms for Optimization}}, Springer.

\bibitem[Bian {\em et~al.}(2017)Bian, Buhmann, Krause and
  Tschiatschek]{Bian_etal_2017}
A.~A. Bian, J.~M. Buhmann, A.~Krause and S.~Tschiatschek  (2017), Guarantees
  for greedy maximization of non-submodular functions with applications, in
  {\em Proceedings of the 34th International Conference on Machine Learning
  (ICML 2017)} (D.~Precup and Y.~W. Teh, eds), Vol.~70 of Proceedings of
  Machine Learning Research, PMLR, pp.~498--507.

\bibitem[Blackwell(1951)Blackwell]{Blackwell_1951}
D.~Blackwell  (1951), {Comparison of experiments}, in {\em Proceedings of the
  2nd Berkeley Symposium on Mathematical Statistics and Probability},
  University of California Press, pp.~93--102.

\bibitem[Blackwell(1953)Blackwell]{Blackwell_1953}
D.~Blackwell  (1953), {Equivalent comparisons of experiments}, {\em The Annals
  of Mathematical Statistics} {\bf 24}(2), 265--272.

\bibitem[Blanchard and Sapsis(2021)Blanchard and Sapsis]{blanchard2021output}
A.~Blanchard and T.~Sapsis  (2021), {Output-weighted optimal sampling for
  Bayesian experimental design and uncertainty quantification}, {\em SIAM/ASA
  Journal on Uncertainty Quantification} {\bf 9}(2), 564--592.

\bibitem[Blau {\em et~al.}(2022)Blau, Bonilla, Chades and Dezfouli]{Blau_2022}
T.~Blau, E.~V. Bonilla, I.~Chades and A.~Dezfouli  (2022), {Optimizing
  sequential experimental design with deep reinforcement learning}, in {\em
  Proceedings of the 39th International Conference on Machine Learning (ICML
  2022)} (K.~Chaudhuri, S.~Jegelka, L.~Song, C.~Szepesvari, G.~Niu and
  S.~Sabato, eds), Vol. 162 of Proceedings of Machine Learning Research, PMLR,
  pp.~2107--2128.

\bibitem[Blum(1954)Blum]{Blum1954}
J.~R. Blum  (1954), {Multidimensional stochastic approximation methods}, {\em
  The Annals of Mathematical Statistics} {\bf 25}(4), 737--744.

\bibitem[Bochkina(2019)Bochkina]{Bochkina2019}
N.~Bochkina  (2019), {Bernstein–von Mises theorem and misspecified models: A
  review}, in {\em Foundations of Modern Statistics} (D.~Belomestny,
  C.~Butucea, E.~Mammen, E.~Moulines, M.~Rei{\ss} and V.~V. Ulyanov, eds), Vol.
  425 of Springer Proceedings in Mathematics \& Statistics, Springer,
  pp.~355--380.

\bibitem[Bogachev {\em et~al.}(2005)Bogachev, Kolesnikov and
  Medvedev]{bogachev2005triangular}
V.~I. Bogachev, A.~V. Kolesnikov and K.~V. Medvedev  (2005), Triangular
  transformations of measures, {\em Sbornik Mathematics} {\bf 196}(3), 309.

\bibitem[Bogunovic {\em et~al.}(2018)Bogunovic, Zhao and
  Cevher]{Bogunovic_etal_2018_SupModRatio}
I.~Bogunovic, J.~Zhao and V.~Cevher  (2018), Robust maximization of
  non-submodular objectives, in {\em Proceedings of the Twenty-First
  International Conference on Artificial Intelligence and Statistics}
  (A.~Storkey and F.~Perez-Cruz, eds), Vol.~84 of Proceedings of Machine
  Learning Research, PMLR, pp.~890--899.

\bibitem[Borges and Biros(2018)Borges and Biros]{borges2018reconstruction}
C.~Borges and G.~Biros  (2018), Reconstruction of a compactly supported sound
  profile in the presence of a random background medium, {\em Inverse Problems}
  {\bf 34}(11), 115007.

\bibitem[Bose(1939)Bose]{bose1939construction}
R.~C. Bose  (1939), On the construction of balanced incomplete block designs,
  {\em Annals of Eugenics} {\bf 9}(4), 353--399.

\bibitem[Bose and Nair(1939)Bose and Nair]{bose1939partially}
R.~C. Bose and K.~R. Nair  (1939), Partially balanced incomplete block designs,
  {\em Sankhy{\=a}} {\bf 4}, 337--372.

\bibitem[Box(1992)Box]{Box_1992}
G.~E.~P. Box  (1992), {Sequential experimentation and sequential assembly of
  designs}, {\em Quality Engineering} {\bf 5}(2), 321--330.

\bibitem[Boyd and Vandenberghe(2004)Boyd and Vandenberghe]{boyd2004convex}
S.~P. Boyd and L.~Vandenberghe  (2004), {\em {Convex Optimization}}, Cambridge
  University Press.

\bibitem[Brockwell and Kadane(2003)Brockwell and Kadane]{Brockwell_2003}
A.~E. Brockwell and J.~B. Kadane  (2003), {A gridding method for Bayesian
  sequential decision problems}, {\em Journal of Computational and Graphical
  Statistics} {\bf 12}(3), 566--584.

\bibitem[Buchbinder {\em et~al.}(2014)Buchbinder, Feldman, Naor and
  Schwartz]{buchbinder2014submodular}
N.~Buchbinder, M.~Feldman, J.~Naor and R.~Schwartz  (2014), Submodular
  maximization with cardinality constraints, in {\em Proceedings of the
  Twenty-Fifth Annual ACM-SIAM Symposium on Discrete Algorithms}, SIAM,
  pp.~1433--1452.

\bibitem[Bui-Thanh {\em et~al.}(2013)Bui-Thanh, Ghattas, Martin and
  Stadler]{BuiThanh2013}
T.~Bui-Thanh, O.~Ghattas, J.~Martin and G.~Stadler  (2013), {A computational
  framework for infinite-dimensional Bayesian inverse problems part I: The
  linearized case, with application to global seismic inversion}, {\em SIAM
  Journal on Scientific Computing} {\bf 35}(6), A2494--A2523.

\bibitem[Caflisch(1998)Caflisch]{caflisch1998monte}
R.~E. Caflisch  (1998), {Monte Carlo and quasi-Monte Carlo methods}, {\em Acta
  numerica} {\bf 7}, 1--49.

\bibitem[Calinescu {\em et~al.}(2011)Calinescu, Chekuri, {P\'{a}l} and
  {Vondr\'{a}k}]{Calinescu_Chekuri_Pal_Vondrak_2011}
G.~Calinescu, C.~Chekuri, M.~{P\'{a}l} and J.~{Vondr\'{a}k}  (2011), Maximizing
  a monotone submodular function subject to a matroid constraint, {\em SIAM
  Journal on Computing} {\bf 40}(6), 1740--1766.

\bibitem[Campbell and Beronov(2019)Campbell and Beronov]{Campbell2019}
T.~Campbell and B.~Beronov  (2019), {Sparse variational inference: Bayesian
  coresets from scratch}, in {\em Advances in Neural Information Processing
  Systems 32} (H.~Wallach, H.~Larochelle, A.~Beygelzimer, F.~d\textquotesingle
  Alch\'{e}-Buc, E.~Fox and R.~Garnett, eds), Curran Associates,
  pp.~11461--11472.

\bibitem[Campbell and Broderick(2018)Campbell and
  Broderick]{campbell2018bayesian}
T.~Campbell and T.~Broderick  (2018), {Bayesian coreset construction via greedy
  iterative geodesic ascent}, in {\em Proceedings of the 35th International
  Conference on Machine Learning (ICML 2018)}, Vol.~80 of Proceedings of
  Machine Learning Research, PMLR, pp.~698--706.

\bibitem[Campbell and Broderick(2019)Campbell and
  Broderick]{campbell2019automated}
T.~Campbell and T.~Broderick  (2019), {Automated scalable Bayesian inference
  via Hilbert coresets}, {\em Journal of Machine Learning Research} {\bf
  20}(1), 551--588.

\bibitem[Carlier {\em et~al.}(2016)Carlier, Chernozhukov and
  Galichon]{carlier2016}
G.~Carlier, V.~Chernozhukov and A.~Galichon  (2016), {Vector quantile
  regression: An optimal transport approach}, {\em The Annals of Statistics}
  {\bf 44}(3), 1165--1192.

\bibitem[Carlin {\em et~al.}(1998)Carlin, Kadane and Gelfand]{Carlin_1998}
B.~P. Carlin, J.~B. Kadane and A.~E. Gelfand  (1998), {Approaches for optimal
  sequential decision analysis in clinical trials}, {\em Biometrics} {\bf
  54}(3), 964--975.

\bibitem[Carlon {\em et~al.}(2020)Carlon {\em et~al.}]{Carlon2020}
A.~G. Carlon, B.~M. Dia, L.~Espath, R.~H. Lopez and R.~Tempone  (2020),
  {Nesterov-aided stochastic gradient methods using Laplace approximation for
  Bayesian design optimization}, {\em Computer Methods in Applied Mechanics and
  Engineering} {\bf 363}, 112909.

\bibitem[Carmona and Nicholls(2020)Carmona and Nicholls]{Carmona2020}
C.~U. Carmona and G.~K. Nicholls  (2020), {Semi-modular inference: Enhanced
  learning in multi-modular models by tempering the influence of components},
  in {\em Proceedings of the 23rd International Conference on Artificial
  Intelligence and Statistics}, Vol. 108 of Proceedings of Machine Learning
  Research, PMLR, pp.~4226--4235.

\bibitem[Caselton and Zidek(1984)Caselton and Zidek]{caselton1984optimal}
W.~F. Caselton and J.~V. Zidek  (1984), Optimal monitoring network designs,
  {\em Statistics \& Probability Letters} {\bf 2}(4), 223--227.

\bibitem[Catanach and Das(2023)Catanach and Das]{catanach2023metrics}
T.~A. Catanach and N.~Das  (2023), Metrics for {B}ayesian optimal experiment
  design under model misspecification, in {\em 2023 62nd IEEE Conference on
  Decision and Control (CDC)}, IEEE, pp.~7707--7714.

\bibitem[Cavagnaro {\em et~al.}(2010)Cavagnaro, Myung, Pitt and
  Kujala]{Cavagnaro_2010}
D.~R. Cavagnaro, J.~I. Myung, M.~A. Pitt and J.~V. Kujala  (2010), {Adaptive
  design optimization: A mutual information-based approach to model
  discrimination in cognitive science}, {\em Neural Computation} {\bf 22}(4),
  887--905.

\bibitem[Chaloner(1984)Chaloner]{chaloner1984optimal}
K.~Chaloner  (1984), {Optimal Bayesian experimental design for linear models},
  {\em The Annals of Statistics} {\bf 12}, 283--300.

\bibitem[Chaloner and Larntz(1989)Chaloner and Larntz]{chaloner1989optimal}
K.~Chaloner and K.~Larntz  (1989), {Optimal Bayesian design applied to logistic
  regression experiments}, {\em Journal of Statistical Planning and Inference}
  {\bf 21}(2), 191--208.

\bibitem[Chaloner and Verdinelli(1995)Chaloner and Verdinelli]{Chaloner_1995}
K.~Chaloner and I.~Verdinelli  (1995), {Bayesian experimental design: A
  review}, {\em Statistical Science} {\bf 10}(3), 273--304.

\bibitem[Chang(2012)Chang]{Chang2012}
K.-H. Chang  (2012), {Stochastic Nelder--Mead simplex method -- A new globally
  convergent direct search method for simulation optimization}, {\em European
  Journal of Operational Research} {\bf 220}(3), 684--694.

\bibitem[Chekuri {\em et~al.}(2014)Chekuri, {Vondr\'{a}k} and
  Zenklusen]{Chekuri_Vondrak_Zenklusen_2014}
C.~Chekuri, J.~{Vondr\'{a}k} and R.~Zenklusen  (2014), Submodular function
  maximization via the multilinear relaxation and contention resolution
  schemes, {\em SIAM Journal on Computing} {\bf 43}(6), 1831--1879.

\bibitem[Chen {\em et~al.}(2021)Chen, Wang, Zhou and Ross]{Chen_2021}
X.~Chen, C.~Wang, Z.~Zhou and K.~Ross  (2021), {Randomized ensembled double
  Q-learning: Learning fast without a model}, in {\em 9th International
  Conference on Learning Representations (ICLR 2021)}.
\newblock Available at
  \href{https://openreview.net/forum?id=AY8zfZm0tDd}{https://openreview.net/forum?id=AY8zfZm0tDd}.

\bibitem[Chevalier and Ginsbourger(2013)Chevalier and
  Ginsbourger]{chevalier2013fast}
C.~Chevalier and D.~Ginsbourger  (2013), {Fast computation of the multi-points
  expected improvement with applications in batch selection}, in {\em Learning
  and Intelligent Optimization}, Vol. 7997 of Lecture Notes in Computer
  Science, Springer, pp.~59--69.

\bibitem[Chowdhary {\em et~al.}(2023)Chowdhary, Tong, Stadler and
  Alexanderian]{Chowdhary2023}
A.~Chowdhary, S.~Tong, G.~Stadler and A.~Alexanderian  (2023), {Sensitivity
  analysis of the information gain in infinite-dimensional Bayesian linear
  inverse problems}.
\newblock Available at
  \href{https://arxiv.org/abs/2310.16906}{arXiv:2310.16906}.

\bibitem[Christen and Nakamura(2003)Christen and Nakamura]{Christen_2003}
J.~A. Christen and M.~Nakamura  (2003), {Sequential stopping rules for species
  accumulation}, {\em Journal of Agricultural, Biological {\&} Environmental
  Statistics} {\bf 8}(2), 184--195.

\bibitem[Clyde(2001)Clyde]{Clyde_2001}
M.~Clyde  (2001), {Experimental design: Bayesian designs}, in {\em
  International Encyclopedia of the Social {\&} Behavioral Sciences} (N.~J.
  Smelser and P.~B. Baltes, eds), Science Direct, pp.~5075--5081.

\bibitem[Cohn {\em et~al.}(1996)Cohn, Ghahramani and Jordan]{Cohn_1996}
D.~A. Cohn, Z.~Ghahramani and M.~I. Jordan  (1996), {Active learning with
  statistical models}, {\em Journal of Artificial Intelligence Research} {\bf
  4}, 129--145.

\bibitem[Conforti and Cornu\'{e}jols(1984)Conforti and
  Cornu\'{e}jols]{Conforti_Cornuejols_1984}
M.~Conforti and G.~Cornu\'{e}jols  (1984), Submodular set functions, matroids
  and the greedy algorithm: Tight worst-case bounds and some generalizations of
  the {R}ado--{E}dmonds theorem, {\em Discrete Applied Mathematics} {\bf 7}(3),
  251--274.

\bibitem[Conn {\em et~al.}(2009)Conn, Scheinberg and Vicente]{Conn2009}
A.~R. Conn, K.~Scheinberg and L.~N. Vicente  (2009), {\em {Introduction to
  Derivative-Free Optimization}}, SIAM.

\bibitem[Cook and Nachtsheim(1980)Cook and Nachtsheim]{cook1980comparison}
R.~D. Cook and C.~J. Nachtsheim  (1980), {A comparison of algorithms for
  constructing exact D-optimal designs}, {\em Technometrics} {\bf 22}(3),
  315--324.

\bibitem[Cotter {\em et~al.}(2013)Cotter, Roberts, Stuart and
  White]{Cotter_etal_2013}
S.~L. Cotter, G.~O. Roberts, A.~M. Stuart and D.~White  (2013), {MCMC methods
  for functions: Modifying old algorithms to make them faster}, {\em
  Statistical Science} {\bf 28}(3), 424--446.

\bibitem[Cover and Thomas(2006)Cover and Thomas]{Cover_2006}
T.~A. Cover and J.~A. Thomas  (2006), {\em {Elements of Information Theory}},
  second edition, Wiley.

\bibitem[Cox(1946)Cox]{Cox_1946a}
R.~T. Cox  (1946), {Probability, frequency and reasonable expectation}, {\em
  American Journal of Physics} {\bf 14}(1), 1--13.

\bibitem[Craig and Fisher(1936)Craig and Fisher]{craig1936design}
C.~C. Craig and R.~A. Fisher  (1936), The design of experiments, {\em The
  American Mathematical Monthly} {\bf 43}(3), 180.

\bibitem[Cui and Tong(2022)Cui and Tong]{Cui2022}
T.~Cui and X.~T. Tong  (2022), {A unified performance analysis of
  likelihood-informed subspace methods}, {\em Bernoulli} {\bf 28}(4),
  2788--2815.

\bibitem[Cui {\em et~al.}(2023)Cui, Dolgov and Zahm]{cui2023scalable}
T.~Cui, S.~Dolgov and O.~Zahm  (2023), {Scalable conditional deep inverse
  Rosenblatt transports using tensor trains and gradient-based dimension
  reduction}, {\em Journal of Computational Physics} {\bf 485}, 112103.

\bibitem[Cui {\em et~al.}(2016)Cui, Law and Marzouk]{Cui_2016}
T.~Cui, K.~J.~H. Law and Y.~M. Marzouk  (2016), {Dimension-independent
  likelihood-informed MCMC}, {\em Journal of Computational Physics} {\bf 304},
  109--137.

\bibitem[Cui {\em et~al.}(2014)Cui {\em et~al.}]{Cui2014a}
T.~Cui, J.~Martin, Y.~M. Marzouk, A.~Solonen and A.~Spantini  (2014),
  {Likelihood-informed dimension reduction for nonlinear inverse problems},
  {\em Inverse Problems} {\bf 30}(11), 114015.

\bibitem[Czy\.{z} {\em et~al.}(2023)Czy\.{z} {\em et~al.}]{Czyz2023}
P.~Czy\.{z}, F.~Grabowski, J.~Vogt, N.~Beerenwinkel and A.~Marx  (2023), Beyond
  normal: On the evaluation of mutual information estimators, in {\em Advances
  in Neural Information Processing Systems 36} (A.~Oh, T.~Neumann,
  A.~Globerson, K.~Saenko, M.~Hardt and S.~Levine, eds), Curran Associates,
  pp.~16957--16990.

\bibitem[Das and Kempe(2011)Das and Kempe]{Das_Kempe_2011}
A.~Das and D.~Kempe  (2011), {Submodular meets spectral: Greedy algorithms for
  subset selection, sparse approximation and dictionary selection}, in {\em
  Proceedings of the 28th International Conference on Machine Learning (ICML
  2011)} (L.~Getoor and T.~Scheffer, eds), ACM, pp.~1057--1064.

\bibitem[DasGupta(1995)DasGupta]{DasGupta_1995}
A.~DasGupta  (1995), {Review of optimal Bayes designs}, Technical report,
  Purdue University, West Lafayette, IN.

\bibitem[Dasgupta(2011)Dasgupta]{dasgupta2011two}
S.~Dasgupta  (2011), Two faces of active learning, {\em Theoretical Computer
  Science} {\bf 412}(19), 1767--1781.

\bibitem[Dashti and Stuart(2017)Dashti and Stuart]{StuartDashtihandbook}
M.~Dashti and A.~M. Stuart  (2017), The {B}ayesian approach to inverse
  problems, in {\em Handbook of Uncertainty Quantification} (R.~Ghanem,
  D.~Higdon and H.~Owhadi, eds), Springer, pp.~311--428.

\bibitem[Dashti {\em et~al.}(2013)Dashti, Law, Stuart and Voss]{dashti2013map}
M.~Dashti, K.~J. Law, A.~M. Stuart and J.~Voss  (2013), {MAP estimators and
  their consistency in Bayesian nonparametric inverse problems}, {\em Inverse
  Problems} {\bf 29}(9), 095017.

\bibitem[Dewaskar {\em et~al.}(2023)Dewaskar, Tosh, Knoblauch and
  Dunson]{dewaskar2023robustifying}
M.~Dewaskar, C.~Tosh, J.~Knoblauch and D.~B. Dunson  (2023), Robustifying
  likelihoods by optimistically re-weighting data.
\newblock Available at
  \href{https://arxiv.org/abs/2303.10525}{arXiv:2303.10525}.

\bibitem[Dick {\em et~al.}(2013)Dick, Kuo and Sloan]{dick2013high}
J.~Dick, F.~Y. Kuo and I.~H. Sloan  (2013), {High-dimensional integration: The
  quasi-Monte Carlo way}, {\em Acta Numerica} {\bf 22}, 133--288.

\bibitem[Dong {\em et~al.}(2024)Dong {\em et~al.}]{Dong2024}
J.~Dong, C.~Jacobsen, M.~Khalloufi, M.~Akramc, W.~Liu, K.~Duraisamy and X.~Huan
   (2024), Variational {B}ayesian optimal experimental design with normalizing
  flows.
\newblock Available at
  \href{https://arxiv.org/abs/2404.13056}{arXiv:2404.13056}.

\bibitem[Donsker and Varadhan(1983)Donsker and Varadhan]{Donsker1983}
M.~D. Donsker and S.~R.~S. Varadhan  (1983), {Asymptotic evaluation of certain
  Markov process expectations for large time. IV}, {\em Communications on Pure
  and Applied Mathematics} {\bf 36}(2), 183--212.

\bibitem[Dror and Steinberg(2008)Dror and Steinberg]{Dror_2008}
H.~A. Dror and D.~M. Steinberg  (2008), {Sequential experimental designs for
  generalized linear models}, {\em Journal of the American Statistical
  Association} {\bf 103}(481), 288--298.

\bibitem[Drovandi {\em et~al.}(2013)Drovandi, McGree and
  Pettitt]{Drovandi_2013}
C.~C. Drovandi, J.~M. McGree and A.~N. Pettitt  (2013), {Sequential Monte Carlo
  for Bayesian sequentially designed experiments for discrete data}, {\em
  Computational Statistics {\&} Data Analysis} {\bf 57}(1), 320--335.

\bibitem[Drovandi {\em et~al.}(2014)Drovandi, McGree and
  Pettitt]{Drovandi_2014}
C.~C. Drovandi, J.~M. McGree and A.~N. Pettitt  (2014), {A sequential Monte
  Carlo algorithm to incorporate model uncertainty in Bayesian sequential
  design}, {\em Journal of Computational and Graphical Statistics} {\bf 23}(1),
  3--24.

\bibitem[Duncan(1970)Duncan]{duncan1970calculation}
T.~E. Duncan  (1970), {On the calculation of mutual information}, {\em SIAM
  Journal on Applied Mathematics} {\bf 19}(1), 215--220.

\bibitem[Duong {\em et~al.}(2023)Duong, Helin and
  Rojo-Garcia]{duong2023stability}
D.-L. Duong, T.~Helin and J.~R. Rojo-Garcia  (2023), {Stability estimates for
  the expected utility in Bayesian optimal experimental design}, {\em Inverse
  Problems} {\bf 39}(12), 125008.

\bibitem[Elfving(1952)Elfving]{elfving1952optimum}
G.~Elfving  (1952), {Optimum allocation in linear regression theory}, {\em The
  Annals of Mathematical Statistics} {\bf 23}(2), 255--262.

\bibitem[Englezou {\em et~al.}(2022)Englezou, Waite and
  Woods]{englezou2022approximate}
Y.~Englezou, T.~W. Waite and D.~C. Woods  (2022), Approximate {L}aplace
  importance sampling for the estimation of expected {S}hannon information gain
  in high-dimensional {B}ayesian design for nonlinear models, {\em Statistics
  and Computing} {\bf 32}(5), 82.

\bibitem[Eskenazis and Shenfeld(2024)Eskenazis and
  Shenfeld]{eskenazis2024intrinsic}
A.~Eskenazis and Y.~Shenfeld  (2024), Intrinsic dimensional functional
  inequalities on model spaces, {\em Journal of Functional Analysis} {\bf
  286}(7), 110338.

\bibitem[Eswar {\em et~al.}(2024)Eswar, Rao and Saibaba]{eswar2024bayesian}
S.~Eswar, V.~Rao and A.~K. Saibaba  (2024), Bayesian {D}-optimal experimental
  designs via column subset selection.
\newblock Available at
  \href{https://arxiv.org/abs/2402.16000}{arXiv:2402.16000}.

\bibitem[Fan(1967)Fan]{Fan_1967}
K.~Fan  (1967), {Subadditive functions on a distributive lattice and an
  extension of Sz\'{a}sz's inequality}, {\em Journal of Mathematical Analysis
  and Applications} {\bf 18}(2), 262--268.

\bibitem[Fan(1968)Fan]{Fan_1968}
K.~Fan  (1968), An inequality for subadditive functions on a distributive
  lattice, with application to determinantal inequalities, {\em Linear Algebra
  and its Applications} {\bf 1}(1), 33--38.

\bibitem[Fedorov(1972)Fedorov]{fedorov1972theory}
V.~V. Fedorov  (1972), {\em Theory of Optimal Experiments}, Academic Press.

\bibitem[Fedorov(1996)Fedorov]{Fedorov_1996}
V.~V. Fedorov  (1996), {Design of spatial experiments: Model fitting and
  prediction}, Technical report, Oak Ridge National Laboratory, Oak Ridge, TN.

\bibitem[Fedorov and Flanagan(1997)Fedorov and Flanagan]{fedorov1997optimal}
V.~V. Fedorov and D.~Flanagan  (1997), {Optimal monitoring network design based
  on Mercer’s expansion of covariance kernel}, {\em Journal of Combinatorics,
  Information and System Sciences} {\bf 23}, 237--250.

\bibitem[Fedorov and Hackl(1997)Fedorov and Hackl]{Fedorov_1997}
V.~V. Fedorov and P.~Hackl  (1997), {\em {Model-Oriented Design of
  Experiments}}, Vol. 125 of Lecture Notes in Statistics, Springer.

\bibitem[Fedorov and M{\"u}ller(2007)Fedorov and
  M{\"u}ller]{fedorov2007optimum}
V.~V. Fedorov and W.~G. M{\"u}ller  (2007), Optimum design for correlated
  fields via covariance kernel expansions, in {\em mODa 8: Advances in
  Model-Oriented Design and Analysis} (J.~L\'{o}pez-Fidalgo, J.~M.
  Rodr\'{i}guez-D\'{i}az and B.~Torsney, eds), Contributions to Statistics,
  Physica, Springer, pp.~57--66.

\bibitem[Feldman and Langberg(2011)Feldman and Langberg]{Feldman2011}
D.~Feldman and M.~Langberg  (2011), {A unified framework for approximating and
  clustering data}, in {\em Proceedings of the 43rd Annual ACM Symposium on
  Theory of Computing (STOC 2011)}, ACM, p.~569–578.

\bibitem[Feng and Marzouk(2019)Feng and Marzouk]{Feng_2019}
C.~Feng and Y.~M. Marzouk  (2019), {A layered multiple importance sampling
  scheme for focused optimal Bayesian experimental design}.
\newblock Available at
  \href{https://arxiv.org/abs/1903.11187}{arXiv:1903.11187}.

\bibitem[Fisher {\em et~al.}(1978)Fisher, Nemhauser and Wolsey]{Fisher_1978}
M.~L. Fisher, G.~L. Nemhauser and L.~A. Wolsey  (1978), {An analysis of
  approximations for maximizing submodular set functions—II}, {\em
  Mathematical Programming} {\bf 8}(1), 73--87.

\bibitem[Fisher(1936)Fisher]{fisher1936design}
R.~A. Fisher  (1936), Design of experiments, {\em British Medical Journal} {\bf
  1}(3923), 554.

\bibitem[Ford {\em et~al.}(1989)Ford, Titterington and Kitsos]{Ford_1989}
I.~Ford, D.~M. Titterington and C.~P. Kitsos  (1989), {Recent advances in
  nonlinear experimental design}, {\em Technometrics} {\bf 31}(1), 49--60.

\bibitem[Foster {\em et~al.}(2021)Foster, Ivanova, Malik and
  Rainforth]{Foster_2021}
A.~Foster, D.~R. Ivanova, I.~Malik and T.~Rainforth  (2021), {Deep adaptive
  design: Amortizing sequential Bayesian experimental design}, in {\em
  Proceedings of the 38th International Conference on Machine Learning (ICML
  2021)} (M.~Meila and T.~Zhang, eds), Vol. 139 of Proceedings of Machine
  Learning Research, PMLR, pp.~3384--3395.

\bibitem[Foster {\em et~al.}(2019)Foster {\em et~al.}]{Foster_2019}
A.~Foster, M.~Jankowiak, E.~Bingham, P.~Horsfall, Y.~W. Teh, T.~Rainforth and
  N.~Goodman  (2019), {Variational Bayesian optimal experimental design}, in
  {\em Advances in Neural Information Processing Systems 32} (H.~Wallach,
  H.~Larochelle, A.~Beygelzimer, F.~d\textquotesingle Alch\'{e}-Buc, E.~Fox and
  R.~Garnett, eds), Curran Associates, pp.~14036--14047.

\bibitem[Foster {\em et~al.}(2020)Foster {\em et~al.}]{Foster2020}
A.~Foster, M.~Jankowiak, M.~O'Meara, Y.~W. Teh and T.~Rainforth  (2020), {A
  unified stochastic gradient approach to designing Bayesian-optimal
  experiments}, in {\em Proceedings of the 23rd International Conference on
  Artificial Intelligence and Statistics}, Vol. 108 of Proceedings of Machine
  Learning Research, PMLR, pp.~2959--2969.

\bibitem[Frazier(2018)Frazier]{frazier2018tutorial}
P.~I. Frazier  (2018), Bayesian optimization, {\em INFORMS TutORials in
  Operations Research} {\bf 2018}, 255--278.

\bibitem[Freund {\em et~al.}(1997)Freund, Seung, Shamir and
  Tishby]{freund1997selective}
Y.~Freund, H.~S. Seung, E.~Shamir and N.~Tishby  (1997), Selective sampling
  using the query by committee algorithm, {\em Machine Learning} {\bf 28},
  133--168.

\bibitem[Fujishige(2005)Fujishige]{fujishige2005}
S.~Fujishige  (2005), {\em {Submodular Functions and Optimization}}, Vol.~58 of
  Annals of Discrete Mathematics, second edition, Elsevier.

\bibitem[Gantmacher and Kre\u{\i}n(1960)Gantmacher and
  Kre\u{\i}n]{Gantmacher_Krein_1960}
F.~R. Gantmacher and M.~G. Kre\u{\i}n  (1960), {\em Oszillationsmatrizen,
  Oszillationskerne und Kleine Schwingungen Mechanischer Systeme}, Vol.~5,
  Akademie.

\bibitem[Gao {\em et~al.}(2018)Gao, Oh and Viswanath]{Gao2018}
W.~Gao, S.~Oh and P.~Viswanath  (2018), {Demystifying fixed $k$-nearest
  neighbor information estimators}, {\em IEEE Transactions on Information
  Theory} {\bf 64}(8), 5629--5661.

\bibitem[Gautier and Pronzato(2000)Gautier and Pronzato]{Gautier_2000}
R.~Gautier and L.~Pronzato  (2000), {Adaptive control for sequential design},
  {\em Discussiones Mathematicae Probability and Statistics} {\bf 20}(1),
  97--113.

\bibitem[Ghattas and Willcox(2021)Ghattas and Willcox]{Ghattas2021}
O.~Ghattas and K.~Willcox  (2021), {Learning physics-based models from data:
  Perspectives from inverse problems and model reduction}, {\em Acta Numerica}
  {\bf 30}, 445--554.

\bibitem[Giles(2015)Giles]{Giles2015}
M.~B. Giles  (2015), {Multilevel Monte Carlo methods}, {\em Acta Numerica} {\bf
  24}, 259--328.

\bibitem[Gin{\'e} and Nickl(2021)Gin{\'e} and Nickl]{gine2021mathematical}
E.~Gin{\'e} and R.~Nickl  (2021), {\em {Mathematical Foundations of
  Infinite-Dimensional Statistical Models}}, Cambridge University Press.

\bibitem[Ginebra(2007)Ginebra]{Ginebra_2007}
J.~Ginebra  (2007), {On the measure of the information in a statistical
  experiment}, {\em Bayesian Analysis} {\bf 2}(1), 167--212.

\bibitem[Giraldi {\em et~al.}(2018)Giraldi, Le~Ma{\^\i}tre, Hoteit and
  Knio]{giraldi2018optimal}
L.~Giraldi, O.~P. Le~Ma{\^\i}tre, I.~Hoteit and O.~M. Knio  (2018), {Optimal
  projection of observations in a Bayesian setting}, {\em Computational
  Statistics \& Data Analysis} {\bf 124}, 252--276.

\bibitem[Gneiting and Raftery(2007)Gneiting and Raftery]{Gneiting_2007}
T.~Gneiting and A.~E. Raftery  (2007), {Strictly proper scoring rules,
  prediction, and estimation}, {\em Journal of the American Statistical
  Association} {\bf 102}(477), 359--378.

\bibitem[Go and Isaac(2022)Go and Isaac]{Go2022}
J.~Go and T.~Isaac  (2022), {Robust expected information gain for optimal
  Bayesian experimental design using ambiguity sets}, in {\em Proceedings of
  the 38th Conference on Uncertainty in Artificial Intelligence}, Vol. 180 of
  Proceedings of Machine Learning Research, PMLR, pp.~718--727.

\bibitem[Goda {\em et~al.}(2020)Goda, Hironaka and Iwamoto]{goda2020multilevel}
T.~Goda, T.~Hironaka and T.~Iwamoto  (2020), {Multilevel Monte Carlo estimation
  of expected information gains}, {\em Stochastic Analysis and Applications}
  {\bf 38}(4), 581--600.

\bibitem[Goda {\em et~al.}(2022)Goda, Hironaka, Kitade and Foster]{Goda2022}
T.~Goda, T.~Hironaka, W.~Kitade and A.~Foster  (2022), {Unbiased MLMC
  stochastic gradient-based optimization of Bayesian experimental designs},
  {\em SIAM Journal on Scientific Computing} {\bf 44}(1), A286--A311.

\bibitem[Gorodetsky and Marzouk(2016)Gorodetsky and Marzouk]{Gorodetsky_2016}
A.~Gorodetsky and Y.~Marzouk  (2016), {Mercer kernels and integrated variance
  experimental design: Connections between Gaussian process regression and
  polynomial approximation}, {\em SIAM/ASA Journal on Uncertainty
  Quantification} {\bf 4}(1), 796--828.

\bibitem[Gramacy(2020)Gramacy]{gramacy2020surrogates}
R.~B. Gramacy  (2020), {\em {Surrogates: Gaussian Process Modeling, Design, and
  Optimization for the Applied Sciences}}, CRC Press.

\bibitem[Gramacy(2022)Gramacy]{GPfullyBayessoftware}
R.~B. Gramacy  (2022), {{plgp:} Particle learning of {G}aussian processes}.
\newblock Available at
  \href{https://cran.r-project.org/package=plgp}{https://cran.r-project.org/package=plgp}.

\bibitem[Gramacy and Apley(2015)Gramacy and Apley]{gramacy2015local}
R.~B. Gramacy and D.~W. Apley  (2015), {Local Gaussian process approximation
  for large computer experiments}, {\em Journal of Computational and Graphical
  Statistics} {\bf 24}(2), 561--578.

\bibitem[Gr{\"{u}}nwald and van Ommen(2017)Gr{\"{u}}nwald and van
  Ommen]{Grunwald2017}
P.~Gr{\"{u}}nwald and T.~van Ommen  (2017), {Inconsistency of Bayesian
  inference for misspecified linear models, and a proposal for repairing it},
  {\em Bayesian Analysis} {\bf 12}(4), 1069--1103.

\bibitem[G{\"{u}}rkan {\em et~al.}(1994)G{\"{u}}rkan, {\"{O}}zge and
  Robinson]{Gurkan1994}
G.~G{\"{u}}rkan, A.~Y. {\"{O}}zge and S.~M. Robinson  (1994), {Sample-path
  optimization in simulation}, in {\em Proceedings of the 1994 Winter
  Simulation Conference (WSC '94)} (J.~D. Tew, M.~S. Manivannan, D.~A. Sadowski
  and A.~F. Seila, eds), ACM, pp.~247--254.

\bibitem[Haber {\em et~al.}(2008)Haber, Horesh and Tenorio]{Haber_2008}
E.~Haber, L.~Horesh and L.~Tenorio  (2008), {Numerical methods for experimental
  design of large-scale linear ill-posed inverse problems}, {\em Inverse
  Problems} {\bf 24}(5), 055012.

\bibitem[Haber {\em et~al.}(2009)Haber, Horesh and Tenorio]{haber2009numerical}
E.~Haber, L.~Horesh and L.~Tenorio  (2009), Numerical methods for the design of
  large-scale nonlinear discrete ill-posed inverse problems, {\em Inverse
  Problems} {\bf 26}(2), 025002.

\bibitem[Hainy {\em et~al.}(2016)Hainy, Drovandi and McGree]{Hainy_2016}
M.~Hainy, C.~C. Drovandi and J.~M. McGree  (2016), {Likelihood-free extensions
  for Bayesian sequentially designed experiments}, in {\em mODa 11: Advances in
  Model-Oriented Design and Analysis} (J.~Kunert, C.~M\"{u}ller and
  A.~Atkinson, eds), Contributions to Statistics, Springer, pp.~153--161.

\bibitem[Hainy {\em et~al.}(2022)Hainy, Price, Restif and Drovandi]{Hainy_2022}
M.~Hainy, D.~J. Price, O.~Restif and C.~Drovandi  (2022), {Optimal Bayesian
  design for model discrimination via classification}, {\em Statistics and
  Computing} {\bf 32}(2), 25.

\bibitem[Hairer {\em et~al.}(2011)Hairer, Stuart and Voss]{HairerStuart}
M.~Hairer, A.~M. Stuart and J.~Voss  (2011), {Signal processing problems on
  function space: Bayesian formulation, stochastic PDEs and effective MCMC
  methods}, in {\em The Oxford Handbook of Nonlinear Filtering} (D.~Crisan and
  B.~Rozovskii, eds), Oxford University Press, pp.~833--873.

\bibitem[Harari and Steinberg(2014)Harari and Steinberg]{harari2014optimal}
O.~Harari and D.~M. Steinberg  (2014), {Optimal designs for Gaussian process
  models via spectral decomposition}, {\em Journal of Statistical Planning and
  Inference} {\bf 154}, 87--101.

\bibitem[He {\em et~al.}(2022)He, Kou and Peng]{He_2022}
X.~D. He, S.~Kou and X.~Peng  (2022), {Risk measures: Robustness,
  elicitability, and backtesting}, {\em Annual Review of Statistics and Its
  Application} {\bf 9}(1), 141--166.

\bibitem[Healy and Schruben(1991)Healy and Schruben]{Healy1991}
K.~Healy and L.~W. Schruben  (1991), {Retrospective simulation response
  optimization}, in {\em Proceedings of the 1991 Winter Simulation Conference
  (WSC '91)} (B.~L. Nelson, W.~D. Kelton and G.~M. Clark, eds), IEEE Computer
  Society, pp.~901--906.

\bibitem[Hedayat(1981)Hedayat]{Hedayat_1981}
A.~Hedayat  (1981), {Study of optimality criteria in design of experiments}, in
  {\em Statistics and Related Topics: International Symposium Proceedings}
  (M.~Csorgo, D.~A. Dawson, J.~N.~K. Rao and A.~K. M.~E. Sahel, eds), Elsevier
  Science, pp.~39--56.

\bibitem[Helin and Kretschmann(2022)Helin and Kretschmann]{helin2022non}
T.~Helin and R.~Kretschmann  (2022), {Non-asymptotic error estimates for the
  Laplace approximation in Bayesian inverse problems}, {\em Numerische
  Mathematik} {\bf 150}(2), 521--549.

\bibitem[Helin {\em et~al.}(2022)Helin, Hyv{\"{o}}nen and Puska]{Helin_2022}
T.~Helin, N.~Hyv{\"{o}}nen and J.-P. Puska  (2022), {Edge-promoting adaptive
  Bayesian experimental design for X-ray imaging}, {\em SIAM Journal on
  Scientific Computing} {\bf 44}(3), B506--B530.

\bibitem[Herrmann {\em et~al.}(2020)Herrmann, Schwab and
  Zech]{herrmann2020deep}
L.~Herrmann, C.~Schwab and J.~Zech  (2020), {Deep neural network expression of
  posterior expectations in Bayesian PDE inversion}, {\em Inverse Problems}
  {\bf 36}(12), 125011.

\bibitem[Hoang {\em et~al.}(2014)Hoang, Low, Jaillet and
  Kankanhalli]{hoang2014nonmyopic}
T.~N. Hoang, B.~K.~H. Low, P.~Jaillet and M.~Kankanhalli  (2014), {Nonmyopic
  $\varepsilon$-Bayes-optimal active learning of Gaussian processes}, in {\em
  Proceedings of the 31st International Conference on Machine Learning (ICML
  2014)}, Vol.~32 of Proceedings of Machine Learning Research, PMLR,
  pp.~739--747.

\bibitem[Hochba(1997)Hochba]{hochba1997approximation}
D.~S. Hochba  (1997), Approximation algorithms for {NP}-hard problems, {\em ACM
  SIGACT News} {\bf 28}(2), 40--52.

\bibitem[Huan(2015)Huan]{Huan_2015}
X.~Huan  (2015), {Numerical approaches for sequential Bayesian optimal
  experimental design}, PhD thesis, Massachusetts Institute of Technology.

\bibitem[Huan and Marzouk(2013)Huan and Marzouk]{huan2013simulation}
X.~Huan and Y.~M. Marzouk  (2013), Simulation-based optimal {B}ayesian
  experimental design for nonlinear systems, {\em Journal of Computational
  Physics} {\bf 232}(1), 288--317.

\bibitem[Huan and Marzouk(2014)Huan and Marzouk]{Huan2014}
X.~Huan and Y.~M. Marzouk  (2014), {Gradient-based stochastic optimization
  methods in Bayesian experimental design}, {\em International Journal for
  Uncertainty Quantification} {\bf 4}(6), 479--510.

\bibitem[Huan and Marzouk(2016)Huan and Marzouk]{Huan_2016}
X.~Huan and Y.~M. Marzouk  (2016), {Sequential Bayesian optimal experimental
  design via approximate dynamic programming}.
\newblock Available at
  \href{https://arxiv.org/abs/1604.08320}{arXiv:1604.08320}.

\bibitem[Huang {\em et~al.}(2020)Huang, Chen, Tsirigotis and
  Courville]{huang2020convex}
C.-W. Huang, R.~T.~Q. Chen, C.~Tsirigotis and A.~Courville  (2020), {Convex
  potential flows: Universal probability distributions with optimal transport
  and convex optimization}.
\newblock Available at
  \href{https://arxiv.org/abs/2012.05942}{arXiv:2012.05942}.

\bibitem[Huggins {\em et~al.}(2016)Huggins, Campbell and
  Broderick]{huggins2016coresets}
J.~Huggins, T.~Campbell and T.~Broderick  (2016), {Coresets for scalable
  Bayesian logistic regression}, in {\em Advances in Neural Information
  Processing Systems 29} (D.~Lee, M.~Sugiyama, U.~Luxburg, I.~Guyon and
  R.~Garnett, eds), Curran Associates, pp.~4080--4088.

\bibitem[Huggins and Miller(2023)Huggins and Miller]{Huggins2023}
J.~H. Huggins and J.~W. Miller  (2023), {Reproducible model selection using
  bagged posteriors}, {\em Bayesian Analysis} {\bf 18}(1), 79--104.

\bibitem[Ivanova {\em et~al.}(2021)Ivanova {\em et~al.}]{Ivanova_2021}
D.~R. Ivanova, A.~Foster, S.~Kleinegesse, M.~U. Gutmann and T.~Rainforth
  (2021), {Implicit deep adaptive design: Policy-based experimental design
  without likelihoods}, in {\em Advances in Neural Information Processing
  Systems 34} (M.~Ranzato, A.~Beygelzimer, Y.~Dauphin, P.~Liang and J.~W.
  Vaughan, eds), Curran Associates, pp.~25785--25798.

\bibitem[Jacob {\em et~al.}(2017)Jacob, Murray, Holmes and Robert]{Jacob2017}
P.~E. Jacob, L.~M. Murray, C.~C. Holmes and C.~P. Robert  (2017), {Better
  together? Statistical learning in models made of modules}.
\newblock Available at
  \href{https://arxiv.org/abs/1708.08719}{arXiv:1708.08719}.

\bibitem[Jagalur-Mohan and Marzouk(2021)Jagalur-Mohan and
  Marzouk]{jagalur2021batch}
J.~Jagalur-Mohan and Y.~Marzouk  (2021), Batch greedy maximization of
  non-submodular functions: Guarantees and applications to experimental design,
  {\em Journal of Machine Learning Research} {\bf 22}(1), 11397--11458.

\bibitem[Jaynes and Bretthorst(2003)Jaynes and Bretthorst]{Jaynes_2003}
E.~T. Jaynes and G.~L. Bretthorst  (2003), {\em {Probability Theory: The Logic
  of Science}}, Cambridge University Press.

\bibitem[Johnson and Barrett(1985)Johnson and Barrett]{Johnson_1985}
C.~R. Johnson and W.~W. Barrett  (1985), {Spanning-tree extensions of the
  Hadamard--Fischer inequalities}, {\em Linear Algebra and its Applications}
  {\bf 66}, 177--193.

\bibitem[Johnson and Nachtsheim(1983)Johnson and Nachtsheim]{johnson1983some}
M.~E. Johnson and C.~J. Nachtsheim  (1983), {Some guidelines for constructing
  exact D-optimal designs on convex design spaces}, {\em Technometrics} {\bf
  25}(3), 271--277.

\bibitem[Johnson {\em et~al.}(1990)Johnson, Moore and
  Ylvisaker]{johnson1990minimax}
M.~E. Johnson, L.~M. Moore and D.~Ylvisaker  (1990), Minimax and maximin
  distance designs, {\em Journal of Statistical Planning and Inference} {\bf
  26}(2), 131--148.

\bibitem[Jones {\em et~al.}(1998)Jones, Schonlau and Welch]{jones1998efficient}
D.~R. Jones, M.~Schonlau and W.~J. Welch  (1998), Efficient global optimization
  of expensive black-box functions, {\em Journal of Global Optimization} {\bf
  13}, 455--492.

\bibitem[Joseph {\em et~al.}(2015)Joseph, Gul and Ba]{Joseph_2015}
V.~R. Joseph, E.~Gul and S.~Ba  (2015), {Maximum projection designs for
  computer experiments}, {\em Biometrika} {\bf 102}(2), 371--380.

\bibitem[Joseph {\em et~al.}(2020)Joseph, Gul and Ba]{Joseph_2020}
V.~R. Joseph, E.~Gul and S.~Ba  (2020), {Designing computer experiments with
  multiple types of factors: The MaxPro approach}, {\em Journal of Quality
  Technology} {\bf 52}(4), 343--354.

\bibitem[Jourdan and Franco(2010)Jourdan and Franco]{jourdan2010optimal}
A.~Jourdan and J.~Franco  (2010), {Optimal Latin hypercube designs for the
  Kullback--Leibler criterion}, {\em AStA Advances in Statistical Analysis}
  {\bf 94}, 341--351.

\bibitem[Kaelbling {\em et~al.}(1998)Kaelbling, Littman and
  Cassandra]{Kaelbling_1998}
L.~P. Kaelbling, M.~L. Littman and A.~R. Cassandra  (1998), {Planning and
  acting in partially observable stochastic domains}, {\em Artificial
  Intelligence} {\bf 101}(1-2), 99--134.

\bibitem[Kaelbling {\em et~al.}(1996)Kaelbling, Littman and
  Moore]{Kaelbling1996}
L.~P. Kaelbling, M.~L. Littman and A.~W. Moore  (1996), {Reinforcement
  learning: A survey}, {\em Journal of Artificial Intelligence Research} {\bf
  4}(1), 237--285.

\bibitem[Kaipio and Kolehmainen(2013)Kaipio and Kolehmainen]{Kaipio2013}
J.~Kaipio and V.~Kolehmainen  (2013), {Approximate marginalization over
  modelling errors and uncertainties in inverse problems}, in {\em Bayesian
  Theory and Applications}, Oxford University Press, pp.~644--672.

\bibitem[Kaipio and Somersalo(2006)Kaipio and Somersalo]{kaipio2006statistical}
J.~Kaipio and E.~Somersalo  (2006), {\em Statistical and Computational Inverse
  Problems}, Vol. 160 of Applied Mathematical Sciences, Springer.

\bibitem[Kaipio and Somersalo(2007)Kaipio and Somersalo]{Kaipio2007}
J.~Kaipio and E.~Somersalo  (2007), {Statistical inverse problems:
  Discretization, model reduction and inverse crimes}, {\em Journal of
  Computational and Applied Mathematics} {\bf 198}(2), 493--504.

\bibitem[Karaca and Kamgarpour(2018)Karaca and
  Kamgarpour]{Karaca_etal_2018_SupModRatio}
O.~Karaca and M.~Kamgarpour  (2018), Exploiting weak supermodularity for
  coalition-proof mechanisms, in {\em 2018 IEEE Conference on Decision and
  Control (CDC)}, IEEE, pp.~1118--1123.

\bibitem[Karhunen(1947)Karhunen]{Karhunen_1947}
K.~Karhunen  (1947), {Uber lineare Methoden in der
  Wahrscheinlichkeitsrechnung}, {\em Am. Acad. Sci. Fennicade, Ser. A, I} {\bf
  37}, 3--79.

\bibitem[Kelmans and Kimelfeld(1983)Kelmans and Kimelfeld]{Kelmans_1983}
A.~K. Kelmans and B.~N. Kimelfeld  (1983), Multiplicative submodularity of a
  matrix's principal minor as a function of the set of its rows and some
  combinatorial applications, {\em Discrete Mathematics} {\bf 44}(1), 113--116.

\bibitem[Kennamer {\em et~al.}(2023)Kennamer, Walton and
  Ihler]{kennamer2023design}
N.~Kennamer, S.~Walton and A.~Ihler  (2023), Design amortization for {B}ayesian
  optimal experimental design, in {\em Proceedings of the 37th AAAI Conference
  on Artificial Intelligence} (B.~Williams, Y.~Chen and J.~Neville, eds),
  number~7, AAAI Press, pp.~8220--8227.

\bibitem[Kennedy and O'Hagan(2001)Kennedy and O'Hagan]{Kennedy_2001}
M.~C. Kennedy and A.~O'Hagan  (2001), {Bayesian calibration of computer
  models}, {\em Journal of the Royal Statistical Society: Series B (Statistical
  Methodology)} {\bf 63}(3), 425--464.

\bibitem[Kiefer(1958)Kiefer]{kiefer1958nonrandomized}
J.~Kiefer  (1958), On the nonrandomized optimality and randomized nonoptimality
  of symmetrical designs, {\em The Annals of Mathematical Statistics} {\bf
  29}(3), 675--699.

\bibitem[Kiefer(1959)Kiefer]{kiefer1959optimum}
J.~Kiefer  (1959), Optimum experimental designs, {\em Journal of the Royal
  Statistical Society: Series B (Methodological)} {\bf 21}(2), 272--304.

\bibitem[Kiefer(1961{\em a})Kiefer]{kiefer1961optimum}
J.~Kiefer  (1961{\em a}), {Optimum designs in regression problems, II}, {\em
  The Annals of Mathematical Statistics} {\bf 32}(1), 298--325.

\bibitem[Kiefer(1961{\em b})Kiefer]{Kiefer_1960}
J.~Kiefer  (1961{\em b}), {Optimum experimental designs V, with applications to
  systematic and rotatable designs}, in {\em Proceedings of the Fourth Berkeley
  Symposium on Mathematical Statistics and Probability}, Vol.~1, University of
  California Press, pp.~381--405.

\bibitem[Kiefer(1974)Kiefer]{kiefer1974general}
J.~Kiefer  (1974), {General equivalence theory for optimum designs (approximate
  theory)}, {\em The Annals of Statistics} {\bf 2}(5), 849--879.

\bibitem[Kiefer and Wolfowitz(1952)Kiefer and Wolfowitz]{Kiefer1952}
J.~Kiefer and J.~Wolfowitz  (1952), {Stochastic estimation of the maximum of a
  regression function}, {\em The Annals of Mathematical Statistics} {\bf
  23}(3), 462--466.

\bibitem[Kiefer and Wolfowitz(1959)Kiefer and
  Wolfowitz]{kieferWolfowitz1959optimum}
J.~Kiefer and J.~Wolfowitz  (1959), Optimum designs in regression problems,
  {\em The Annals of Mathematical Statistics} {\bf 30}(2), 271--294.

\bibitem[Kiefer and Wolfowitz(1960)Kiefer and Wolfowitz]{kiefer1960equivalence}
J.~Kiefer and J.~Wolfowitz  (1960), The equivalence of two extremum problems,
  {\em Canadian Journal of Mathematics} {\bf 12}, 363--366.

\bibitem[Kim {\em et~al.}(2014)Kim {\em et~al.}]{Kim_2014}
W.~Kim, M.~A. Pitt, Z.-L. Lu, M.~Steyvers and J.~I. Myung  (2014), {A
  hierarchical adaptive approach to optimal experimental design}, {\em Neural
  Computation} {\bf 26}, 2565--2492.

\bibitem[King and Wong(2000)King and Wong]{King_2000}
J.~King and W.-K. Wong  (2000), {Minimax D-optimal designs for the logistic
  model}, {\em Biometrics} {\bf 56}(4), 1263--1267.

\bibitem[Kleijn and van~der Vaart(2012)Kleijn and van~der Vaart]{Kleijn2012}
B.~J.~K. Kleijn and A.~W. van~der Vaart  (2012), {The Bernstein--von-Mises
  theorem under misspecification}, {\em Electronic Journal of Statistics} {\bf
  6}, 354--381.

\bibitem[Kleinegesse and Gutmann(2020)Kleinegesse and Gutmann]{Kleinegesse2020}
S.~Kleinegesse and M.~U. Gutmann  (2020), {B}ayesian experimental design for
  implicit models by mutual information neural estimation, in {\em Proceedings
  of the 37th International Conference on Machine Learning (ICML 2020)}
  (H.~Daum\'{e} and A.~Singh, eds), Vol. 119 of Proceedings of Machine Learning
  Research, PMLR, pp.~5316--5326.

\bibitem[Kleinegesse and Gutmann(2021)Kleinegesse and
  Gutmann]{Kleinegesse2021a}
S.~Kleinegesse and M.~U. Gutmann  (2021), {Gradient-based Bayesian experimental
  design for implicit models using mutual information lower bounds}.
\newblock Available at
  \href{https://arxiv.org/abs/2105.04379}{arXiv:2105.04379}.

\bibitem[Kleinegesse {\em et~al.}(2021)Kleinegesse, Drovandi and
  Gutmann]{Kleinegesse_2021}
S.~Kleinegesse, C.~Drovandi and M.~U. Gutmann  (2021), {Sequential Bayesian
  experimental design for implicit models via mutual information}, {\em
  Bayesian Analysis} {\bf 16}(3), 773--802.

\bibitem[Kleywegt {\em et~al.}(2002)Kleywegt, Shapiro and Homem-de
  Mello]{Kleywegt2002}
A.~J. Kleywegt, A.~Shapiro and T.~Homem-de Mello  (2002), The sample average
  approximation method for stochastic discrete optimization, {\em SIAM Journal
  on Optimization} {\bf 12}(2), 479--502.

\bibitem[Knapik {\em et~al.}(2011)Knapik, van~der Vaart and van
  Zanten]{vanZanten}
B.~T. Knapik, A.~W. van~der Vaart and J.~H. van Zanten  (2011), {Bayesian
  inverse problems with Gaussian priors}, {\em The Annals of Statistics} {\bf
  39}(5), 2626--2657.

\bibitem[Knothe(1957)Knothe]{knothe1957contributions}
H.~Knothe  (1957), Contributions to the theory of convex bodies, {\em Michigan
  Mathematical Journal} {\bf 4}(1), 39--52.

\bibitem[Ko {\em et~al.}(1995)Ko, Lee and Queyranne]{Ko_1995}
C.-W. Ko, J.~Lee and M.~Queyranne  (1995), An exact algorithm for maximum
  entropy sampling, {\em Operations Research} {\bf 43}(4), 684--691.

\bibitem[Kobyzev {\em et~al.}(2020)Kobyzev, Prince and
  Brubaker]{kobyzev2020normalizing}
I.~Kobyzev, S.~J. Prince and M.~A. Brubaker  (2020), {Normalizing flows: An
  introduction and review of current methods}, {\em IEEE Transactions on
  Pattern Analysis and Machine Intelligence} {\bf 43}(11), 3964--3979.

\bibitem[Konda and Tsitsiklis(1999)Konda and Tsitsiklis]{Konda_1999}
V.~R. Konda and J.~N. Tsitsiklis  (1999), {Actor-critic algorithms}, in {\em
  Advances in Neural Information Processing Systems 12} (S.~Solla, T.~Leen and
  K.~M\"{u}ller, eds), MIT Press, pp.~1008--1014.

\bibitem[Korkel {\em et~al.}(1999)Korkel, Bauer, Bock and
  Schloder]{Korkel_1999}
S.~Korkel, I.~Bauer, H.~G. Bock and J.~P. Schloder  (1999), {A sequential
  approach for nonlinear optimum experimental design in DAE systems}, in {\em
  Scientific Computing in Chemical Engineering II} (F.~Keil, W.~Mackens,
  H.~Voss and J.~Werther, eds), Springer, pp.~338--345.

\bibitem[Kotelyanski\u{\i}(1950)Kotelyanski\u{\i}]{Kotelyanskii_1950}
D.~M. Kotelyanski\u{\i}  (1950), On the theory of nonnegative and oscillating
  matrices, {\em Ukrains'kyi Matematychnyi Zhurnal} {\bf 2}(2), 94--101.

\bibitem[Kouri {\em et~al.}(2022)Kouri, Jakeman and {Gabriel
  Huerta}]{Kouri_2022}
D.~P. Kouri, J.~D. Jakeman and J.~{Gabriel Huerta}  (2022), Risk-adapted
  optimal experimental design, {\em SIAM/ASA Journal on Uncertainty
  Quantification} {\bf 10}(2), 687--716.

\bibitem[Koval {\em et~al.}(2020)Koval, Alexanderian and
  Stadler]{koval2020optimal}
K.~Koval, A.~Alexanderian and G.~Stadler  (2020), {Optimal experimental design
  under irreducible uncertainty for linear inverse problems governed by PDEs},
  {\em Inverse Problems} {\bf 36}(7), 075007.

\bibitem[Koval {\em et~al.}(2024)Koval, Herzog and
  Scheichl]{koval2024tractable}
K.~Koval, R.~Herzog and R.~Scheichl  (2024), Tractable optimal experimental
  design using transport maps.
\newblock Available at
  \href{https://arxiv.org/abs/2401.07971}{arXiv:2401.07971}.

\bibitem[Kozachenko and Leonenko(1987)Kozachenko and Leonenko]{Kozachenko1987}
L.~F. Kozachenko and N.~N. Leonenko  (1987), {A statistical estimate for the
  entropy of a random vector}, {\em Problems of Information Transmission} {\bf
  23}(2), 9--16.

\bibitem[Kraskov {\em et~al.}(2004)Kraskov, St{\"{o}}gbauer and
  Grassberger]{Kraskov2004}
A.~Kraskov, H.~St{\"{o}}gbauer and P.~Grassberger  (2004), {Estimating mutual
  information}, {\em Physical Review E - Statistical Physics, Plasmas, Fluids,
  and Related Interdisciplinary Topics} {\bf 69}(6), 16.

\bibitem[Krause and Golovin(2014)Krause and Golovin]{Krause_Golovin_2014}
A.~Krause and D.~Golovin  (2014), Submodular function maximization, in {\em
  Tractability, Practical Approaches to Hard Problems} (B.~Lucas, H.~Youssef
  and K.~Pushmeet, eds), Cambridge University Press, pp.~71--104.

\bibitem[Krause {\em et~al.}(2008)Krause, Singh and Guestrin]{Krause_2008}
A.~Krause, A.~Singh and C.~Guestrin  (2008), {Near-optimal sensor placements in
  Gaussian processes: Theory, efficient algorithms and empirical studies}, {\em
  Journal of Machine Learning Research} {\bf 9}, 235--284.

\bibitem[Kuhn {\em et~al.}(2019)Kuhn, Esfahani, Nguyen and
  Shafieezadeh-Abadeh]{kuhn2019wasserstein}
D.~Kuhn, P.~M. Esfahani, V.~A. Nguyen and S.~Shafieezadeh-Abadeh  (2019),
  Wasserstein distributionally robust optimization: Theory and applications in
  machine learning, in {\em Operations Research \& Management Science in the
  Age of Analytics}, INFORMS, pp.~130--166.

\bibitem[Kushner and Yin(2003)Kushner and Yin]{Kushner2003}
H.~J. Kushner and G.~G. Yin  (2003), {\em {Stochastic Approximation and
  Recursive Algorithms and Applications}}, second edition, Springer.

\bibitem[Lam and Willcox(2017)Lam and Willcox]{lam2017lookahead}
R.~Lam and K.~Willcox  (2017), Lookahead {B}ayesian optimization with
  inequality constraints, in {\em Advances in Neural Information Processing
  Systems 30} (I.~Guyon, U.~V. Luxburg, S.~Bengio, H.~Wallach, R.~Fergus,
  S.~Vishwanathan and R.~Garnett, eds), Curran Associates, pp.~1890--1900.

\bibitem[Larson {\em et~al.}(2019)Larson, Menickelly and Wild]{Larson2019}
J.~Larson, M.~Menickelly and S.~M. Wild  (2019), {Derivative-free optimization
  methods}, {\em Acta Numerica} {\bf 28}, 287--404.

\bibitem[Lau and Zhou(2020)Lau and Zhou]{lau2020spectral}
L.~C. Lau and H.~Zhou  (2020), A spectral approach to network design, in {\em
  Proceedings of the 52nd Annual ACM SIGACT Symposium on Theory of Computing
  (STOC 2020)}, ACM, pp.~826--839.

\bibitem[Lau and Zhou(2022)Lau and Zhou]{lau2022local}
L.~C. Lau and H.~Zhou  (2022), A local search framework for experimental
  design, {\em SIAM Journal on Computing} {\bf 51}(4), 900--951.

\bibitem[{Le~Cam}(1964){Le~Cam}]{LeCam_1964}
L.~{Le~Cam}  (1964), Sufficiency and approximate sufficiency, {\em The Annals
  of Mathematical Statistics} {\bf 35}(4), 1419--1455.

\bibitem[Lehmann and Casella(1998)Lehmann and Casella]{Lehmann_1998}
E.~L. Lehmann and G.~Casella  (1998), {\em {Theory of Point Estimation}},
  Springer Texts in Statistics, Springer.

\bibitem[Leskovec {\em et~al.}(2007)Leskovec {\em et~al.}]{leskovec2007cost}
J.~Leskovec, A.~Krause, C.~Guestrin, C.~Faloutsos, J.~VanBriesen and N.~Glance
  (2007), Cost-effective outbreak detection in networks, in {\em Proceedings of
  the 13th ACM SIGKDD International Conference on Knowledge Discovery and Data
  Mining}, ACM, pp.~420--429.

\bibitem[Letizia {\em et~al.}(2023)Letizia, Novello and
  Tonello]{letizia2023variational}
N.~A. Letizia, N.~Novello and A.~M. Tonello  (2023), Variational $f$-divergence
  and derangements for discriminative mutual information estimation.
\newblock Available at
  \href{https://arxiv.org/abs/2305.20025}{arXiv:2305.20025}.

\bibitem[Lewis(1995)Lewis]{lewis1995sequential}
D.~D. Lewis  (1995), A sequential algorithm for training text classifiers:
  Corrigendum and additional data, {\em SIGIR Forum} {\bf 29}(2), 13--19.

\bibitem[Li {\em et~al.}(2024{\em a})Li, Baptista and
  Marzouk]{fengyi2024forthcoming}
F.~Li, R.~Baptista and Y.~Marzouk  (2024{\em a}), Expected information gain
  estimation via density approximations: Sample allocation and dimension
  reduction.
\newblock Forthcoming.

\bibitem[Li {\em et~al.}(2024{\em b})Li, Marzouk and Zahm]{li2023principal}
M.~T.~C. Li, Y.~Marzouk and O.~Zahm  (2024{\em b}), Principal feature detection
  via $\phi$-{S}obolev inequalities.
\newblock To appear in \textit{Bernoulli}. Available at
  \href{https://bernoullisociety.org/publications/bernoulli-journal/bernoulli-journal-papers}{https://bernoullisociety.org/publications/
  bernoulli-journal/bernoulli-journal-papers}.

\bibitem[Liepe {\em et~al.}(2013)Liepe, Filippi, Komorowski and
  Stumpf]{Liepe_2013}
J.~Liepe, S.~Filippi, M.~Komorowski and M.~P.~H. Stumpf  (2013), Maximizing the
  information content of experiments in systems biology, {\em PLoS
  Computational Biology} {\bf 9}(1), e1002888.

\bibitem[Lillicrap {\em et~al.}(2016)Lillicrap {\em et~al.}]{Lillicrap_2015}
T.~P. Lillicrap, J.~J. Hunt, A.~Pritzel, N.~Heess, T.~Erez, Y.~Tassa, D.~Silver
  and D.~Wierstra  (2016), {Continuous control with deep reinforcement
  learning}, in {\em Proceedings of the 4th International Conference on
  Learning Representations (ICLR 2016)} (Y.~Bengio and Y.~LeCun, eds).

\bibitem[Lindley(1956)Lindley]{Lindley_1956}
D.~V. Lindley  (1956), On a measure of the information provided by an
  experiment, {\em The Annals of Mathematical Statistics} {\bf 27}(4),
  986--1005.

\bibitem[Lo{\`{e}}ve(1948)Lo{\`{e}}ve]{Loeve_1948}
M.~Lo{\`{e}}ve  (1948), {Fonctions al{\'{e}}atoires du second ordre}, in {\em
  Processus Stochastique et Mouvement Brownien} (P.~L{\'{e}}vy, ed.), Gauthier
  Villars.

\bibitem[Long {\em et~al.}(2013)Long, Scavino, Tempone and Wang]{Long_2013}
Q.~Long, M.~Scavino, R.~Tempone and S.~Wang  (2013), {Fast estimation of
  expected information gains for Bayesian experimental designs based on Laplace
  approximations}, {\em Computer Methods in Applied Mechanics and Engineering}
  {\bf 259}, 24--39.

\bibitem[Loredo(2011)Loredo]{Loredo_2010}
T.~J. Loredo  (2011), Rotating stars and revolving planets: {B}ayesian
  exploration of the pulsating sky, in {\em Bayesian Statistics 9: Proceedings
  of the Ninth Valencia International Meeting} (J.~M. Bernardo, M.~J. Bayarri
  and J.~O. Berger, eds), Oxford University Press, Benidorm, Spain,
  pp.~361--392.

\bibitem[Lov{\'a}sz(1983)Lov{\'a}sz]{Lovasz1983}
L.~Lov{\'a}sz  (1983), Submodular functions and convexity, in {\em Mathematical
  Programming The State of the Art: Bonn 1982} (A.~Bachem, B.~Korte and
  M.~Gr{\"o}tschel, eds), Springer, pp.~235--257.

\bibitem[Lov{\'a}sz(2007)Lov{\'a}sz]{lovasz_book}
L.~Lov{\'a}sz  (2007), {\em {Combinatorial Problems and Exercises}}, second
  edition, American Mathematical Society.

\bibitem[MacKay(1992)MacKay]{Mackay_1992}
D.~J.~C. MacKay  (1992), Information-based objective functions for active data
  selection, {\em Neural Computation} {\bf 4}(4), 590--604.

\bibitem[Madan {\em et~al.}(2020)Madan, Nikolov, Singh and
  Tantipongpipat]{madan2020maximizing}
V.~Madan, A.~Nikolov, M.~Singh and U.~Tantipongpipat  (2020), Maximizing
  determinants under matroid constraints, in {\em 2020 IEEE 61st Annual
  Symposium on Foundations of Computer Science (FOCS)}, IEEE, pp.~565--576.

\bibitem[Madan {\em et~al.}(2019)Madan, Singh, Tantipongpipat and
  Xie]{madan2019combinatorial}
V.~Madan, M.~Singh, U.~Tantipongpipat and W.~Xie  (2019), Combinatorial
  algorithms for optimal design, in {\em Proceedings of the 32nd Conference on
  Learning Theory}, Vol.~99 of Proceedings of Machine Learning Research, PMLR,
  pp.~2210--2258.

\bibitem[Mak {\em et~al.}(1999)Mak, Morton and Wood]{Mak1999}
W.-K. Mak, D.~P. Morton and R.~K. Wood  (1999), {Monte Carlo bounding
  techniques for determining solution quality in stochastic programs}, {\em
  Operations Research Letters} {\bf 24}(1-2), 47--56.

\bibitem[Manole {\em et~al.}(2021)Manole, Balakrishnan, Niles-Weed and
  Wasserman]{manole2021plugin}
T.~Manole, S.~Balakrishnan, J.~Niles-Weed and L.~Wasserman  (2021), Plugin
  estimation of smooth optimal transport maps.
\newblock Available at
  \href{https://arxiv.org/abs/2107.12364}{arXiv:2107.12364}.

\bibitem[Markowitz(1952)Markowitz]{Markowitz_1952}
H.~Markowitz  (1952), Portfolio selection, {\em The Journal of Finance} {\bf
  7}(1), 77--91.

\bibitem[Marzouk and Xiu(2009)Marzouk and Xiu]{marzouk2009stochastic}
Y.~Marzouk and D.~Xiu  (2009), A stochastic collocation approach to {B}ayesian
  inference in inverse problems, {\em Communications in Computational Physics}
  {\bf 6}, 826--847.

\bibitem[Marzouk {\em et~al.}(2016)Marzouk, Moselhy, Parno and
  Spantini]{Marzouk_2016}
Y.~Marzouk, T.~Moselhy, M.~Parno and A.~Spantini  (2016), Sampling via measure
  transport: {A}n introduction, in {\em Handbook of Uncertainty Quantification}
  (R.~Ghanem, D.~Higdon and H.~Owhadi, eds), Springer, pp.~1--41.

\bibitem[McAllester and Stratos(2020)McAllester and Stratos]{McAllester2020}
D.~McAllester and K.~Stratos  (2020), Formal limitations on the measurement of
  mutual information, in {\em Proceedings of the 23rd International Conference
  on Artificial Intelligence and Statistics}, Vol. 108 of Proceedings of
  Machine Learning Research, PMLR, pp.~875--884.

\bibitem[McGree {\em et~al.}(2012)McGree, Drovandi and Pettitt]{McGree_2012}
J.~McGree, C.~Drovandi and A.~Pettitt  (2012), {A sequential Monte Carlo
  approach to the sequential design for discriminating between rival continuous
  data models}, Technical report, Queensland University of Technology,
  Brisbane, Australia.

\bibitem[McKay {\em et~al.}(1979)McKay, Beckman and
  Conover]{mckay1979comparison}
M.~D. McKay, R.~J. Beckman and W.~J. Conover  (1979), A comparison of three
  methods for selecting values of input variables in the analysis of output
  from a computer code, {\em Technometrics} {\bf 21}(2), 239.

\bibitem[Melendez {\em et~al.}(2021)Melendez {\em
  et~al.}]{melendez2021designing}
J.~A. Melendez, R.~J. Furnstahl, H.~W. Grie{\ss}hammer, J.~A. McGovern, D.~R.
  Phillips and M.~T. Pratola  (2021), Designing optimal experiments: {A}n
  application to proton {C}ompton scattering, {\em The European Physical
  Journal A} {\bf 57}, 1--24.

\bibitem[Mertikopoulos {\em et~al.}(2020)Mertikopoulos, Hallak, Kavis and
  Cevher]{mertikopoulos2020almost}
P.~Mertikopoulos, N.~Hallak, A.~Kavis and V.~Cevher  (2020), On the almost sure
  convergence of stochastic gradient descent in non-convex problems, in {\em
  Advances in Neural Information Processing Systems 33} (H.~Larochelle,
  M.~Ranzato, R.~Hadsell, M.~Balcan and H.~Lin, eds), Curran Associates,
  pp.~1117--1128.

\bibitem[Meyer and Nachtsheim(1995)Meyer and Nachtsheim]{meyer1995coordinate}
R.~K. Meyer and C.~J. Nachtsheim  (1995), The coordinate-exchange algorithm for
  constructing exact optimal experimental designs, {\em Technometrics} {\bf
  37}(1), 60--69.

\bibitem[Miller and Dunson(2019)Miller and Dunson]{Miller2019}
J.~W. Miller and D.~B. Dunson  (2019), Robust {B}ayesian inference via
  coarsening, {\em Journal of the American Statistical Association} {\bf
  114}(527), 1113--1125.

\bibitem[Mirzasoleiman {\em et~al.}(2015)Mirzasoleiman {\em
  et~al.}]{lazier_than_lazy_greedy}
B.~Mirzasoleiman, A.~Badanidiyuru, A.~Karbasi, J.~Vondr{\'a}k and A.~Krause
  (2015), Lazier than lazy greedy, in {\em Proceedings of the 29th {AAAI}
  Conference on Artificial Intelligence (AAAI 2015)}, p.~1812–1818.

\bibitem[Mirzasoleiman {\em et~al.}(2013)Mirzasoleiman, Karbasi, Sarkar and
  Krause]{Mirzasoleiman_etal_2013}
B.~Mirzasoleiman, A.~Karbasi, R.~Sarkar and A.~Krause  (2013), Distributed
  submodular maximization: {I}dentifying representative elements in massive
  data, in {\em Advances in Neural Information Processing Systems 26} (C.~J.
  Burges, L.~Bottou, M.~Welling, Z.~Ghahramani and K.~Q. Weinberger, eds),
  Curran Associates, pp.~2049--2057.

\bibitem[Mnih {\em et~al.}(2015)Mnih {\em et~al.}]{Mnih_2015}
V.~Mnih, K.~Kavukcuoglu, D.~Silver, A.~A. Rusu, J.~Veness, M.~G. Bellemare,
  A.~Graves, M.~Riedmiller, A.~K. Fidjeland, G.~Ostrovski, S.~Petersen,
  C.~Beattie, A.~Sadik, I.~Antonoglou, H.~King, D.~Kumaran, D.~Wierstra,
  S.~Legg and D.~Hassabis  (2015), {Human-level control through deep
  reinforcement learning}, {\em Nature} {\bf 518}(7540), 529--533.

\bibitem[Mo{\v{c}}kus(1975)Mo{\v{c}}kus]{movckus1975bayesian}
J.~Mo{\v{c}}kus  (1975), {On Bayesian methods for seeking the extremum}, in
  {\em Optimization Techniques IFIP Technical Conference} (G.~I. Marchuk, ed.),
  Springer, pp.~400--404.

\bibitem[Mohamed {\em et~al.}(2020)Mohamed, Rosca, Figurnov and
  Mnih]{mohamed2020monte}
S.~Mohamed, M.~Rosca, M.~Figurnov and A.~Mnih  (2020), Monte {C}arlo gradient
  estimation in machine learning, {\em Journal of Machine Learning Research}
  {\bf 21}(1), 5183--5244.

\bibitem[Morrison {\em et~al.}(2018)Morrison, Oliver and Moser]{Morrison2018}
R.~E. Morrison, T.~A. Oliver and R.~D. Moser  (2018), {Representing model
  inadequacy: A stochastic operator approach}, {\em SIAM/ASA Journal on
  Uncertainty Quantification} {\bf 6}(2), 457--496.

\bibitem[{Moselhy} and Marzouk(2012){Moselhy} and Marzouk]{ElMoselhy_2012}
T.~A. {Moselhy} and Y.~M. Marzouk  (2012), {Bayesian inference with optimal
  maps}, {\em Journal of Computational Physics} {\bf 231}(23), 7815--7850.

\bibitem[M{\"{u}}ller {\em et~al.}(2007)M{\"{u}}ller {\em et~al.}]{Muller_2007}
P.~M{\"{u}}ller, D.~A. Berry, A.~P. Grieve, M.~Smith and M.~Krams  (2007),
  {Simulation-based sequential Bayesian design}, {\em Journal of Statistical
  Planning and Inference} {\bf 137}(10), 3140--3150.

\bibitem[M{\"{u}}ller {\em et~al.}(2022)M{\"{u}}ller, Duan and {Garcia
  Tec}]{Muller_2022}
P.~M{\"{u}}ller, Y.~Duan and M.~{Garcia Tec}  (2022), {Simulation‐based
  sequential design}, {\em Pharmaceutical Statistics} {\bf 21}(4), 729--739.

\bibitem[Murphy(2003)Murphy]{Murphy_2003}
S.~A. Murphy  (2003), {Optimal dynamic treatment regimes}, {\em Journal of the
  Royal Statistical Society: Series B (Statistical Methodology)} {\bf 65}(2),
  331--366.

\bibitem[Myung and Pitt(2009)Myung and Pitt]{Myung_2009}
J.~I. Myung and M.~A. Pitt  (2009), Optimal experimental design for model
  discrimination, {\em Psychological Review} {\bf 116}(3), 499--518.

\bibitem[Nelder and Mead(1965)Nelder and Mead]{Nelder1965}
J.~A. Nelder and R.~Mead  (1965), {A simplex method for function minimization},
  {\em The Computer Journal} {\bf 7}(4), 308--313.

\bibitem[Nemhauser and Wolsey(1978)Nemhauser and Wolsey]{Nemhauser1978b}
G.~L. Nemhauser and L.~A. Wolsey  (1978), Best algorithms for approximating the
  maximum of a submodular set function, {\em Mathematics of Operations
  Research} {\bf 3}(3), 177--188.

\bibitem[Nemhauser {\em et~al.}(1978)Nemhauser, Wolsey and
  Fisher]{Nemhauser1978}
G.~L. Nemhauser, L.~A. Wolsey and M.~L. Fisher  (1978), {An analysis of
  approximations for maximizing submodular set functions---I}, {\em
  Mathematical Programming} {\bf 14}(1), 265--294.

\bibitem[Nesterov(1983)Nesterov]{Nesterov1983}
Y.~Nesterov  (1983), {A method of solving a convex programming problem with
  convergence rate $\mathcal{O}\left(1/k^2\right)$}, {\em Soviet Mathematics
  Doklady} {\bf 27}(2), 372--376.

\bibitem[Ng and Russell(2000)Ng and Russell]{Ng_2000}
A.~Ng and S.~Russell  (2000), Algorithms for inverse reinforcement learning, in
  {\em Proceedings of the 17th International Conference on Machine Learning
  (ICML 2000)}, Morgan Kaufmann, pp.~663--670.

\bibitem[Nguyen {\em et~al.}(2010)Nguyen, Wainwright and Jordan]{Nguyen2010}
X.~Nguyen, M.~J. Wainwright and M.~I. Jordan  (2010), {Estimating divergence
  functionals and the likelihood ratio by convex risk minimization}, {\em IEEE
  Transactions on Information Theory} {\bf 56}(11), 5847--5861.

\bibitem[Niederreiter(1992)Niederreiter]{Niederreiter_1992}
H.~Niederreiter  (1992), {\em {Random Number Generation and Quasi-Monte Carlo
  Methods}}, SIAM.

\bibitem[Nikolov and Singh(2016)Nikolov and Singh]{nikolov2016maximizing}
A.~Nikolov and M.~Singh  (2016), Maximizing determinants under partition
  constraints, in {\em Proceedings of the 48th Annual ACM Symposium on Theory
  of Computing (STOC 2016)}, ACM, pp.~192--201.

\bibitem[Nikolov {\em et~al.}(2022)Nikolov, Singh and
  Tantipongpipat]{nikolov2022proportional}
A.~Nikolov, M.~Singh and U.~Tantipongpipat  (2022), Proportional volume
  sampling and approximation algorithms for {A}-optimal design, {\em
  Mathematics of Operations Research} {\bf 47}(2), 847--877.

\bibitem[Nocedal and Wright(2006)Nocedal and Wright]{Nocedal2006}
J.~Nocedal and S.~J. Wright  (2006), {\em {Numerical Optimization}}, Springer.

\bibitem[Norkin {\em et~al.}(1998)Norkin, Pflug and Ruszczynski]{Norkin1998}
V.~Norkin, G.~Pflug and A.~Ruszczynski  (1998), {A branch and bound method for
  stochastic global optimization}, {\em Mathematical Programming} {\bf
  83}(1-3), 425--450.

\bibitem[O'Hagan {\em et~al.}(2006)O'Hagan {\em et~al.}]{OHagan_2006a}
A.~O'Hagan, C.~E. Buck, A.~Daneshkhah, J.~R. Eiser, P.~H. Garthwaite, D.~J.
  Jenkinson, J.~E. Oakley and T.~Rakow  (2006), {\em {Uncertain Judgements:
  Eliciting Experts' Probabilities}}, Wiley.

\bibitem[Orozco {\em et~al.}(2024)Orozco, Herrmann and Chen]{Chen2024}
R.~Orozco, F.~J. Herrmann and P.~Chen  (2024), Probabilistic {B}ayesian optimal
  experimental design using conditional normalizing flows.
\newblock Available at
  \href{https://arxiv.org/abs/2402.18337}{arXiv:2402.18337}.

\bibitem[Overstall and McGree(2022)Overstall and McGree]{overstall2022bayesian}
A.~Overstall and J.~McGree  (2022), Bayesian decision-theoretic design of
  experiments under an alternative model, {\em Bayesian Analysis} {\bf 17}(4),
  1021--1041.

\bibitem[Overstall(2022)Overstall]{Overstall_2022}
A.~M. Overstall  (2022), {Properties of Fisher information gain for Bayesian
  design of experiments}, {\em Journal of Statistical Planning and Inference}
  {\bf 218}, 138--146.

\bibitem[Overstall and Woods(2017)Overstall and Woods]{Overstall2017}
A.~M. Overstall and D.~C. Woods  (2017), Bayesian design of experiments using
  approximate coordinate exchange, {\em Technometrics} {\bf 59}(4), 458--470.

\bibitem[Overstall {\em et~al.}(2018)Overstall, McGree and
  Drovandi]{overstall2018approach}
A.~M. Overstall, J.~M. McGree and C.~C. Drovandi  (2018), An approach for
  finding fully {B}ayesian optimal designs using normal-based approximations to
  loss functions, {\em Statistics and Computing} {\bf 28}, 343--358.

\bibitem[Owen(1992)Owen]{Owen_1992}
A.~B. Owen  (1992), Orthogonal arrays for computer experiments, integration and
  visualization, {\em Statistica Sinica} {\bf 2}, 439--452.

\bibitem[Owen(2013)Owen]{mcbook}
A.~B. Owen  (2013), Monte {C}arlo theory, methods and examples.
\newblock Available at \href{https://artowen.su.domains/mc/}{https://
  artowen.su.domains/mc/}.

\bibitem[Papadimitriou and Steiglitz(1998)Papadimitriou and
  Steiglitz]{Papadimitriou_Steiglitz_book}
C.~H. Papadimitriou and K.~Steiglitz  (1998), {\em Combinatorial Optimization:
  Algorithms and Complexity}, Courier Corporation.

\bibitem[Papamakarios {\em et~al.}(2021)Papamakarios {\em
  et~al.}]{papamakarios2021normalizing}
G.~Papamakarios, E.~Nalisnick, D.~J. Rezende, S.~Mohamed and
  B.~Lakshminarayanan  (2021), Normalizing flows for probabilistic modeling and
  inference, {\em Journal of Machine Learning Research} {\bf 22}(1),
  2617--2680.

\bibitem[Pardo-Ig{\'u}zquiza(1998)Pardo-Ig{\'u}zquiza]{pardo1998maximum}
E.~Pardo-Ig{\'u}zquiza  (1998), Maximum likelihood estimation of spatial
  covariance parameters, {\em Mathematical Geology} {\bf 30}, 95--108.

\bibitem[Peters and Schaal(2008)Peters and Schaal]{Peters_2008}
J.~Peters and S.~Schaal  (2008), Natural actor-critic, {\em Neurocomputing}
  {\bf 71}(7--9), 1180--1190.

\bibitem[Pilz(1991)Pilz]{pilz1991bayesian}
J.~Pilz  (1991), {\em {Bayesian Estimation and Experimental Design in Linear
  Regression Models}}, Wiley.

\bibitem[Polyak and Juditsky(1992)Polyak and Juditsky]{Polyak1992}
B.~T. Polyak and A.~B. Juditsky  (1992), {Acceleration of stochastic
  approximation by averaging}, {\em SIAM Journal on Control and Optimization}
  {\bf 30}(4), 838--855.

\bibitem[Pompe and Jacob(2021)Pompe and Jacob]{Pompe2021}
E.~Pompe and P.~E. Jacob  (2021), {Asymptotics of cut distributions and robust
  modular inference using posterior bootstrap}.
\newblock Available at
  \href{https://arxiv.org/abs/2110.11149}{arXiv:2110.11149}.

\bibitem[Pooladian and Niles-Weed(2021)Pooladian and
  Niles-Weed]{pooladian2021entropic}
A.-A. Pooladian and J.~Niles-Weed  (2021), Entropic estimation of optimal
  transport maps.
\newblock Available at
  \href{https://arxiv.org/abs/2109.12004}{arXiv:2109.12004}.

\bibitem[Poole {\em et~al.}(2019)Poole {\em et~al.}]{poole2019variational}
B.~Poole, S.~Ozair, A.~Van Den~Oord, A.~Alemi and G.~Tucker  (2019), On
  variational bounds of mutual information, in {\em Proceedings of the 36th
  International Conference on Machine Learning (ICML 2019)}, Vol.~97 of
  Proceedings of Machine Learning Research, PMLR, pp.~5171--5180.

\bibitem[Powell(2011)Powell]{Powell2011}
W.~B. Powell  (2011), {\em {Approximate Dynamic Programming: Solving the Curses
  of Dimensionality}}, second edition, Wiley.

\bibitem[Prangle {\em et~al.}(2023)Prangle, Harbisher and
  Gillespie]{Prangle_2023}
D.~Prangle, S.~Harbisher and C.~S. Gillespie  (2023), Bayesian experimental
  design without posterior calculations: {A}n adversarial approach, {\em
  Bayesian Analysis} {\bf 18}(1), 133--163.

\bibitem[Pronzato and M{\"u}ller(2012)Pronzato and
  M{\"u}ller]{pronzato2012design}
L.~Pronzato and W.~G. M{\"u}ller  (2012), Design of computer experiments:
  {S}pace filling and beyond, {\em Statistics and Computing} {\bf 22},
  681--701.

\bibitem[Pronzato and Thierry(2002)Pronzato and Thierry]{Pronzato_2002}
L.~Pronzato and {\'{E}}.~Thierry  (2002), {Sequential experimental design and
  response optimisation}, {\em Statistical Methods and Applications} {\bf
  11}(3), 277--292.

\bibitem[Pronzato and Walter(1985)Pronzato and Walter]{Pronzato_1985}
L.~Pronzato and E.~Walter  (1985), {Robust experiment design via stochastic
  approximation}, {\em Mathematical Biosciences} {\bf 75}(1), 103--120.

\bibitem[Pukelsheim(2006)Pukelsheim]{Pukelsheim_2006}
F.~Pukelsheim  (2006), {\em {Optimal Design of Experiments}}, SIAM.

\bibitem[Rahimian and Mehrotra(2019)Rahimian and
  Mehrotra]{rahimian2019distributionally}
H.~Rahimian and S.~Mehrotra  (2019), Distributionally robust optimization: {A}
  review.
\newblock Available at
  \href{https://arxiv.org/abs/1908.05659}{arXiv:1908.05659}.

\bibitem[Raiffa and Schlaifer(1961)Raiffa and Schlaifer]{Raiffa_1961}
H.~Raiffa and R.~Schlaifer  (1961), {\em {Applied Statistical Decision
  Theory}}, Wiley.

\bibitem[Rainforth {\em et~al.}(2018)Rainforth {\em
  et~al.}]{rainforth2018nesting}
T.~Rainforth, R.~Cornish, H.~Yang, A.~Warrington and F.~Wood  (2018), On
  nesting {M}onte {C}arlo estimators, in {\em Proceedings of the 35th
  International Conference on Machine Learning (ICML 2018)}, Vol.~80 of
  Proceedings of Machine Learning Research, PMLR, pp.~4267--4276.

\bibitem[Rainforth {\em et~al.}(2023)Rainforth, Foster, Ivanova and
  Smith]{Rainforth_2023}
T.~Rainforth, A.~Foster, D.~R. Ivanova and F.~B. Smith  (2023), Modern
  {B}ayesian experimental design, {\em Statistical Science} {\bf 39}(1),
  100--114.

\bibitem[Rasmussen and Williams(2006)Rasmussen and Williams]{Rasmussen_2006}
C.~E. Rasmussen and C.~K.~I. Williams  (2006), {\em {Gaussian Processes for
  Machine Learning}}, MIT Press.

\bibitem[Rhee and Glynn(2015)Rhee and Glynn]{Rhee2015}
C.~H. Rhee and P.~W. Glynn  (2015), {Unbiased estimation with square root
  convergence for SDE models}, {\em Operations Research} {\bf 63}(5),
  1026--1043.

\bibitem[Riis {\em et~al.}(2022)Riis {\em et~al.}]{riis2022bayesian}
C.~Riis, F.~Antunes, F.~H{\"u}ttel, C.~Lima~Azevedo and F.~Pereira  (2022),
  Bayesian active learning with fully {B}ayesian {G}aussian processes, in {\em
  Advances in Neural Information Processing Systems 35} (S.~Koyejo, S.~Mohamed,
  A.~Agarwal, D.~Belgrave, K.~Cho and A.~Oh, eds), Curran Associates,
  pp.~12141--12153.

\bibitem[Riley {\em et~al.}(2019)Riley {\em et~al.}]{Riley_2019}
Z.~B. Riley, R.~A. Perez, G.~W. Bartram, S.~M. Spottswood, B.~P. Smarslok and
  T.~J. Beberniss  (2019), {Aerothermoelastic experimental design for the
  AEDC/VKF Tunnel C: Challenges associated with measuring the response of
  flexible panels in high-temperature, high-speed wind tunnels}, {\em Journal
  of Sound and Vibration} {\bf 441}, 96--105.

\bibitem[Robbins and Monro(1951)Robbins and Monro]{Robbins1951}
H.~Robbins and S.~Monro  (1951), A stochastic approximation method, {\em The
  Annals of Mathematical Statistics} {\bf 22}(3), 400--407.

\bibitem[Robertazzi and Schwartz(1989)Robertazzi and
  Schwartz]{Robertazzi_Schwartz_1989}
T.~Robertazzi and S.~Schwartz  (1989), An accelerated sequential algorithm for
  producing {D}-optimal designs, {\em SIAM Journal on Scientific and
  Statistical Computing} {\bf 10}(2), 341--358.

\bibitem[Rockafellar and Royset(2015)Rockafellar and Royset]{Rockafellar_2015}
R.~T. Rockafellar and J.~O. Royset  (2015), Measures of residual risk with
  connections to regression, risk tracking, surrogate models, and ambiguity,
  {\em SIAM Journal on Optimization} {\bf 25}(2), 1179--1208.

\bibitem[Rockafellar and Uryasev(2002)Rockafellar and
  Uryasev]{Rockafellar_2002}
R.~T. Rockafellar and S.~Uryasev  (2002), {Conditional value-at-risk for
  general loss distributions}, {\em Journal of Banking \& Finance} {\bf 26}(7),
  1443--1471.

\bibitem[Rockafellar and Uryasev(2013)Rockafellar and
  Uryasev]{Rockafellar_2013}
R.~T. Rockafellar and S.~Uryasev  (2013), {The fundamental risk quadrangle in
  risk management, optimization and statistical estimation}, {\em Surveys in
  Operations Research and Management Science} {\bf 18}(1--2), 33--53.

\bibitem[Rosenblatt(1952)Rosenblatt]{rosenblatt1952remarks}
M.~Rosenblatt  (1952), Remarks on a multivariate transformation, {\em The
  Annals of Mathematical Statistics} {\bf 23}(3), 470--472.

\bibitem[Royset(2022)Royset]{Royset_2022}
J.~O. Royset  (2022), Risk-adaptive approaches to learning and decision making:
  {A} survey.
\newblock Available at
  \href{https://arxiv.org/abs/2212.00856}{arXiv:2212.00856}.

\bibitem[Rudolf and Sprungk(2018)Rudolf and Sprungk]{rudolf2018generalization}
D.~Rudolf and B.~Sprungk  (2018), {On a generalization of the preconditioned
  Crank--Nicolson Metropolis algorithm}, {\em Foundations of Computational
  Mathematics} {\bf 18}, 309--343.

\bibitem[Ruppert(1988)Ruppert]{Ruppert1988}
D.~Ruppert  (1988), {Efficient estimations from a slowly convergent
  Robbins-Monro process}, Technical report, Cornell University.
\newblock Available at
  \href{http://ecommons.cornell.edu/bitstream/handle/1813/8664/TR000781.pdf?sequence=1}{http://ecommons.cornell.edu/
  bitstream/handle/1813/8664/TR000781.pdf?sequence=1}.

\bibitem[Ruthotto {\em et~al.}(2018)Ruthotto, Chung and Chung]{Ruthotto2018}
L.~Ruthotto, J.~Chung and M.~Chung  (2018), Optimal experimental design for
  inverse problems with state constraints, {\em SIAM Journal on Scientific
  Computing} {\bf 40}(4), B1080--B1100.

\bibitem[Ryan {\em et~al.}(2016)Ryan, Drovandi, McGree and Pettitt]{Ryan_2016}
E.~G. Ryan, C.~C. Drovandi, J.~M. McGree and A.~N. Pettitt  (2016), A review of
  modern computational algorithms for {B}ayesian optimal design, {\em
  International Statistical Review} {\bf 84}(1), 128--154.

\bibitem[Ryan(2003)Ryan]{ryan2003estimating}
K.~J. Ryan  (2003), Estimating expected information gains for experimental
  designs with application to the random fatigue-limit model, {\em Journal of
  Computational and Graphical Statistics} {\bf 12}(3), 585--603.

\bibitem[Sacks {\em et~al.}(1989)Sacks, Welch, Mitchell and Wynn]{Sacks_1989}
J.~Sacks, W.~J. Welch, T.~J. Mitchell and H.~P. Wynn  (1989), Design and
  analysis of computer experiments, {\em Statistical Science} {\bf 4}(4),
  118--128.

\bibitem[Santambrogio(2015)Santambrogio]{santambrogio2015optimal}
F.~Santambrogio  (2015), {\em Optimal Transport for Applied Mathematicians:
  Calculus of Variations, PDEs, and Modeling}, Vol.~87 of Progress in Nonlinear
  Differential Equations and their Applications, Springer.

\bibitem[Santner {\em et~al.}(2018)Santner, Williams and Notz]{Santner_2018}
T.~J. Santner, B.~J. Williams and W.~I. Notz  (2018), {\em {The Design and
  Analysis of Computer Experiments}}, second edition, Springer.

\bibitem[Sargsyan {\em et~al.}(2019)Sargsyan, Huan and Najm]{Sargsyan2019}
K.~Sargsyan, X.~Huan and H.~N. Najm  (2019), Embedded model error
  representation for {B}ayesian model calibration, {\em International Journal
  for Uncertainty Quantification} {\bf 9}(4), 365--394.

\bibitem[Sargsyan {\em et~al.}(2015)Sargsyan, Najm and Ghanem]{Sargsyan2015}
K.~Sargsyan, H.~N. Najm and R.~G. Ghanem  (2015), On the statistical
  calibration of physical models, {\em International Journal of Chemical
  Kinetics} {\bf 47}(4), 246--276.

\bibitem[Schein and Ungar(2007)Schein and Ungar]{schein2007active}
A.~I. Schein and L.~H. Ungar  (2007), Active learning for logistic regression:
  {A}n evaluation, {\em Machine Learning} {\bf 68}, 235--265.

\bibitem[Schillings and Schwab(2016)Schillings and
  Schwab]{schillings2016scaling}
C.~Schillings and C.~Schwab  (2016), Scaling limits in computational {B}ayesian
  inversion, {\em ESAIM: Mathematical Modelling and Numerical Analysis} {\bf
  50}(6), 1825--1856.

\bibitem[Schillings {\em et~al.}(2020)Schillings, Sprungk and
  Wacker]{schillings2020convergence}
C.~Schillings, B.~Sprungk and P.~Wacker  (2020), {On the convergence of the
  Laplace approximation and noise-level-robustness of Laplace-based Monte Carlo
  methods for Bayesian inverse problems}, {\em Numerische Mathematik} {\bf
  145}, 915--971.

\bibitem[Schrijver(2003)Schrijver]{Schrijver_book}
A.~Schrijver  (2003), {\em Combinatorial Optimization: Polyhedra and
  Efficiency}, Springer.

\bibitem[Sebastiani and Wynn(2000)Sebastiani and Wynn]{Sebastiani_2000}
P.~Sebastiani and H.~P. Wynn  (2000), {Maximum entropy sampling and optimal
  Bayesian experimental design}, {\em Journal of the Royal Statistical Society:
  Series B (Statistical Methodology)} {\bf 62}(1), 145--157.

\bibitem[Seo {\em et~al.}(2000)Seo, Wallat, Graepel and
  Obermayer]{seo2000gaussian}
S.~Seo, M.~Wallat, T.~Graepel and K.~Obermayer  (2000), Gaussian process
  regression: {A}ctive data selection and test point rejection, in {\em
  Proceedings of the IEEE-INNS-ENNS International Joint Conference on Neural
  Networks (IJCNN 2000)}, Springer, pp.~27--34.

\bibitem[Settles(2009)Settles]{settles.tr09}
B.~Settles  (2009), Active learning literature survey, Computer Sciences
  Technical Report 1648, University of Wisconsin--Madison.

\bibitem[Shah and Sinha(1989)Shah and Sinha]{Shah_1989}
K.~R. Shah and B.~K. Sinha  (1989), {\em {Theory of Optimal Designs}}, Vol.~54
  of Lecture Notes in Statistics, Springer.

\bibitem[Shahriari {\em et~al.}(2016)Shahriari {\em
  et~al.}]{shahriari2015taking}
B.~Shahriari, K.~Swersky, Z.~Wang, R.~P. Adams and N.~de~Freitas  (2016),
  Taking the human out of the loop: {A} review of {B}ayesian optimization, {\em
  Proceedings of the IEEE} {\bf 104}(1), 148--175.

\bibitem[Shapiro(1991)Shapiro]{Shapiro1991}
A.~Shapiro  (1991), Asymptotic analysis of stochastic programs, {\em Annals of
  Operations Research} {\bf 30}(1), 169--186.

\bibitem[Shapiro {\em et~al.}(2021)Shapiro, Dentcheva and
  Ruszczynski]{Shapiro_2021}
A.~Shapiro, D.~Dentcheva and A.~Ruszczynski  (2021), {\em {Lectures on
  Stochastic Programming: Modeling and Theory}}, third edition, SIAM.

\bibitem[Shashaani {\em et~al.}(2018)Shashaani, Hashemi and
  Pasupathy]{Shashaani2018}
S.~Shashaani, F.~S. Hashemi and R.~Pasupathy  (2018), {ASTRO-DF: A} class of
  adaptive sampling trust-region algorithms for derivative-free stochastic
  optimization, {\em SIAM Journal on Optimization} {\bf 28}(4), 3145--3176.

\bibitem[Shen(2023)Shen]{Shen_2023b}
W.~Shen  (2023), Reinforcement learning based sequential and robust {B}ayesian
  optimal experimental design, PhD thesis, University of Michigan.

\bibitem[Shen and Huan(2021)Shen and Huan]{Shen_2021}
W.~Shen and X.~Huan  (2021), Bayesian sequential optimal experimental design
  for nonlinear models using policy gradient reinforcement learning.
\newblock Available at
  \href{https://arxiv.org/abs/2110.15335}{arXiv:2110.15335}.

\bibitem[Shen and Huan(2023)Shen and Huan]{Shen_2023}
W.~Shen and X.~Huan  (2023), {Bayesian sequential optimal experimental design
  for nonlinear models using policy gradient reinforcement learning}, {\em
  Computer Methods in Applied Mechanics and Engineering} {\bf 416}, 116304.

\bibitem[Shen {\em et~al.}(2023)Shen, Dong and Huan]{Shen_2023a}
W.~Shen, J.~Dong and X.~Huan  (2023), Variational sequential optimal
  experimental design using reinforcement learning.
\newblock Available at
  \href{https://arxiv.org/abs/2306.10430}{arXiv:2306.10430}.

\bibitem[Shewry and Wynn(1987)Shewry and Wynn]{Shewry_1987}
M.~C. Shewry and H.~P. Wynn  (1987), {Maximum entropy sampling}, {\em Journal
  of Applied Statistics} {\bf 14}(2), 165--170.

\bibitem[Siade {\em et~al.}(2017)Siade, Hall and Karelse]{Siade_2017}
A.~J. Siade, J.~Hall and R.~N. Karelse  (2017), A practical, robust methodology
  for acquiring new observation data using computationally expensive
  groundwater models, {\em Water Resources Research} {\bf 53}(11), 9860--9882.

\bibitem[Silver {\em et~al.}(2014)Silver {\em et~al.}]{Silver_2014}
D.~Silver, G.~Lever, N.~Heess, T.~Degris, D.~Wierstra and M.~Riedmiller
  (2014), Deterministic policy gradient algorithms, in {\em Proceedings of the
  31st International Conference on Machine Learning (ICML 2014)}, Vol.~32 of
  Proceedings of Machine Learning Research, PMLR, pp.~387--395.

\bibitem[Singh and Xie(2020)Singh and Xie]{singh2020approximation}
M.~Singh and W.~Xie  (2020), {Approximation algorithms for D-optimal design},
  {\em Mathematics of Operations Research} {\bf 45}(4), 1512--1534.

\bibitem[Solonen {\em et~al.}(2012)Solonen, Haario and Laine]{Solonen_2012}
A.~Solonen, H.~Haario and M.~Laine  (2012), Simulation-based optimal design
  using a response variance criterion, {\em Journal of Computational and
  Graphical Statistics} {\bf 21}(1), 234--252.

\bibitem[Song and Ermon(2020)Song and Ermon]{Song2020}
J.~Song and S.~Ermon  (2020), Understanding the limitations of variational
  mutual information estimators, in {\em Proceedings of the 8th International
  Conference on Learning Representations (ICLR 2020)}.
\newblock Available at
  \href{https://openreview.net/forum?id=B1x62TNtDS}{https://openreview.net/forum?id=B1x62TNtDS}.

\bibitem[Song {\em et~al.}(2019)Song, Chen and Yue]{song2019general}
J.~Song, Y.~Chen and Y.~Yue  (2019), A general framework for multi-fidelity
  {B}ayesian optimization with {G}aussian processes, in {\em Proceedings of the
  22nd International Conference on Artificial Intelligence and Statistics},
  Vol.~89 of Proceedings of Machine Learning Research, PMLR, pp.~3158--3167.

\bibitem[Spall(1998{\em a})Spall]{Spall1998a}
J.~C. Spall  (1998{\em a}), {An overview of the simultaneous perturbation
  method for efficient optimization}, {\em Johns Hopkins APL Technical Digest}
  {\bf 19}(4), 482--492.

\bibitem[Spall(1998{\em b})Spall]{Spall1998}
J.~C. Spall  (1998{\em b}), Implementation of the simultaneous perturbation
  algorithm for stochastic optimization, {\em IEEE Transactions on Aerospace
  and Electronic Systems} {\bf 34}(3), 817--823.

\bibitem[Spantini {\em et~al.}(2018)Spantini, Bigoni and
  Marzouk]{spantini2018inference}
A.~Spantini, D.~Bigoni and Y.~Marzouk  (2018), Inference via low-dimensional
  couplings, {\em Journal of Machine Learning Research} {\bf 19}(1),
  2639--2709.

\bibitem[Spantini {\em et~al.}(2017)Spantini {\em et~al.}]{spantini2017goal}
A.~Spantini, T.~Cui, K.~Willcox, L.~Tenorio and Y.~Marzouk  (2017),
  Goal-oriented optimal approximations of {B}ayesian linear inverse problems,
  {\em SIAM Journal on Scientific Computing} {\bf 39}(5), S167--S196.

\bibitem[Spantini {\em et~al.}(2015)Spantini {\em et~al.}]{Spantini2015}
A.~Spantini, A.~Solonen, T.~Cui, J.~Martin, L.~Tenorio and Y.~Marzouk  (2015),
  Optimal low-rank approximations of {B}ayesian linear inverse problems, {\em
  SIAM Journal on Scientific Computing} {\bf 37}(6), A2451--A2487.

\bibitem[Spielman and Srivastava(2008)Spielman and
  Srivastava]{spielman2008graph}
D.~A. Spielman and N.~Srivastava  (2008), Graph sparsification by effective
  resistances, in {\em Proceedings of the 40th Annual ACM Symposium on Theory
  of Computing (STOC 2008)}, ACM, pp.~563--568.

\bibitem[Sp{\"o}ck(2012)Sp{\"o}ck]{spock2012spatial}
G.~Sp{\"o}ck  (2012), Spatial sampling design based on spectral approximations
  to the random field, {\em Environmental Modelling \& Software} {\bf 33},
  48--60.

\bibitem[Sp{\"o}ck and Pilz(2010)Sp{\"o}ck and Pilz]{spock2010spatial}
G.~Sp{\"o}ck and J.~Pilz  (2010), Spatial sampling design and covariance-robust
  minimax prediction based on convex design ideas, {\em Stochastic
  Environmental Research and Risk Assessment} {\bf 24}, 463--482.

\bibitem[Spokoiny(2023)Spokoiny]{spokoiny2023dimension}
V.~Spokoiny  (2023), Dimension free nonasymptotic bounds on the accuracy of
  high-dimensional {L}aplace approximation, {\em SIAM/ASA Journal on
  Uncertainty Quantification} {\bf 11}(3), 1044--1068.

\bibitem[Sprungk(2020)Sprungk]{sprungk2020local}
B.~Sprungk  (2020), {On the local Lipschitz stability of Bayesian inverse
  problems}, {\em Inverse Problems} {\bf 36}(5), 055015.

\bibitem[Sreekumar and Goldfeld(2022)Sreekumar and
  Goldfeld]{sreekumar2022neural}
S.~Sreekumar and Z.~Goldfeld  (2022), Neural estimation of statistical
  divergences, {\em Journal of Machine Learning Research} {\bf 23}(126), 1--75.

\bibitem[Sriver {\em et~al.}(2009)Sriver, Chrissis and Abramson]{Sriver2009}
T.~A. Sriver, J.~W. Chrissis and M.~A. Abramson  (2009), {Pattern search
  ranking and selection algorithms for mixed variable simulation-based
  optimization}, {\em European Journal of Operational Research} {\bf 198}(3),
  878--890.

\bibitem[Steinberg and Hunter(1984)Steinberg and Hunter]{Steinberg_1984}
D.~M. Steinberg and W.~G. Hunter  (1984), Experimental design: {R}eview and
  comment, {\em Technometrics} {\bf 26}(2), 71--97.

\bibitem[Stone(1959)Stone]{stone1959application}
M.~Stone  (1959), Application of a measure of information to the design and
  comparison of regression experiments, {\em The Annals of Mathematical
  Statistics} pp.~55--70.

\bibitem[Strutz and Curtis(2024)Strutz and Curtis]{Strutz_2023}
D.~Strutz and A.~Curtis  (2024), {Variational Bayesian experimental design for
  geophysical applications: Seismic source location, amplitude versus offset
  inversion, and estimating CO$_2$ saturations in a subsurface reservoir}, {\em
  Geophysical Journal International} {\bf 236}(3), 1309--1331.

\bibitem[Stuart and Teckentrup(2018)Stuart and Teckentrup]{stuart2018posterior}
A.~Stuart and A.~Teckentrup  (2018), {Posterior consistency for Gaussian
  process approximations of Bayesian posterior distributions}, {\em Mathematics
  of Computation} {\bf 87}(310), 721--753.

\bibitem[Stuart(2010)Stuart]{Stuart_2010}
A.~M. Stuart  (2010), {Inverse problems: A Bayesian perspective}, {\em Acta
  Numerica} {\bf 19}, 451--559.

\bibitem[Sun and Yeh(2007)Sun and Yeh]{Sun_2007}
N.-Z. Sun and W.~W. Yeh  (2007), {Development of objective-oriented groundwater
  models: 2. Robust experimental design}, {\em Water Resources Research} {\bf
  43}(2), 1--14.

\bibitem[Sutton and Barto(2018)Sutton and Barto]{Sutton2018}
R.~S. Sutton and A.~G. Barto  (2018), {\em {Reinforcement Leaning}}, second
  edition, MIT Press.

\bibitem[Sutton {\em et~al.}(1999)Sutton, McAllester, Singh and
  Mansour]{Sutton_2000}
R.~S. Sutton, D.~McAllester, S.~P. Singh and Y.~Mansour  (1999), Policy
  gradient methods for reinforcement learning with function approximation, in
  {\em Advances in Neural Information Processing Systems 12} (S.~Solla, T.~Leen
  and K.~M\"{u}ller, eds), MIT Press, pp.~1057--1063.

\bibitem[Sviridenko(2004)Sviridenko]{sviridenko2004note}
M.~Sviridenko  (2004), A note on maximizing a submodular set function subject
  to a knapsack constraint, {\em Operations Research Letters} {\bf 32}(1),
  41--43.

\bibitem[Sviridenko {\em et~al.}(2017)Sviridenko, {Vondr\'{a}k} and
  Ward]{Sviridenko_etal_2017}
M.~Sviridenko, J.~{Vondr\'{a}k} and J.~Ward  (2017), Optimal approximation for
  submodular and supermodular optimization with bounded curvature, {\em
  Mathematics of Operations Research} {\bf 42}(4), 1197--1218.

\bibitem[Tang(1993)Tang]{Tang_1993}
B.~Tang  (1993), Orthogonal array-based {L}atin hypercubes, {\em Journal of the
  American Statistical Association} {\bf 88}(424), 1392--1397.

\bibitem[Tec {\em et~al.}(2023)Tec, Duan and M{\"{u}}ller]{Tec_2023}
M.~Tec, Y.~Duan and P.~M{\"{u}}ller  (2023), A comparative tutorial of
  {B}ayesian sequential design and reinforcement learning, {\em The American
  Statistician} {\bf 77}(2), 223--233.

\bibitem[Terejanu {\em et~al.}(2012)Terejanu, Upadhyay and Miki]{Terejanu2012}
G.~Terejanu, R.~R. Upadhyay and K.~Miki  (2012), {Bayesian experimental design
  for the active nitridation of graphite by atomic nitrogen}, {\em Experimental
  Thermal and Fluid Science} {\bf 36}, 178--193.

\bibitem[Torczon(1997)Torczon]{Torczon1997}
V.~Torczon  (1997), On the convergence of pattern search algorithms, {\em SIAM
  Journal on Optimization} {\bf 7}(1), 1--25.

\bibitem[Tzoumas {\em et~al.}(2021)Tzoumas, Carlone, Pappas and
  Jadbabaie]{Tzoumas_etal_2017_SupModRatio}
V.~Tzoumas, L.~Carlone, G.~J. Pappas and A.~Jadbabaie  (2021), {LQG} control
  and sensing co-design, {\em IEEE Transactions on Automatic Control} {\bf
  66}(4), 1468--1483.

\bibitem[van~den Oord {\em et~al.}(2018)van~den Oord, Li and
  Vinyals]{VandenOord2018}
A.~van~den Oord, Y.~Li and O.~Vinyals  (2018), Representation learning with
  contrastive predictive coding.
\newblock Available at
  \href{https://arxiv.org/abs/1807.03748}{arXiv:1807.03748}.

\bibitem[Vazirani(2001)Vazirani]{vazirani2001approximation}
V.~V. Vazirani  (2001), {\em Approximation Algorithms}, Springer.

\bibitem[Villa {\em et~al.}(2021)Villa, Petra and Ghattas]{PetraTOMS}
U.~Villa, N.~Petra and O.~Ghattas  (2021), {HIPPYlib: An extensible software
  framework for large-scale inverse problems governed by PDEs: Part I:
  Deterministic inversion and linearized Bayesian inference}, {\em ACM
  Transactions on Mathematical Software} {\bf 47}(2), 1--34.

\bibitem[Villani(2009)Villani]{villani2009optimal}
C.~Villani  (2009), {\em Optimal Transport: Old and New}, Springer.

\bibitem[{Von Toussaint}(2011){Von Toussaint}]{vonToussaint_2011}
U.~{Von Toussaint}  (2011), {Bayesian inference in physics}, {\em Reviews of
  Modern Physics} {\bf 83}, 943--999.

\bibitem[Vondr\'{a}k(2008)Vondr\'{a}k]{Vondrak_2008}
J.~Vondr\'{a}k  (2008), Optimal approximation for the submodular welfare
  problem in the value oracle model, in {\em Proceedings of the Fortieth Annual
  ACM Symposium on Theory of Computing (STOC 2008)}, ACM, pp.~67--74.

\bibitem[{Vondr\'{a}k}(2010){Vondr\'{a}k}]{Vondrak_2011}
J.~{Vondr\'{a}k}  (2010), Submodularity and curvature: {T}he optimal algorithm,
  {\em RIMS Kokyuroku Bessatsu} {\bf B23}, 253--266.

\bibitem[Vondr\'{a}k {\em et~al.}(2011)Vondr\'{a}k, Chekuri and
  Zenklusen]{Vondrak_Chekuri_Zenklusen_2011}
J.~Vondr\'{a}k, C.~Chekuri and R.~Zenklusen  (2011), Submodular function
  maximization via the multilinear relaxation and contention resolution
  schemes, in {\em Proceedings of the Forty-third Annual ACM Symposium on
  Theory of Computing (STOC 2011)}, ACM, pp.~783--792.

\bibitem[Wald(1943)Wald]{wald1943efficient}
A.~Wald  (1943), On the efficient design of statistical investigations, {\em
  The Annals of Mathematical Statistics} {\bf 14}(2), 134--140.

\bibitem[Walker(2016)Walker]{Walker_2016}
S.~G. Walker  (2016), {Bayesian information in an experiment and the Fisher
  information distance}, {\em Statistics {\&} Probability Letters} {\bf 112},
  5--9.

\bibitem[Wang {\em et~al.}(2020)Wang, Clark, Liu and Frazier]{wang2020parallel}
J.~Wang, S.~C. Clark, E.~Liu and P.~I. Frazier  (2020), {Parallel Bayesian
  global optimization of expensive functions}, {\em Operations Research} {\bf
  68}(6), 1850--1865.

\bibitem[Wang and Marzouk(2022)Wang and Marzouk]{wang2022minimax}
S.~Wang and Y.~Marzouk  (2022), On minimax density estimation via measure
  transport.
\newblock Available at
  \href{https://arxiv.org/abs/2207.10231}{arXiv:2207.10231}.

\bibitem[Wang {\em et~al.}(2023)Wang, Jin, Schmitt and Olhofer]{wang2023recent}
X.~Wang, Y.~Jin, S.~Schmitt and M.~Olhofer  (2023), {Recent advances in
  Bayesian optimization}, {\em ACM Computing Surveys} {\bf 55}(13s), 1--36.

\bibitem[Wathen and Christen(2006)Wathen and Christen]{Wathen_2006}
J.~K. Wathen and J.~A. Christen  (2006), Implementation of backward induction
  for sequentially adaptive clinical trials, {\em Journal of Computational and
  Graphical Statistics} {\bf 15}(2), 398--413.

\bibitem[Weaver and Meeker(2021)Weaver and Meeker]{Weaver_2021}
B.~P. Weaver and W.~Q. Meeker  (2021), Bayesian methods for planning
  accelerated repeated measures degradation tests, {\em Technometrics} {\bf
  63}(1), 90--99.

\bibitem[Weaver {\em et~al.}(2016)Weaver, Williams, Anderson-Cook and
  Higdon]{Weaver2016}
B.~P. Weaver, B.~J. Williams, C.~M. Anderson-Cook and D.~M. Higdon  (2016),
  Computational enhancements to {B}ayesian design of experiments using
  {G}aussian processes, {\em Bayesian Analysis} {\bf 11}(1), 191--213.

\bibitem[Whittle(1973)Whittle]{whittle1973some}
P.~Whittle  (1973), Some general points in the theory of optimal experimental
  design, {\em Journal of the Royal Statistical Society: Series B
  (Methodological)} {\bf 35}(1), 123--130.

\bibitem[Wolsey and Nemhauser(1999)Wolsey and Nemhauser]{Wolsey_Nemhauser_book}
L.~A. Wolsey and G.~L. Nemhauser  (1999), {\em {Integer and Combinatorial
  Optimization}}, Wiley.

\bibitem[Wu and Hamada(2011)Wu and Hamada]{wu2011experiments}
C.~F.~J. Wu and M.~S. Hamada  (2011), {\em Experiments: Planning, Analysis, and
  Optimization}, Wiley.

\bibitem[Wu and Frazier(2019)Wu and Frazier]{wu2019practical}
J.~Wu and P.~Frazier  (2019), Practical two-step lookahead {B}ayesian
  optimization, in {\em Advances in Neural Information Processing Systems 32}
  (H.~Wallach, H.~Larochelle, A.~Beygelzimer, F.~d\textquotesingle
  Alch\'{e}-Buc, E.~Fox and R.~Garnett, eds), Curran Associates,
  pp.~9813--9823.

\bibitem[Wu {\em et~al.}(2023{\em a})Wu, Chen and Ghattas]{wu2023offline}
K.~Wu, P.~Chen and O.~Ghattas  (2023{\em a}), An offline-online decomposition
  method for efficient linear {B}ayesian goal-oriented optimal experimental
  design: {A}pplication to optimal sensor placement, {\em SIAM Journal on
  Scientific Computing} {\bf 45}(1), B57--B77.

\bibitem[Wu {\em et~al.}(2023{\em b})Wu, O’Leary-Roseberry, Chen and
  Ghattas]{wu2023large}
K.~Wu, T.~O’Leary-Roseberry, P.~Chen and O.~Ghattas  (2023{\em b}),
  Large-scale {B}ayesian optimal experimental design with derivative-informed
  projected neural network, {\em Journal of Scientific Computing} {\bf 95}(1),
  30.

\bibitem[Wynn(1972)Wynn]{wynn1972results}
H.~P. Wynn  (1972), Results in the theory and construction of {D}-optimum
  experimental designs, {\em Journal of the Royal Statistical Society: Series B
  (Methodological)} {\bf 34}(2), 133--147.

\bibitem[Wynn(1984)Wynn]{wynn1984jack}
H.~P. Wynn  (1984), {Jack Kiefer's contributions to experimental design}, {\em
  The Annals of Statistics} {\bf 12}(2), 416--423.

\bibitem[Xu and Liao(2020)Xu and Liao]{Xu2020}
Z.~Xu and Q.~Liao  (2020), Gaussian process based expected information gain
  computation for {B}ayesian optimal design, {\em Entropy} {\bf 22}(2), 258.

\bibitem[Yates(1933)Yates]{yates1933principles}
F.~Yates  (1933), The principles of orthogonality and confounding in replicated
  experiments, {\em The Journal of Agricultural Science} {\bf 23}(1), 108--145.

\bibitem[Yates(1937)Yates]{yates1937design}
F.~Yates  (1937), The design and analysis of factorial experiments.
\newblock Technical Communication no. 35, Imperial Bureau of Soil Science.

\bibitem[Yates(1940)Yates]{yates1940lattice}
F.~Yates  (1940), Lattice squares, {\em The Journal of Agricultural Science}
  {\bf 30}(4), 672--687.

\bibitem[Zahm {\em et~al.}(2022)Zahm {\em et~al.}]{Zahm2022}
O.~Zahm, T.~Cui, K.~Law, A.~Spantini and Y.~Marzouk  (2022), {Certified
  dimension reduction in nonlinear Bayesian inverse problems}, {\em Mathematics
  of Computation} {\bf 91}(336), 1789--1835.

\bibitem[Zhan and Xing(2020)Zhan and Xing]{zhan2020expected}
D.~Zhan and H.~Xing  (2020), Expected improvement for expensive optimization:
  {A} review, {\em Journal of Global Optimization} {\bf 78}(3), 507--544.

\bibitem[Zhang {\em et~al.}(2021)Zhang, Bi and Zhang]{Zhang2021}
J.~Zhang, S.~Bi and G.~Zhang  (2021), A scalable gradient-free method for
  {B}ayesian experimental design with implicit models, in {\em Proceedings of
  the 24th International Conference on Artificial Intelligence and Statistics},
  Vol. 130 of Proceedings of Machine Learning Research, PMLR, pp.~3745--3753.

\bibitem[Zhang {\em et~al.}(2022)Zhang {\em et~al.}]{zhang2022wasserstein}
X.~Zhang, J.~Blanchet, Y.~Marzouk, V.~A. Nguyen and S.~Wang  (2022),
  Distributionally robust {G}aussian process regression and {B}ayesian inverse
  problems.
\newblock Available at
  \href{https://arxiv.org/abs/2205.13111}{arXiv:2205.13111}.

\bibitem[Zheng {\em et~al.}(2020)Zheng, Hayden, Pacheco and
  Fisher]{zheng2020sequential}
S.~Zheng, D.~Hayden, J.~Pacheco and J.~W. Fisher  (2020), Sequential {B}ayesian
  experimental design with variable cost structure, in {\em Advances in Neural
  Information Processing Systems 33} (H.~Larochelle, M.~Ranzato, R.~Hadsell,
  M.~Balcan and H.~Lin, eds), Curran Associates, pp.~4127--4137.

\bibitem[Zhong {\em et~al.}(2024)Zhong, Shen, Catanach and Huan]{Zhong_2024}
S.~Zhong, W.~Shen, T.~Catanach and X.~Huan  (2024), Goal-oriented {B}ayesian
  optimal experimental design for nonlinear models using {M}arkov chain {M}onte
  {C}arlo.
\newblock Available at
  \href{https://arxiv.org/abs/2403.18072}{arXiv:2403.18072}.

\bibitem[Zhu and Stein(2005)Zhu and Stein]{zhu2005spatial}
Z.~Zhu and M.~L. Stein  (2005), Spatial sampling design for parameter
  estimation of the covariance function, {\em Journal of Statistical Planning
  and Inference} {\bf 134}(2), 583--603.

\bibitem[Zhu and Stein(2006)Zhu and Stein]{zhu2006spatial}
Z.~Zhu and M.~L. Stein  (2006), Spatial sampling design for prediction with
  estimated parameters, {\em Journal of Agricultural, Biological, and
  Environmental Statistics} {\bf 11}, 24--44.

\bibitem[Zong(1999)Zong]{zong2008sphere}
C.~Zong  (1999), {\em {Sphere Packings}}, Springer.

\end{thebibliography}
\label{lastpage}
\end{document}